\documentclass[11pt,a4paper]{article}
\usepackage{amssymb}
 \newcommand{\cameraready}{}

\usepackage{amsthm}

\usepackage[mathscr]{euscript}
\usepackage{amsmath}
\usepackage[english]{babel} 
\usepackage[a4paper,margin=2.25cm]{geometry}

\usepackage[T1]{fontenc}
\usepackage[utf8]{inputenc}
 \usepackage{lmodern}
\usepackage{fancyhdr}
\usepackage{graphicx}
 \PassOptionsToPackage{names,dvipsnames}{xcolor}
 
\usepackage{tikz}
\usetikzlibrary{calc}
\usetikzlibrary{automata,positioning}

\usepackage{microtype}%if unwanted, comment out or use option "draft"
\usepackage{amsmath}
\usepackage{mathtools}
\usepackage{proof}
\usepackage{hyperref}
\usepackage{mdframed}
\mdfdefinestyle{alttight}{innertopmargin=0pt,innerleftmargin=2pt,innerrightmargin=2pt}

\usepackage{xspace}
\usepackage{enumerate}
\usepackage{mathpartir}
\usepackage{bussproofs}
\usepackage{multirow}

\usepackage{cleveref}

\usepackage{colortbl}

\usepackage[T1]{fontenc}

\usepackage{todonotes}

\usepackage{stmaryrd}
\usepackage{float}
\restylefloat{table}

\usepackage{arydshln}
\newcommand{\newadd}[1]{\ifx\cameraready\undefined \textcolor{blue!70}{#1} \else #1\fi}

\providecommand{\thesisalt}[2]{\ifx\thesis\undefined#1\else#2\fi}
\providecommand{\cralt}[2]{\ifx\cameraready\undefined#1\else#2\fi}

\newenvironment{newaddenv}
{
  \ifx\cameraready\undefined \color{blue!70} \else \fi
}
{} 

\usepackage{tikz}
\usepackage{comment}
\usepackage{wrapfig}
\usetikzlibrary{decorations.pathreplacing}
\usetikzlibrary{decorations.pathmorphing}

\usepackage{thmtools}
\usepackage{thm-restate}

\declaretheorem[name=Proposition, numberwithin=section]{proposition}

\usepackage{cleveref}
\newcommand{\ghot}[1]{#1\! \leadsto\! \diamond}
\newcommand{\linecondit}[3]{\ensuremath{(#1)\,\mathsf{::}\,#2\mathbf{:}\,#3}}

\newcommand{\balan}[1]{\ensuremath{\mathsf{balanced}(#1)}}
\newcommand{\subst}[2]{\{^{#1}\!/{\scriptstyle #2}\}}

\newcommand{\lrangle}[1]{\langle #1 \rangle}
\newcommand{\blrangle}[1]{\big\langle #1 \big\rangle}

\newcommand{\relS}{\ensuremath{\mathcal{S}\xspace}}

\newcommand{\brtext}[1]{[\textrm{\small #1}]}

\newcommand{\ltsrule}[1]{{\footnotesize \ensuremath{\lrangle{\textsc{#1}}}}}
\newcommand{\eltsrule}[1]{{\footnotesize [\textsc{#1}]}}
\newcommand{\trule}[1]{{\footnotesize\brtext{#1}}}
\newcommand{\orule}[1]{{\scriptsize{\brtext{#1}}}}

\newcommand{\noi}{\noindent}

\newcommand{\bnfis}{\;::=\;}
\newcommand{\bnfbar}{\;\;{|}\;\;}
\newcommand{\sbnfbar}{\;\;|\;\;}

\newcommand{\set}[1]{\{#1\}}
\newcommand{\es}{\emptyset}

\newcommand{\dom}[1]{\mathsf{dom}(#1)}
\newcommand{\codom}[1]{\mathsf{codom}(#1)}

\newcommand{\freev}[1]{\lrangle{#1}}
\newcommand{\boundv}[1]{(#1)}

\newcommand{\inact}{\mathbf{0}}

\newcommand{\Par}{\;|\;}
\newcommand{\news}[1]{(\nu\, #1)\,}
\newcommand{\newsp}[2]{(\nu\, #1)(#2)}

\newcommand{\varp}[1]{#1}
\newcommand{\rvar}[1]{#1}
\newcommand{\recp}[2]{\mu \rvar{#1}. #2}

\newcommand{\Def}{\sessionfont{def}\ }

\newcommand{\defeq}{\stackrel{\Def}{=}}

\newcommand{\repl}{\ast\,}
\newcommand{\parcomp}[2]{\prod_{#1}{#2}}

\newcommand{\myfontformat}[1]{\mathsf{#1}}
\newcommand{\bn}[1]{\myfontformat{bn}(#1)}
\newcommand{\fn}[1]{\myfontformat{fn}(#1)}
\newcommand{\rfn}[1]{\myfontformat{rn}(#1)}

\newcommand{\fv}[1]{\myfontformat{fv}(#1)}

\newcommand{\fs}[1]{\myfontformat{fs}(#1)}

\newcommand{\subj}[1]{\mathtt{subj}(#1)}

\newcommand{\scong}{\equiv}

\newcommand{\fwb}{\approx^\mathtt{C}}

\newcommand{\red}{\longrightarrow}

\def\subst#1#2{\{\raisebox{.5ex}{\small$#1$}\! / \mbox{\small$#2$}\}}

\newcommand{\sessionfont}[1]{\mathtt{#1}}
\newcommand{\vart}[1]{\mathsf{#1}}

\newcommand{\shsep}{.}
\newcommand{\outses}{!}
\newcommand{\inpses}{?}
\newcommand{\selses}{\triangleleft}
\newcommand{\brases}{\triangleright}
\newcommand{\dual}[1]{\overline{#1}}
\newcommand{\cat}{,}

\newcommand{\bout}[2]{#1 \outses \freev{#2} \shsep}
\newcommand{\about}[2]{#1 \outses \freev{#2}}
\newcommand{\boutt}[2]{#1 \outses \freev{#2}}
\newcommand{\bbout}[2]{#1 \outses \blrangle{#2} \shsep}
\newcommand{\abbout}[2]{#1 \outses \blrangle{#2}}
\newcommand{\binpt}[2]{#1 \inpses \boundv{#2}}
\newcommand{\abinp}[2]{#1 \inpses \boundv{#2}}
\newcommand{\binp}[2]{#1 \inpses \boundv{#2} \shsep}
\newcommand{\bsel}[2]{#1 \selses #2 \shsep}

\newcommand{\bbra}[2]{#1 \brases \set{#2}}

\newcommand{\tfont}[1]{\mathtt{#1}}
\newcommand{\tsep}{\textcolor{darkgray}{;}}

\newcommand{\chtype}[1]{\lrangle{#1}}

\newcommand{\outtype}{\outses}
\newcommand{\inptype}{\inpses}

\newcommand{\trec}[2]{\mu\vart{#1}.#2}
\newcommand{\tvar}[1]{\vart{#1}}

\newcommand{\tinact}{\tfont{\textcolor{darkgray}{end}}}

\newcommand{\btoutt}[1]{\outtype \lrangle{#1}}
\newcommand{\btinpt}[1]{\inptype (#1) }

\newcommand{\btout}[1]{\outtype \lrangle{#1} \tsep}
\newcommand{\bbtout}[1]{\outtype \big\langle{#1}\big\rangle \tsep}
\newcommand{\btinp}[1]{\inptype (#1) \tsep}
\newcommand{\bbtinp}[1]{\inptype \big({#1}\big) \tsep}
\newcommand{\btsel}[1]{\oplus \set{#1}}

\newcommand{\btbra}[1]{\& \set{#1}}

\newcommand{\proves}{\vdash}
\newcommand{\hastype}{\triangleright}

\newcommand{\outlts}{\outses}
\newcommand{\inplts}{\inpses}

\newcommand{\bactout}[2]{#1 \outlts \freev{#2}}

\newcommand{\bactinp}[2]{#1 \inplts \freev{#2}}

\makeatletter
\newcommand{\xMapsto}[2][]{\ext@arrow 0599{\Mapstofill@}{#1}{#2}}
\def\Mapstofill@{\arrowfill@{\Mapstochar\Relbar}\Relbar\Rightarrow}
\makeatother

\newcommand{\by}[1]{\mathrel{\xrightarrow{#1}}}
\newcommand{\By}[1]{\mathrel{\xRightarrow{#1}}}

\newcommand{\hby}[1]{\mathrel{\xmapsto{#1}}}

\newcommand{\bistyp}{\rightleftharpoons}

\newcommand{\oldiecolor}{OliveGreen}
\definecolor{pcolor}{rgb}{0.36, 0.54, 0.66}
\definecolor{tcolor}{rgb}{0.57, 0.36, 0.51}

\newcommand{\map}[1]{\ensuremath{\textcolor{\oldiecolor}{\llbracket}#1\textcolor{\oldiecolor}{\rrbracket}}}
\newcommand{\tmap}[2]{\ensuremath{\textcolor{\oldiecolor}{(\!\!\langle}#1\textcolor{\oldiecolor}{\rangle\!\!)}^{#2}}}

\newcommand{\auxmapp}[3]{\ensuremath{\big\lfloor\!\!\big\lfloor#1\big\rfloor\!\!\big\rfloor^#2_#3}}
\newcommand{\vtmap}[2]{{\ensuremath{\big\lfloor #1\big\rfloor^{#2}}}}

\newcommand{\dualof}{\ \mathsf{dual}\ }

\newcommand{\HOp}{\ensuremath{\mathsf{HO}\pi}\xspace}
\newcommand{\sessp}{\ensuremath{\pi}\xspace}

\newcommand{\HO}{\ensuremath{\mathsf{HO}}\xspace}
\newcommand{\mHO}{\ensuremath{\mu\mathsf{HO}}\xspace}

\newcommand{\msts}{MSTs\xspace}

\newcommand{\Proc}{\ensuremath{\diamond}}

\newcommand{\appl}[2]{#1\, {#2}}
\newcommand{\abs}[2]{\lambda #1.\,#2}

\newcommand{\lollipop}{\multimap}
\newcommand{\sharedop}{\rightarrow}

\newcommand{\lhot}[1]{#1\! \lollipop\! \diamond}
\newcommand{\shot}[1]{#1\! \sharedop\! \diamond}

\newcommand{\vmap}[1]{|\!|#1|\!|}

\newcommand{\fotrigger}[5]{\binp{#1}{#2} \newsp{#3}{\binp{#3}{y}  \mapchar{#4}{y} \Par \bout{\dual{#3}}{#5} \inact}}

\newcommand{\horel}[6]{#1; #2 \proves #3 #4 #5 \proves #6}
\newcommand{\bool}%{\sessionfont{bool}}
{\tfont{\textcolor{darkgray}{bool}}}
\newcommand{\nat}%{\sessionfont{nat}}
{\tfont{\textcolor{darkgray}{nat}}}
\newcommand{\tint}{\tfont{\textcolor{darkgray}{int}}}

\newcommand{\y}{\ensuremath{y}}

\newcommand{\X}{\varp{X}}

\newtheorem{notation}{Notation}[section]

\newcommand{\nonhosyntax}[1]{\colorbox{lightgray}{\ensuremath{#1}}}
\newcommand{\nonpisyntax}[1]{\fcolorbox{black}{white}{\ensuremath{#1}}}

\newcommand{\newc}[1]{#1}

\newif\ifny\nyfalse
\newif\ifdm\dmtrue
\newif\ifrhu\rhutrue
\newif\ifjp\jptrue
\newif\ifjp\jptrue
\newcommand{\newjb}[1]{{#1}}

\newcommand{\AT}[2]{#1 \! : \! #2}

\newcommand{\defref}[1]{\Cref{#1}}

\newcommand{\thmref}[1]{\Cref{#1}}

\newcommand{\appref}[1]{\Cref{#1}}
\newcommand{\lemref}[1]{\Cref{#1}}

\definecolor{lightgray}{gray}{0.75}

\newcommand{\mapchar}[2]{\ensuremath{[\!\!(#1)\!\!]^{#2}}}
\newcommand{\omapchar}[1]{\ensuremath{[\!\!(#1)\!\!]_{\mathsf{c}}}}

\newcommand{\ftrigger}[3]{#1 \Leftarrow_{\mathtt{C}} #2{\,:\,}#3}

\makeatletter
\newcommand*{\rom}[1]{({\expandafter\romannumeral #1})}
\makeatother

\newcommand{\defas}{\triangleq}

\usepackage{fixme}
\fxsetup{theme=color,mode=multiuser,draft,inline,marginclue}
\definecolor{fxtarget}{rgb}{0.8300,0.1400,0.1400}
\FXRegisterAuthor{dan}{adan}{\colorbox{green!70}{\color{black}DF:}}
\newcommand{\deccolor}{BrickRed}
\newcommand{\decoptcolor}{MidnightBlue}
\newcommand{\AmeliaMod}[3]{\textcolor{\deccolor}{\mathcal{A}}^{#1}_{#2}\textcolor{\deccolor}{\big(}#3\textcolor{\deccolor}{\big)}}
\newcommand{\F}[1]{\textcolor{\deccolor}{\mathcal{F}(}\textcolor{black}{#1}\textcolor{\deccolor}{)}}
\newcommand{\Do}[1]{\textcolor{\decoptcolor}{\mathcal{F}^\ast(}\textcolor{black}{#1}\textcolor{\decoptcolor}{)}}
\newcommand{\msessp}{\ensuremath{\mu\pi}\xspace}

\newcommand{\mGt}[1]{\textcolor{\deccolor}{\mathcal{H}(}#1\textcolor{\deccolor}{)}}
\newcommand{\Gtopt}[1]{\textcolor{\decoptcolor}{\mathcal{H}}^{\textcolor{\decoptcolor}{*}}\textcolor{\decoptcolor}{(}#1\textcolor{\decoptcolor}{)}}

\newcommand{\mRt}[1]{\textcolor{\deccolor}{\mathcal{R}(}#1\textcolor{\deccolor}{)}}
\newcommand{\mRts}[3]{\textcolor{\deccolor}{\mathcal{R}_{\circ}(}#3\textcolor{\deccolor}{)}}
\newcommand{\hidejp}[1]{}

\newcommand{\B}[3]{\textcolor{\oldiecolor}{\mathcal{B}}^{#1}_{#2}\textcolor{\oldiecolor}{\big(\textcolor{black}{#3}\big)}}
\newcommand{\V}[3]{\textcolor{\oldiecolor}{\mathcal{V}}^{#1}_{#2}\textcolor{\oldiecolor}{\big(\textcolor{black}{#3}\big)}}

\newcommand{\Gt}[1]{\textcolor{\oldiecolor}{\mathcal{G}(\textcolor{black}{#1})}}
\newcommand{\Rt}[1]{\textcolor{\oldiecolor}{\mathcal{R}(\textcolor{black}{#1})}}
\newcommand{\Rts}[3]{\textcolor{\oldiecolor}{\mathcal{R}^{\star}(\textcolor{black}{#3})}}
\newcommand{\envR}{\Delta_\mu}
\newcommand{\thetaR}{\Theta_\mu}
\newcommand{\thetax}{\Theta_{\rvar{X}}}
\newcommand{\envPropR}{\Phi}
\newcommand{\D}[1]{\textcolor{\oldiecolor}{\mathcal{D}(\textcolor{black}{#1})}}

\newcommand{\len}[1]{|#1|}
\newcommand{\incrname}[2]{\subst{#1_{#2+1}}{#1_#2}}

\newcommand{\prop}{c}
\newcommand{\apropinp}[1]{\abinp{\prop_{#1}}}
\newcommand{\propinp}[1]{\binp{\prop_{#1}}}

\newcommand{\propout}[1]{\bout{\dual{\prop_{#1}}}}
\newcommand{\apropout}[1]{\about{\dual{\prop_{#1}}}}
\newcommand{\proprinp}[1]{\binp{\prop^{#1}}}
\newcommand{\proprout}[1]{\bout{\prop^{#1}}}
\newcommand{\proprinpk}[2]{\binp{\prop^{#1}_{#2}}}
\newcommand{\proproutk}[2]{\bout{\dual {\prop^{#1}_{#2}}}}

\newcommand{\apropbout}[1]{\abbout{\dual{\prop_{#1}}}}

\newcommand{\propinprec}[1]{\binp{\prop^r_{#1}}}

\newcommand{\propoutrec}[1]{\bout{\dual{\prop^r_{#1}}}}

\newcommand{\propinprecsh}[1]{\binp{\prop^r_{#1}}}

\newcommand{\propoutrecsh}[1]{\bout{{\prop^r_{#1}}}}
\newcommand{\apropoutrecsh}[1]{\about{\prop^r_{#1}}}

\newcommand{\unaryint}[1]{\ensuremath{\textsf{odd}(#1)}}

\newcommand{\lin}[1]{\myfontformat{lin}(#1)}

\newcommand{\slhotup}[1]{{#1}^{\leadsto}}

\newcommand{\degree}{l}

\newcommand{\mugamma}{\gamma}

\newcommand{\fnb}[2]{\myfontformat{fnb}(#1,#2)}

\newcommand{\wtd}[1]{\widetilde{#1}}
\newcommand{\li}[1]{\widetilde{#1}}

\newcommand{\lenHO}[1]{\textcolor{\deccolor}{\lbag} #1 \textcolor{\deccolor}{\rbag}}
\newcommand{\lenHOopt}[1]{\textcolor{\decoptcolor}{\lbag} #1 \textcolor{\decoptcolor}{\rbag^*}}

\newcommand{\indexed}[3]{\mathsf{indexed}_{#3}(#1,#2)}

\newcommand{\Cbs}[3]{\textcolor{\behcolor}{\widehat{\mathscr{C}}}^{#1}_{#2}\textcolor{\behcolor}{\big(}#3\textcolor{\behcolor}{\big)}}
\newcommand{\fpn}[1]{\ensuremath{\myfontformat{fpn}(#1)}}
\newcommand{\nextn}[1]{\ensuremath{\mathsf{next}(#1)}}

\newcommand{\indices}[1]{\ensuremath{\mathsf{index}(#1)}}
\newcommand{\bname}[1]{\ensuremath{\myfontformat{nbd}(#1)}}

\newcommand{\mstindex}{\ensuremath{l}}

\newcommand{\bns}[1]{\ensuremath{\tilde {#1_*}}}
\newcommand{\wbns}[1]{\ensuremath{\widetilde {#1_*}}}

\newcommand{\iname}[1]{\ensuremath{\breve{#1}}}

\newcommand{\highlighta}[1]{\colorbox{lightgray}{\ensuremath{#1}}}

\newcommand{\horelm}[5]{#1 \proves #2 #3 #4 \proves #5}
\newcommand{\mstb}{\approx^{\mathtt{M}}}

\newcommand{\numpropam}[1]{\textcolor{\deccolor}{\#(}#1\textcolor{\deccolor}{)}}
\newcommand{\numprop}[1]{\textcolor{\decoptcolor}{\#^*(}#1\textcolor{\decoptcolor}{)}}

\newcommand{\mapcharm}[2]{\langle #1 \rangle^{#2}}
\newcommand{\omapcharm}[1]{\langle #1 \rangle_{\texttt{c}}}

\newcommand{\Rtopt}[1]{\textcolor{\decoptcolor}{\mathcal{R}^*(}#1\textcolor{\decoptcolor}{)}}
\newcommand{\Rtsopt}[3]{\textcolor{\decoptcolor}{\mathcal{R}^{*}_\circ(}#3\textcolor{\decoptcolor}{)}}

\newcommand{\tr}{\myfontformat{tr}}
\newcommand{\traux}[1]{\myfontformat{tr}(#1)}

\newcommand{\initname}[2]{\ensuremath{\subst{#1_{#2}}{#1}}}

\newcommand{\subl}[1]{\textsf{sub}(#1)}

\newcommand{\ftriggerm}[3]{#1 \Leftarrow_{\mathtt{m}} #2{\,:\,}#3}
\newcommand{\fotriggerm}[6]{\binp{#1}{#2} \newsp{#3}{\binp{#3}{\wtd y}  \mapcharm{#4}{y}_{#6} \Par \bout{\dual{#3}}{#5} \inact}}

\newcommand{\processrelate}{\diamond}

\newcommand{\rhom}{\ensuremath{\rho_{\mathtt m}}}

\newcommand{\sigmav}{\ensuremath{\sigma_v}}

\newcommand{\empg}{{}}

\newcommand{\recpx}[1]{\ensuremath{\myfontformat{frv}(#1)}}

\newcommand{\lenrecpx}[1]{\ensuremath{\#_X(#1)}}

\newcommand{\recdep}{\ensuremath{d}}

\newcommand{\lentr}[1]{\ensuremath{\mathsf{brn}(#1)}}

\newcommand{\arxivlink}{https://arxiv.org/abs/2107.12345}

\newcommand{\longversion}[2]{\ifx\ppdp\undefined#1\else#2\fi}

\newcommand{\appnote}{\ifx\ppdp\undefined 
      \noindent \emph{Note:}   
      Due to space limits, 
      omitted  material (definitions, proofs, additional examples) can be found in the appendices. 
  \else 
    \noindent \emph{Note:}   
    Due to space limits, omitted material (definitions, proofs, additional examples) can 
    be found at \url{\arxivlink}. 
\fi 
}

\newcommand{\appnoteEx}
{\ifx\ppdp\undefined 
      \noindent The breakdown of $B'$ is similar and given in 
      \appref{ex:ex1-decomp}.
  \else 
    \noindent The breakdown of $B'$ is similar. 
\fi 
}

\newcommand{\indT}[1]{{[\textcolor{black}{#1}\rangle}}
\newcommand{\indTaux}[2]{{[\textcolor{black}{#1}\rangle^{\star}_{\textcolor{black}{#2}}}}

\newcommand{\plen}[1]{{\textcolor{\oldiecolor}{\lbag \textcolor{black}{#1}\rbag}}}
\newcommand{\propoutt}[1]{\boutt{\dual{\prop_{#1}}}}

\newcommand{\recprov}[3]{\binp{\prop^{#1}}{#2}\appl{#2}{#3}}

{}

\excludecomment{revisionrem}

\newcommand{\ltenc}[1]{\map{#1}}

\newcommand{\ltab}[1]{l_{\#}[#1]}
\newcommand{\lti}[1]{l_{\mathtt{i}}[#1]}
\newcommand{\lto}[1]{l_{\mathtt{o}}[#1]}
\newcommand{\ltend}{\varnothing[]}

\newcommand{\ltsel}[3]{\langle #1 : #2 \rangle_{#3}}
\newcommand{\ltbra}{\langle l_i : \tau_i\rangle_{i \in I}}
\newcommand{\ltcase}[5]{\textbf{case} \ #1 \ \textbf{of}
\ \{#2\_(#3) \triangleright #4\}_{#5}}
\newcommand{\ltcasep}[2]{\textbf{case} \ #1 \ \textbf{of}
\ \{#2\}}

\newcommand{\ltc}[1]{\#[#1]}

\makeatletter
\@ifpackageloaded{stix}{%
}{%
  \DeclareFontEncoding{LS2}{}{\noaccents@}
  \DeclareFontSubstitution{LS2}{stix}{m}{n}
  \DeclareSymbolFont{stix@largesymbols}{LS2}{stixex}{m}{n}
  \SetSymbolFont{stix@largesymbols}{bold}{LS2}{stixex}{b}{n}
  \DeclareMathDelimiter{\lBrace}{\mathopen} {stix@largesymbols}{"E8}%
                                            {stix@largesymbols}{"0E}
  \DeclareMathDelimiter{\rBrace}{\mathclose}{stix@largesymbols}{"E9}%
                                            {stix@largesymbols}{"0F}
}
\makeatother

\newcommand{\Boptsb}[3]{\textcolor{\decoptcolor}{\mathscr{A}^{\boxplus}}^{#1}_{#2}\textcolor{\decoptcolor}{\big(}#3\textcolor{\decoptcolor}{\big)}}

\newcommand{\behcolor}{RoyalBlue}

\newcommand{\Brecpi}[3]{\textcolor{\decoptcolor}{\widehat{\mathscr{A}}}^{\,#1}_{#2}\textcolor{\decoptcolor}{\big(}#3\textcolor{\decoptcolor}{\big)}}

\newcommand{\Bopt}[3]{\textcolor{\decoptcolor}{\mathscr{A}}^{#1}_{#2}\textcolor{\decoptcolor}{\big(}#3\textcolor{\decoptcolor}{\big)}}

\newcommand{\Dletpi}{\mathscr{J}}

\newcommand{\Dbpi}[3]{\textcolor{\behcolor}{\Dletpi}^{#1}_{#2}\textcolor{\behcolor}{\big(}#3\textcolor{\behcolor}{\big)}}
\newcommand{\Dbrecpi}[3]{\textcolor{\behcolor}{\widehat{\Dletpi}}^{\,#1}_{#2}\textcolor{\behcolor}{\big(}#3\textcolor{\behcolor}{\big)}}
\newcommand{\Cbpi}[3]{\textcolor{\behcolor}{\mathcal{C}}^{#1}_{#2}\textcolor{\behcolor}{\big(}#3\textcolor{\behcolor}{\big)}}

\newcommand{\Cbrecpi}[3]{\textcolor{\behcolor}{\widehat{\mathcal{C}}}^{\,#1}_{#2}\textcolor{\behcolor}{\big(}#3\textcolor{\behcolor}{\big)}}

\newcommand{\rvardepth}[1]{\ensuremath{\delta(#1)}}
\newcommand{\rvardepthaux}[1]{\ensuremath{\widehat{\delta}(#1)}}

\newcommand{\indf}[1]{\ensuremath{\iota(#1)}}
\newcommand{\indfaux}[2]{\ensuremath{\widehat{\iota}_{#1}(#2)}}

\newtheorem{theorem}{Theorem}[section]

\newtheorem{lemma}{Lemma}[section]
\theoremstyle{definition}
\newtheorem{definition}{Definition}[section]

\theoremstyle{definition}
\newtheorem{example}{Example}[section]

\newtheorem{remark}{Remark}[section]

\title{Minimal Session Types for the $\pi$-calculus}
\usepackage{authblk}
\author{Alen Arslanagi\'{c}}
\author{Jorge A. P\'{e}rez}
\author{Anda-Amelia Palamariuc}
\affil{University of Groningen, The Netherlands}
\date{\today}                     %% if you don't need date to appear
\newcommand{\fixed}[1]{#1}

\usepackage{rotating}
\usepackage{adjustbox}
\usepackage{import}

\begin{document}
\maketitle

\begin{abstract}
Session types are a discipline for the static verification of message-passing programs.
A session type specifies a channel's protocol as \emph{sequences} of exchanges.
It is most relevant to investigate session-based concurrency by identifying the \emph{essential notions} that enable program specification and verification. 
Following that perspective, prior work identified \emph{minimal} session types (\msts), a sub-class of session types without the sequentiality construct, which specifies the structure of communication actions.
This formulation of session types led to establish a \emph{minimality result}:  every process typable with standard session types can be compiled down to a process typable using \msts, which mimics sequentiality in types via additional process synchronizations.
Such a minimality result is significant because it justifies session types in terms of themselves, without resorting to external notions; it was proven for a {higher-order} session $\pi$-calculus, in which values are abstractions (functions from names to processes). 

In this paper, we study \msts and their associated minimality result but now for the session $\pi$-calculus, the (first-order) language in which values are names and for which session types have been more widely studied.
We first show that this new minimality result can be obtained by {composing} known results.
Then, we develop optimizations of this new minimality result and prove that the associated transformation into processes with \msts satisfies dynamic correctness.
\end{abstract}

% \chapter{MST}
\section{Introduction}

Session types are a type-based approach to statically ensure that message-passing programs correctly implement some predefined protocols~\cite{HondaK:typdyi,honda.vasconcelos.kubo:language-primitives}. 
A session type stipulates the sequence and payload of the messages exchanged along a channel. 
Sequentiality, denoted by the action prefix~`$\,;\,$', is arguably the most distinctive construct of session types, as it allows to specify structured communications. 
For instance, in the session type $S = \btinp{\tint} \btinp{\tint} \btout{\bool} \tinact$,  this  construct conveniently specifies a channel protocol that \emph{first} receives~(`$?$') two integers, \emph{then} sends~(`$!$') a Boolean, and \emph{finally} ends.
 
%In this paper,  we investigate  for the $\pi$-calculus, the paradigmatic calculus of concurrency. This elementary formulation is called .

 Because sequentiality is so useful for protocol specification and verification, a natural question is whether it could be recovered by other, simpler means. 
 To investigate this question, Arslanagi\'{c} et al.~\cite{APV19} defined \emph{minimal session types} (\msts), an {alternative formulation} of session types.
 The difference between standard and minimal session types concerns {sequentiality}  in types:  \msts is the sub-class of session types without sequentiality.
To see the difference, consider session types for input and output, denoted `$\btout{U} S$' and `$\btinp{U} S$', respectively:  in standard session types, the type $S$ denotes an arbitrary (session) protocol; in \msts, the type $S$ can only be `$\tinact$', the type of the terminated protocol.
Arslanagi\'{c} et al.~established a \emph{minimality result}: every well-typed session process can be \emph{decomposed} into a process typable with \msts. Their approach to decomposition is inspired by Parrow's work on \emph{trios processes} for the untyped $\pi$-calculus~\cite{DBLP:conf/birthday/Parrow00}. The minimality result justifies session types in terms of themselves, without resorting to external notions, and shows that sequentiality in types is useful but not indispensable, because it can be precisely mimicked by the process decomposition.  

  The minimality result based on \msts in~\cite{APV19} was proven for \HO, a \emph{higher-order} process calculus in which values are  abstractions (functions from names to processes). \HO does not include name passing nor process recursion, but it can encode them precisely~\cite{DBLP:conf/esop/KouzapasPY16,DBLP:journals/iandc/KouzapasPY19}. An important question left open in~\cite{APV19} is whether the minimality result holds for a \emph{first-order} session $\pi$-calculus, i.e., a language in which values are names. This is a relevant question, as session types have been more widely studied in the first-order setting, from foundational and practical angles. Unlike \HO, the session $\pi$-calculus we consider, dubbed \sessp in the following, natively supports recursion and recursive types.

In this paper, we report \emph{two positive answers} to this question.
  Our \emph{first answer} is simple, perhaps even deceptively so.  In order to establish the minimality result for \sessp, we compose the decomposition in~\cite{APV19} with the mutual encodings between \HO and \sessp  given in~\cite{DBLP:conf/esop/KouzapasPY16,DBLP:journals/iandc/KouzapasPY19}. 
  We call this the \emph{decompose by composing} approach.
  
  \begin{figure}[!t]
    %     \begin{wrapfigure}
          %  \begin{wrapfigure}{r}{0.25\textwidth}
          %  \vspace{-4mm}
    \begin{mdframed}%[style=alttight]
    \centering
          \begin{tikzpicture}[scale = 1.25, every node/.style = { scale = 1.25}] % make it bigger
            \tikzset{node distance = 15ex}
            \tikzset{line width = 1pt}
            \tikzstyle{snakeme} = [decorate, decoration = snake ]
            \tikzstyle{smaller} = [scale = 0.65]
            \tikzstyle{small} = [scale = 0.75]
            
            \node (PI) {\sessp};
            \node [right of=PI] (MPI) {\msessp};
            \node [below of=PI] (HO) {\HO};
            \node [right of=HO] (MHO) {\mHO};
            
            %\draw[dotted, ->] (HO) to (MHO);
            \draw[->, dotted] (HO) -- node[above, small] {$\D{\cdot}$} (MHO);
            \draw[->, dotted] (PI) -- node [right, small] {$\map{\cdot}^1_{g}$} (HO);
            \draw[->, dotted] (MHO) -- node [right, small] {$\map{\cdot}^2$} (MPI);
            \draw[->] (PI) -- node[below, scale=0.85] {$\F{\cdot}$} (MPI);
        \end{tikzpicture}
        \end{mdframed}
        \caption{First approach to a minimality result for \sessp: Decompose by Composing. \label{pi:enc_drawing}}
        % \vspace{-5mm}
    % \end{wrapfigure}
    %\vspace{-3mm}
    \end{figure}
%         \begin{figure}[!h]
% %     \begin{wrapfigure}
%       % \begin{wrapfigure}{r}{0.25\textwidth}
%       % \vspace{-4mm}
% \begin{mdframed}[style=alttight]
% \centering
% 	    \begin{tikzpicture}[scale = 0.5, every node/.style = { scale = 1.25}] % make it bigger
%         \tikzset{node distance = 11ex}
%         \tikzset{line width = 1pt}
%         \tikzstyle{snakeme} = [decorate, decoration = snake ]
%         \tikzstyle{smaller} = [scale = 0.65]
%         \tikzstyle{small} = [scale = 0.75]
        
%         \node (PI) {\sessp};
%         \node [right of=PI] (MPI) {\msessp};
%         \node [below of=PI] (HO) {\HO};
%         \node [right of=HO] (MHO) {\mHO};
        
%         %\draw[dotted, ->] (HO) to (MHO);
%         \draw[->, dotted] (HO) -- node[above, smaller] {$\D{\cdot}$} (MHO);
%         \draw[->, dotted] (PI) -- node [right, smaller] {$\map{\cdot}^1_{g}$} (HO);
%         \draw[->, dotted] (MHO) -- node [right, smaller] {$\map{\cdot}^2$} (MPI);
%         \draw[->] (PI) -- node[below, small] {$\F{\cdot}$} (MPI);
%     \end{tikzpicture}
%     \end{mdframed}
%     \caption{Decomposition by composition. \label{enc_drawing}}
%     %\vspace{-5mm}
% %\end{wrapfigure}
% %\vspace{-3mm}
% \end{figure}
More precisely, let  \msessp and \mHO denote the process languages \sessp and \HO   with \msts (rather than with standard session types). 
Also, let  $\D{\cdot}$ denote the  decomposition function from $\HO$ to $\mHO$ defined in~\cite{APV19}.
Kouzapas et al.~\cite{DBLP:conf/esop/KouzapasPY16,DBLP:journals/iandc/KouzapasPY19} gave typed encodings from \sessp to \HO (denoted $\map{\cdot}^1_{g}$)  and from $\HO$ to $\sessp$ (denoted $\map{\cdot}^2$). 
   Therefore, given $\D{\cdot}$, $\map{\cdot}^1_{g}$\,, and  $\map{\cdot}^2$, to define a decomposition function $\F{\cdot}: \sessp \to \msessp$, it suffices to compose these three functions following    \Cref{pi:enc_drawing}.
This is sound for our purposes, because $\map{\cdot}^1_{g}$  and $\map{\cdot}^2$ preserve sequentiality in processes and their types. %---\jcheck{that is, both encodings preserve \msts}.

The correctness of the decomposition function $\F{\cdot}$ follows from that of its constituent functions. $\F{\cdot}$ is significant, as it provides an elegant, positive answer to the question of whether the minimality result in~\cite{APV19} holds for $\sessp$ and all its constructs, including recursion.  Indeed, it proves that the kind of values exchanged do not influence sequentiality in session types: the minimality result of~\cite{APV19} is not specific to the abstraction-passing language $\HO$.

Unfortunately, $\F{\cdot}$ is not entirely satisfactory. A side effect of composing  $\D{\cdot}$, $\map{\cdot}^1_{g}$\,, and  $\map{\cdot}^2$ is that the resulting decomposition of $\sessp$ into $\msessp$ exhibits some \emph{suboptimal features}, in particular redundant synchronizations. 
%is \emph{inefficient}, as it induces
These shortcomings are particularly noticeable in the treatment of processes with recursion and recursive types. The \emph{second answer} to the open question from~\cite{APV19} is an optimized variant of  $\F{\cdot}$, dubbed $\Do{\cdot}$, which avoids redundant synchronizations and treats recursive processes and variables directly, exploiting the fact that $\sessp$ supports recursion natively.

  \paragraph{Contributions.}
The main contributions of this paper are:
\begin{enumerate}
	\item Two new minimality results for \sessp. The first leverages the function  $\F{\cdot}$, obtained by following the ``decompose by composing'' approach in \Cref{pi:enc_drawing} (\Cref{pi:t:amtyprecdec});  the second exploits $\Do{\cdot}$, an optimized variant of $\F{\cdot}$ without redundant communications (\Cref{pi:t:decompcore}). Also, $\Do{\cdot}$ provides direct support for   recursion and recursive types.
	We provide metrics for the improvements in moving from $\F{\cdot}$ to  $\Do{\cdot}$ (\Cref{pi:p:propopt}).
	
	\item The minimality results are, in essence, a guarantee of \emph{static correctness} for our decompositions. In the case of the optimized decomposition $\Do{\cdot}$, we complement this static guarantee with a \emph{dynamic correctness} guarantee: following~\cite{DBLP:journals/corr/abs-2301-05301}, we prove that well-typed processes $P$ and $\Do{P}$ are behaviorally equivalent (\Cref{pi:t:mainbsthm}).
	\item Examples that illustrate the workings of $\F{\cdot}$ and $\Do{\cdot}$.
\end{enumerate}

\paragraph{Organization.}
   The rest of this paper is organized as follows.
Next, \Cref{pi:s:prelim} collects definitions and results from prior works, useful to our developments. 
\Cref{pi:s:dbc} develops the ``decompose by composing'' approach, and presents the first minimality result for \sessp. 
\Cref{pi:s:optimizations} reports the optimized decomposition function, our second minimality result, and its dynamic correctness guarantee. 
\Cref{pi:sec:sb} discusses by example the extension of $\Do{\cdot}$ with constructs for labeled choices (selection and branching).
 \Cref{pi:s:rw} discusses related works, and in particular compares our approach and results against the work by Dardha et al.~\cite{DBLP:conf/ppdp/DardhaGS12,DBLP:journals/iandc/DardhaGS17,DBLP:journals/corr/Dardha14}.
 Finally, \Cref{pi:s:concl} concludes.

  Omitted  technical material appears \longversion{in the appendices.}{in ~\cite{AAP21-full}.} 
      Throughout the paper, we use colors: they are meant to facilitate reading, but are not indispensable for following the text. In particular, we use 
      \textcolor{\oldiecolor}{green} color for notions from prior works, and 
      \textcolor{\deccolor}{red} and \textcolor{\decoptcolor}{blue} colors to distinguish elements of our first and second decompositions, respectively.
      
      \paragraph{Origin of the Results.} This paper is an extended and revised version of a conference paper that appeared at PPDP'21~\cite{DBLP:conf/ppdp/ArslanagicP021} and was presented as a short paper at EXPRESS/SOS'21. With respect to~\cite{DBLP:conf/ppdp/ArslanagicP021}, the current presentation includes complete technical details and extended examples; in particular,  \Cref{pi:sec:sb} is new and \Cref{pi:s:rw} has been substantially extended.

% \noindent \emph{Note:}   
% Due to space limits, 
% %throughout the paper we consider languages without constructs for labeled choice, which we discuss towards the end of the paper.
% %Also, 
% omitted  material (definitions, proofs, additional examples) can be found in the appendices.

\section{Preliminaries}
\label{pi:s:prelim}

In this section, we recall the syntax, semantics, and session type system of \HOp~\cite{KouzapasPY17,DBLP:journals/iandc/KouzapasPY19}.\footnote{We consider \HOp without labeled choices (selection and branching); these constructs will be discussed in \Cref{pi:sec:sb}.}
As mentioned above, we are concerned with \sessp, which is the first-order sub-language of \HOp.
We also recall the mutual encodings between \sessp and
  \HO, the other relevant sub-language of \HOp~\cite{DBLP:conf/esop/KouzapasPY16,DBLP:journals/iandc/KouzapasPY19}. 
 Finally, we briefly discuss \msts for \HO and overview the minimality result in~\cite{APV19,DBLP:journals/corr/abs-2301-05301}. 

\subsection{\HOp (and its Sub-languages \HO and \sessp)}
%\noindent For the purpose of keeping this subsection as a short overview, we introduce the details defining \HOp, as they include both \HO and \sessp-calculus. The syntax and operational semantics %and the type systemare identical to those defined in \cite{MinimalSessionTypes}. We proceed by borrowing the presented information.\\

	\begin{figure}[!t]
	\begin{mdframed}[style=alttight]
				\begin{align*}
			n   \bnfis &  a,b \bnfbar s, \dual{s} 
			\\
			u,w   \bnfis & n \bnfbar x,y,z 
			\\
			V,W   \bnfis &\nonhosyntax{u} \bnfbar \nonpisyntax{\abs{x}{P}} \sbnfbar \nonpisyntax{x,y,z}
			\\[1mm]
			P,Q
			  \bnfis &
			\bout{u}{V}{P}  \sbnfbar  \binp{u}{x}{P} %\sbnfbar			\bsel{u}{l} P \sbnfbar \bbra{u}{l_i:P_i}_{i \in I}  
%			\\[1mm]
			  \sbnfbar  
			  \nonpisyntax{\appl{V}{u}} \sbnfbar  P\Par Q \sbnfbar \news{n} P 
			\sbnfbar \inact \sbnfbar \nonhosyntax{\rvar{X} \sbnfbar \recp{X}{P}}
%			\\%[1mm]
%			 \bnfbar &
%			\nonhosyntax{\rvar{X} \bnfbar \recp{X}{P}}
		\end{align*}
			\end{mdframed}
	\caption{\newjb{Syntax of \HOp. The calculus \sessp is the sub-language of \HOp that lacks \nonpisyntax{\text{boxed}} constructs, whereas \HO is the sub-language that lacks \nonhosyntax{\text{shaded}} constructs.}}
	\label{top:fig:syntax}
%	\vspace{-1mm}
%	\Hlinefig
%\vspace{-3mm}
\end{figure}
%We start by presenting the \textbf{syntax} of \HOp and its sub-languages \HO and \sessp, as studied in~\cite{DBLP:journals/iandc/KouzapasPY19,KouzapasPY17}. 
\paragraph{Syntax.}
\Cref{top:fig:syntax} gives the  {syntax} of processes $P, Q, \dots$, values $V, W, \ldots$, and conventions for names.  Identifiers $a, b, c, \dots$ denote \emph{shared names}, while $s, \dual{s}, \dots $ are used for \emph{session names}.  \emph{Names} (shared or sessions) are denoted by $m, n \dots$, and $x, y, z, \dots$ range over variables.  A notion of duality applies to names:  the dual of $n$ is denoted $\dual{n}$. Duality is defined only on session names; this way, $\dual{\dual{s}} = s$, but $\dual{a} = a$.

We write $\widetilde x$ to denote a tuple $(x_1, \dots, x_n)$, and use $\epsilon$ to denote the empty tuple. Given a tuple  $\widetilde x$, we use $|\widetilde x|$ to denote its length. 
With a slight abuse of notation, we sometimes write $\widetilde x$ to refer to its associated finite set of elements (and vice versa).

%The notations presented so far are common for both \sessp and \HO. Values are simply channel names in \sessp, while in \HO  they can be abstractions and variables. 
An abstraction $\abs{x}{P}$ binds $x$ to its body $P$. In processes, sequentiality is specified via prefixes. The \emph{output} prefix, $\bbout{u}{V}P$, sends value $V$ on name $u$, then continues as $P$. Its dual is the \emph{input} prefix, $\binp{u}{x}P$, in which variable $x$ is bound. %Labelled (deterministic) choice consists of \emph{selection} and \emph{branching}, denoted as $\bsel{u}{l}P$ and $\bbra{u}{l_i:P_i}_{i \in I}$.
In the following we shall consider \emph{polyadic} communication (i.e.,  communication of tuples of values), and so we have output and input prefixes of the form $\bbout{u}{\widetilde{V}}P$ and $\binp{u}{\widetilde{x}}P$, respectively.
Parallel composition, $P \Par Q$, specifies the combined behavior of two processes running simultaneously. Restriction $\news{n}P$ binds the endpoints $n, \dual{n}$ in process $P$. 
Process $\inact$ denotes inaction. 
Recursive variables and recursive processes are denoted $\rvar{X} $ and $\recp{X}{P}$, respectively. 
Replication is denoted by the shorthand notation~$\repl{P}$, which stands for $\recp{X}(P \Par X)$.

%Although not specifically mentioned before, the process calculi used for the scope of this paper are considered polyadic.

The sets of free variables, sessions, and names of a process are denoted $\fv{P}$, $\fs{P}$,  and $\fn{P}$. A process $P$ is \emph{closed} if  $\fv{P} = \emptyset$.
We write  $\bbout{u}{}P$ and $\binp{u}{}P$ when the communication objects are not relevant. Also, we omit trailing occurrences of  $\inact$.
%\fn

As shown in \Cref{top:fig:syntax}, the  {sub-languages} \sessp and \HO of \HOp differ as follows:
application $V u$ is only present in \HO; constructs for recursion $\recp{X}{P}$ are present in \sessp but not in \HO.

\begin{figure}[!t]
\begin{mdframed}[style=alttight]
%\[
%	\begin{array}{rcllcrcll}
	\begin{align*}
		\appl{(\abs{x}{P})}{u}   & \red  P \subst{u}{x} 
		& \orule{App} &
		\\[1mm]
		\bout{n}{V} P \Par \binp{\dual{n}}{x} Q & \red  P \Par Q \subst{V}{x} 
		& \orule{Pass} &
		\\[1mm]
%		\bsel{n}{l_j} Q \Par \bbra{\dual{n}}{l_i : P_i}_{i \in I} & \red  Q \Par P_j ~~(j \in I)~~ 
%		& \orule{Sel} & 
%		\\[1mm]
%		&&
		P \red P'\Rightarrow  \news{n} P   & \red    \news{n} P' 
		& \orule{Res} & 
		\\[1mm]
		P \red P'  \Rightarrow    P \Par Q  & \red   P' \Par Q  
		& \orule{Par} & 
		\\[1mm]
%		&&
		P \scong Q \red Q' \scong P'  \Rightarrow  P  & \red  P'
		& \orule{Cong} &
	\end{align*}
%\]
	\begin{align*}
	&	
		P_1 \Par P_2 \scong P_2 \Par P_1
		\quad
		P_1 \Par (P_2 \Par P_3) \scong (P_1 \Par P_2) \Par P_3
		\\[1mm]
	&	P \Par \inact \scong P
		\quad 
		P \Par \news{n} Q \scong \news{n}(P \Par Q)
		~~ (n \notin \fn{P})
		\\[1mm] 
	&	\news{n} \inact \scong \inact
		\qquad  
		\recp{X}{P} \scong P\subst{\recp{X}{P}}{\rvar{X}}
		\qquad
		P \scong Q \textrm{ if } P \scong_\alpha Q
	\end{align*}
%\vspace{-3mm}
\end{mdframed}
\caption{Reduction Semantics of $\HOp$. 
\vspace{-3mm}
\label{fig:reduction}}
%\Hlinefig
\end{figure}

\paragraph{Reduction Semantics.}
  The  {operational semantics} of \HOp, enclosed in \Cref{fig:reduction}, is expressed through a \emph{reduction relation}, denoted $\red$. 
Reduction is closed under \emph{structural congruence}, $\scong$, which identifies equivalent processes from a structural perspective. 
 We write $P\subst{V}{x}$ to denote the capture-avoiding substitution of variable $x$ with value $V$ in $P$. 
As customary, the capture-avoiding, simultaneous substitution in the polyadic setting is denoted $\subst{\widetilde V}{\widetilde x}$, with the assumption that $|\widetilde V| = |\widetilde x|$.
 We write $\sigma, \sigma', \ldots$ to range over substitutions, and 
 write `$\{ \}$'     to denote the empty substitution. 
In~\Cref{fig:reduction}, Rule~$\orule{App}$ denotes application, which only concerns names.
Rule~$\orule{Pass}$ defines a shared or session interaction on channel $n$'s endpoints. %Labelled (deterministic) choice is defined by Rule~$\orule{Sel}$. 
The remaining rules are standard.

We shall often write $P \red^k P'$ to denote a sequence of $k > 0$ reductions between $P$ and $P'$.

\thesisalt{
  \paragraph{Labeled Transition System (LTS).}
  \label{top:ss:lts}
  % \subimport{sources/}{untyped-lts.tex}
  % \section{Labelled Transition System for Processes}
% \label{top:ss:lts}

We define the interaction of processes with their environment using action labels $\ell$:
\begin{center}
	\begin{tabular}{l}
		$\ell
			\bnfis  \tau 
			\bnfbar	\news{\widetilde{m}} \bactout{n}{V}
			\bnfbar	\bactinp{n}{V}$ 
			%\bnfbar	\bactsel{n}{l} 
			%\bnfbar	\bactbra{n}{l}$
	\end{tabular}
\end{center}
\noi 
Label $\tau$ defines internal actions.
Action
$\news{\widetilde{m}} \bactout{n}{V}$
denotes the sending of value $V$
over channel $n$ with a possible empty set of restricted names
$\widetilde{m}$ 
(we may write $\bactout{n}{V}$ when $\widetilde{m}$ is empty).
Dually, the action for value reception is 
$\bactinp{n}{V}$.
%Actions for select and branch on a label~$l$ are denoted $\bactsel{n}{l}$ and $\bactbra{n}{l}$, respectively.
We write $\fn{\ell}$ and $\bn{\ell}$ to denote the
sets of free/bound names in $\ell$, respectively.
%and set $\mathsf{n}(\ell)=\bn{\ell}\cup \fn{\ell}$. 
Given $\ell \neq \tau$, we 
say $\ell$ is a \emph{visible action}; we
write $\subj{\ell}$
to denote its \emph{subject}.
\newc{This way, we have:  
$\subj{\news{\widetilde{m}} \bactout{n}{V}} = 
\subj{\bactinp{n}{V}} = 
%\subj{\bactsel{n}{l}} = 
%\subj{\bactbra{n}{l}} = 
n$}.
\fixed{\emph{Dual actions}
occur on subjects that are dual between them and carry the same
object; thus, output is dual to input. % and selection is dual to branching.
%\begin{definition}[Dual Actions]\label{def:dualact}
We define duality on actions
as the least symmetric relation $\asymp$ on action labels that satisfies:
\[
	%\bactsel{n}{l} \asymp \bactbra{\dual{n}}{l}
	%\qquad \qquad 
	\news{\widetilde{m}} \bactout{n}{V} \asymp \bactinp{\dual{n}}{V}
\]}

%%%%%%%%%%%%%%%%%%%%%%%%%% UNTYPED LTS %%%%%%%%%%%%%%%%%%%%%%%%%%%%%%%%%%%%%%%%%%%%%%%%%%%%%

\begin{figure}[t!]
    \begin{mdframed}
        	\begin{mathpar}
		\inferrule[\ltsrule{App}]{
		}{
			\appl{(\abs{x}{P})}{V} \by{\tau} P \subst{V}{x}
		}
		\and
		\inferrule[\ltsrule{Snd}]{
		}{
			\bout{n}{V} P \by{\bactout{n}{V}} P
		}
		\and
		\inferrule[\ltsrule{Rv}]{
		}{
			\binp{n}{x} P \by{\bactinp{n}{V}} P\subst{V}{x}
		}
		\and
%		\inferrule[\ltsrule{Sel}]{
%		}{
%			\bsel{s}{l}{P} \by{\bactsel{s}{l}} P
%		}
%		\and
%		\inferrule[\ltsrule{Bra}]{j\in I
%		}{
%			\bbra{s}{l_i:P_i}_{i \in I} \by{\bactbra{s}{l_j}} P_j 
%		}
%		\and
		\inferrule[\ltsrule{Alpha}]{
			P \scong_\alpha Q
			\and
			Q\by{\ell} P'
		}{
			P \by{\ell} P'
		}
		\and
		\inferrule[\ltsrule{Res}]{
			P \by{\ell} P'
			\and
			n \notin \fn{\ell}
		}{
			\news{n} P \by{\ell} \news{n} P'
		}
		\and
		\inferrule[\ltsrule{New}]{
			P \by{\news{\widetilde{m}} \bactout{n}{V}} P'
			\and
			m_1 \in \fn{V}
		}{
			\news{m_1} P \by{\news{m_1\cat\widetilde{m}} \bactout{n}{V}} P'
		}
		\and
		\inferrule[\ltsrule{Par${}_L$}]{
			P \by{\ell} P'
			\and
			\bn{\ell} \cap \fn{Q} = \es
		}{
			P \Par Q \by{\ell} P' \Par Q
		}
		\and
		\inferrule[\ltsrule{Tau}]{
			P \by{\ell_1} P'
			\and
			Q \by{\ell_2} Q'
			\and
			\ell_1 \asymp \ell_2
		}{
			P \Par Q \by{\tau} \newsp{\bn{\ell_1} \cup \bn{\ell_2}}{P' \Par Q'}
		}
		\and
		\inferrule[\ltsrule{Rec}]{
			P\subst{\recp{X}{P}}{\rvar{X}} \by{\ell} P'
		}{
			\recp{X}{P}  \by{\ell} P'
		}
	\end{mathpar}
\end{mdframed}
\caption[The Untyped LTS for \HOp processes.]{The Untyped LTS for \HOp processes. We omit Rule $\ltsrule{Par${}_R$}$.  \label{top:fig:untyped_LTS}}
\end{figure}

%%%%%%%%%%%%%%%%%%%%%%%%%% UNTYPED LTS %%%%%%%%%%%%%%%%%%%%%%%%%%%%%%%%%%%%%%%%%%%%%%%%%%%%%

%%%%%%%%%%%%%%%%%%%%%%%%%% UNTYPED LTS %%%%%%%%%%%%%%%%%%%%%%%%%%%%%%%%%%%%%%%%%%%%%%%%%%%%%

The (early) labeled transition system
(LTS) %LTS
for \emph{untyped} processes
is given in
\Cref{top:fig:untyped_LTS}. 
We write $P_1 \by{\ell} P_2$ with the usual meaning.
The rules are standard~\cite{KYHH2015,KY2015}; we comment on some of them. 
A process with an output prefix can
interact with the environment with an output action that carries a value
$V$ (Rule~$\ltsrule{Snd}$).  Dually, in Rule~$\ltsrule{Rv}$ a
receiver process can observe an input of an arbitrary value $V$.
%Select and branch processes observe the select and branch actions in Rules~$\ltsrule{Sel}$ and $\ltsrule{Bra}$, respectively.
Rule $\ltsrule{Res}$ 
%closes the LTS under restriction 
enables an observable action from a process with an outermost restriction, provided that 
 the restricted name does not occur free in the action. 
If a restricted name occurs free in
the carried value of an output action,
the process performs scope opening (Rule~$\ltsrule{New}$).  
Rule~$\ltsrule{Rec}$ handles recursion unfolding.
Rule~$\ltsrule{Tau}$ 
states that two parallel processes which perform
dual actions can synchronise by an internal transition.
Rules~$\ltsrule{Par${}_L$}$/$\ltsrule{Par${}_R$}$ 
and $\ltsrule{Alpha}$ 
%close the LTS under parallel composition and $\alpha$-renaming. 
define standard treatments for actions under parallel composition and $\alpha$-renaming.

In defining (typed) behavioral equivalences for processes, later on we shall consider a \emph{typed} LTS, i.e., an enhancement of the untyped LTS in
\Cref{top:fig:untyped_LTS} with session types. 
}{}

\subsection{Session Types for \HOp}
\label{pi:ss:types}
	\begin{figure}[!t]
	\begin{mdframed}%[style=alttight]
				\begin{align*}
						%\textrm{(value)} &
			U \bnfis & C \bnfbar L
			& %\\[1mm]
			%\textrm{(name)} & 
			C \bnfis & S \bnfbar \chtype{S} \bnfbar \chtype{L}
			\\%[1mm]
			%\textrm{(abstractions)}~~ & 
			L \bnfis & \shot{U} \bnfbar \lhot{U}
			& %\\[1mm]		
			%\textrm{(session)} & 
			S \bnfis &  \btout{U} S \bnfbar \btinp{U} S  
			\bnfbar  							
			 \trec{t}{S} \bnfbar \vart{t}  \bnfbar \tinact
			\\%[1mm]					
			% \bnfbar &  \btsel{l_i:S_i}_{i \in I}  \bnfbar \btbra{l_i:S_i}_{i \in I}
			%\\
	%	\end{align*}
%		      \vspace{1em}
      \noalign{\hbox to \textwidth{\leaders\hbox to 3pt{\hss . \hss}\hfil}}
%		      \vspace{1em}
%	\begin{align*}
U  \bnfis	&	\shot{\widetilde{C}} \bnfbar \lhot{\widetilde{C}}
		&
		C   \bnfis &		M  \bnfbar  {\chtype{U}}
\\
		% M  \bnfis  & 
    	% \tinact  \bnfbar  \btout{\widetilde{U}} \tinact \bnfbar 
    % \btinp{\widetilde{U}} \tinact \qquad \qquad  
    % \\
        \mugamma  \bnfis & \tinact \bnfbar \vart{t}
&
    M \bnfis & \mugamma  \bnfbar  \btout{\widetilde{U}} \mugamma 		
    	\bnfbar \btinp{\widetilde{U}} \mugamma \bnfbar \trec{t}{M }	
	%	\\
   % &  \bnfbar \btsel{l_i:M_i}_{i \in I}					\bnfbar \btbra{l_i:M_i}_{i \in I} \\
	\end{align*}
			\end{mdframed}
		\caption{STs for \HOp (top) and MSTs for \HO (bottom). \label{top:f:sts}}
		%\vspace{-5mm}
	\end{figure}
	\Cref{top:f:sts} (top) gives the syntax of types.
Value types $U$ include
the first-order types $C$ and the higher-order
types $L$. 
In \fixed{our} examples we use other value (base) types, such as $\tint$ and $\bool$.

Session types are denoted with $S$ and
shared types with $\chtype{S}$ and $\chtype{L}$.
\newc{We write $\Proc$ to denote the \emph{process type}.}
%Note that we dissallow type $\chtype{U}$, thus
%in the type discipline shared names cannot carry shared names.
%In name types, $\chtype{U}$ is shared name types 
%which are sent via shared names. 
\newc{The functional} types $\shot{U}$ and $\lhot{U}$ denote
{\em shared} and {\em linear} higher-order 
%\jpc{functional}
types, respectively.
The {\em output type}
$\btout{U} S$ is assigned to a name that 
first sends a value of
type $U$ and then follows the type described by $S$.  
Dually, $\btinp{U} S$ denotes an {\em input type}. 
%The {\em selection type}
%$\btsel{l_i:S_i}_{i \in I}$ 
%and the 
%{\em branching type}
%$\btbra{l_i:S_i}_{i \in I}$ 
%define labelled choice, \newc{implemented at the level of processes by internal and external choice mechanisms, respectively.}
Type $\tinact$ is the termination type. 
We assume the {\em recursive type} $\trec{t}{S}$ is guarded,
i.e., the type variable $\vart{t}$ only appears under prefixes. 
This way, types such as, 
e.g.,  the type $\trec{t}{\vart{t}}$ are not allowed. 
The sets of free/bound variables of a   type $S$ are defined as usual; 
the sole binder is $\trec{t}{S}$.
Closed session types do not have free type variables.

\begin{figure}[t!]
\begin{mdframed}
\[
	\begin{array}{c}
	\inferrule[(Sess)]{}{\Gamma; \emptyset; \set{u:S} \proves u \hastype S} 
		\qquad
		\inferrule[(Sh)]{}{\Gamma \cat u : U; \emptyset; \emptyset \proves u \hastype U}
		\\ \\ 
		\inferrule[(LVar)]{}{\Gamma; \set{x: \lhot{C}}; \emptyset \proves x \hastype \lhot{C}}
						\qquad
		\inferrule[(RVar)]{}{\Gamma \cat \rvar{X}: \Delta; \emptyset; \Delta  \proves \rvar{X} \hastype \Proc}
				\\  \\
		\inferrule[(Abs)]{
			\Gamma; \Lambda; \Delta_1 \proves P \hastype \Proc
			\quad
			\Gamma; \es; \Delta_2 \proves x \hastype C
		}{
			\Gamma\backslash x; \Lambda; \Delta_1 \backslash \Delta_2 \proves \abs{{x}}{P} \hastype \lhot{{C}}
		}
		\quad
		\inferrule[(App)]{
			%\begin{array}{c}
				%U = \hot{C} \lor \shot{C}
				%\\
				{\Gamma; \Lambda; \Delta_1 \proves V \hastype \ghot{C} ~~
				\leadsto\, \in \{\lollipop, \sharedop\} \quad
				\Gamma; \es; \Delta_2 \proves u \hastype C
				}
		%	\end{array}
		}{
			\Gamma; \Lambda; \Delta_1 \cat \Delta_2 \proves \appl{V}{u} \hastype \Proc
		} 

		\\  \\
		\inferrule[(Prom)]{
			\Gamma; \emptyset; \emptyset \proves V \hastype 
                         \lhot{C}
		}{
			\Gamma; \emptyset; \emptyset \proves V \hastype 
                         \shot{C}
		} 
		\quad
		\inferrule[(EProm)]{
		\Gamma; \Lambda \cat x : \lhot{C}; \Delta \proves P \hastype \Proc
		}{
			\Gamma \cat x:\shot{C}; \Lambda; \Delta \proves P \hastype \Proc
		}
				\quad
		\inferrule[(End)]{
			\Gamma; \Lambda; \Delta  \proves P \hastype \Proc  \quad u \not\in \dom{\Gamma, \Lambda,\Delta}
		}{
			\Gamma; \Lambda; \Delta \cat u: \tinact  \proves P \hastype \Proc
		}
		\\  \\
		\inferrule[(Rec)]{
			\Gamma \cat \rvar{X}: \Delta; \emptyset; \Delta  \proves P \hastype \Proc
		}{
			\Gamma ; \emptyset; \Delta  \proves \recp{X}{P} \hastype \Proc
		}
		\qquad
			\inferrule[(Par)]{
			\Gamma; \Lambda_{i}; \Delta_{i} \proves P_{i} \hastype \Proc \quad i=1,2
		}{
			\Gamma; \Lambda_{1} \cat \Lambda_2; \Delta_{1} \cat \Delta_2 \proves P_1 \Par P_2 \hastype \Proc
		}
		\qquad
				\inferrule[(Nil)]{ }{\Gamma; \emptyset; \emptyset \proves \inact \hastype \Proc}
		\\  \\
		\inferrule[(Send)]{
					%\begin{array}{c}
					u:S \in \Delta_1 \cat \Delta_2 %\\
					\quad
			\Gamma; \Lambda_1; \Delta_1 \proves P \hastype \Proc
			\quad
			\Gamma; \Lambda_2; \Delta_2 \proves V \hastype U
			%\end{array}
		}{
			\Gamma; \Lambda_1 \cat \Lambda_2; ((\Delta_1 \cat \Delta_2) \setminus u:S) \cat u:\btout{U} S \proves \bout{u}{V} P \hastype \Proc
		}
		\\  \\
		\inferrule[(Req)]{
			%\begin{array}{c}
%				\Gamma; \es; \es \proves u \hastype U_1
%				\qquad
%				\Gamma; \Lambda; \Delta_1 \proves P \hastype \Proc
%				%\\
%				\qquad
%				\Gamma; \es; \Delta_2 \proves V \hastype U_2
%				\\
%				(U_1 = \chtype{S} 
%                                \land  
%                                U_2 = S)
%				\lor
%				 (U_1 = \chtype{L} 
%                                \land  
%                                 U_2 = L)
%                                 \\
                                 				{\Gamma; \Lambda; \Delta_1 \proves P \hastype \Proc
				\quad
                                 \Gamma; \es; \es \proves u \hastype \chtype{{U}}
				\quad
				\Gamma; \es; \Delta_2 \proves V \hastype {U}
				%\quad \mathcal{U} \in \{S, L\}
				\quad (U = S) \lor (U = L)
				}
			%\end{array}
		}{
			\Gamma; \Lambda; \Delta_1 \cat \Delta_2 \proves \bout{u}{V} P \hastype \Proc
		}
		~~
		\\ \\
				\inferrule[(Rcv)]{
		%\begin{array}{c}
			\Gamma; \Lambda_1; \Delta_1 \cat u: S \proves P \hastype \Proc
			\quad
			\Gamma; \Lambda_2; \Delta_2 \proves {x} \hastype {U}
		%	\end{array}
		}{
			\Gamma \backslash x; \Lambda_1\backslash \Lambda_2; \Delta_1\backslash \Delta_2 \cat u: \btinp{U} S \vdash \binp{u}{{x}} P \hastype \Proc
		}
		~~
		\inferrule[(Acc)]{
			\begin{array}{c}
%				\Gamma; \emptyset; \emptyset \proves u \hastype U_1 
%				\quad
%				\Gamma; \Lambda_1; \Delta_1 \proves P \hastype \Proc
%				\\
%				\Gamma; \Lambda_2; \Delta_2 \proves x \hastype U_2\\
%				(U_1 = \chtype{S} 
%                                \land
%                                U_2 = S)
%				\lor
%				 (U_1 = \chtype{L} 
%                                \land 
%                                 U_2 = L)
%                                 \\
                                 {
                                 				\Gamma; \Lambda_1; \Delta_1 \proves P \hastype \Proc
				\quad
                                 	\Gamma; \emptyset; \emptyset \proves u \hastype \chtype{{U}} 
				}
				\\
				{\Gamma; \Lambda_2; \Delta_2 \proves x \hastype  {U}
				\quad %\mathcal{U} \in \{S, L\}
				\quad (U = S) \lor (U = L)
				}
	               \end{array}
		}{
			\Gamma\backslash x; \Lambda_1 \backslash \Lambda_2; \Delta_1 \backslash \Delta_2 \proves \binp{u}{x} P \hastype \Proc
		}	
		\\  \\
%				\inferrule[(Bra)]{
%			 \forall i \in I \quad \Gamma; \Lambda; \Delta \cat u:S_i \proves P_i \hastype \Proc
%		}{
%			\Gamma; \Lambda; \Delta \cat u: \btbra{l_i:S_i}_{i \in I} \proves \bbra{u}{l_i:P_i}_{i \in I}\hastype \Proc
%		}
%		\qquad
%	 	\inferrule[(Sel)]{
%			\Gamma; \Lambda; \Delta \cat u: S_j  \proves P \hastype \Proc \quad j \in I
%		}{
%			\Gamma; \Lambda; \Delta \cat u:\btsel{l_i:S_i}_{i \in I} \proves \bsel{u}{l_j} P \hastype \Proc
%		}
%		\\  \\
		\inferrule[(ResS)]{
			\Gamma; \Lambda; \Delta \cat s:S_1 \cat \dual{s}: S_2 \proves P \hastype \Proc \quad S_1 \dualof S_2
		}{
			\Gamma; \Lambda; \Delta \proves \news{s} P \hastype \Proc
		}
		\qquad
		\inferrule[(Res)]{
			\Gamma\cat a:\chtype{S} ; \Lambda; \Delta \proves P \hastype \Proc
		}{
			\Gamma; \Lambda; \Delta \proves \news{a} P \hastype \Proc
		}
		\end{array}
\]
\end{mdframed}
%\vspace{-3mm}
\caption{Typing Rules for $\HOp$. % (including selection and branching constructs).
%See \appref{app:types} for a full account.
\label{top:fig:typerulesmys}}
%\Hline
%\vspace{-1mm}
\end{figure}
%\myparagraph{Typing System of \HOp}

Session types for \sessp exclude  $L$ from value types $U$ and $\chtype{L}$ from  $C$; session types for \HO exclude $C$ from value types $U$.

We write $S_1 \dualof S_2$ if 
$S_1$ is the \emph{dual} of $S_2$.   
Intuitively, 
duality
converts $!$ into $?$ %and $\oplus$ into $\&$ 
(and vice-versa).
This intuitive definition is enough for our purposes; the formal definition is co-inductive, see~\cite{DBLP:journals/iandc/KouzapasPY19,KouzapasPY17}.
Typing \emph{environments} are defined below:
%\[
\begin{align*}
	\Gamma  & \bnfis  \emptyset \bnfbar \Gamma \cat \varp{x}: \shot{U} \bnfbar \Gamma \cat u: \chtype{S} \bnfbar \Gamma \cat u: \chtype{L} 
        \bnfbar \Gamma \cat \rvar{X}: \Delta 
        \\
	\Lambda &\bnfis  \emptyset \bnfbar \Lambda \cat \AT{x}{\lhot{U}} 
	\\
	%\qquad\qquad
	%\hfill 
	\Delta   & \bnfis   \emptyset \bnfbar \Delta \cat \AT{u}{S}
\end{align*}
%]
We shall refer to $\Delta$ as the \emph{session environment}.
 $\Gamma$, $\Lambda$, and $\Delta$ satisfy different structural principles.
%Intuitively, the \emph{exchange} principle indicates that the ordering of type assignments does not matter. 
%\emph{Weakening} says that type assignments need not be used. 
%Finally, \emph{contraction}  says that type assignments may be duplicated.}
 $\Gamma$ maps variables and shared names to value types, and recursive 
variables to session environments;  
it admits weakening, contraction, and exchange principles.
$\Lambda$ maps variables to 
%the
 linear %functional 
higher-order
types, and so it is relevant only for processes featuring passing of abstractions.
$\Delta$ maps
session names to session types. 
Both $\Lambda$ and $\Delta$ %behave linearly: they 
are
only subject to exchange.  
%We require that t
The domains of $\Gamma$,
$\Lambda$ and $\Delta$ 
(denoted $\dom{\Gamma}$, $\dom{\Lambda}$, and  $\dom{\Delta}$, respectively)
are assumed pairwise distinct.

Given $\Gamma$, we write $\Gamma \backslash x$ to denote the environment obtained from 
 $\Gamma$ by removing the assignment $x:\shot{U}$, for some $U$.
This notation applies similarly to $\Delta$ and $\Lambda$; 
we write $\Delta \backslash \Delta'$ (and $\Lambda \backslash \Lambda'$) with the expected meaning.
Notation
$\Delta_1\cdot \Delta_2$ means 
the disjoint union of $\Delta_1$ and $\Delta_2$.  
We define \emph{typing judgements} for values $V$
and processes $P$:
%\begin{center}
%\begin{tabular}{c}
	$$\Gamma; \Lambda; \Delta \proves V \hastype U \qquad \qquad \qquad \qquad \qquad \Gamma; \Lambda; \Delta \proves P \hastype \Proc$$
%\end{tabular}
%\end{center}
The judgement on the left
says that under environments $\Gamma$, $\Lambda$, and $\Delta$ value $V$
has type $U$; the  judgement on the right says that under
environments $\Gamma$, $\Lambda$, and $\Delta$ process $P$ has the process type~$\Proc$.
The typing rules are presented in~\Cref{top:fig:typerulesmys}. 
% \longversion{The typing rules are presented in Appendix~\ref{top:app:typing}.}{
%   The typing rules are presented in \cite{AAP21-full}. 
% }

% \input{type_system.tex}

Type soundness   for \HOp  
relies on two auxiliary notions: 
%that contain dual endpoints typed with dual types.
%The following definition ensures two session endpoints 
%are dual each other. 

%\smallskip

\begin{definition}[Session Environments: Balanced/Reduction]\label{top:d:wtenvred}%\rm
	Let $\Delta$ be a session environment.
	\begin{enumerate}[$\bullet$]
	\item  $\Delta$ is {\em balanced}, written $\balan{\Delta}$, if whenever
	$s: S_1, \dual{s}: S_2 \in \Delta$ then $S_1 \dualof S_2$.
	\item We define  reduction  $\red$ on session environments as: %\\ %[-2mm]
\begin{align*}
	\Delta \cat s: \btout{U} S_1 \cat \dual{s}: \btinp{U} S_2  & \red  
	\Delta \cat s: S_1 \cat \dual{s}: S_2  
%	\\
%	\Delta \cat s: \btsel{l_i: S_i}_{i \in I} \cat \dual{s}: \btbra{l_i: S_i'}_{i \in I} &\red \Delta \cat s: S_k \cat \dual{s}: S_k' \ (k \in I)
\end{align*}
\end{enumerate}
\end{definition}

%We state the type soundness result.
\begin{theorem}[Type Soundness~\cite{KouzapasPY17}]\label{t:sr}%\rm
			Suppose $\Gamma; \es; \Delta \proves P \hastype \Proc$
			with
			$\balan{\Delta}$. 
			Then $P \red P'$ implies $\Gamma; \es; \Delta'  \proves P' \hastype \Proc$
			and $\Delta = \Delta'$ or $\Delta \red \Delta'$
			with $\balan{\Delta'}$. 
\end{theorem}

\subsection{Typed Encodings between \sessp and \HO}
\label{s:encoding}
The encodings $\map{\cdot}^1_g: \sessp \to \HO$
and  $\map{\cdot}^2:  \HO\to\sessp $ are \emph{typed}: each consists of a translation on processes and a translation on types. 
This way, $\tmap{\cdot}{1}$ translates   types for first-order processes into   types for higher-order processes, while $\tmap{\cdot}{2}$ operates in the opposite direction---see Figures~\ref{top:f:enc:hopi_to_ho}
and~\ref{top:f:enc:ho_to_sessp}, respectively.
{Remarkably, these translations on processes and types do not alter their sequentiality.} 
%We briefly discuss these encodings.

% The first encoding is $\map{\cdot}^1: \sessp \to \HO$, presented in Figure~\ref{f:enc:hopi_to_ho}. Its opposing counterpart, the second encoding, is denoted by $\map{\cdot}^2$ and depicted in Figure
% ~\ref{top:f:enc:ho_to_sessp}. Both figures contain the respective type mappings as well, which are used to cast session types into the desired process calculi. So, $\tmap{\cdot}{1}$ maps a first-order session type into a higher-order session type, while $\tmap{\cdot}{2}$ achieves the opposite result. Remarkably, these mappings do not alter the sequentiality property of a process' type; they \emph{preserve minimality}. 
% Hence, by combining them with the type decomposition function defined in \cite{MinimalSessionTypes}, we can obtain a method which translates a first-order process into a minimally-typed first-order process. Let us briefly discuss each figure. \\

\paragraph{From \sessp to \HO.}
To mimic the sending of name $w$, the encoding $\map{\cdot}^1_g$  encloses $w$
within the body of an input-guarded abstraction. The corresponding input process
receives this higher-order value, applies it on a restricted session, and sends
the encoded continuation through the session's dual.

Several auxiliary notions are used to encode recursive processes; we describe
them intuitively (see \cite{DBLP:journals/iandc/KouzapasPY19} for full details).
The key idea is to encode recursive processes in \sessp using a ``duplicator''
process in \HO, circumventing linearity by replacing free names with variables.
The parameter $g$ is a map from process variables to sequences of name
variables. 
% Also, $\vmap{{\cdot}}$ maps sequences of session names into sequences
% of variables, and $\auxmapp{\cdot}{{}}{\es}$ maps processes with free names to
% processes without free names (but with free variables instead).
\newadd{To handle linearity, auxiliary mappings are defined: 
$\vmap{{\cdot}}$ maps sequences of session names into sequences
of variables, and $\auxmapp{\cdot}{{}}{\es}$ maps processes with free names to
processes without free names (but with free variables instead): 
}

\begin{newaddenv}
    \begin{definition}[Auxiliary Mappings] \label{top:d:trabs}
        \label{top:d:auxmap}
        We define mappings $\vmap{\cdot}$ and $\auxmapp{\cdot}{{}}{\Psi}$ as follows:
        \begin{itemize}
        \item  \fixed{$\vmap{\cdot}:\mathcal{N}^\omega  \longrightarrow \mathcal{V}^\omega$}
          is a map of sequences of lexicographically ordered names to sequences of variables, defined
          inductively 
          as: 
          \begin{align*}
          \vmap{\epsilon} & = \epsilon 
          \\
          \vmap{n \cat \tilde{m}} & = x_n \cat \vmap{\tilde{m}} \quad \text{($x$ fresh)}
          \end{align*}
        
        \item Given a set of session names and variables $\Psi$, 
           the   map
          $\auxmapp{\cdot}{{}}{\Psi}: \HO \to \HO$
          is as in \Cref{top:f:auxmap}.
          \end{itemize}
        \end{definition}
    \end{newaddenv}

The encoding \tmap{\cdot}{1} depends on the auxiliary function
$\vtmap{\cdot}{1}$, defined on value types. Following the encoding on processes,
this mapping on values takes a first-order value type and encodes it into a
linear higher-order value type, which encloses an input type that expects to
receive another higher-order type. Notice how the innermost higher-order value
type is either shared or linear, following the nature of the given type.

\begin{newaddenv}
  \begin{figure}[t!]
  \begin{mdframed}
    \begin{align*}
        \auxmapp{\bout{w}{\abs{x}{Q}} P}{{}}{\Psi} & \defeq \bout{u}{\abs{x}{\auxmapp{Q}{{}}{{\Psi \cat x}}}} \auxmapp{P}{{}}{\Psi}
        &
        \auxmapp{\bbra{w}{l_i:P_i}_{i \in I}}{{}}{\Psi} & \defeq \bbra{u}{l_i:\auxmapp{P_i}{{}}{\Psi}}_{i \in I}		
        \\
        \auxmapp{\binp{w}{x} P}{{}}{\Psi} & \defeq \binp{u}{x} \auxmapp{P}{{}}{\Psi} 
        &
        \auxmapp{\bsel{w}{l} P}{{}}{\Psi} & \defeq \bsel{u}{l} \auxmapp{P}{{}}{\Psi} 
        \\
        \auxmapp{\news{n} P}{{}}{\Psi} & \defeq \news{n} \auxmapp{P}{{}}{{\Psi \cat n}}
        &
        \auxmapp{\appl{(\lambda x.Q)}{w}}{{}}{\Psi}  & \defeq \appl{(\lambda x.\auxmapp{Q}{{}}{{\Psi \cat x}})}{u}
        \\
        \auxmapp{P \Par Q}{{}}{\Psi} & \defeq \auxmapp{P}{{}}{\Psi} \Par \auxmapp{Q}{{}}{\Psi} 
        &
        \auxmapp{\appl{x}{w}}{{}}{\Psi} & \defeq \appl{x}{u}
        \\
        \auxmapp{\inact}{{}}{\Psi}  & \defeq  \inact	
        & 
\end{align*}
\begin{center}
    {In all cases: $u = 
    \begin{cases} 
    x_n & \text{if $w$ is a name $n$ and  $n \not\in \Psi$ ($x$ fresh)}
    \\
    w & \text{otherwise: $w$ is a variable or a name $n$ and $n \in \Psi$} 
         \end{cases}$}
\end{center}
\end{mdframed}
%\vspace{-3mm}
\caption[Auxiliary mapping used to encode \HOp into \HO.]{\label{top:f:auxmap} Auxiliary mapping used to encode \HOp into \HO (\Cref{top:d:auxmap}).}
%\vspace{-1mm}
%\Hlinefig 
\end{figure}
\end{newaddenv}

\begin{figure}[!t]
\begin{mdframed}%[style=alttight]
\noi{\bf Terms:} 
\begin{align*}
	\map{\bout{u}{w} P}^{1}_ {g}	&\defeq	\!\bout{u}{ \abs{z}{\,\binp{z}{x} (\appl{x}{w})} } \map{P}^{1}_{g}
	\\
	\map{\binp{u}{\AT{x}{C}} Q}^{1}_{g}	&\defeq	\!\binp{u}{y} \newsp{s}{\appl{y}{s} \!\Par\! \bout{\dual{s}}{\abs{x}{\map{Q}^{1}_{g}}} \inact}
	\\
%		\map{\bsel{u}{l} P}^{1}_{g} &\defeq \bsel{u}{l} \map{P}^{1}_{g}
%	\\
%	\map{\bbra{u}{l_i{:}P_i}_{i \in I}}^{1}_{g} &\defeq \!\bbra{u}{l_i: \map{P_i}^{1}_{g}}_{i \in I}
%	\\
		\map{P \Par Q}^{1}_{g} & \defeq \map{P}^{1}_{g} \Par \map{Q}^{1}_{g}
	\\
	\map{\news{n} P}^{1}_{g} &\defeq \!\news{n} \map{P}^{1}_{g}
		\\
	\map{\inact}^{1}_{g} &\defeq \inact \\
  \map{\recp{X}{P}}^{1}_{g} &\defeq
		\newsp{s}{\bbout{\dual{s}}{V} \inact  \Par  \binp{s}{z_\X} \map{P}^{1}_{{g,\{\rvar{X}\to \tilde{n}\}}}}
		\\
    & \qquad  \text{where } (\tilde{n} = \fn{P}) \\
    & \qquad V = \abs{(\vmap{\tilde{n}}, y)} \,{\binp{y}{z_\X} 
    \auxmapp{\map{P}^{1}_{{g,\{\rvar{X}\to \tilde{n}\}}}}{{}}{\es}} \\
    \map{\rvar{X}}^{1}_{g} & \defeq 
		\newsp{s}{\appl{z_X}{(\tilde{n}, s)} \Par \bout{\dual{s}}{ z_X} \inact}  \quad (\tilde{n} = g(\rvar{X}))
%\\
\\[1mm]
\noalign{\bf Types:}
%\begin{align*}
	\vtmap{{S}}{1} & \defeq	\lhot{(\btinp{\lhot{\tmap{S}{1}}} \tinact)}
	\\ 
	\vtmap{\chtype{S}}{1} & \defeq	\lhot{(\btinp{\shot{\chtype{\tmap{S}{1}}}} \tinact)}
	\\[1mm]
	\tmap{\btout{U} S}{1} & \defeq \btout{{\vtmap{U}{1}}} \tmap{S}{1}
	\\
	\tmap{\btinp{U} S}{1} & \defeq\btinp{{\vtmap{U}{1}}} \tmap{S}{1}
	\\
%	\tmap{\btsel{l_i: S_i}_{i \in I}}{1} & \defeq \btsel{l_i: \tmap{S_i}{1}}_{i \in I}
%	\\
%	\tmap{\btbra{l_i: S_i}_{i \in I}}{1} & \defeq \btbra{l_i: \tmap{S_i}{1}}_{i \in I}
%	\\
	\tmap{\chtype{S}}{1} & \defeq	\chtype{\tmap{S}{1}} 
		\qquad \quad 
		\tmap{\trec{t}{S}}{1}   \defeq \trec{t}{\tmap{S}{1}}
	\\
		\tmap{\tinact}{1}  & \defeq  \tinact
		\qquad \qquad \quad 
		\tmap{\vart{t}}{1}  \defeq \vart{t}
%    \\
%    \tmap{\trec{t}{S}}{1}  & \defeq \trec{t}{\tmap{S}{1}} \\
%    \tmap{\vart{t}}{1} & \defeq \vart{t} 
  \end{align*}
\end{mdframed}
%
%\vspace{-1mm}
\caption{\label{top:f:enc:hopi_to_ho}Typed encoding of \sessp into \HO, selection from \cite{DBLP:journals/iandc/KouzapasPY19}. 
Above,
$\fn{P}$ is a lexicographically ordered sequence  of free names in $P$.
\longversion{Maps $\vmap{\cdot}$ and 
$\auxmapp{\cdot}{{}}{\Psi}$ are in 
\defref{top:d:auxmap} and \Cref{top:f:auxmap}.}{
  Maps $\vmap{\cdot}$ and 
  $\auxmapp{\cdot}{{}}{\Psi}$ can be found in \cite{AAP21-full}. 
}
} 
%While $\map{\cdot}^{1}_{g}$ encodes processes,  \tmap{\cdot}{1} encodes session types.
% }
%\vspace{-1mm}
%(cf.~\defref{d:enc:fotohorec}).
%Mappings 
%$\map{\cdot}^2$,
%$\mapt{\cdot}^2$, 
%and 
%$\mapa{\cdot}^2$
%are homomorphisms for the other processes/types/labels. 
%\Hlinefig
%\vspace{-4mm}
\end{figure}

\paragraph{From \HO to \sessp.}
%\todo[]{Shouldn't this be "Back to \sessp from \HO"?}
The encoding $\map{\cdot}^2$ simulates higher-order communication using first-order constructs, following  Sangiorgi~\cite{SangiorgiD:expmpa}. The idea is to use \emph{trigger names}, which point towards copies of input-guarded server processes that should be activated. 
The encoding of abstraction sending distinguishes two cases: if the abstraction body does not contain any free session names (which are linear), then the server can be  replicated. 
Otherwise, if the value contains session names then its corresponding server name must be used exactly once. 
The encoding of abstraction receiving proceeds inductively, noticing that the variable is now a placeholder for a first-order name. 
The encoding of application is also in two cases; both of them depend on the creation of a fresh session, which is used to pass around the applied name. 
%The encoding on types $\tmap{\cdot}{2}$  maps a higher-order value type  to a shared, first-order input value type. 

\begin{figure}[!t]
\begin{mdframed}
{\bf Terms:} 
\begin{align*}
	 \map{\bout{u}{\abs{x}{Q}} P}^{2}  & \defeq  
	\begin{cases}
		\newsp{a}{\bout{u}{a} (\map{P}^{2} \Par \repl{\binp{a}{y} \binp{y}{x} \map{Q}^{2}})}
		& \text{if $\fs{Q} = \emptyset$}
		\\ 
		& \text{~with  $*P = \recp{X}{(P \Par \rvar{X})}$}
		\\
		\newsp{a}{\bout{u}{a} (\map{P}^{2} \Par \binp{a}{y} \binp{y}{x} \map{Q}^{2})}
		& \text{otherwise} %\dk{Q \textrm{ linear}} \\
	\end{cases}
\\
	 \map{\binp{u}{x} P}^{2} & \defeq  \binp{u}{x} \map{P}^{2}
	%\quad \quad \pmap{\appl{(\abs{x}{P})}{u}}{2} \!\! \defeq \!\! \pmap{P\subst{u}{x}}{2}
	 \\
	 \map{\appl{x}{u}}^{2} & \defeq \newsp{s}{\bout{x}{s} \bout{\dual{s}}{u} \inact}
	\\
	 \map{\appl{(\abs{x}{P})}{u}}^{2}  & \defeq  %\newsp{s}{\bout{a}{s} \bout{\dual{s}}{u} \inact} \\
	\newsp{s}{\binp{s}{x} \map{P}^{2} \Par \bout{\dual{s}}{u} \inact}
 % \\
%  \qquad \quad \text{where: $*P = \recp{X}{(P \Par \rvar{X})}$}
\end{align*}
{\bf Types:}
\begin{align*}
		\tmap{\btout{\lhot{S}}S_1}{2} & \defeq \bbtout{\chtype{\btinp{\tmap{S}{2}}\tinact}}\tmap{S_1}{2}
		\\
		\tmap{\btinp{\lhot{S}}S_1}{2} & \defeq \bbtinp{\chtype{\btinp{\tmap{S}{2}}\tinact}}\tmap{S_1}{2}
%\noindent{\bf Labels:}\ 
%		\mapa{(\nu \tilde{m})\bactout{n}{\abs{ x}{P}} }^2  & \defeq & \news{m} \bactout{n}{m} \\
%		\mapa{\bactinp{n}{\abs{ x}{P}} }^2 &  \defeq & \bactinp{n}{m}
%\quad \quad m \text{ fresh}
%\\[1mm]
%\hline
\end{align*}
\end{mdframed}

\caption{Typed encoding of \HO into \sessp~\cite{DBLP:journals/iandc/KouzapasPY19}. 
% {Here, $*P$ stands for $$}
%While $\map{\cdot}^2$ translates processes, \tmap{\cdot}{2} encodes types. 
\label{top:f:enc:ho_to_sessp}
}
%\vspace{-1mm}
%\Hlinefig
\vspace{-4mm}
\end{figure}

% \thesisalt{
%   \subsection{Labelled Transition System for Processes}
%   \label{top:ss:lts}
%   % \subimport{sources/}{untyped-lts.tex}
%   \input{untyped-lts.tex}
% }{}

\subsection{Minimal Session Types and the Minimality Result for \HO}
As mentioned above, MSTs are session types without sequencing: in session types such as 
$\btout{U} S$ and $\btinp{U} S$, the type $S$ can only correspond to $\tinact$.
The syntax of \msts for \HO  is in \Cref{top:f:sts} (bottom).
Recall that we write \mHO to denote \HO with \msts.
	The decomposition $\D{\cdot}$ in~\cite{APV19} relies crucially on polyadic communication.
Hence, following~\cite{DBLP:conf/esop/KouzapasPY16,DBLP:journals/iandc/KouzapasPY19}, value types are of the form $\shot{\widetilde{C}}$ and $\lhot{\widetilde{C}}$. Similarly, minimal session types for output and input are of the form $\btout{\widetilde{U}} \tinact$ and $\btinp{\widetilde{U}}\tinact$: they communicate tuples of values but lack a continuation.

\paragraph{Trios Processes.}
 Following Parrow~\cite{DBLP:conf/birthday/Parrow00}, we define the decomposition $\D{\cdot}$  on \HO processes in terms of a \emph{breakdown function} $\B{k}{\tilde x}{\cdot}$, which translates a process into a composition of \emph{trios processes} (or simply \emph{trios}).
A {trio} is a process with exactly three nested prefixes.
If $P$ is a sequential process with $k$ nested actions, then $\D{P}$ will contain $k$  trios running in parallel: each trio in $\D{P}$ will enact exactly one prefix from $P$; the breakdown function must be carefully designed to ensure that trios trigger each other in such a way that $\D{P}$ preserves the prefix sequentiality in $P$.
While trios decompositions elegantly induce processes typable with \msts, they are not a goal in themselves; rather, they offer one possible path to better understand sequentiality in session types.

We use  some useful terminology from~\cite{DBLP:conf/birthday/Parrow00}.
The \textit{context} of a trio is a tuple of variables $\widetilde x$, possibly empty, which  makes variable bindings explicit.
We use a reserved set of \textit{propagator names} (or simply \emph{propagators}), denoted by $\prop_k, \prop_{k+1}, \ldots$, to carry contexts and trigger the
next trio.
Propagators with recursive types are denoted by $\prop_k^r, \prop_{k+1}^r, \ldots$.
We say that  a \emph{leading trio} is the one that receives a context, performs an action, and triggers the next one;
 a \emph{control trio} only activates other trios.

\paragraph{The Breakdown Function: Preliminaries}
The breakdown function works on both processes and values.
The breakdown of process $P$ is denoted by $\B{k}{\tilde x}{P}$, where $k$ is the
index for the propagator $\prop_k$, and $\widetilde x$ is the context to be received by the previous trio.
Similarly,
the breakdown  of a value $V$ is denoted by
$\V{k}{\tilde x}{V}$. 

\Cref{mst:t:bdowncore-ho} gives   $\B{k}{\tilde x}{\cdot}$ and $\V{k}{\tilde x}{\cdot}$ as defined in \cite{APV19}.
\fixed{As we explain below, the decompositions exploit type information. Hence, in the table we include side conditions derived from typing; notice that for session types we have either $C = S$ or $C = \chtype{S}$.}

\fixed{Before commenting on the entries in \Cref{mst:t:bdowncore-ho}
and formally defining $\D{\cdot}$, 
 let us introduce some auxiliary notations and notions.
Let $\eta_1, \eta_2$, and $\eta_3$ denote some mathematical objects (say, a name or a finite sequence of names). We shall use the one-line conditional
$$
\eta_1 = \linecondit{c}{\eta_2}{\eta_3}
$$ %\line
to express that the equality $\eta_1 = \eta_2$ holds if the Boolean condition $c$ is true, and that the equality $\eta_1 = \eta_3$ holds otherwise.}

Let $\widetilde u = (a,b,s,s',\ldots)$ be a finite tuple of names.
We shall write $\mathsf{init}(\widetilde u)$ to denote the tuple
$(a_1,b_1,s_1,s'_1,\ldots)$.
We say that a process has been \emph{initialized} if it only involves \emph{indexed names} (i.e., all of its names have some index).
\begin{table}[!t]
 
% \begin{table}[!t]
    \begin{tabular}{ |l|l|l|}
      \rowcolor{gray!25}
      \hline
      $P$ &
        \multicolumn{2}{l|}{
      \begin{tabular}{l}
        \noalign{\smallskip}
        $\B{k}{\tilde x}{P}$
        \smallskip
      \end{tabular}
    }  \\
      \hline
    
    $\bout{u_i}{V}{Q}$ &
        \begin{tabular}{l}
          \noalign{\smallskip}
          $\bullet~ \neg\traux{S}$: %& 
          \\
          \quad $\propinp{k}{\wtd x}
                \bbout{u_i}{\V{k+1}{\tilde y}{V\sigma}}
                \apropout{k+1}{\wtd w}  \Par \B{k+1}{\tilde w}{Q\sigma}$ 
          \smallskip 
          \\
          \hdashline 
          \noalign{\smallskip}
          $\bullet~ \traux{S}$:  \\
          \quad $\propinp{k}{\wtd x}
                  \abbout{\prop^u}{N_V}   \Par 
            \B{k+1}{\tilde w}{Q}$
            \\
            \quad where:
            \\
            \qquad $N_V = \abs{\wtd z}
                  {\bbout{z_{\indT{S}}}{\V{k+1}{\tilde y}{V}}}{} \big(\apropout{k+1}{\wtd w} \Par 
              \recprov{u}{x}{\wtd z}\big) 
          $
         \smallskip
      \end{tabular}
      &
      %side-conditions
      \begin{tabular}{l}
        \noalign{\smallskip}
        $u_i:S$ \\
        $\widetilde y = \fv{V}$\\
        $\widetilde w = \fv{Q}$ \\
        % $\degree = \len{V}$ \\
        $\sigma = \nextn{u_i}$ \\
        $\wtd z = (z_1,\ldots,z_{\len{\Rts{}{s}{S}}})$
        \smallskip
      \end{tabular}
      \\
    \hline 
    $\binp{u_i}{y}Q$
    &
    
      \begin{tabular}{l}
          \noalign{\smallskip}
          $\bullet~ \traux{S}$: 
          \\
          \quad $\propinp{k}{\wtd x}\binp{u_i}{y} \apropout{k+1}{\wtd w}
         \Par \B{k+1}{\tilde w}{Q\sigma}$ %& (if $\neg\traux{S}$)
         \smallskip 
          \\
          \hdashline 
          \noalign{\smallskip} 
             $\bullet~ \neg\traux{S}$: 
          \\ 
         \quad $\propinp{k}{\wtd x}\abbout{\prop^u}
         {N_y} 
         \Par \B{k+1}{\tilde w}{Q}$ %& (if $\traux{S}$)
         \\
         \quad where:
         \\
         \qquad $N_y = \abs{\wtd z}{\binp{z_{\indT{S}}}{y} \big(\apropout{k+1}{\wtd w} \Par 
         \recprov{u}{x}{\wtd z}} \big)$
          \smallskip
      \end{tabular}
      &
      %side-conditions
      \begin{tabular}{l}
        \noalign{\smallskip}
       $u_i:S$ \\
        $\wtd w = \fv{Q} $ \\
      $\sigma = \nextn{u_i}$ \\
      $\wtd z = (z_1,\ldots,z_{\len{\Rts{}{s}{S}}})$
        \smallskip
      \end{tabular}
        \\
      \hline
    
        $\appl{V}{(\wtd r, u_i)}$ &
      \begin{tabular}{l}
        \noalign{\smallskip}
        $\propinp{k}{\widetilde x}\overbracket{\prop^{r_1}!\big\langle 
          \lambda \widetilde z_1. \prop^{r_2}!\langle\lambda \widetilde z_2.\cdots. 
       \prop^{r_n}!\langle \lambda \widetilde z_n.}^{n = |\tilde r|} 
       Q \rangle \,\rangle \big\rangle$ \\
      where:
      \\
       \quad $Q = \appl{\V{k+1}{\tilde x}{V}}{(\widetilde z_1,\ldots,
       \widetilde z_n, \widetilde m)}$
        \smallskip
      \end{tabular}
      &
      %side-conditions
      \begin{tabular}{l}
          \noalign{\smallskip}
          $u_i:C$ \\
        $\forall r_i \in \widetilde r.(r_i: S_i \wedge \traux{S_i} \wedge$\\
        \qquad $\wtd{z_i} = (z^i_1,\ldots,z^i_{\len{\Rts{}{s}{S_i}}}))$\\
        $\wtd m = (u_i, \ldots, u_{i+\len{\Gt{C}}-1})$
        \smallskip
      \end{tabular}
      \\
      \hline
      $\news{s:C}{P'}$ &
      \begin{tabular}{l}
        \noalign{\smallskip}
         $\bullet \neg\traux{C}:$
         \\
         $\news{\widetilde{s}:\Gt{C}}{\,\B{k}{\tilde x}
                 {P'\sigma}}$ %& (if $\neg\tr(C)$)
           \smallskip 
           \\
           \hdashline
           \noalign{\smallskip}
           $\bullet ~\traux{C}:$
           \\
       $\news{\widetilde{s}:\Gt{C}}
             \news{c^s}
        \recprov{s}{x}{\wtd s}
        % \binp{\prop^s}{b}(\appl{b}{\wtd s}) 
        \Par \news{c^{\bar{s}}}
        \recprov{\bar s}{x}{\wtd {\dual s}}
          \Par \B{k}{\tilde x}{P'\sigma}$
        \smallskip
      \end{tabular}
      &
      %side-conditions
      \begin{tabular}{l}
        \noalign{\smallskip}
        $\wtd x = \fv{P'}$ \\
        $\wtd{s} = (s_1,\ldots,s_{\len{\Gt{C}}})$ \\
        $\wtd {\dual{s}} = (\dual{s_1},\ldots,\dual{s_{\len{\Gt{C}}}})$ \\
        $\sigma=\subst{s_1 \dual{s_1}}{s \dual{s}}$
        \smallskip
      \end{tabular}
      \\
      \hline
      $Q \Par R$ &
      \begin{tabular}{l}
        \noalign{\smallskip}
        $\propinp{k}{\wtd x} \propout{k+1}{\wtd y}
        \apropout{k+\degree+1}{\wtd w}   \Par \B{k+1}{\tilde y}{Q} \Par \B{k+\degree+1}{\tilde w}{R}$
        \smallskip
      \end{tabular}
      &
      %side-conditions
      \begin{tabular}{l}
        \noalign{\smallskip}
        $\wtd y  = \fv{Q}$ \\ 
        $\wtd w = \fv{R}$ \\
        $\degree = \plen{Q}$
        \smallskip
      \end{tabular}
          \\
      \hline
    
      $\inact$ &
    \begin{tabular}{l}
      \noalign{\smallskip}
      $\propinp{k}{}\inact$
     \smallskip
    \end{tabular}
    &
    \begin{tabular}{l}
        \noalign{\smallskip}
    %$\widetilde x = \es$ 
    \smallskip
    \end{tabular}
    %$\widetilde x = \epsilon$
    %side-conditions
      \\
      \hline

    \rowcolor{gray!25}
    $V$ &
      \multicolumn{2}{l|}{
    \begin{tabular}{l}
      \noalign{\smallskip}
      $\V{}{\tilde x}{V}$
      \smallskip
    \end{tabular} } 
    \\
    \hline
    $y$ &
    \begin{tabular}{l}
      \noalign{\smallskip}
      $y$
      \smallskip
    \end{tabular}
    &
    %side-conditions\\
    \begin{tabular}{l}
      \noalign{\smallskip}
    %  $\widetilde x = \fv{V}$ \\
    %  $\widetilde{y} = (y_1,\ldots,y_{\len{\Gt{C}}})$
      \smallskip
    \end{tabular}
    \\
    \hline
    
    % $\abs{(\widetilde y z):\slhotup{(\wtd S C)}}P$ 
    $\abs{(\wtd y z)}P$ 
    & 
    \begin{tabular}{l}
      \noalign{\smallskip}
    $\abs{(\widetilde{y^1},\ldots,\widetilde{y^n}, \widetilde z):
    \slhotup{(\widetilde{M})}}{N}$ 
    \\
    \smallskip
      where:
      \\
      $\widetilde{M} = (\Gt{S_1},\ldots,\Gt{S_n}, \Gt{C})$
      \\
             $N = \news{\widetilde \prop} \news{\widetilde \prop_r}
         \prod_{i \in \len{\widetilde y}}
         (\recprov{y_i}{x}{\wtd y^i})
         \Par \apropout{1}{\widetilde x}
     \Par \B{1}{\tilde x}{P \subst{z_1}{z}}$
    \smallskip 
    \end{tabular}
    &
    %side-conditions\\
    \begin{tabular}{l}
      % \noalign{\smallskip}
      \noalign{\smallskip}
      $\wtd y z : \wtd S C$ \\
      $\forall y_i \in \widetilde y.(y_i: S_i \wedge \traux{S_i} \wedge$\\
       \qquad $\widetilde{y^i} = (y^i_1,\ldots,y^i_{\len{\Gt{S_i}}}))$\\
       $\widetilde z = (z_1,\ldots,z_{\len{\Gt{C}}})$ \\
       $\wtd \prop = (\prop_1,\ldots,\prop_{\plen{P}})$ \\ 
    $\widetilde{\prop_r} = \bigcup_{r\in \tilde{y}}\prop^r$ 
      \smallskip
    \end{tabular}
    \\
    \hline
    \end{tabular}
    % \caption{The breakdown function for processes and values. \label{mst:t:bdowncore}}
    % \end{table}
  \caption{The breakdown function for \HO processes and values~\cite{APV19,DBLP:journals/corr/abs-2301-05301}. \label{mst:t:bdowncore-ho}}
  \end{table}

  \begin{definition}[Predicates on Types (and Names)]
  \label{d:predtypes}
    Given a session type $C$, we write $\traux{C}$  to indicate that $C$ is a
    tail-recursive session type, and $\lin{C}$ otherwise.
    %  that is if $C \equiv \trec{t}S$.
%Also, given $u:C$, we write $\lin{u}$ if $C = S$ and  $\neg \tr(S)$.
Given $u:C$, we write $\lin{u}$ if $C = S$ and  $\neg \traux{S}$
and $\traux{u}$ otherwise.
  \end{definition}
  
    \begin{definition}[Subsequent index substitution]
    \label{pi:d:nextn}
    Let $n_i$ be an indexed name. We define $\nextn{n_i} = \linecondit{\lin{n_i}}{\incrname{n}{i}}{\{\}}$.
  \end{definition}

\begin{figure}[!t]
\begin{mdframed}
\vspace{-4mm}
	\begin{align*}
  \Gt{\btout{U}{S}} &=
  \begin{cases}
  \btoutt{\Gt{U}}   &  \text{if $S = \tinact$} \\
  \btoutt{\Gt{U}}\, ,\Gt{S} & \text{otherwise}  
\end{cases} 
 \\
 \Gt{\btinp{U}{S}} &=
  \begin{cases}
  \btinpt{\Gt{U}}  &  \text{if $S = \tinact$} \\
  \btinpt{\Gt{U}} \, , \Gt{S} & \text{otherwise}
  \end{cases}   
  \\
  \Gt{\trec{t}{S}} & = 
  \begin{cases} 	
    \Rt{S} & \text{if $\traux{\trec{t}{S}}$} \\ 
    \trec{t}{\Gt{S}} & \text{if $\neg\traux{\trec{t}{S}}$ and $\Gt{S}$ is a singleton}
 \end{cases} 
 \\
  \Gt{\tinact} &= \tinact 
  \\
   \Gt{\vart{t}} & = \vart{t} 
  \\
  \Gt{\lhot{C}} &= \lhot{\Gt{C}} 
  \\
  \Gt{\shot{C}} &= \shot{\Gt{C}}
  \\
  % \Gt{S_1,\ldots,S_n} &=
  % \Gt{S_1},\ldots,\Gt{S_n} 
  % \\
    \Gt{\chtype{U}} &= \chtype{\Gt{U}}
   \\
   \Rt{\vart{t}} & = \epsilon
  \\
    \Rt{\btout{U}S} &= \trec{t}{\btout{\Gt{U}}} \tvar{t}, \Rt{S}
   \\
   \Rt{\btinp{U}S} & = \trec{t}{\btinp{\Gt{U}}} \tvar{t}, \Rt{S} 
  % & 
  % \Rt{\btout{U}S} = \trec{t}{\btout{\Gt{U}}} \tvar{t}, \Rt{S}
  % \Rt{\btinp{U}S} = \trec{t}{\btinp{\Gt{U}}} \tvar{t}, \Rt{S} 
  % \\ 
  % \Rts{}{s}{\btinp{U}S} &= \Rts{}{s}{S}
  % \\
  % \Rts{}{s}{\btout{U}S} &= \Rts{}{s}{S}
\end{align*}
\end{mdframed}	
\caption[Decomposing session types into minimal session types.]{Decomposing session types into minimal session types (\Cref{mst:def:typesdecomp})   \label{mst:fig:typesdecomp}}
\end{figure}

In order to determine the required number of propagators ($\prop_k, \prop_{k+1}, \ldots$) required in the breakdown of processes and  values, we  define the \emph{degree} of a process: 
\begin{definition}[Degree of a Process]
  \label{mst:def:sizeproc}
	Let $P$ be an \HO process.
	The \emph{degree} of $P$, denoted $\plen{P}$, is defined as
	follows:
	$$ %\plen
	\plen{P} =
	\begin{cases}
    \plen{Q} + 1 & \text{if $P = \bout{u_i}{V}Q$ or $P=\binp{u_i}{y}Q$}
    \\
	% \len{V} + \len{Q} + 1 & \text{if $P = \bout{u_i}{V}Q$}
	% \\
	% \len{Q} + 1 & \text{if $P =\bout{u_i}{y}Q$ or $P=\binp{u_i}{y}Q$}
	% \\
	% 0 & \text{if $P = \appl{V}{u_i}$}
	% \\
	\plen{P'} & \text{if $P = \news{s:S}P'$}
	\\
	\plen{Q} + \plen{R} + 1 & \text{if $P = Q \Par R$}
	\\
	1 & \text{if $P = \appl{V}{u_i}$ or $P = \inact$}
	\end{cases}
	$$

\end{definition}

\fixed{\paragraph{The Breakdown Function: Intutions}
The breakdown function relies on type information: names are decomposed based on their session types; also, the shape of decomposed process depends on whether the associated session type is tail-recursive or not.
Moreover, we shall rely on decompositions/functions on types, denoted $\Gt{\cdot}$, $\Rt{\cdot}$, and $\Rts{}{s}{\cdot}$, which will be defined later on, in
  \Cref{mst:def:typesdecomp}.
Next, we describe the most interesting cases of \Cref{mst:t:bdowncore-ho}; the reader is referred to~\cite{APV19,DBLP:journals/corr/abs-2301-05301} for details and examples. }

\paragraph{Output:}
The decomposition of $\bout{u_i}{V}Q$ requires decomposing both the sent value $V$ and the continuation $Q$. 
We distinguish two cases, depending on the type $S$ of $u_i$:
\begin{itemize}
  \item If $\neg\tr(S)$ then  $u_i$ is linear or shared, and so the decomposition consists of a leading trio that
mimics an output action in parallel with the breakdown of  $Q$.
The context  $\wtd x$ must include the free variables of $V$ and $Q$, which are 
denoted $\wtd y$ and $\wtd w$, respectively. These tuples are not
necessarily disjoint: variables with shared types can appear free in both $V$
and $Q$. The value $V$ is then broken down with parameters $\wtd
y$ and $k+1$; the latter serves to consistently generate propagators for the
trios in the breakdown of $V$, denoted $\V{}{\tilde y}{V\sigma}$. 
The substitution $\sigma$  increments the index of session
names; it is applied to both $V$ and $Q$ before they are broken down. 
By taking $\sigma = \nextn{u_i}$ we 
distinguish two cases: 
\begin{itemize}
\item If name $u_i$ is linear (i.e., it has a session type) then its future
occurrences are renamed into $u_{i+1}$, and $\sigma = \subst{u_{i+1}}{u_i}$; 
\item Otherwise, if $u_i$ is {shared}, then $\sigma = \{\}$.
\end{itemize}
Note that if $u_i$ is linear then it appears either in $V$ or $Q$  and $\sigma$
affects only one of them. The last prefix   activates the
breakdown of  $Q$ with its corresponding context $\wtd w$.

\smallskip

In case $V  = y$, the same strategy applies; 
notice that variable $y$ is not propagated further if it does not appear free in $Q$. 

\item 
If $\tr(S)$ then $u_i$ is tail-recursive, and so the decomposition consists of a leading trio that mimics the output action
running in parallel with the breakdown of  $Q$. After receiving the context
$\widetilde x$, the leading trio sends an abstraction $N_V$ along $\prop^u$, which 
performs several tasks. First, $N_V$ collects the sequence of names $\tilde u$;
then, it mimics the output action of $P$ along one of such names (denoted $u_{\indT{S}}$, see below) and
triggers the next trio, with context $\wtd w$; finally, it reinstates the
server on $\prop^u$ for the next trio that uses~$u$.  
Indexing is not relevant in this case. 
\end{itemize}

\paragraph{Input:}
To decompose a process $\binp{u_i}{y}Q$ we distinguish two cases, as before: 
\rom{1}  name $u_i$ is linear or shared or \rom{2} tail-recursive.
In case~\rom{1}, the breakdown consists of a leading trio that mimics the input action and possibly extends the context with the
received variable $y$. In case~\rom{2}, when $u_i$ has tail-recursive session type $S$, the decomposition is as in the output case.
%In this case we need to receive $\tilde u$ using one of the received names ($u_{f(S)}$) as a subject for the input action.

\paragraph{Application:}
For simplicity we consider the breakdown of applications of the form $\appl{V}{(\wtd r, u_i)}$, where
every $r_i \in \wtd r$ is such that $\tr(r_i)$ and only  $u_i$ is such that $\neg\tr(u_i)$.
The general case involves different orders in names and multiple names with non-recursive types, and is defined similarly.      
%\begin{align*}
% \B{k}{\tilde x}{\appl{V}{(\widetilde r, u_i)}} = &
% \propinp{k}{\widetilde x}\overbracket{\prop^{r_1}!\big\langle 
%        \lambda \widetilde z_1. \prop^{r_2}!\langle\lambda 
%        \widetilde z_2.\cdots. 
%	 	\prop^{r_n}!\langle \lambda \widetilde z_n.}^{n = |\tilde r|} 
%	 	\appl{\V{}{\tilde x}{V}}{(\widetilde z_1,\ldots,
%	 	\widetilde z_n, \widetilde m)} \rangle \,\rangle \big\rangle
%\end{align*}

Before commenting on process $ \B{k}{\tilde x}{\appl{V}{(\widetilde r, u_i)}}$, 
let us discuss how names in $(\wtd r, u_i)$ are decomposed using types.
Letting $|\tilde
r| = n$ and $i \in \{1,\ldots,n\}$, for each $r_i \in \widetilde r$ (with
$r_i:S_i$) we generate a sequence $\wtd
z_i=(z^i_1,\ldots,z^i_{\len{\Rts{}{s}{S_i}}})$ as in the output case. 
We decompose name $u_i$ (with $u_i:C$) as $\wtd m = (u_i,\ldots,u_{i+\len{\Gt{C}}-1})$. 

The decomposition first receives a context $\wtd x$ for value $V$: we break down $V$
   with $\wtd x$ as a context since these variables need to be propagated to the
   abstracted process.
Subsequently, an output on $\prop^{r_1}$ sends a value containing $n$ abstractions that occur nested
within output prefixes---as explained in~\cite{DBLP:journals/corr/abs-2301-05301}, this is similar to a partial instantiation mechanism.
For each $j \in \{1,\ldots,n-1\}$, each abstraction
binds $\widetilde z_j$ and sends the next abstraction along $\prop^{r_{j+1}}$.
The innermost abstraction abstracts over $\widetilde z_n$ and encloses
the process $\appl{\V{}{\tilde x}{V}}{(\widetilde z_1,\ldots, \widetilde z_n,
\widetilde m)}$, which effectively mimics the  application. 
This abstraction nesting binds all variables $\widetilde z_i$, the 
decompositions of all tail-recursive names ($\wtd r$). 
%This structure can be seen as an encoding of partial application: by virtue of a
%single synchronization on $\prop^{r_i}$ part of variables (i.e., $\widetilde
%z_i$) will be instantiated.   

\paragraph{Composition:}
 The breakdown of a process $Q \Par R$ consists of a control trio that triggers the breakdowns of $Q$ and $R$; it does not mimic any
 action of the source process. 
 In the decomposed process $\B{k}{\tilde x}{Q \Par R}$, the tuple $\wtd y \subseteq \wtd x$ (resp. $\wtd
 w \subseteq \wtd x$) collects the free variables in  $Q$ (resp. $R$). To avoid
 name conflicts, the trigger for the breakdown of  $R$ is $\dual
 {c_{k+\degree+1}}$, with $\degree = \plen{Q}$ (cf. \Cref{mst:def:sizeproc}).

\paragraph{Restriction:}
To decompose $\news{s:C}{P'}$ we examine $C$;
the interesting case is when $\tr(C)$.
In that case, 
    we decompose $s$ into $\wtd{s} = (s_1,\ldots,s_{\len{\Gt{S}}})$ and
    $\dual{s}$ into $\widetilde{\dual{s}} =
    (\dual{s_1},\ldots,\dual{s_{\len{\Gt{S}}}})$. Notice
    that as $\tr(C)$ we have  
    $C \equiv \trec{t}S$, therefore $\Gt{C} = \Rt{S}$.  
     The breakdown introduces two servers in parallel with the breakdown of
      $P'$; they provide names for $s$ and $\dual{s}$ along $\prop^s$
      and $\prop^{\dual{s}}$, respectively. The server on $\prop^s$ (resp.
      $\prop^{\dual{s}}$) receives a value   
      and applies it to the sequence $\widetilde s$ (resp.
      $\widetilde{\dual{s}}$). We restrict over $\widetilde s$ and propagators
      $\prop^s$ and $\prop^{\dual{s}}$. 

\paragraph{Inaction:}
To breakdown $\inact$, we define a so-called \emph{degenerate} trio with only one input prefix
that receives a context that by construction will always be empty (i.e., $\widetilde x = \epsilon$).

\paragraph{Value:}
Let us discuss the particular case of values of the form 
$\abs{(\widetilde y, z):\slhotup{(\widetilde S, C)}}P$, 
where $\tr(y_i)$ holds for every $y_i \in \widetilde y$ and $\neg \tr(z)$, and 
$\leadsto \in \{\lollipop, \sharedop\}$.
The general case is defined similarly. 
  
In the process for the decomposed value, denoted
$\V{}{\tilde x}{\abs{(\widetilde y, z):\slhotup{(\widetilde S, C)}}P}$, 
 every $y_i$ (with $y_i : S_i$) is
decomposed into $\widetilde y^i=(y_1,\ldots,y_{\len{\Gt{S_i}}})$. We use
$C$ to decompose $z$ into $\widetilde{z}$. 
We abstract over $\wtd
y^1,\ldots,\wtd y^n, \wtd z$; 
%  $\widetilde{z}$; 
 the body of the abstraction (i.e. $N$) is the composition of recursive names
 propagators, the control trio, and the breakdown of $P$, with name index
 initialized with the substitution $\subst{z_1}{z}$.  
 For every $y_i \in \widetilde y$ there is a server
$\recprov{y_i}{x}{\wtd y^i}$
as a subprocess in the
abstracted composition---the rationale for these servers is as in previous cases.
  We restrict the
 propagators 
 $\wtd \prop = (\prop_1,\ldots,\prop_{\plen{P}})$: this
 enables us to type the value in a shared environment when $\leadsto =
 \rightarrow$. 
% Variable $z$ is decomposed as in \Cref{mst:t:bdowncore}.
% The breakdown is similar to the (monadic) shared value given in \Cref{mst:t:bdowncore}. 
 Also, we restrict special propagator names $\wtd \prop_r
= \bigcup_{r\in \tilde{v}}\prop^r$.

\paragraph{Decomposing Types}
We may now formally introduce the decompositions on types (and some associated notions), which have been informally used above.

\begin{definition}[Decomposing Session Types] 
  \label{mst:def:typesdecomp}
  Given the session, higher-order, and shared types of \Cref{top:f:sts}, the \emph{type decomposition function}  $\Gt{\cdot}$ is defined using the auxiliary function $\Rt{\cdot}$ as in \Cref{mst:fig:typesdecomp}. 
We write $\len{\Gt{S}}$ to denote the length of  $\Gt{S}$ (and similarly for $\Rt{\cdot}$).
\end{definition}

\fixed{Notice that $\Gt{\cdot}$ is a partial function, which operates on a sub-class of non-tail-recursive session types: in \Cref{mst:fig:typesdecomp}, the condition `$\Gt{S}$ is a singleton' allows us to decompose  types such as `$
  \trec{t}{\btinp{\shot{(\btinp{\mathsf{str}}\btout{\mathsf{str}}\tinact,
  \vart{t})}}\tinact}$', which is non-tail-recursive and features sequencing as an argument but not at the top-level.}
  
  To handle the unfolding of recursive types, we shall use the following auxiliary function, which decomposes guarded recursive types, by first ignoring all the actions until the recursion.
\begin{definition}[Decomposing an Unfolded Recursive Type]
Let $S$ be a session type. The function $\Rts{}{s}{\cdot}$: is defined as follows:
  \begin{align*}
    \Rts{}{s}{\trec{t}{S}} & = \Rt{S}
    \\
    \Rts{}{s}{\btinp{U}S} & = \Rts{}{s}{S} 
    \\
    \Rts{}{s}{\btout{U}S} & = \Rts{}{s}{S}
  \end{align*} 
\end{definition}
 
\begin{newaddenv}
  Given an unfolded recursive session type $S$, the auxiliary function $\indT{S}$ returns the position of the top-most prefix of $S$ within its body.
  \begin{definition}[Index function]
  \label{mst:def:indexfunction}
  Let $S$ be an (unfolded) recursive session type. The function $\indT{S}$ is defined as follows:  
  \begin{align*}
    \indT{S} = \begin{cases}
     \indTaux{S'\subst{S}{\vart{t}}}{0} & \text{if} \ S =\trec{t}{S'} \\
     \indTaux{S}{0} & \text{otherwise}
   \end{cases}
   \qquad
   \indTaux{T}{l} = \begin{cases}
   	\len{\Rt{S}} - l + 1 & \text{if $T=\trec{t}{S}$}
   	\\
   		\indTaux{S}{l+1} & \text{if $T=\btout{U}S$ or $T=\btinp{U}S$}
   \end{cases}
  \end{align*}
%  \noindent where $ \indTaux{S}{l}$:
%  \begin{align*}
%  \indTaux{\trec{t}{S}}{l} & = \len{\Rt{S}} - l + 1
%    \\
% \indTaux{\btout{U}S}{l}  & = \indTaux{S}{l+1}
%  \\
%   \indTaux{\btinp{U}S}{l} & = \indTaux{S}{l+1}
%  \end{align*}
  \end{definition}
\end{newaddenv}

\begin{example}
	\fixed{Let  $S = \trec{t}\btinp{\tint}\btinp{\bool}\btout{\bool}\vart{t}$
      and 
      $T = \btinp{\bool}\btout{\bool}S$.
      Then $\indT{T} = 2$ since the top-most prefix of $T$ (`$\btinp{\bool}$') is 
      the second prefix in the body of $S$.
  }
\end{example}

\paragraph{The Process Decomposition and the Minimality Result.}
Having introduced the breakdown function, we can now define the decomposition of \HO processes:

\begin{definition}[Decomposing Processes~\cite{APV19,DBLP:journals/corr/abs-2301-05301}]
  \label{mst:def:hodecomp}
  \label{mst:def:decomp}
	Let $P$ be a closed \HO process such that $\widetilde u = \fn{P}$.
  The \emph{decomposition} of $P$, denoted $\D{P}$, is
defined as:
  $$
  \D{P} = \news{\widetilde \prop}\big(\propout{k}{} \inact \Par \B{k}{\epsilon}{P\sigma}\big)
  $$
  where:
  $k > 0$;
  $\widetilde \prop = (\prop_k,\ldots,\prop_{k+\len{P}-1})$;
  $\sigma = \subst{\mathsf{init}(\tilde u)}{\widetilde u}$;
  and
  the  {breakdown function} $\B{k}{\tilde x}{\cdot}$, %where $\widetilde{x}$ is a tuple of variables, 
  is as defined  in Table~\ref{mst:t:bdowncore-ho}.
  \end{definition}

% \todo[inline]{We need to recall the decomposition of types from ECOOP (core fragment is enough)}

% \begin{definition}[Decomposing Session Types] 
%   \label{top:def:ho-typesdecomp}
% %  Let $S$ be a session type, $U$ be a higher-order type, $C$ be a name type, and
% %  $\chtype{U}$ be a shared type, all as in \Cref{f:sts}. 
%    The \emph{decomposition function} on the types of \Cref{top:f:sts}, denoted $\Gt{\cdot}$, is defined in 
%    \Cref{top:f:def:ho-typesdecomp}.
      
%    \begin{figure}
%    \begin{mdframed}[style=alttight]
%   \begin{align*}
%     \Gt{\btout{U}{S}} &=
%     \begin{cases}
%     \btout{\Gt{U}}{\tinact}  &  \text{if $S = \tinact$} \\
%     \btout{\Gt{U}}{\tinact}\, ,\Gt{S} & \text{otherwise}
%     \end{cases} \\
%     \Gt{\btinp{U}{S}} &=
%     \begin{cases}
%     \btinp{\Gt{U}}{\tinact}  &  \text{if $S = \tinact$} \\
%     \btinp{\Gt{U}}{\tinact}\, , \Gt{S} & \text{otherwise}
%     \end{cases} \\
% %      \Gt{\tinact} &= \tinact \\
%     \Gt{\lhot{C}} &= \lhot{\Gt{C}} \quad   \Gt{\tinact} = \tinact \\ 
%     \Gt{\shot{C}} &= \shot{\Gt{C}}  \quad \Gt{\chtype{U}} = \chtype{\Gt{U}}\\ 
%    % \Gt{\chtype{U}} &= \chtype{\Gt{U}}\\
%     \Gt{S_1,\ldots,S_n} &=
%     \Gt{S_1},\ldots,\Gt{S_n}
%   \end{align*}
%      \end{mdframed}	
%      \caption{Decomposition of types (cf. \defref{top:def:ho-typesdecomp})}
% \label{top:f:def:ho-typesdecomp}
%    \end{figure}
%   \end{definition}

As already mentioned, the \emph{minimality result}  in \cite{APV19} ensures that if $P$ is a well-typed \HO process then $\D{P}$ is a well-typed \mHO process. It attests that the sequentiality in the session types for $P$ is appropriately accommodated by the decomposition $\D{P}$. Its proof relies on an auxiliary result establishing the typability of $\B{k}{\tilde x}{P}$. 

In the following, given a session environment $\Delta = \Delta_1, \Delta_2$,   we shall write $\Delta_1 \circ \Delta_2$ to indicate the split of $\Delta$ into an environment $\Delta_1$ containing non-recursive names and a (disjoint) environment $\Delta_2$ containing recursive names.  
Also, we will use $\Theta, \Theta', \ldots$ to denote session environments induced by our decompositions.

\begin{restatable}[Typability of $\B{\text{-}}{\text{-}}{\cdot}$~\cite{APV19}]{lemm}{typerec}
  \label{mst:t:typecore}
  Let $P$ be an indexed \HO process and $V$ be a value.
  \begin{enumerate}
  \item
  If $\Gamma;\Lambda;\Delta \circ  \envR \proves P \hastype \Proc$ then
  $\Gt{\Gamma_1}\cat \envPropR;\es;\Gt{\Delta} \cat \Theta \proves \B{k}{\tilde x}{P} \hastype \Proc$, 
  where: 
  \begin{itemize}
  \item $k > 0$
  \item $\widetilde r = \dom{\envR}$
  \item $\envPropR = \prod_{r \in \tilde r} c^r:\chtype{\lhot{\Rts{}{s}{\envR(r )}}}$
  \item $\widetilde x = \fv{P}$ 
  \item $\Gamma_1=\Gamma \setminus \widetilde x$ 
  \item $\dom{\Theta} =
  \{\prop_k,\ldots,\prop_{k+\plen{P}-1}\} \cup
  \{\dual{\prop_{k+1}},\ldots,\dual{\prop_{k+\plen{P}-1}}\}$
  \item $\Theta(\prop_k)=
  \btinpt{U_1, \ldots, U_n} $, where $(\Gt{\Gamma}\cat\Gt{\Lambda})(\widetilde x) = (x_1 : U_1, \ldots, x_n : U_n)$
  \item $\balan{\Theta}$
  \end{itemize}

  \item If $\Gamma;\Lambda;\Delta \circ  \envR \proves V \hastype 
  \lhot{\wtd T}$ then
  $\Gt{\Gamma} \cat \envPropR; \Gt{\Lambda};\Gt{\Delta} \proves 
  \V{k}{\tilde x}{V} \hastype \lhot{\Gt{\wtd T}}$, 
  where: 
  \begin{itemize}
  	\item 
$\wtd x = \fv{V}$ 
\item 
  $\envPropR = \prod_{r \in \tilde r} c^r:\chtype{\lhot{\Rts{}{s}{\envR(r)}}}$
    \end{itemize}

\end{enumerate}
\end{restatable} 
% \end{newaddenv}

% \begin{theorem}[Minimality Result~\cite{APV19}]
% \label{t:decompcore}
% 	Let $P$ be a closed \HO process with $\widetilde u = \fn{P}$ and
% 	$\sigma = \subst{\mathsf{init}(\widetilde u)}{\widetilde u}$.
% 	If $\Gamma;\es;\Delta \proves P \hastype \Proc$ then
% 	$$\Gt{\Gamma \sigma};\es;\Gt{\Delta \sigma} \proves \D{P} \hastype \Proc$$
% \end{theorem}

\begin{restatable}[Minimality Result for \HO~\cite{APV19}]{theorem}{decompcore}
  \label{mst:t:decompcore}
	Let $P$ be a closed \HO process (i.e. $\fv{P} = \emptyset$) with $\widetilde u = \fn{P}$.
  % and $\widetilde v
	% = \rfn{P}$. 
	% \\
	If $\Gamma;\es;\Delta \circ  \envR \proves P \hastype \Proc$, 
	%where $\envR$ only involves recursive session types,  
	then $\Gt{\Gamma
	\sigma};\es;\Gt{\Delta\sigma}  \cat \Gt{\envR \sigma} \proves \D{P} \hastype
	\Proc$, where $\sigma = \subst{\mathsf{init}(\widetilde u)}{\widetilde u}$.
\end{restatable}

Having summarized the results on which our developments stand (in particular, the minimality result for \HO), we now move on to establish the minimality result for \sessp. 
%We shall follow \cite{APV19} in first defining a breakdown function and then defining a decomposition function for \sessp processes.

% \input{sources/preliminaries.tex}

\section{Decompose by Composing}
\label{pi:s:dbc}

In this section, our goal is to obtain a minimality result for $\sessp$ following the ``decompose by composing'' approach (cf. \Cref{pi:enc_drawing}). 
We start by defining $\msessp$, i.e., the language $\sessp$ equipped with \msts. 
Then, we define a decomposition function $\F{\cdot}: \sessp \to \msessp$,  given  in terms of a breakdown function  denoted $\AmeliaMod{k}{\tilde x}{\cdot}_g$ (cf.~\Cref{pi:table:repo}).
As anticipated, this breakdown function will result from the composition of  $\map{\cdot}^1_g$, $\B{k}{\tilde x}{\cdot}$, and $\map{\cdot}^2$, that is, 
$
\AmeliaMod{k}{\tilde x}{\cdot}_g = 
 \map{ \B{k}{\tilde x}{ \map{\cdot}^1_g } }^2
$.
Using $\F{\cdot}$, we shall obtain the desired minimality result for \sessp, which will be given by \Cref{pi:t:amtyprecdec}. 
% \jcheck{Reference?}

\begin{remark} 
\label{rem:typing}
\fixed{In the following we focus on $\sessp$, and so we sometimes use typing judgments of the form $\Gamma; \Delta \proves P \hastype \Proc$, i.e., the judgment introduced in \Cref{pi:ss:types} but without the environment $\Lambda$, which is not relevant for first-order processes.
Based on the rules in~\Cref{top:fig:typerulesmys}, we can derive typing rules for  \sessp with polyadic communication. 
Such rules, given in~\Cref{top:fig:typerulesmyspoly}, use the shortened typing judgments for $\sessp$.}
	\begin{figure}[t!]
		\begin{mdframed}
	\[
		\begin{array}{c}
			\inferrule[(PolyVar)]{}{\Gamma \cat \wtd x : \wtd M_x;\wtd y : \wtd M_y \proves 
			\wtd x \wtd y \hastype \wtd M_x \wtd M_y} \\[4mm]
	%		~~\inferrule[(Sh)]{}{\Gamma \cat u:C;\es \proves u \hastype C}
	%		\\[4mm]
	%		\inferrule[(RVar)]{}
	%		{\Gamma \cdot X : \Delta;\Delta \proves X \hastype \Proc}
	%		~~\inferrule[(Nil)]
	%		{}
	%		{\Gamma;\emptyset \proves \inact \hastype \Proc}
	%		~~\inferrule[(Rec)]
	%		{\Gamma \cdot X : \Delta;\Delta \proves P \hastype \Proc}
	%		{\Gamma;\Delta \proves \recp{X}P \hastype \Proc}
	%		\\[4mm]
			\inferrule[(PolySend)]{
					%	\begin{array}{c}
	%					u:S \in \Delta_1 
	%					\quad
				\Gamma; \Delta_1 \cat u:S \proves P \hastype \Proc
				\quad
				\Gamma; \Delta_2 \proves \wtd x \hastype \wtd C
				%\end{array}
			}{
				\Gamma; \Delta_1 \cat \Delta_2 \cat u:\btout{\wtd C} S \proves \bout{u}{\wtd x} P \hastype \Proc
			}
	%		\\[4mm]
			\qquad 
			\inferrule[(PolyRcv)]{
			\begin{array}{c}
				\Gamma; \Delta_1 \cat u: S \proves P \hastype \Proc
				\quad
				\Gamma; \Delta_2 \proves {\wtd x} \hastype {\wtd C}
				\end{array}
			}{
				\Gamma \backslash \wtd x; \Delta_1 \setminus \Delta_2 \cat u: \btinp{\wtd C} S \vdash \binp{u}{{\wtd x}} P \hastype \Proc
			}
			\\[4mm]
			\inferrule[(PolyResS)]
			{\Gamma;\Delta \cat \wtd s:\wtd S_1 \cat  \wtd{\dual{s}}:\wtd S_2 \proves P \hastype \Proc
			\quad S_1 \dualof S_2}
			{\Gamma;\Delta \proves \news{\wtd s}P \hastype \Proc}
			\qquad 
			\inferrule[(PolyRes)]
			{\Gamma;\Delta \cat \wtd a:\wtd C  \proves P \hastype \Proc}
			{\Gamma;\Delta \proves \news{\wtd a}P \hastype \Proc}
			~~ 
			\\[4mm]
	%		\inferrule[(Sel)]
	%		{\Gamma;\Delta \cat u:S_j \proves P \hastype \Proc 
	%		\qquad j \in I}
	%		{\Gamma;\Delta \cat u: \btsel{l_i:S_i}_{i \in I} \proves 
	%		\bsel{u}{l_j}P \hastype \Proc}
	%		\\[4mm]
	%		\inferrule[(Bra)]
	%		{ \forall i \in I \qquad \Gamma;\Delta \cat u:S_i \proves P_i \hastype \Proc}
	%		{\Gamma;\Delta \cat u: \btbra{l_i:S_i}_{i \in I} \proves 
	%		\bbra{u}{l_i:P_i}_{i \in I}}
	%		~~\inferrule[(Par)]
	%		{\Gamma;\Delta_i \proves P_i \hastype \Proc \qquad i = 1,2}
	%		{\Gamma;\Delta_1 \cat \Delta_2 \proves P_1 \Par P_2 \hastype \Proc}
	%		\\[4mm] 
			\inferrule[(Req)]{
				\begin{array}{c}
					\Gamma;  \es \proves u \hastype \chtype{\wtd S}
					\quad
					\Gamma; \Delta_1 \proves P \hastype \Proc
					\\
					\Gamma;  \Delta_2 \proves \wtd x \hastype \wtd S
				\end{array}
			}{
				\Gamma;  \Delta_1 \cat \Delta_2 \proves \bout{u}{\wtd x} P \hastype \Proc
			}
			\qquad 
			\inferrule[(Acc)]{
				\begin{array}{c}
					\Gamma; \emptyset \proves u \hastype \chtype{\wtd S}
					\quad
					\Gamma;  \Delta_1 \proves P \hastype \Proc
					\\
					\Gamma; \Delta_2 \proves \wtd x \hastype \wtd S\\
					   \end{array}
			}{
				\Gamma\backslash \wtd x; \Delta_1 \setminus \Delta_2 \proves \binp{u}{\wtd x} P \hastype \Proc
			}
			\end{array}
	\]
	%\vspace{-3mm}
		\end{mdframed}
	\caption{Derived typing rules for the polyadic variant of $\pi$ (cf. \Cref{rem:typing})
	%See~\cite{KouzapasPY17} for a full account.
	\label{top:fig:typerulesmyspoly}}
	%\Hline
	%\vspace{-1mm}
	\end{figure}
	\end{remark}
	
\subsection{Key Ideas}
\label{pi:ss:idea}
  Conceptually, the breakdown function $\AmeliaMod{k}{\tilde x}{\cdot}_g$ can be obtained in two steps: 
  \begin{enumerate}
  	\item  First, we use the composition $\B{k}{\tilde x}{ \map{\cdot}^1_g }$, which given a \sessp process returns a \mHO process. 
  	%Let us write $\Amelia{k}{\tilde x}{\cdot}_g$ to denote this composition.
  	\item  Second, we use $\map{\cdot}^2$ to transform the \mHO process obtained in (1) into a \msessp process. 

  \end{enumerate}
  We illustrate these two steps in detail, considering output, input, and recursive processes.

\subsubsection{Output and Input Processes}

\paragraph{Output.} 
 Given $P = \bout{u_i}{w_j}{Q}$, we first obtain:
$$
    \B{k}{\tilde x}{ \map{\bout{u_i}{w_j}{Q}}^1_g }  = 
%        \Amelia{k}{\tilde x}{\bout{u_i}{w_j}{Q}}_g = 
        \propinp{k}{\tilde x} \bbout{u_i}{W} \apropout{k+3}{\tilde x}  \Par
        \B{k+3}{\tilde x}{ \map{Q\sigma}^1_g} 
%        \Amelia{k+3}{\tilde x}{Q\sigma}_g 
        $$
        where 
        \begin{align*}
        \sigma & = \linecondit{u_i:S}{\incrname{u}{i}}{\{\} }
        \\
        W & = \abs{z_1}
        \big( \apropout{k+1}{} \Par \propinp{k+1}{} \binp{z_1}{x} 
        \apropout{k+2}{x} \Par \propinp{k+2}{x}\big(\appl{x}{\widetilde w}\big)\big)
                \end{align*}
    Then,  we transform 
    %$\Amelia{k}{\tilde x}{\bout{u_i}{w_j}{Q}}_g$ 
    $\B{k}{\tilde x}{ \map{\bout{u_i}{w_j}{Q}}^1_g } $
 into the following \msessp process, using $\map{\cdot}^2$:
    \begin{align*}
               \AmeliaMod{k}{\tilde x}{\bout{u_i}{w_j}{Q}}_g 
               = & 
~\map{ \B{k}{\tilde x}{ \map{\bout{u_i}{w_j}{Q}}^1_g } }^2 
\\ 
   =  &	~   \propinp{k}{\widetilde x} \news{a} \big(\bout{u_i}{a}
    \big( \apropout{k+3}{\widetilde x} \Par
          \AmeliaMod{k+3}{\tilde x}{Q \sigma}_g \Par
        \\
        & \qquad \qquad \qquad \binp{a}{y} \binp{y}{z_1} \apropout{k+1}{z_1} \Par
        \propinp{k+1}{z_1} \binp{z_1}{x} \apropout{k+2}{x} \Par
        \\
        & \qquad \qquad \qquad \qquad \propinp{k+2}{x} 
        \news{s} \big(\bout{x}{s} \about{ \dual{s} }{\widetilde{w}}\big) \big) \big)
      \end{align*}
We briefly describe the resulting process.
The subprocess mimicking the output action on $u_i$ is guarded by an input on $c_k$.
  Then, the output of $w_j$ on $u_i$ is mimicked via several steps: first, a private name $a$ is sent along $u_i$; 
    then, after several redirections involving local trios, the breakdown of $w_j$ 
    is sent along name $s$. 
    We can see that the output action on $u_i$ enables the forwarding of the context $\wtd x$ to the breakdown of $Q$, the continuation of the output. 

    Another form of output is when both $u_i$ and/or $w_j$  have recursive types. 
We shall refer to names with tail-recursive types as  
    \emph{recursive names}.
This output case with recursive names is similar to the one just discussed, and given in~\Cref{pi:table:repo}. 
%In essence, it has the same structure but with an additional mechanism to obtaied     an object name in the same way a subject name is obtained via $c^r$. 

    \paragraph{Input.} The breakdown of $\binp{u_i}{w}Q$ is 
    as follows:
          \begin{align*}
          \AmeliaMod{k}{\tilde x}{\binp{u_i}{w}Q}_g = & 
~\map{ \B{k}{\tilde x}{ \map{\binp{u_i}{w}Q}^1_g } }^2 
\\
           =  &~\propinp{k}{\widetilde x}\binp{u_i}{y} \apropout{k+1}{\widetilde x, y} \Par
            \\
            &\quad \news{s_1} \big(\propinp{k+1}{\widetilde x, y} \propout{k+2}{y} \apropout{k+3}{\widetilde x} \Par
            \\
            &\qquad   \qquad  \propinp{k+2}{y} \news{s}\big(\bout{y}{s} \about{\dual{s}}{s_1}   ) \Par
            \\
            &\qquad  \qquad \qquad \propinp{k+3}{\widetilde x} \news{a}  \big(\bout{\dual{s_1}}{a} 
            \big( \apropout{k+l+4}{} \Par \propinp{k+l+4}{} \inact \Par %\widehat{Q}_{\tilde{x}}             
            \\
            &\qquad \qquad \qquad \qquad \binp{a}{y'} \binp{y'}
        {\widetilde{w}} \big( \apropout{k+4}{\widetilde x} \Par
        \AmeliaMod{k+4}{\widetilde x}{Q \subst{w_1}{w} \sigma }_g \big) \big)\big)\big)
        \end{align*}
%  where 
%        $\widehat{Q}_{\tilde{x}} = \binp{a}{y'} \binp{y'}
%        {\widetilde{w}} \big( \apropout{k+4}{\widetilde x} \Par
%        \AmeliaMod{k+4}{\widetilde x}{Q \subst{w_1}{w} \sigma }_g \big)$.
        
 In this case, the  activation on $\prop_k$ enables        the input   on $u_i$. 
         After several redirections, 
           the actual input of variables $\wtd w$ is 
         on a name received for $y'$, which binds them in the breakdown of ${Q}$. 
         Hence, the context $\wtd x$ does not get extended for an inductive call: 
         it only gets extended locally (it is propagated by $\prop_{k+1}$). 
         Indeed, the context is always empty  and propagators only enable subsequent actions. 
         The context does play a role in breaking down input actions with recursive names:  in that case, the variables $z_X$ (generated in the encoding of  recursion, cf. \Cref{top:f:enc:hopi_to_ho}) get propagated as context. 

        %  \thesisalt{
\subsubsection{Recursion}
\begin{newaddenv}
  Now, we illustrate the resulting decomposition for
  processes involving tail-recursive names. 
  \paragraph{Output.} The breakdown of 
  $\bout{r}{w_j} Q$ when $\traux{r}$ is as follows: 
\begin{align*}
  \AmeliaMod{k}{\tilde x}{\bout{r}{w_j} Q} =&\propinp{k}{\widetilde x}
  \news{a_1}{}\proprout{r}{a_1}
  % \\  \qquad \qquad \qquad 
  \big(\AmeliaMod{k+3}{\tilde x}{P}_g
  \Par \binp{a_1}{y_1}\binp{y_1}{\widetilde z}W \big) 
  \\
  &\quad \text{where:} 
  \\
  &\qquad W =  \news{a_2}\big(\bout{z_{\indf{S}}}{a_2} 
  % \\ 
  % \qquad \qquad 
  \big(\propout{k+3}{\widetilde x}\proprinp{r}{b}
        \news{s} \big(\bout{b}{s}\about{\dual{s}}{\widetilde z} \big) \Par 
         \\ 
         &\qquad \qquad  \qquad 
         \binp{a_2}{y_2}\binp{y_2}{z'_1}
        % \news{\widetilde \prop}
        % \\ \qquad \qquad  \qquad 
        \big(\apropout{k+1}{} \Par 
        % \\ 
        \propinp{k+1}{}\binp{z'_1}{x}\apropout{k+2}{x} \Par 
        \\ 
        &\qquad \qquad  \qquad \qquad 
        \propinp{k+2}{x}\news{s'} \big(\bout{x}{s'}\about{\dual s'}{\widetilde w} \big) \big) \big) \big)
\end{align*}
This breakdown follows the essential ideas of 
the case  $\neg \traux{r}$, discussed above. 
The difference is the following: before mimicking the action on $r$, the process has to obtain 
the decomposition of name $r$ by a communication on $\prop^r$. 
Again, as a consequence of composing encodings, this communication is carried out via several channel redirections: first, 
the fresh name $a_1$ is sent along $\prop^r$; then, 
a name is received on $a_1$, along which the  breakdown of  $r$, denoted by 
$\wtd z$, is finally received.
Notice that here we obtain the entire decomposition of $r$. 
However, to properly mimic the original 
output action in $W$, we need one specific channel from $\wtd z$, which is appropriately selected 
% we pick appropriate index in 
based on its type $S$ by the index function 
$\indf{\cdot}$
given in~\Cref{pi:def:indexfunction-comp}---this is the first action of $W$, i.e., $z_{\indf{S}}$. 
Finally, the entire decomposition of  $r$ is again made 
available for future trios by communication on $\prop^r$. 
This way, we are able to send back 
 recursive names so they can be used in the next instance of a recursive body. 
%  all recursive names 
% used in a body of recursion 
% in the same way as for corresponding \HO process.

% Next, in subprocess $W$, 
% the original output action is mimicked on $z_{f(S)}$, 
% by $f(S)$ appropriate index of 
% channel $r$ is picked based on its tail-recursive type $S$: 
% in the same way as for corresponding \HO process. 
% Finally, 

\paragraph*{Recursive process.}
The process $\recp{X}P$ is broken down as given below. Recall that the mapping $ \vmap{\cdot}$
has been defined in \Cref{top:d:trabs}. 
\begin{align*}
  &\news{s_1} \big(  \propinp{k}{\widetilde x}\propout{k+1}{\wtd x}
      \apropout{k+3}{\wtd x} \Par
      \\
      &
      \qquad \quad \propinp{k+1}{\wtd x} \news{a_1}\big( \bout{\dual s_1}{a_1} \big( \apropout{k+2}{}
      \Par \apropinp{k+2}{} \Par
      \\ 
      &
      \qquad \qquad \qquad  \qquad  \qquad
      \propinp{k+3}{\widetilde x}\binp{s_1}{z_x}
      \apropout{k+4}{\wtd x, z_x} 
      \Par
      \\  
      & 
      \qquad \qquad \qquad  \qquad  \qquad
      \AmeliaMod{k+4}{\tilde x, z_x}{P}_{g, \{X \rightarrow {\tilde n}\}} \Par 
      \\
      &
     \qquad  \qquad \qquad \qquad \qquad  
      \repl\binp{a_1}{y'_1}
      \binp{y'_1}{\vmap{\widetilde n^1}, \ldots, \vmap{\widetilde n^m}, y_1}\widehat{P}\,\big)\big)\big)
      \\
      &
      \text{where:} 
      \\
      &
     \widehat{P} = \news{\widetilde \prop}\big( \prod_{0 < i \leq m}
        \proprinp{n_i}{b}\news{s'}\big(\bout{b}{s'}\about{\dual s'}
        {\vmap{\widetilde n^i}}\big)
        \Par 
        \\ 
        &
        \qquad \qquad \qquad  \qquad
        \apropout{k+2}{\widetilde x} \Par
        % \\ \quad \quad 
        \propinp{k+2}{\widetilde x}\binp{y_1}{z_x}\apropout{k+3}{\widetilde x, z_x} 
        \Par
        \auxmapp{ \AmeliaMod{k+3}{\tilde{x}, z_x}
        {P}_{g, \{X \rightarrow {\tilde n}\}}  }{{}}{{\tilde \prop, \tilde \prop_r}}\big)
\end{align*}
\noindent 
Moreover, above we assume that $\wtd n = \fn{P}$, $m = \len{\wtd n}$, $\vmap{\widetilde n} = (\vmap{n_1}, \dots, \vmap{n_m})$, $\vmap{n_i} : S_i$   
and $\vmap{\widetilde n^i} = (\vmap{n_1^i}, \dots, \vmap{n_{\len{\mGt{S_i}}}^i })$ 
for $i \in \{1,\ldots,m \}$. 

The breakdown of this process works in coordination with the breakdown of the recursive variable $\rvar{X}$ (described below). 
The main mechanism here is concerned with controlling 
\rom{1} the activation of 
new instances of recursive body (i.e., $P$) and
\rom{2} the propagation of recursive names to a subsequent instance 
(i.e., $\wtd n = \fn{P}$). 
The first instance is given by $\AmeliaMod{k+4}{\tilde x, z_x}{P}_{g, \{X \rightarrow {\tilde n}\}}$. Notice that by the definition of 
other cases, this instance will first collect all the 
recursive names by the communication with top-level providers on 
names $\prop^r$ for $r \in \wtd n$. 
The mechanism for generating subsequent instances is given by the replicated process
$\repl\binp{a_1}{y'_1}
\binp{y'_1}{\vmap{\widetilde n^1}, \ldots, \vmap{\widetilde n^m}, y_1}\widehat{P}$. 
Recall that replication is supported via the encoding $*P = \recp{X}(P \Par X)$.
Here, the intention is that this process again receives 
decompositions of recursive names in $\wtd n$ on shared name $a_1$ 
from the breakdown of $\rvar{X}$: to see this, 
notice that link to $a_1$ is propagated to 
the breakdown of $P$ via name $z_x$. 
% ... 
% This mechanism is responsible for 
% sending back recursive names to a new instance of recursive body 
% when this instance is activated. 
% ... 

\paragraph*{Recursive Variable.}
Following the previous description, the 
variable $\rvar{X}$ is broken down as follows: 
\begin{align*}
  &\news{s_1}\big( \propinp{k}{z_x}\propout{k+1}{z_x}\apropout{k+2}{z_x} 
    \Par
    \\ 
    &
    \quad \qquad \qquad 
    \propinp{k+2}{z_x}\bout{\dual s_1}{z_x}\apropout{k+3}{} 
    \Par \apropinp{k+3}{} \Par 
     \\ 
     &
     \qquad \qquad  \qquad \qquad 
     \propinp{k+1}{z_x}\news{a_1}
     \big( \proprout{n_1}{a_1}
    \big(
    \binp{a_1}{y_1}\binp{y_1}{\widetilde z_1} \ldots \news{a_j}Q\big)\big)
    % \Par \propinp{k+2}{z_x}\bout{\dual s_1}{z_x}\apropout{k+3}{} 
    % \Par \propinp{k+3}{}\inact 
    \big) 
    \\ 
    &
    \text{where:} 
    \\
    & \qquad 
    Q = \big(\proprout{n_j}{a_j}\big(\binp{a_j}{y_j}\binp{y_j}{\widetilde z_j}
    \news{s'}\big(\bout{z_x}{s'}\about{\dual s'}{\widetilde z_1, \ldots, \widetilde z_j, s_1}\big)\big)\big)
\end{align*}
\noindent 
and  $\wtd n = g(X)$, $\len{\widetilde n} = j$, 
$n_i : S$, and 
$\widetilde{z_i} = (z_1^i, \dots, z_{\len{\mRts{}{}{S_i}}}^i)$ 
for $i \in \{1, \dots, j\}$. 

As described above, the main role of this breakdown is to 
send back decomposition of all recursive names used in a recursive body (given by 
$\wtd n = g(X)$) to a next instance of a recursive body. 
This breakdown accomplishes this by \rom{1} first  
collecting recursive names by communication on 
$\prop^r$ for $r \in \wtd n$ with trios of a breakdown of the recursive body, 
and \rom{2} the propagating those names to a subsequent instance 
by communication on $z_x$. 
As explained in the previous case, 
this link is established at the entry of recursive process, 
and propagated throughout trios of its decomposition up to this process. 
% The link $z_x$ is propagated from previous trios. 
% ...
\end{newaddenv} 
\cralt{
  %\todo[inline]{TODO: Give key ideas for composition of recursion}
}{}
% }{} % thesisalt
\subsection{Formal Definitions}
 We start by defining \msts for \sessp:
\begin{definition}[Minimal Session Types, \msts]
  \label{pi:d:mtypesi}
  \emph{Minimal session types} for \sessp are defined in~\Cref{pi:f:mtypesi}.

  %		\begin{align*}
  %		C  & \bnfis		M  \bnfbar  {\chtype{M}}
  %\\
  %		M & \bnfis 	\tinact  \bnfbar  \btout{\widetilde{C}} \tinact \bnfbar \btinp{\widetilde{C}} \tinact
  %	\end{align*}
  \end{definition}

\begin{figure}[t]
\begin{mdframed}[style=alttight]
	    \begin{align*}
      C  & \bnfis		M  \bnfbar  {\chtype{M}}
  % \\
      % M  &\bnfis 	\tinact  \bnfbar  \btout{\widetilde{C}} \tinact \bnfbar \btinp{\widetilde{C}} \tinact
      \\ 
      \mugamma  &\bnfis \tinact \bnfbar \vart{t}  \\
    M  &\bnfis 	\mugamma  \bnfbar  \btout{\widetilde{C}} \mugamma \bnfbar \btinp{\widetilde{C}} \mugamma \bnfbar \trec{t}{M }		 
    \end{align*}
    \end{mdframed}
    \vspace{-3mm}
    \caption[Minimal Session Types for \sessp]{Minimal Session Types for \sessp (cf.~\Cref{pi:d:mtypesi})}\label{pi:f:mtypesi}
 \end{figure}
 
  % \input{ameliatable}
  % \begin{sidewaystable}[t!]
  %   % \caption{Decompose by composing: Breakdown function $\AmeliaMod{k}{\tilde x}{\cdot}_g$ for $\sessp$ processes 
  %   % (cf.~\defref{pi:def:am-decomp}). \label{pi:table:repo}}
  %   \caption{Simple caption}
  %   % \input{ameliatable1.tex}
  %   % \vspace{-4mm}
  % \end{sidewaystable}
  % \begin{adjustbox}{angle=90}
  \begin{table*}[!t]
   \thesisalt{  % \begin{table*}[!ht]
\begin{tabular}{ |l|l|l|}
  \rowcolor{gray!25}
  \hline
  $P$ &
    \multicolumn{2}{l|}{
  \begin{tabular}{l}
    \noalign{\smallskip}
    $\AmeliaMod{k}{\tilde x}{P}_g$
    \smallskip
  \end{tabular}
}  \\
  \hline
%%% OUTPUT
$\bout{u_i}{w_j} Q$ &
      \begin{tabular}{l}
      \noalign{\smallskip}
      $\propinp{k}{\widetilde x} \news{a} \big(\bout{u_i}{a}
       \big( \apropout{k+3}{\widetilde x} \Par
          \AmeliaMod{k+3}{\tilde x}{Q \sigma}_g \Par$
        \\
        \qquad 
        $\binp{a}{y} \binp{y}{z_1} \apropout{k+1}{z_1} \Par$
        \\
        \qquad \quad 
        $\propinp{k+1}{z_1} \binp{z_1}{x} \apropout{k+2}{x} \Par$
        \\
        $\qquad \qquad \qquad \qquad \propinp{k+2}{x} 
        \news{s} \big(\bout{x}{s} \about{ \dual{s} }{\widetilde{w}} \big) \big) \big)$
     \smallskip
  \end{tabular}
  &
  %side-conditions
  \begin{tabular}{l}
    \noalign{\smallskip}
    $w_j:C $ \\
    $k = j+\len{\mGt{C}}-1$ \\
    $\widetilde w = (w_j,\ldots,w_{k})$ \\ 
    % $\widetilde w = (w_j,\ldots,w_{j+\len{\mGt{C}}-1})$ \\
   
    % $\widetilde z = \fv{Q}$ \\
		% $\sigma = \linecondit{u_i:S}{\incrname{u}{i}}{\{\} }$
    $\sigma = \nextn{u_i}$
%		\begin{cases}
%		\incrname{u}{i} & \text{if } u_i:S \\
%		\{\} & \text{otherwise}
%		\end{cases}
		% $
		%$\sigma$
		% $$\sigma = \begin{cases}
		% \incrname{u}{i} & if \\
		% \{\} &
		% \end{cases}$$
    \smallskip
  \end{tabular}
  \\
%  \hline
%	$\bout{u_i}{y}Q$
%	&
%	\begin{tabular}{l}
%	\noalign{\smallskip}
%	$\propinp{k}{\widetilde x} \bout{u_i}{y}
%	\propout{k+1}{\widetilde x'} \inact \Par \B{k+1}{\tilde x'}{Q\sigma}$ \\
%
%	\smallskip
%\end{tabular}
%	&
%	\begin{tabular}{l}
%		\noalign{\smallskip}
%		%$\widetilde x \eqprm \widetilde z y$ \\
%%		$\widetilde x' = \begin{cases}
%%		\widetilde z y, & \text{if} \ y \in \fv{Q}  \\
%%		\widetilde z, & \text{otherwise}
%%		\end{cases}$ \\
%	$\widetilde x' = \fv{Q}$ \\
%	%$y \in \widetilde x$ \\
%	$\sigma = \begin{cases}
%	\incrname{u}{i} & \text{if } u_i:S \\
%	\{\} & \text{otherwise}
%	\end{cases}$
%	\smallskip
%\end{tabular}
\hline 

%%% INPUT
$\binp{u_i}{w}Q$
&

  \begin{tabular}{l}
      \noalign{\smallskip}
            $\propinp{k}{\widetilde x}\binp{u_i}{y} \apropout{k+1}{\widetilde x, y} \Par$ 
            \\
            \quad 
            $\news{s_1} \big(\propinp{k+1}{\widetilde x, y} 
            \propout{k+2}{y} \apropout{k+3}{\widetilde x} \Par$ 
            \\
            \qquad \qquad 
            $\propinp{k+2}{y} \news{s} \big(\bout{y}{s} \about{\dual{s}}{s_1}   ) \Par$ 
            \\
            \quad \qquad \qquad
            $\propinp{k+3}{\widetilde x} 
            \news{a}  \big(\bout{\dual{s_1}}{a}$ 
            \\ \qquad \qquad \qquad \quad 
            $\big( 
              \apropout{k+l+4}{} \Par \propinp{k+l+4}{} \inact \Par \widehat{Q}_{\tilde{x}}~  \big) \big) \big)$
            %%%%
            \smallskip 
            \\
      where:\\
      $ \!\begin{aligned} 
        \widehat{Q}_{\tilde{x}} &= \binp{a}{y'} \binp{y'}
        {\widetilde{w}} \big( \apropout{k+4}{\widetilde x} \Par
        \AmeliaMod{k+4}{\widetilde x}{Q \subst{w_1}{w} \sigma }_g \big) 
      \end{aligned}$ 
%      \smallskip
%        \\
  \smallskip
  \end{tabular}
  &
  %side-conditions
  \begin{tabular}{l}
    \noalign{\smallskip}
%    	$\widetilde x' = \begin{cases}
%		\widetilde x y & \text{if } y \in \fv{Q} \\
%		\widetilde x & \text{otherwise}
%		\end{cases}$ \\
%    $w:\slhotup{C}$\\
    $w : C$\\
    $\widetilde w = (w_1, \ldots, w_{\len{\mGt{C}}}) $\\
% 	$\widetilde z = \fv{Q} $ \\
	$l = \lenHO{Q}$ \smallskip \\
		%$\sigma = \begin{cases}
		%\incrname{u}{i} & \text{if } u_i:S \\
		%\{\} & \text{otherwise}
		%\end{cases}$
    
    % $\widetilde{\prop} = \begin{cases} 
    %         \epsilon & \text{if} \ \leadsto = \multimap \\
    %      (\prop_k,\ldots,\prop_{k+\lenHO{P}-1}) & \text{if} \ \leadsto = \rightarrow
    %  \end{cases}$ \smallskip \\
     
    %  $\sigma = \linecondit{u_i:S}{\incrname{u}{i}}{\epsilon}$
     $\sigma = \nextn{u_i}$
%     \begin{cases}
%		\incrname{u}{i} & \text{if } u_i:S \\
%	    \epsilon & \text{otherwise}
%		\end{cases}

\smallskip
  \end{tabular}
    \\
  \hline

    %$\appl{V}{u_i}$ &
  %\begin{tabular}{l}
    %\noalign{\smallskip}
    %$\propinp{k}{\widetilde x}\appl{\V{k+1}{\tilde x}{V}}{\widetilde m}$
    %\smallskip
  %\end{tabular}
  %&
  %%side-conditions
  %\begin{tabular}{l}
  %	\noalign{\smallskip}
  %	$u_i:C$ \\
  %	$\widetilde x = \fv{V}$ \\
  %  $\widetilde{m} = (u_i,\ldots,u_{i+\len{\Gt{C}}-1})$
  %  \smallskip
  %\end{tabular}
  %\\
  %\hline

%  $\appl{y}{u_i}$ &
%  \begin{tabular}{l}
%    \noalign{\smallskip}
%    $\propinp{k}{\widetilde x}\appl{y}{\widetilde m}$
%    \smallskip
%  \end{tabular}
%  &
%  %side-conditions \\
%  \begin{tabular}{l}
%    \noalign{\smallskip}
%    $u_i:C$ \\
%    %$\widetilde x = y$ \\
%    $\widetilde m = (u_i,\ldots,u_{i+\len{\Gt{C}}-1})$
%    \smallskip
%  \end{tabular}
%\\

%%% INPUT WITH RECURSIVE NAME 
$\bout{r_i}{w_j} P$ &
\begin{tabular}{l}
  \noalign{\smallskip}
  $\propinp{k}{\widetilde x}
  \news{a_1}{}\proprout{r}{a_1}$
  \\  \qquad \qquad \qquad 
  $\big(\AmeliaMod{k+3}{\tilde x}{P}_g
  \Par \binp{a_1}{y_1}\binp{y_1}{\widetilde z}W \big)$ \\
  \text{where:} \\
  $W =  \news{a_2}\big(\bout{z_{\indf{S}}}{a_2}$ 
  \\ \qquad \qquad 
  $\big(\propout{k+3}{\widetilde x}\proprinp{r}{b}
        \news{s} \big(\bout{b}{s}\about{\dual{s}}{\widetilde z} \big) \Par$
         \\ \qquad \qquad  \quad 
        $ \binp{a_2}{y_2}\binp{y_2}{z'_1}$
        % \news{\widetilde \prop}
        \\ \qquad \qquad  \qquad 
        $\big(\apropout{k+1}{} \Par$ 
        % \\ 
        $\propinp{k+1}{}\binp{z'_1}{x}\apropout{k+2}{x} \Par$  
        \\ \qquad \qquad  \qquad \quad 
        $\propinp{k+2}{x}\news{s'} \big(\bout{x}{s'}\about{\dual s'}{\widetilde w} \big) \big) \big) \big)$
 \smallskip
\end{tabular}
&
\begin{tabular}{l}
	\noalign{\smallskip}
   $r:S \wedge \traux{S}$ \\
   $\wtd z = (z_1,\ldots,z_{\len{\mRts{}{s}{S}}})$\\
   $\widetilde{c} = (c_{k+1}, c_{k+2})$ \\
  %  $\widetilde{y} = \fn{P}$ \\
   $w:C$ 
   \\
   $k = j+\len{\mGt{C}}-1$ \\
   $\widetilde w = (w_j,\ldots,w_{k})$ 
  %  $\wtd{w} = (w_j, \ldots, w_{j+ \len{\mGt{C}} - 1})$
\smallskip
\end{tabular}
  \\
  \hline
  $\binp{r_i}{w} P$ &
\begin{tabular}{l}
  \noalign{\smallskip}
    $\propinp{k}{\widetilde x}\news{a_1}$ 
    \\ \quad 
    $\big(\proprout{r}{a_1}
    \big(\news{s_1}\big(\propinp{k+1}{y}\propout{k+2}{y}\apropout{k+3}{}
    \Par$
    \\  \quad 
    $\propinp{k+2}{y}\news{s}\big(\bout{y}{s}\about{\dual s}{s_1}) \Par $
    \\ \quad 
    $\propinp{k+3}{}\news{a_2} \big(\bout{s_1}{a_2}\big(
    \apropout{k+l+4}{} \Par \propinp{k+l+4}{}\inact \Par$
    \\ \qquad 
    $\binp{a_2}{y_2}\binp{y_2}{\widetilde w}
    % \news{\widetilde \prop}
    \big(\apropout{k+4}{\widetilde x}
    \Par \AmeliaMod{k+4}{\tilde x}{P\subst{w_1}{w}}_g
    \big)\big)\big)  \Par$
    \\ \qquad \qquad 
   $
     \binp{a_1}{y_1}\binp{y_1}{\widetilde z}\binp{z_{\indf{S}}}{y}$
    \\ \qquad \qquad \qquad 
    $ 
    \propout{k+1}{y}\proprinp{r}{b}
    \news{s'}\big(\bout{b}{s'}
    \about{\dual s'}{\widetilde z}\big)
    \big)\big) \big)$
 \smallskip
\end{tabular}
& 
\begin{tabular}{l}
	\noalign{\smallskip}
   $r:S \wedge \traux{S}$ \\
   $\wtd z = (z_1,\ldots,z_{\len{\mRts{}{s}{S}}})$\\
  %  $\widetilde{x}' = \fn{P}$\\
   $l = \lenHO{P}$ \\
  %  $l = \begin{cases} 
  %           \lenHO{P} & \text{if} \ \leadsto = \multimap \\
  %        0 & \text{if} \ \leadsto = \rightarrow
  %    \end{cases}$\\

   $w:C$ 
   \\
   $\wtd{w} = (w_1, \ldots, w_{\len{\mGt{C}}})$
  %  $\wtd{\prop} =  (\prop_{k+4},\ldots,\prop_{k+\lenHO{P}+3})$ \\
  %  $\widetilde{\prop} = \begin{cases} 
  %           \epsilon & \text{if} \ \leadsto = \multimap \\
  %        (\prop_{k+4},\ldots,\prop_{k+\lenHO{P}+3}) & \text{if} \ \leadsto = \rightarrow
  %    \end{cases}$
\smallskip
\end{tabular}
  \\
  \hline
\end{tabular} 
}{
    % \begin{table}[!t]
      \resizebox{1.01\textwidth}{!}{
        }}
  \caption[Decompose by composing: Breakdown function $\AmeliaMod{k}{\tilde x}{\cdot}_g$ (Part I)]{Decompose by composing (Part 1/2): Breakdown function $\AmeliaMod{k}{\tilde x}{\cdot}_g$
  (cf.~\Cref{pi:def:am-decomp}). \label{pi:table:repo}}
  % \vspace{-4mm}
  \end{table*}

  \begin{table*}[!t]
    \thesisalt{
      % \begin{table*}[!ht]
    \begin{tabular}{ |l|l|l|}
      \rowcolor{gray!25}
      \hline
      $P$ &
        \multicolumn{2}{l|}{
      \begin{tabular}{l}
        \noalign{\smallskip}
        $\AmeliaMod{k}{\tilde x}{P}_g$
        \smallskip
      \end{tabular}
    }  \\
      \hline
  %% RESTRICTION
  % \begin{tabular} 
  $\news{s}{P'}$ &
  \begin{tabular}{l}
    \noalign{\smallskip}
 	$\news{\widetilde{s}:\mGt{C}}{\,\AmeliaMod{k}{\tilde x}
 			{P'\sigma}_g}$
    \smallskip
  \end{tabular}
  &
  %side-conditions
  \begin{tabular}{l}
    \noalign{\smallskip}
    % $\widetilde x = \fv{P'}$ \\
    $s : C$ \\ 
    $\widetilde{s} = (s_1,\ldots,s_{\len{\mGt{C}}})$  
    \\ 
    $\sigma = 	\subst{s_1 \dual{s_1}}{s \dual{s}}$
% 	$\sigma = \begin{cases}
% 				\subst{s_1 \dual{s_1}}{s \dual{s}} & \text{if} \ C = S \\
% 				\subst{s_1}{s} & \text{if} \ C = \chtype{U}
%             \end{cases}
% $
    \smallskip
  \end{tabular}
  \\
  \hline
   %%% RESTRICTION TAIL RECURSIVE 
   $\news{r}{P'}$ &
   \begin{tabular}{l}
     \noalign{\smallskip}
    $\news{\widetilde{r}:\mRt{S}}\,
    \proprinp{r}{b}\news{s'}\big(\bout{b}{s'}\about{\dual s'}
        {\widetilde r}\big) \Par$ 
        \\  \qquad   
        $\proprinp{\dual r}{b}\news{s'}\big(\bout{b}{s'}\about{\dual s'}
        {\widetilde {\dual r}}\big)
   \Par \AmeliaMod{k}{\tilde x}  
        {P'\sigma}_g$
     \smallskip
   \end{tabular}
   &
   %side-conditions
   \begin{tabular}{l}
     \noalign{\smallskip}
     % $\widetilde x = \fv{P'}$ \\
     $r:\trec{t}{S}$
     \\ 
     $\traux{\trec{t}{S}}$  
     \\ 
     $\sigma = 	\subst{r_1 \dual{r_1}}{r \dual{r}}$ 
    \\
     $\widetilde{r} = (r_1,\ldots,r_{\len{\mRt{S}}})$ \\
     $\widetilde{\dual r} = (\dual r_1,\ldots,\dual r_{\len{\mRt{S}}})$
    %  $\sigma = 	\subst{s_1 \dual{s_1}}{s \dual{s}}$
 % 	$\sigma = \begin{cases}
 % 				\subst{s_1 \dual{s_1}}{s \dual{s}} & \text{if} \ C = S \\
 % 				\subst{s_1}{s} & \text{if} \ C = \chtype{U}
 %             \end{cases}
 % $
     \smallskip
   \end{tabular}
   \\
   \hline
%%% PARALLEL
  $Q_1 \Par Q_2$ &
  \begin{tabular}{l}
    \noalign{\smallskip}
    $\propinp{k}{\tilde x} \propout{k+1}{\tilde y}
    \apropout{k+\degree+1}{\tilde z}   \Par \AmeliaMod{k+1}{\tilde y}{Q_1}_g \Par \AmeliaMod{k+\degree+1}{\tilde z}{Q_2}_g$
    \smallskip
  \end{tabular}
  &
  %side-conditions
  \begin{tabular}{l}
    \noalign{\smallskip}
    $\widetilde y  = \fv{Q_1}$  
    \\ 
    $\widetilde z = \fv{Q_2}$  
    \\ 
    $\degree = \lenHO{Q}$
    \smallskip
  \end{tabular}
      \\
  \hline

%%% INACTION
  $\inact$ &
\begin{tabular}{l}
  \noalign{\smallskip}
  $\propinp{k}{}\inact$
 \smallskip
\end{tabular}
&
\begin{tabular}{l}
	\noalign{\smallskip}
\smallskip
\end{tabular}
  \\
  % \hline
  \hline 
  %%%% RECURSIVE PROCESS
  $\recp{X}{P}$ &
\begin{tabular}{l}
  \noalign{\smallskip}
      $\news{s_1} \big(  \propinp{k}{\widetilde x}\propout{k+1}{\wtd x}
      \apropout{k+3}{\wtd x} \Par$ 
      \\
      $\qquad \quad \propinp{k+1}{\wtd x} \news{a_1}\big( \bout{s_1}{\dual a_1} \big( \apropout{k+2}{}
      \Par \apropinp{k+2}{} \Par$
      \\ \qquad \qquad \qquad 
      $\propinp{k+3}{\widetilde x}\binp{s_1}{z_x}
      \apropout{k+4}{\wtd x, z_x} 
      \Par$
      \\  \qquad \qquad \qquad  
      $
      \AmeliaMod{k+4}{\tilde x, z_x}{P}_{g, \{X \rightarrow {\tilde n}\}} \Par $\\
      \qquad \qquad \qquad  
      $
      \repl\binp{a_1}{y'_1}
      \binp{y'_1}{\vmap{\widetilde n^1}, \ldots, \vmap{\widetilde n^m}, y_1}\widehat{P}\,\big)\big)\big)$ 
      \\
      \text{where:} 
      \\
     $\widehat{P} = \news{\widetilde \prop}\big( \prod_{0 < i \leq m}
        \proprinp{n_i}{b}\news{s'}\big(\bout{b}{s'}\about{\dual s'}
        {\vmap{\widetilde n^i}}\big)
        \Par$ 
        \\ \qquad \qquad 
        $\apropout{k+2}{\widetilde x} \Par$
        % \\ \quad \quad 
        $ 
        \propinp{k+2}{\widetilde x}\binp{y_1}{z_x}\apropout{k+3}{\widetilde x, z_x} 
        \Par$
        \\    \qquad \qquad 
        $\auxmapp{ \AmeliaMod{k+3}{\tilde{x}, z_x}
        {P}_{g, \{X \rightarrow {\tilde n}\}}  }{{}}{{\tilde \prop, \tilde \prop_r}}\big)$
 \smallskip
\end{tabular}
&
\begin{tabular}{l}
	\noalign{\smallskip}
    $\wtd n = \fn{P}$ \\
    $m = \len{\wtd n}$ \\
    $\vmap{\widetilde n} = (\vmap{n_1}, \dots, \vmap{n_m})$ \\
    $i \in \{1,\ldots,m \}.$ \\
    $\   \vmap{n_i} : S_i$ \\
    $\   \vmap{\widetilde n^i} = (\vmap{n_1^i}, \dots, \vmap{n_{\len{\mGt{S_i}}}^i })$\\
    $\widetilde{\prop} =(\prop_{k+2},\ldots,$ 
    \\ \qquad \qquad 
    $\prop_{k+\lenHO{ \vmap{P}^1_{g, \{X \rightarrow \tilde n \}}{{}} }+1})$ \\
    $\wtd c_r = \bigcup_{v \in \wtd n} c^v$
\smallskip
\end{tabular}
\\
  \hline
  %%%% RECURSIVE VARIABLE
  $\rvar{X}$ &
\begin{tabular}{l}
  \noalign{\smallskip}
    $\news{s_1}\big( \propinp{k}{z_x}\propout{k+1}{z_x}\apropout{k+2}{z_x} 
    \Par$
    \\ \qquad \qquad 
    $\propinp{k+2}{z_x}\bout{\dual s_1}{z_x}\apropout{k+3}{} 
    \Par \apropinp{k+3}{} \Par$ 
     \\ \qquad \qquad \qquad
     $\propinp{k+1}{z_x}\news{a_1}
     \big( \proprout{n_1}{a_1}$
     \\ \qquad \qquad \qquad \qquad
    $\big(
    \binp{a_1}{y_1}\binp{y_1}{\widetilde z_1} \ldots \news{a_j}Q\big)\big)
    % \Par \propinp{k+2}{z_x}\bout{\dual s_1}{z_x}\apropout{k+3}{} 
    % \Par \propinp{k+3}{}\inact 
    \big)$ 
    \\ 
    \text{where:} 
    \\
    $Q = \big(\proprout{n_j}{a_j}\big(\binp{a_j}{y_j}\binp{y_j}{\widetilde z_j}$
    \\ 
    $\qquad \qquad \news{s'}\big(\bout{z_x}{s'}\about{\dual s'}{\widetilde z_1, \ldots, \widetilde z_j, s_1}\big)\big)\big)$
 \smallskip
\end{tabular}
&
\begin{tabular}{l}
	\noalign{\smallskip}
      $\wtd n = g(X)$ \\
      $\len{\widetilde n} = j$ \\
      $i \in \{1, \dots, j\}$ \\
      $\quad n_i : S \wedge \traux{S_i}$ \\
      $\quad \widetilde{z_i} = (z_1^i, \dots, z_{\len{\mRts{}{}{S_i}}}^i)$
\smallskip
\end{tabular}
% \\
%   \hline
%   $\inact$ &
% \begin{tabular}{l}
%   \noalign{\smallskip}
%   $\propinp{k}{}\inact$
%  \smallskip
% \end{tabular}
% &
% \begin{tabular}{l}
% 	\noalign{\smallskip}
% \smallskip
% \end{tabular}
  \\
  \hline
\end{tabular}
% \caption{Decompose by composition: Breakdown function $\AmeliaMod{k}{\tilde x}{\cdot}_g$ for \sessp processes 
% (cf. Definition~\ref{def:am-decomp}). \label{table:repo}}
% \vspace{-4mm}
% \end{table*}
}
    {\resizebox{1.01\textwidth}{!}{}}
  \caption[Decompose by composing: Breakdown function $\AmeliaMod{k}{\tilde x}{\cdot}_g$ (Part II)]{
    Decompose by composing (Part 2/2): Breakdown function $\AmeliaMod{k}{\tilde x}{\cdot}_g$ 
  (cf.~\Cref{pi:def:am-decomp}). \label{pi:table:repo2}}
  % \vspace{-4mm}
  \end{table*}
% \end{adjustbox} 

The breakdown function $\AmeliaMod{k}{\tilde x}{\cdot}_g$ for all constructs of \sessp is given in~\Cref{pi:table:repo,pi:table:repo2}; it  relies on several auxiliary definitions, most notably:
\begin{itemize}
	\item The \emph{degree} of a process $P$, denoted  $\lenHO{P}$;
%	\item The predicate $\tr(C)$, which indicates that $C$ is a
%    \item Tail-recursive session types;
	\item The functions $\mGt{\cdot}$ and $\mRt{\cdot}$, which decompose session types into \msts.
\end{itemize}
%\jcheck{[Why does it mention HO?]}
We now formally define these notions.
\begin{definition}[Degree of a Process]
  \label{pi:def:am-sizeproc}
  The \emph{degree} of a process $P$, denoted  $\lenHO{P}$, is defined as:
%    $$
%    \lenHO{P} =
%    \begin{cases}
%    \lenHO{Q} + 3 & \text{if $P = \bout{u_i}{w_j}Q$}
%    \\
%    \lenHO{Q} + 5 & \text{if $P = \binp{u_i}{x:C}Q$}
%    \\
%    \lenHO{Q} & \text{if $P = \news{s:S}Q$}
%    \\
%    \lenHO{Q} + \lenHO{R} + 1 & \text{if $P = Q \Par R$}
%    \\
%    1 & \text{if $P = \inact$ } \\
%    \lenHO{Q} + 4 & \text{if } P = \recp{X}Q \\
%    4 & \text{if } P = \rvar{X} 
%    \end{cases}
%    $$
    \begin{align*}
      \lenHO{\bout{u_i}{w_j}Q} & =  \lenHO{Q} + 3 & 
   \lenHO{\news{s:S}Q} & = \lenHO{Q}
   & \lenHO{ \inact} & =  1   
    \\
    \lenHO{\binp{u_i}{x:C}Q} & =  \lenHO{Q} + 5 
    &
    \lenHO{Q \Par R} & = \lenHO{Q} + \lenHO{R} + 1      \\
   \lenHO{\rvar{X} } & = 4  
   &
   \lenHO{\recp{X}Q} & = \lenHO{Q} + 4 
    \end{align*}
  \end{definition}
  
%  \begin{definition}[A Predicate on Names]
%    Recall that we write $\traux{C}$  to indicate that $C$ is a
%    tail-recursive session type (\Cref{d:predtypes}).
%    %  that is if $C \equiv \trec{t}S$.
%Given $u:C$, we write $\lin{u}$ if $C = S$ and  $\neg \traux{S}$.
%  \end{definition}

  % \begin{definition}[Predicates on Types $\lin{\cdot}$]
  %   Let $u$ be a session type $C$. We define the
  %   predicate $\lin{C}$ if $C = S$ and  $\neg \tr(S)$.
  % \end{definition}

  % \acheck{We overload predicate $\lin{\cdot}$ (resp. $\tr(\cdot)$)
  % to work on names: let $u$ be a with session type $C$ (i.e. $u:C$), 
  % then  $\lin{u}=\lin{C}$ (resp. $\tr(u) = \tr(C)$).}

We define how to obtain \msts for \sessp from standard session types:

  % Type Decomposition Function G'(): Core Fragment
\begin{definition}[Decomposing First-Order Types]
  \label{pi:decomp:firstordertypes}
%  Let %$P$ be a \sessp process, 
%  $S$ be a session type, $C$ be a name type and $\chtype{S}$ be a shared channel type. 
  The decomposition function $\mGt{\cdot}$ on finite types, obtained by
  combining the mappings $\tmap{\cdot}{1}, \Gt{\cdot}$, and $ \tmap{\cdot}{2}$,
  is defined in~\Cref{pi:f:decomp:firstordertypes} (top, where omitted cases are
  defined homomorphically).
  It is extended to account for recursive session types in~\Cref{pi:f:decomp:firstordertypes} (center).
  
  The auxiliary function $\mRts{}{}{\cdot}$,  given in~\Cref{pi:f:decomp:firstordertypes} (bottom), is used in~\Cref{pi:table:repo} to 
decompose \emph{guarded} tail-recursive types:
  it skips session prefixes until a type of form $\trec{t}S$
  is encountered; when that occurs, the recursive type is decomposed using 
  $\mRt{\cdot}$. 
  \end{definition}
  
\begin{newaddenv}
  
We give the definition of the index function 
  for $\sessp$ recursive types by composing the index function~$\indT{\cdot}$ (for \HO recursive types, cf.~\Cref{mst:def:indexfunction}) 
  and the  encoding of types from~\Cref{top:f:enc:hopi_to_ho,top:f:enc:ho_to_sessp}. 
This composition is straightforward:  notice that $\indT{\cdot}$  counts prefixes 
of session types, and that the encodings of types from~\Cref{top:f:enc:hopi_to_ho,top:f:enc:ho_to_sessp} 
do not alter the number of prefixes. 

%   that 
%   given an unfolded recursive session type $S$ returns the position of the top-most prefix of $S$ within its body.
%   We compose 
% the index function 
% for \HO recursive types~\Cref{mst:def:indexfunction} and 
% types encoding from~\Cref{top:f:enc:hopi_to_ho} 
% and~\Cref{top:f:enc:ho_to_sessp}. 

% \Cref{top:f:enc:hopi_to_ho}

% counts 
% prefixes in session type and 
% follows directly from the~\Cref{mst:def:indexfunction}
% and ... 

% Given an unfolded recursive session type $S$, the auxiliary function $\indT{S}$ returns the position of the top-most prefix of $S$ within its body.
% \subsubsection{Core fragment} 
\begin{definition}[Index function]
  \label{pi:def:indexfunction-comp}
  Let $S$ be an (unfolded) recursive session type. The function $\indf{S}$ is defined as follows:  
  \begin{align*}
   \indf{S} = \begin{cases}
     \indfaux{0}{S'\subst{S}{\vart{t}}} & \text{if $S =\trec{t}{S'}$} \\
     \indfaux{0}{S} & \text{otherwise}
   \end{cases}
\qquad 
\indfaux{l}{T} = 
\begin{cases}
	\indfaux{l+1}{S} & \text{if $T = \btout{U}S$ or $T = \btinp{U}S$} 
	\\
	\len{\mRt{S}} - l + 1 & \text{if $T =\trec{t}{S}$}
\end{cases}
    \end{align*}
  \end{definition}
\end{newaddenv}

  \begin{figure}[!t]
  	\begin{mdframed}[style=alttight]
  \begin{align*}
      \mGt{\btout{C}{S}} &=
          \begin{cases}
              M &  \text{if $S = \tinact$}\\
             M , \mGt{S} & \text{otherwise}
          \end{cases}\\
          &\qquad  \text{where }  
          M =  \bbtout{ \lrangle{ \btinp{ \btinp{ \lrangle{ \btinp{ \mGt{C} }{\tinact} } }{\tinact} }{\tinact} } }{\tinact}  \\
      \mGt{\btinp{C}{S}} &=
          \begin{cases}
              M  &  \text{if $S = \tinact$} \\
              M, \mGt{S} & \text{otherwise}
          \end{cases}\\
          &\qquad \text{where } M = \btinp{ \lrangle{ \btinp{ \btinp{ \lrangle{\btinp{\mGt{C}}{\tinact}} }
          {\tinact} }{\tinact} } }{\tinact}
          \\
      \mGt{\tinact} &= \tinact \\
      \mGt{S_1,\ldots,S_n} &= \mGt{S_1},\ldots,\mGt{S_n}
%        \vspace{-2mm}
\\
      \mGt{\chtype{S}} &= \chtype{\mGt{S}}
  \end{align*}
      ~~~{\hbox to \textwidth{\leaders\hbox to 3pt{\hss . \hss}\hfil}}
        %\\
      \begin{align*}
       \mGt{\trec{t}{S}} &= 
                      \begin{cases}
                          \mRt{S} & \text{if $\trec{t}{S}$ is tail-recursive}\\
                          \trec{t}{\mGt{S}} & \text{otherwise}
                      \end{cases} \\
       \mRt{\btout{S}{S'}} &= 
      \trec{t}{\bbtout{\lrangle{\btinp{ \btinp{\lrangle{\btinp{\mGt{S}}{\tinact}}}
      {\tinact} }{\tinact}}}{\vart{t}}}, \mRt{S'}\\
       \mRt{\btinp{S}{S'}} &= 
      \trec{t}{\bbtinp{\lrangle{\btinp{ \btinp{\lrangle{\btinp{\mGt{S}}{\tinact}}}{\tinact} }
      {\tinact}}}{\vart{t}}}, \mRt{S'} \\
      \mGt{\vart{t}} &= \vart{t}  \\
      \mRt{\vart{t}} & = \epsilon  
  \end{align*}
%  We shall also define $\mRts{}{}{\cdot}$ as follows:
      ~~~{\hbox to \textwidth{\leaders\hbox to 3pt{\hss . \hss}\hfil}}
  \begin{align*}
    \mRts{}{}{\btinp{S}{S'}} = \mRts{}{}{S'} 
  \qquad
   \mRts{}{}{\btout{S}{S'}}  = \mRts{}{}{S'} 
   \qquad 
   \mRts{}{}{\trec{t}{S}} = \mRts{}{}{S} 
  \end{align*}
    	\end{mdframed}
    	\caption[Decomposition of session types $\mGt{\cdot}$]{Decomposition of session types $\mGt{\cdot}$ (cf.~\Cref{pi:decomp:firstordertypes})}
    	\label{pi:f:decomp:firstordertypes}
    	\vspace{-3mm}
  \end{figure}

  Given a typed process $P$, we write $\rfn{P}$ to denote the set of free names of $P$ whose types are recursive.
  We are finally ready to define the decomposition function $\F{\cdot}$, the analog 
  of~\Cref{mst:def:decomp} but for processes in \sessp:
  \begin{definition}[Process Decomposition]
    \label{pi:def:am-decomp}
    Let $P$ be a closed \sessp process
    with $\wtd u = \fn{P}$ and $\wtd v = \rfn{P}$. 
    Given the breakdown function $\AmeliaMod{k}{\tilde x}{\cdot}_g$  in~\Cref{pi:table:repo,pi:table:repo2}, the decomposition $\F{P}$ is defined as:
      $$
      \F{P} = \news{\widetilde \prop}
      \news{\widetilde \prop_r}
      \Big(  
      \propout{k}{} \inact 
      \Par \AmeliaMod{k}{\epsilon}{P\sigma}_g 
%      \Par        \prod_{r \in \tilde{v} } P^r  
      \Par        \prod_{r \in \tilde{v} } \proprinp{r}{b}\news{s} (\bout{b}{s}\about{\dual s}{\wtd r})  
      \Big)
      $$ 
      where: $k > 0$, $\widetilde \prop =
      (\prop_k,\ldots,\prop_{k+\lenHO{P}-1})$;
      $\widetilde{\prop_r} = \bigcup_{r\in \tilde{v}}\prop^r$;
      $\sigma =
      \subst{\mathsf{init}(\tilde u)}{\widetilde u}$; 
%      $P^r = \proprinp{r}{b}\news{s} \big(\bout{b}{s}\about{\dual s}{\wtd r} \big)$ 
      for each $r \in \tilde{v}$, we have $r:S$ and $\wtd{r} = r_1, \ldots,  r_{\len{\mGt{S}}}$. 
     % Also, the breakdown function $\AmeliaMod{k}{\tilde x}{\cdot}_g$ is as defined in Table~\ref{table:repo}.
      % Note that the function $\mathsf{init(\cdot)}$
      % applied on tuple $\widetilde u$ initializes the set of free names by
      % adding an index to each element.
    \end{definition}

\newcommand{\subpr}{R}
\newcommand{\subpq}{Q}
\subsection{Examples}
We illustrate our constructions by means of two representative examples, involving delegation and recursive processes/types.

\begin{example}[A Process with Delegation]
  \label{pi:ex:ex1-decomp}
  \label{pi:ex:decomp-cr} 
  % Let $P$ be a \sessp process which incorporates name-passing and $w$ be a
  % channel through which first an integer value, then a boolean value is sent. 
  Let $P$ be a \sessp process which incorporates name-passing 
  and implements channels $u$ and $\dual w$ with types 
    $S = \btout{\dual T}\tinact$ 
  and $T = \btinp{\tint}\btout{\bool}\tinact$, respectively:
  % The degree of $P$ is $\lenHO{P} = 25$.
  \begin{align}
    P =~ &\news{u:S}\big(\underbrace{\bout{u}{w} \binp{\dual w}{t} \bout{\dual w}{\unaryint{t}} 
    \inact}_{\subpr} \Par \underbrace{\binp{\dual u}{x} \bout{x}{\textsf{5}}\binp{x}{b} \inact}_{\subpq}\big) \label{ex:procpi}
     \\ 
     \red~  & \binp{\dual w}{t} \bout{\dual w}{\unaryint{t}} 
    \inact  \Par {\bout{w}{\textsf{5}}\binp{w}{b} \inact}  \label{ex:procpii}
%     & \qquad \qquad \quad 
%       \binp{\dual u}{x} \bout{x}{\textsf{5}}\binp{x}{b} \inact) = 
%       \news{u}(A \Par \subpq)	
  \end{align}
  % \begin{align*}
  %   P &= \news{u:S}(\bout{u}{w} \binp{\dual w}{t} \bout{\dual w}{\unaryint{t}} \inact \Par
  %     \binp{\dual u}{x} \bout{x}{\textsf{5}}\binp{x}{b} \inact) = \news{u}(A \Par \subpq)	
  % \end{align*}
  
  \noindent 
 We have that $\lenHO{P} = 25$.
  Then, the decomposition of $P$ into a collection of first-order processes
  typed with minimal session types is:
  \begin{align*}
    \F{P} &= %\news{ \prop_1, \dots, c_{25}}\big(\propout{1}{} \inact \Par \AmeliaMod{1}{\epsilon}{P\sigma}_\empg \big)\\
%          &= \news{ \prop_1, \dots, c_{25}}\big(\propout{1}{} \inact \Par 
%          \AmeliaMod{1}{\epsilon}{\news{u}(\subpr \Par \subpq)\sigma}_\empg \big)\\
%          &= 
          \news{ \prop_1, \dots, c_{25}}\big(\propout{1}{} \inact 
          \Par \news{u_1} \AmeliaMod{1}{\epsilon}{(\subpr \Par \subpq)\sigma'}_\empg \big)
          % , 
          % \\
          % &\text{ where } \sigma = \mathsf{init}(\fn{P}),\ 
          % \sigma' = \sigma \cdot \subst{u_1 \dual{u_1}}{u\dual{u}}
  \end{align*}

  \noindent where $\sigma = \mathsf{init}(\fn{P})$ and $ 
          \sigma' = \sigma \cdot \subst{u_1 \dual{u_1}}{u\dual{u}}$. 
          We omit parameter $g$, as it is empty. 
 We have:
  \begin{align*}
      \AmeliaMod{1}{\epsilon}{(\subpr \Par \subpq)\sigma'}_\empg \big) &= 
      \propinp{1}{} \propout{2}{}  \apropout{13}{}  \Par 
      \AmeliaMod{2}{\epsilon}{\subpr \sigma'}_\empg \Par \AmeliaMod{13}{\epsilon}{\subpq \sigma'}_\empg
  \end{align*}

  \noindent We use some abbreviations for 
  subprocesses of $\subpr$ and $\subpq$ : 
  \begin{align*}
    \subpr' &= \binp{\dual{w}_1}{t}\subpr''   
    \qquad \subpr'' = \bout{\dual w_1}
    {\unaryint{t}}\inact  \\
    \subpq' &= \bout{x_1}{5} \subpq''  
    \qquad \subpq'' = \binp{x_1}{b}\inact 
  \end{align*}

  \noindent The breakdown of $\subpr$ is:
  \begin{align*}
    \AmeliaMod{2}{\epsilon}{\subpr}_\empg = & ~\propinp{2}{} \news{a_1}
   \big(\bout{u_1}{a_1} \big( 
        \apropout{5}{} \Par \AmeliaMod{5}{\epsilon}
        {\subpr'}_\empg 
        \Par 
        \\
        & \qquad \qquad \qquad \binp{a_1}{y_1} \binp{y_1}{z_1} \apropout{3}{z_1} \Par 
         \propinp{3}{z_1} \binp{z_1}{x} \apropout{4}{x} \Par 
         \\
         & \qquad \qquad \qquad \quad 
        \propinp{4}{x} \news{s} \big(\bout{x}{s} \about{\dual s}{w_1, w_2} \big) \big) \big)\\
    % next row
    \AmeliaMod{5}{\epsilon}{\subpr'}_\empg = &
    ~\propinp{5}{} \binp{\dual{w_1}}{y_2} \apropout{6}{y_2} \Par \\ 
    & \news{s_1}\big( \propinp{6}{y_2} \propout{7}{y_2} \apropbout{8}{} \Par \\
        & \qquad \qquad \propinp{7}{y_2} \news{s'}\big( \bout{y_2}{s'}  \bout{\dual{s'}}{s_1} \inact \big) \big) 
        \Par 
        \\
        & 
       \qquad \qquad  \qquad \propinp{8}{} \news{a_2} \big( \bout{\dual{s_1}}{a_2} \big(\apropout{10}{} 
        \Par \propinp{10}{} \inact \Par \\
        &\qquad \qquad \qquad \qquad \qquad \binp{a_2}{y_3} \binp{y_3}{t_1} 
        \big( \apropout{9}{} \Par
        \AmeliaMod{9}{\epsilon}{\subpr''}_\empg  \big) \big) \big)  \\
        %  third row
        \AmeliaMod{9}{\epsilon}{\subpr''}_\empg =  &
      ~\propinp{9}{} \news{a} \big(\bout{\dual w_2}{a}
      \big( \apropout{12}{} \Par \propinp{12}{}\inact \Par 
      \\
      & \qquad \qquad \qquad \binp{a}{y} \binp{y}{z_1} \apropout{11}{z_1} \Par
      \propinp{10}{z_1} \binp{z_1}{x} \apropout{11}{x}     \Par
      \\
      &
      \qquad
      %  \\
      %  &
      %  \qquad \qquad \qquad \qquad 
       \qquad \qquad \qquad \qquad \propinp{11}{x} 
       \news{s} \big(\bout{x}{s} \about{ \dual{s} }{\unaryint{t}} \big) \big) \big)
    \end{align*}

  \noindent The breakdown of $\subpq$ is:
  \begin{align*}
    \AmeliaMod{13}{\epsilon}{\subpq}_\empg &= 
    \propinp{13}{}\binp{\dual u_1}{y_4}\apropout{14}{y_4}  \Par 
    \news{s_1}\big( \propinp{14}{y}\propout{15}{y}\apropout{16}{} 
    \Par \\
    & \qquad \quad \propinp{15}{y_4} \news{s''} \big(\bout{y_4}{s''} 
      \bout{\dual{s''}}{s_1} \inact \big) \Par 
      \propinp{16}{} \news{a_3} \big(\bout{s_1}{a_3} \big( \apropout{21}{} \Par \\
    & \qquad \qquad \propinp{21}{} \inact \Par 
    \binp{a_3}{y_5} \binp{y_5}{x_1, x_2} 
    \big(\apropout{17}{} \Par \AmeliaMod{17}{\epsilon}{\subpq'}_\empg \big) \big) \big) \big) 
    \smallskip \bigskip \\
    % new row
  \AmeliaMod{17}{\epsilon}{\subpq'}_\empg &=  
      \propinp{17}{} \news{a_4} \big( \bout{x_1}{a_4} \big( \apropout{20}{}  
      \Par \AmeliaMod{20}{\epsilon}{\subpq''}_\empg  
      \Par \binp{a_4}{y_6} \binp{y_6}{z_1} 
      \apropout{18}{z_1} \Par \\
      &\qquad  \quad \propinp{18}{z_1} \binp{z_1}{x} \apropout{19}{x} \Par 
      \propinp{19}{x} \news{s'''} \big(\bout{x}{s'''} \about{\dual{s'''}}{\textsf{5} } \big) \big) \big)
      \\
      % new row 
      \AmeliaMod{20}{\epsilon}{\subpq''}_\empg &= 
      \propinp{20}{}\binp{x_2}{y} \apropout{21}{y} \Par
      \\
      &\qquad \news{s_1} \big(\propinp{21}{y} \propout{22}{y} \apropout{23}{} \Par
       \propinp{22}{y} \news{s}\big(\bout{y}{s} \about{\dual{s}}{s_1}   \big) \Par
      \\
      &\qquad \qquad \qquad \propinp{23}{\wtd x} \news{a}  \big(\bout{\dual{s_1}}{a} 
      \big( \apropout{25}{} \Par \propinp{25}{} \inact \Par \\
      & \qquad \qquad \qquad \quad \binp{a}{y'} \binp{y'}
        {b_1} \big( \apropout{24}{} \Par
        \propinp{24}{}\inact \big)   \big) \big) \big) 
  \end{align*}

% \begin{align*}
%   \AmeliaMod{20}{\epsilon}{\binp{x_2}{b} \inact}_\empg &= 
%   \propinp{20}{}\binp{x_2}{y} \apropout{21}{y} \Par
%   \\
%   &\qquad \news{s_1} \big(\propinp{21}{y} \propout{22}{y} \apropout{23}{} \Par
%    \propinp{22}{y} \news{s}(\bout{y}{s} \about{\dual{s}}{s_1}   ) \Par
%   \\
%   &\qquad \qquad \qquad \propinp{23}{\wtd x} \news{a}  (\bout{\dual{s_1}}{a} 
%   ( \apropout{25}{} \Par \propinp{25}{} \inact \Par \\
%   & \qquad \qquad \qquad \binp{a}{y'} \binp{y'}
%     {b_1} \big( \apropout{24}{} \Par
%     \propinp{24}{}\inact \big)   ) ) \big) 
%   % &\text{where}
%   % \\
%   %   \widehat{Q}_{\tilde{x}} &= \binp{a}{y'} \binp{y'}
%   %   {b_1} \big( \apropout{k+4}{} \Par
%   %   \AmeliaMod{k+4}{}{Q \subst{w_1}{w} \sigma }_\empg \big) 
% \end{align*}
\begin{comment} 
  \noindent Names $\dual w_1$ and $\dual w_2$ are typed, respectively, with 
  \begin{align*}
  M_1 & = \bbtinp{ \lrangle{ \btinp{ \btinp{ \lrangle{ \btinp{ \textsf{Int} }{\tinact} } }{\tinact} }{\tinact} } }{\tinact}
  \\
  M_2 & = \bbtout{ \lrangle{ \btinp{ \btinp{ \lrangle{ \btinp{ \textsf{Bool}}{\tinact} } }{\tinact} }{\tinact} } }{\tinact}
    \end{align*}
    Name $u_1$ is typed with 
  $\bbtout{ \lrangle{ \btinp{ \btinp{ \lrangle{ \btinp{ \dual M_1, \dual M_2 }{\tinact} } }{\tinact} }{\tinact} } }{\tinact}$. 
  \end{comment} 

  % \noindent The breakdown of $\subpq'$ is similar and given in 
  %       \Cref{pi:ex:ex1-decomp}.  % \noindent The breakdown of $\subpq'$ is similar and given in 
  % % \appref{ex:ex1-decomp}.
  \noindent Type $S$ is broken down into \msts $M_1$ and $M_2$, as follows: 
  \begin{align*} 
    M_1 &= \bbtinp{ \lrangle{ \btinp{ \btinp{ \lrangle{ \btinp{ \tint }{\tinact} } }{\tinact} }{\tinact} } }{\tinact} \\
    M_2 &= \bbtout{ \lrangle{ \btinp{ \btinp{ \lrangle{ \btinp{ \bool}{\tinact} } }{\tinact} }{\tinact} } }{\tinact}
  \end{align*}  

  \noindent Names $\dual w_1$ and $\dual w_2$ are typed with $M_1$ and $M_2$, respectively. Then, 
  name $u_1$ is typed with $M$, given by: 
  \begin{align*} 
    M = \bbtout{ \lrangle{ \btinp{ \btinp{ \lrangle{ \btinp{ \dual M_1, \dual M_2 }{\tinact} } }{\tinact} }{\tinact} } }{\tinact}
  \end{align*}

  % Now let us observe the reduction chain. The collection of processes
  % synchronizes on $\prop_1, \prop_2, \prop_{13}$ after three reductions.

  \noindent Consider the reductions of $\F{P}$ that mimic the exchange of $w$ along $u$ in $P$. We first have three  
  synchronizations on $\prop_1, \prop_2, \prop_{13}$:
  \begin{align*}
    \F{P} \red^{3}~ & \news{\widetilde c} \big( \ \news{a_1} \big( \highlighta{\bout{u_1}{a_1}} \big( 
    \apropout{5}{} \Par 
    \AmeliaMod{5}{\epsilon}{\subpr' }_\empg \Par
    \\
    & \quad  
    \binp{a_1}{y_1} \binp{y_1}{z_1} \apropout{3}{z_1} \Par 
     \propinp{3}{z_1} \binp{z_1}{x} \apropout{4}{x} \Par 
     \\
    & \quad
    \propinp{4}{x} \news{s} \big(\bout{x}{s} \about{\dual s}{w_1, w_2}\big) \big) \big) \Par 
   \highlighta{\binp{\dual u_1}{y_4}}\apropout{14}{y_4}    \Par 
   \\
    & \quad 
  \news{s_1}\big( \propinp{14}{y}\propout{15}{y}\apropout{16}{} 
  \Par  
  \\
  & \qquad 
  \propinp{15}{y_4} \news{s''} \big(\bout{y_4}{s''} 
    \bout{\dual{s''}}{s_1} \inact \big) \Par 
    \\
  & \qquad
    \propinp{16}{} \news{a_3} \big(\bout{s_1}{a_3} \big( \apropout{21}{} \Par \\
  & \qquad \quad \propinp{21}{} \inact \Par 
  \binp{a_3}{y_5} \binp{y_5}{x_1, x_2} 
  \big(\apropout{17}{} \Par 
  \AmeliaMod{17}{\epsilon}{\subpq'}_\empg \big) \big) \big) \big) 
%  \smallskip \bigskip  
%    \\ \bigskip & \quad \quad 
%    \text{where $\widetilde c = (\prop_3, \dots, \prop_{12}, \prop_{14}, \dots, c_{25})$}
\end{align*}
where $\widetilde c = (\prop_3, \dots, \prop_{12}, \prop_{14}, \dots, c_{25})$.
  \begin{comment} 
  \begin{align*}
    \F{P} &\red^{3} \news{\widetilde c} \big( \ \news{a_1} \big( \highlighta{\bout{u_1}{a_1}} \big( 
      \apropout{5}{} \Par 
      \AmeliaMod{5}{\epsilon}{A' }_\empg \Par
      % \\
      % & \quad  
      \binp{a_1}{y_1} \binp{y_1}{z_1} \apropout{3}{z_1} \Par 
       \propinp{3}{z_1} \binp{z_1}{x} \apropout{4}{x} \Par 
       \\
      & \quad
      \propinp{4}{x} \news{s} \big(\bout{x}{s} \about{\dual s}{w_1, w_2}\big) \big) \big) \Par 
     \highlighta{\binp{\dual u_1}{y_4}}\apropout{14}{y_4}    \Par 
    %  \\
    %   & \quad 
    \news{s_1}\big( \propinp{14}{y}\propout{15}{y}\apropout{16}{} 
    \Par  
    \\
    & \qquad 
    \propinp{15}{y_4} \news{s''} \big(\bout{y_4}{s''} 
      \bout{\dual{s''}}{s_1} \inact \big) \Par 
    %   \\
    % & \qquad
      \propinp{16}{} \news{a_3} \big(\bout{s_1}{a_3} \big( \apropout{21}{} \Par \\
    & \qquad \quad \propinp{21}{} \inact \Par 
    \binp{a_3}{y_5} \binp{y_5}{x_1, x_2} 
    \big(\apropout{17}{} \Par 
    \AmeliaMod{17}{\epsilon}{\subpq'}_\empg \big) \big) \big) \big)
    \smallskip \bigskip  
      \\ \bigskip & \quad \quad 
      \text{where $\widetilde c = (\prop_3, \dots, \prop_{12}, \prop_{14}, \dots, c_{25})$}
  \end{align*}
\end{comment} 
Then, the broken down process $\subpr$ communicates with process $\subpq$
  through channel $u_1$ by passing name $a_1$ (highlighted above).
  % which is
  % further sent on $\prop_{14}$
  % and $\prop_{15}$. 
  Here we notice that the original transmission of value $w$ 
  is not immediately mimicked on channel $u$, but it is delegated to 
  some other channel through a series of channel redirections 
  starting with the transmission of name $a_1$. 
  % Further, by breaking down sequential prefixes involved in channel redirections 
  % redundant communications on propagators are introduced. 
  Further, the received name $a_1$ is locally propagated by $\prop_{14}$
  and $\prop_{15}$. This represents redundant communications on propagators 
  induced by breaking down sequential prefixes produced by 
  two encodings $\map{\cdot}^1_g$ and $\map{\cdot}^{2}$
  (i.e., those communications are not present in $P$). 
  Another synchronization occurs on $\prop_{16}$.
  \begin{align*}
    \F{P} \red^{7}~&  \news{\widetilde c_*}\news{a_1} \big( \  
    \apropout{5}{} \Par  
    \AmeliaMod{5}{\epsilon}{\subpr'}_\empg  \Par 
    \highlighta{\binp{a_1}{y_1}} \binp{y_1}{z_1} \apropout{3}{z_1}
     \Par 
     \\
    & 
     \propinp{3}{z_1} \binp{z_1}{x} \apropout{4}{x} \Par   
     \propinp{4}{x} \news{s} \big(\bout{x}{s} \about{\dual s}{w_1, w_2} \big) 
    \Par 
    \\
    &
    \news{s_1}\big(  \news{s''} \big(\highlighta{\bout{a_1}{s''}} 
    \bout{\dual{s''}}{s_1} \inact \big) \Par 
     \news{a_3} \big(\bout{s_1}{a_3} \big( \apropout{21}{} \Par 
    \\
    & 
    \qquad 
    \propinp{21}{} \inact \Par 
    \binp{a_3}{y_5} \binp{y_5}{x_1, x_2} \big(\apropout{17}{} 
    \Par \AmeliaMod{17}{\epsilon}{\subpq'}_\empg \big) \big) \big) \big)  \big)  
%    \\ \bigskip & \quad \quad 
\end{align*}
where 
    $\widetilde c_* = (\prop_3, \dots, c_{12}, c_{17}, \dots, c_{25})$.
    
  The next step involves a communication on $a_1$: session name $s''$ is passed and substitutes variable $y_1$.
  \begin{align*}
    \F{P} \red^{8}~ & \news{\widetilde c_*} \news{s''} \big( \ 
    \apropout{5}{} \Par 
    \AmeliaMod{5}{\epsilon}{\subpr'}_\empg       \Par 
    \\
    &
    \highlighta{ \binp{s''}{z_1}} \apropout{3}{z_1} \Par \propinp{3}{z_1} \binp{z_1}{x} \apropout{4}{x} \Par \\
    &  \propinp{4}{x} \news{s} \big(\bout{x}{s} \about{\dual s}{w_1, w_2} \big) 
    \Par 
    \\
    &
 \news{s_1}\big(  \highlighta{\bout{\dual{s''}}{s_1}} \inact \Par 
    \news{a_3} \big(\bout{s_1}{a_3} \big( \apropout{21}{} \Par \propinp{21}{} \inact \Par \\
    & \quad \binp{a_3}{y_5} \binp{y_5}{x_1, x_2} \big( \ \apropout{17}{} \Par 
    \AmeliaMod{17}{\epsilon}{\subpq' }_\empg  \big) \big) \big) \big) \big) \smallskip 
 %   \\ \bigskip & \quad \quad 
 %   \text{where $\wtd \prop = (\prop_3, \dots, \prop_{12}, \prop_{17}, \dots, \prop_{25})$}
\end{align*}

  \noindent 
  After the synchronization on channel $s''$, name $z_1$ is further sent to the next parallel process through the propagator $c_3$:
  \begin{align*}
      \F{P} \red^{10}~& \news{\widetilde c_{**}} \news{s_1} \big( \ 
      \apropout{5}{} \Par
       \AmeliaMod{5}{\epsilon}{\subpr' }_\empg 
      \Par 
      \\
      &
      \highlighta{\binp{s_1}{x}} \apropout{4}{x} \Par \propinp{4}{x} 
      \news{s} \big(\bout{x}{s} \about{\dual s}{w_1, w_2}\big)  \Par \\
      & \news{a_3} \big(\highlighta{\bout{s_1}{a_3}} \big( \apropout{21}{} 
      \Par \propinp{21}{} \inact \Par  
      \binp{a_3}{y_5} \binp{y_5}{x_1, x_2} 
      \big(\apropout{17}{} \Par 
      \AmeliaMod{17}{\epsilon}{\subpq'}_\empg \big) \big) \big)  \big) 
  \end{align*}

where 
      $\widetilde c_{**} = (\prop_4, \dots, \prop_{12}, \prop_{17}, \dots, \prop_{25})$.

  Communication on $s_1$ leads to variable $x$ being substituted by name $a_3$,
  which is then passed on $c_{4}$ to the next process. In addition, inaction is
  simulated by synchronization on $\prop_{21}$.
  \begin{align*}
      \F{P} \red^{13}~& \news{\widetilde c_{\bullet}} \news{a_3} ( \
      \apropout{5}{} \Par 
      \AmeliaMod{5}{\epsilon}{\subpr' }_\empg \Par 
      \news{s} (
        \highlighta{\bout{a_3}{s}} \about{\dual s}{w_1, w_2})  \Par \\
      & \quad 
      \highlighta{\binp{a_3}{y_5}} \binp{y_5}{x_1, x_2} (\apropout{17}{} 
      \Par \AmeliaMod{17}{\epsilon}{\subpq'  }_\empg ) \ ) 
  \end{align*}
  where $\widetilde c_{\bullet} = (\prop_5, \dots, c_{12}, c_{17}, \dots, c_{25})$.
  
  Now, the distribution of the decomposition of $w$ from one process to another
  can finally be simulated by two reductions: first, a synchronization on $a_3$
  sends the endpoint of session $s$, which replaces variable $y_5$; afterwards,
  the dual endpoint is used to send the names $w_1,w_2$, substituting the
  variables $x_1,x_2$. 
  % Here, we remark that prefix $\abinp{s}{x_1,x_2}$ bounds 
  % variables $x_1,x_2$ in the breakdown of the continuation. 
  \begin{align*}
    \F{P} \red^{14}~&  \news{\widetilde c_{\bullet\bullet}} \news{s} ( \ 
    \apropout{5}{} \Par 
    \AmeliaMod{5}{\epsilon}{\subpr' }_\empg 
    \Par 
    \highlighta{\about{\dual s}{w_1, w_2}} \Par 
    \highlighta{\binp{s}{x_1, x_2}}
     \big(\apropout{17}{} \Par 
     \AmeliaMod{17}{\epsilon}{\subpq' }_\empg \big)  \big) \smallskip 
 %   \\ \bigskip & \quad \quad \text{where 
    %$\widetilde c = (\prop_5, \dots, c_{12}, c_{17}, \dots, c_{25})$}
    \\
    %next reduction
      \red~ & \news{\widetilde c_{\bullet\bullet}}\big( \ 
        \apropout{5}{} \Par \AmeliaMod{5}{\epsilon}{\subpr' }_\empg \Par   
                 \apropout{17}{} \Par 
        \AmeliaMod{17}{\epsilon}{\subpq'}_\empg \subst{w_1 w_2}{x_1 x_2}  \big) 
        = V 
  \end{align*}
where 
      $\widetilde c_{\bullet \bullet} = (\prop_5, \dots, c_{12}, c_{17}, \dots, c_{25})$.

  Here, we remark that prefix $\abinp{s}{x_1,x_2}$ binds 
  variables $x_1,x_2$ in the breakdown of the continuation 
  (i.e., $\AmeliaMod{17}{\epsilon}{\subpq' }_\empg$). 
  Thus, there is no need for propagators to pass contexts: 
  propagators here only serve to enforce the ordering of actions. 
  On the other hand, this relies on a process nesting
  that is induced by the application of encoding 
  $\map{\cdot}^2$ in the composition. 
   Thus, the trio structure is lost. 
  % ( which is by the final application of encoding 
  % $\map{\cdot}^2$ in the composition.  )

  Undoubtedly, the first reduction of  process $P$ in \eqref{ex:procpi}
  has been simulated. We may notice that in $V$ names $w_1$, $w_2$ substitute 
  $x_1$, $x_2$ and the subsequent reduction on $w$ can be simulated on name $w_1$. 
  The subsequent reductions follow the same pattern. 
  % Futher, we remark that prefix 
  % Undoubtedly, the behaviour of the initial first-order process has been 
  % simulated. 
  Thus, the outcome of our decomposition function is a behaviorally
  equivalent process that is typed with \msts.

\end{example}

%%% > ONE COLUMN STARTS 
\begin{figure}[!t]
  \begin{mdframed}
  \begin{align*}
  &\AmeliaMod{1}{\epsilon}{P\subst{r_1}{r}}_\es =
       \news{s_1} \big(  \propinp{1}{}\propout{2}{}
       \apropout{4}{} \Par \\
       & \qquad \qquad \qquad \qquad \qquad
       \propinp{2}{} \news{a_1}\big( \bout{s_1}{a_1} \big( \apropout{3}{}
       \Par \propinp{3}{}\inact \Par
       \\
       & \qquad  \qquad \qquad \qquad \qquad \qquad \qquad \qquad \propinp{4}{}\binp{s_1}{z_x}
       \apropout{5}{z_x} 
       \Par
       \\
       &\qquad \qquad \qquad \qquad \qquad \qquad \qquad \qquad \qquad 
       % \AmeliaMod{5}{z_x}{P'\subst{r_1}{r}}_{\{X \rightarrow {r}\}} 
       R^5
       \Par 
       % \\
       % &\qquad \qquad \qquad \qquad \qquad
       \repl\binp{a_1}{y'_1}
       \binp{y'_1}{x_{r_1}, x_{r_2}, y_1}\widehat{P}\, \big) \big )\big) 
  \\
  &\text{where:} 
  \\
  &\widehat{P} =
         \news{\widetilde \prop}
             \big(\proprinp{r}{b}\news{s'} \big(\bout{b}{s'}\about{\dual s'}
             {x_{r_1}, x_{r_2}} \big)
             \Par \apropout{1}{} \Par 
              \\
              &\quad \quad \quad \quad \quad \qquad  
             \propinp{1}{}\binp{y_1}{z_x}\apropout{2}{z_x} 
             \Par  
             R^2 \subst{x_{r_1}, x_{r_2}}{r_1,r_2}
             \big) 
  \\
  & R^{k}  =
       \propinp{k}{z_x} 
       \\
       &\qquad \qquad 
       \news{a_1}\big(\proprout{r}{a_1}
       (\news{s_1} \big( \propinp{k+1}{y}\propout{k+2}{y}\apropout{k+3}{}
       \Par
       \\
       &\qquad \qquad \propinp{k+2}{y}\news{s} \big(\bout{y}{s}\about{\dual s}{s_1}) \Par 
        \\
        & \qquad \qquad \qquad \quad \quad 
       \propinp{k+3}{}\news{a_2} \big(\bout{s_1}{a_2} \big(
       \apropout{k+l+4}{} \Par 
       \\
       &\qquad \qquad \qquad \qquad\qquad \propinp{k+l+4}{}\inact \Par 
       {\binp{a_2}{y_2}}\highlighta{\binp{y_2}{w_1}} 
       \\
       & \qquad \qquad \qquad \qquad \qquad \quad
        \news{\widetilde \prop} \big(\apropout{k+4}{z_x}
       \Par \AmeliaMod{k+4}{z_x}{\bout{r_2}{-w_1}X}_g
       \big) \big) \big)  \Par
       \\
       & \qquad \qquad \qquad\qquad \qquad \qquad \qquad \binp{a_1}{y_1}\binp{y_1}{z_1,z_2}
       \highlighta{\binp{z_{1}}{y}}
       \\
       & \qquad \qquad\qquad\qquad \qquad \qquad \qquad
       \propout{k+1}{y}\proprinp{r}{b}
       \news{s'} \big(\bout{b}{s'}
       \about{\dual s'}{z_1,z_2} \big) \big) \big) \big)
       \\
  & 
  \AmeliaMod{k+4}{z_x}{\bout{r_2}{-w_1}X}_g =
  \propinp{k}{z_x}
  \news{a_1}{}\proprout{r}{a_1}
  \\
  & \qquad \qquad \qquad \qquad \qquad \qquad 
  (\AmeliaMod{k+7}{z_x}{X}_g
  \Par \binp{a_1}{y_1}\binp{y_1}{\widetilde z}W \big) 
  \\
  %  &
  %  \text{where:} 
  %  \\ 
  &
  W =  \news{a_2}\big(
    \highlighta{\bout{z_{2}}{a_2}}(\propout{k+7}{z_x}\proprinp{r}{b}
        \news{s}(\bout{b}{s}\about{\dual{s}}{\widetilde z} \big) \Par
         \\
         &
        \qquad \qquad \qquad  \binp{a_2}{y_2}\binp{y_2}{z'_1}\news{\widetilde \prop}
        (\apropout{k+5}{} \Par 
        \\
        & 
        \qquad \qquad \qquad \qquad \qquad \qquad \qquad \qquad \propinp{k+5}{}\binp{z'_1}{x}\apropout{k+6}{x} \Par 
        \\ 
        & \qquad \qquad \qquad \qquad \qquad \qquad \qquad \qquad \qquad
        \propinp{k+6}{x}\news{s'}(\bout{x}{s'}
        \highlighta{\about{\dual s'}{-w1}} \big) \big) \big) \big)
  \\
  & \AmeliaMod{k+7}{z_x}{X}_g = 
  \news{s_1}\big( \propinp{k+7}{z_x}\propout{k+8}{z_x}
  \apropout{k+9}{z_x} 
  \Par 
  \\ 
  & \qquad \qquad \qquad \propinp{k+8}{z_x}\news{a_1}
  \big( \proprout{r}{a_1} \big(\propinp{k+9}{z_x}
  \bout{\dual s_1}{z_x}\apropout{k+10}{}
  \Par 
  % \\
  % & \qquad \qquad \qquad \qquad \qquad \quad 
  % \apropout{k+10}{}\propinp{k+10}{}\inact \Par
  \apropinp{k+10}{} \Par
  \\
  & \qquad \qquad \qquad \qquad \qquad \quad 
  \binp{a_1}{y_1}\binp{y_1}{r_1,r_2} \news{s'}(\bout{z_x}{s'}
  \about{\dual s'}{r_1,r_2, s_1} \big) \big) \big) \big) \big) \big)\\
  & \text{with } g = \{ X \mapsto r_1,r_2 \}
  \end{align*}
  \end{mdframed}
  \caption[Breakdown of a recursive process (\Cref{pi:ex:recproc}) ]{Breakdown of a recursive process (\Cref{pi:ex:recproc}) \label{pi:ex:am-rec}}
  \end{figure}

\begin{example}[A Recursive Process]
  \label{pi:ex:recproc}
Let $r$ be a channel with the tail-recursive session type $S = \trec{t}\btinp{\tint}\btout{\tint}\tvar{t}$.  
      We decompose $r$ using $S$ and obtain two channels 
    typed with \msts as in~\Cref{pi:f:decomp:firstordertypes}:
    \begin{align*}
     r_1:\trec{t}{\bbtinp{\lrangle{\btinp{ \btinp{\lrangle{\btinp{\tint}{\tinact}}}{\tinact} }
     {\tinact}}}{\vart{t}}}
     \\
     r_2 : 
     \trec{t}{\bbtout{\lrangle{\btinp{ \btinp{\lrangle{\btinp{\tint}{\tinact}}}
     {\tinact} }{\tinact}}}{\vart{t}}}
    \end{align*}
  Consider now the process $P=\recp{X}\binp{r}{w}\bout{r}{-w}X$. Let us write $P'$ to denote the ``body'' of $P$, i.e., $P' = \binp{r}{w}\bout{r}{-w}X$.
  Then, process $\F{P}$ is
     \begin{align*}
    %   &= \news{\widetilde \prop}
      \news{\widetilde \prop}\news{\prop^r}
      \big(
        \proprinp{r}{b}\news{s}\big(\bout{b}{s}\about{\dual s}{r_1,r_2}\big)
        \Par  
        \apropout{1}{}
      \Par \AmeliaMod{1}{\epsilon}{P\subst{r_1}{r}}_\es \big) 
      % P^r = \proprinp{r}{b}\news{s}(\bout{b}{s}\about{\dual s}{\wtd r})
    \end{align*}

    \noindent where $\wtd \prop=(\prop_1,\ldots,\prop_{\lenHO{P}})$ and 
     $\AmeliaMod{1}{\epsilon}{P\subst{r_1}{r}}_\es $ is in~\Cref{pi:ex:am-rec}. 
    
    In~\Cref{pi:ex:am-rec}, $\AmeliaMod{1}{\epsilon}{P\subst{r_1}{r}}$ simulates
    recursion in $P$ using replication. 
	Given some index $k$, process $R^k$ mimics actions of the recursive body. 
    It first gets a decomposition of $r$ by interacting 
    with the process providing recursive names on $c^r$ (for the first instance, 
    this is a top-level process in $\F{P}$). 
    Then, it mimics the first input action on the 
    channel received for $z_1$ (that is, $r_1$): the input of actual names for 
    $w_1$ is delegated through channel redirections to name $y_2$ (both prefixes are highlighted
    in~\Cref{pi:ex:am-rec}). Once the recursive name is used, the 
    decomposition of recursive name is made available for the 
    breakdown of the continuation by a communication on $c^r$. 
    Similarly, in the continuation, the second action on $r$, output, is mimicked by
    $r_2$ (received for $z_2$), with the output of actual name $w_1$ delegated to
    $\dual s'$ (both prefixes are highlighted in~\Cref{pi:ex:am-rec}). 

    Subprocess $R^5$ is a breakdown of the first instance of the recursive body.
    % Intuitively, $R^5$ will send recursive names along name $a_1$ for an use in
    % a next instance.  
    The replication guarded by $a_1$ produces a next instance,
    i.e.,  
    process $R^2\subst{x_{r_1}, x_{r_2}}{r_1,r_2}$ in $\widehat{P}$. By
     communication on $a_1$ and a few reductions on propagators, it gets
     activated: along $a_1$ it first receives a name for $y'_1$ along which it
     also receives: \rom{1} recursive names $r_1,r_2$ for variables $x_{r_1},
     x_{r_2}$, and \rom{2} a name for $y_1$ along which it will receive      $a_1$ again,  for future
     instances, as it can be seen in $\AmeliaMod{k+7}{z_x}{X}_g$.

    \end{example}

% \begin{example}[Recursive process]
% \label{pi:ex:recproc}
%     Let $S = \trec{t}\btinp{\tint}\btout{\tint}\tvar{t}$ be a tail-recursive session type. 
%       Also, let $P=\recp{X}P'$ be a process implementing a channel $r$ with type
%       with $S$, with $P' = \binp{r}{w}\bout{r}{-w}X$.  
%     We decompose $r$ using $S$ and obtain two channels 
%   typed with \msts as in  \figref{f:decomp:firstordertypes}. We have:
%   \begin{align*}
%    r_1:\trec{t}{\bbtinp{\lrangle{\btinp{ \btinp{\lrangle{\btinp{\tint}{\tinact}}}{\tinact} }
%    {\tinact}}}{\vart{t}}}
%    \\
%    r_2 : 
%    \trec{t}{\bbtout{\lrangle{\btinp{ \btinp{\lrangle{\btinp{\tint}{\tinact}}}
%    {\tinact} }{\tinact}}}{\vart{t}}}
%   \end{align*}
%    The process $\AmeliaMod{k}{\tilde x}{P}_g$ is given in \figref{ex:am-rec} (\appref{app:ss:exrec}).
%    \end{example}

\subsection{Results}
\label{ss:results}
\cralt{
\todo[inline]{I don't like that we refer to notions in ``in Theorem 3.10 and Theorem 3.27 of \cite{APV19}'' - it is not self-contained. If we really need something we should have it in Section 2 (and then point to that section, not to \cite{APV19}). Please revise.}}{}

We close this section by establishing the minimality result for \sessp using the typability of  $\F{\cdot}$. We need some auxiliary definitions to characterize
the propagators required to decompose recursive processes. 
    % \todo[inline]{We need to give some intuitions about the environments,
    %  otherwise the next definition is not self-contained.}

\thesisalt{
  \Cref{mst:t:typecore}}{\Cref{mst:t:typecore}} states typability results by 
introducing two typing environments, denoted $\Theta$ and $\Phi$. 
While environment $\Theta$ is used to type linear propagators (e.g., $\prop_{k},
\prop_{k+1}, \ldots$) generated by the breakdown function $\B{-}{-}{\cdot}$,
environment $\Phi$ types shared propagators used in trios that propagate
breakdown of recursive names (e.g., $c^r, c^v,\ldots$ where $r$ and $v$ are
recursive names). 
% generated
% by $\B{-}{-}{\cdot}$
% when a process has free recursive names. 

\begin{definition}[Session environment for propagators]
  \label{pi:ThetaPrimeDef}
  Let $\Theta$ be the session environment and $\Phi$ be the recursive propagator
  environment defined 
  \thesisalt{in~\Cref{mst:t:typecore}}{in~\Cref{mst:t:typecore}}. Then, by applying the encoding $\tmap{\cdot}{2}$, we define
  $\Theta'$ and $ \Phi'$ as follows: $\Theta' = \tmap{\Theta}{2}, \Phi' =
  \tmap{\Phi}{2}$.
  \end{definition}

We can use $\Theta' = \tmap{\Theta}{2}$ in the following statement, where we 
state the typability result for the breakdown function.
The proof composes previously established results:

\begin{restatable}[Typability of Breakdown]{lemm}{thmamtyprec}
  \label{pi:t:thmamtyprec}
  \label{pi:t:am-typ-rec}
    Let P be an initialized \sessp process. If \ $\Gamma; \Delta, \envR \proves P \hastype \Proc$, then 
    $\mGt{\Gamma}, \Phi';\mGt{\Delta} \cat
      \Theta' \proves \AmeliaMod{k}{\epsilon}{P}_g \hastype \Proc $, 
      where: 
      \begin{itemize} 
      	\item $k > 0;$ 
      	\item  $\widetilde{r} = \dom{\envR}$; 
      	\item  $\Phi' = \parcomp{r \in \tilde{r}}{\prop^r : \chtype{\chtype{\btinp{\mRts{}{}{\envR(r)}}\tinact}}  };$
      	\item $\balan{\Theta'}$ with 
      	 $$\dom{\Theta'} = \{\prop_k,\prop_{k+1},\ldots,\prop_{k+\lenHO{P}-1}\} 
       \cup \{\dual{\prop_{k+1}},\ldots,\dual{\prop_{k+\lenHO{{P}}-1}}\}$$
        such that $ \Theta'(\prop_k)=
      \btinp{\cdot} \tinact $. 
      \end{itemize}
%
%
%      
%      \ and \  
%       with \ 
%       $$\dom{\Theta'} = \{\prop_k,\prop_{k+1},\ldots,\prop_{k+\lenHO{P}-1}\} 
%       \cup \{\dual{\prop_{k+1}},\ldots,\dual{\prop_{k+\lenHO{{P}}-1}}\}$$
%       such that $ \Theta'(\prop_k)=
%      \btinp{\cdot} \tinact $. 
\end{restatable}

\begin{proof}
  Directly from
  \thesisalt{\Cref{mst:t:typecore}}{\Cref{mst:t:typecore}}
  and from Theorems 5.1 and 5.2 from~\cite{DBLP:journals/iandc/KouzapasPY19}. 
See \longversion{\Cref{pi:app:thmamtyprec}}{\cite{AAP21-full}} for details.
\end{proof}

  %%% Decomposition %%%%
  We now consider typability for the decomposition function, using 
$\Phi' =  \tmap{\Phi}{2}$ as in~\Cref{pi:ThetaPrimeDef}. 
% The proof follows from~\Cref{pi:t:am-typ-rec}; see~\longversion{\Cref{pi:app:thmamtyprecdec}}{\cite{AAP21-full}}.

  \begin{restatable}[Minimality Result for \sessp]{theorem}{thmamtyprecdec}
    \label{pi:t:amtyprecdec}
    Let $P$ be a closed \sessp process, with $\widetilde u = \fn{P}$ and $\widetilde v = \rfn{P} $.
    If \ $\Gamma;\Delta, \Delta_\mu \proves P \hastype \Proc$, where $\Delta_\mu$ only involves recursive session types, then \\
    $\mGt{\Gamma \sigma};\mGt{\Delta \sigma}, \mGt{\Delta_\mu \sigma} \proves \F{P} \hastype \Proc$, where
    $\sigma = \subst{\mathsf{init}(\widetilde u)}{\widetilde u}$.
  \end{restatable}

  \begin{proof}
    Directly by using \thmref{pi:t:am-typ-rec};    
    see~\Cref{pi:app:thmamtyprecdec} for details.
  \end{proof}

While \Cref{pi:t:amtyprecdec} gives a useful \emph{static} guarantee about the correctness of our decomposition of $P$ into $\F{P}$, a \emph{dynamic} guarantee that confirms that $P$ and $\F{P}$ are behaviorally equivalent is most relevant. Before establishing such a dynamic guarantee, we explore to what extent $\F{P}$ can be optimized, i.e., whether redundancies induced by the ``decompose by composing'' approach can be safely eliminated.

\section{An Optimized Decomposition}
\label{pi:s:optimizations}

% \section{Optimizations}
% \label{pi:s:optimizations}

%%%% BEGIN MOTIVATION FOR OPTIMIZATION
Although conceptually simple, the ``decompose by composing'' approach to the minimality result induces some suboptimal features. Considering this, in this section we propose $\Do{\cdot}$, an optimization of $\F{\cdot}$ with less synchronizations and direct support for recursion, 
%breakdown function $\AmeliaMod{k}{\tilde x}{\cdot}_g$ 
and establish its correctness properties,  in terms of its corresponding minimality result (static correctness, \Cref{pi:t:decompcore}, Page~\pageref{pi:t:decompcore}) but also \emph{dynamic correctness} (\Cref{pi:t:mainbsthm}, Page~\pageref{pi:t:mainbsthm}).

\subsection{Motivation: Suboptimal Features of ``Decompose by Composing''}
To motivate our insights,  consider  the process $\AmeliaMod{k}{\tilde
x}{\binp{u_i}{w}{Q}}_g$ as presented in~\Cref{pi:ss:idea} and~\Cref{pi:table:repo}. We identify some \emph{suboptimal features} of a  
decomposition based on $\AmeliaMod{k}{\tilde
x}{\cdot}_g$: 

\begin{description}
	\item[Channel redirections] 
While the given process receives a name for variable $w$ along $u_i$, 
its breakdown does not input a breakdown of $w$ directly, but does so through a series of 
channel redirections:  $u_i$ receives a name along which 
it sends the restricted name $s$, along which it sends the restricted name $s_1$ 
and so on. Finally, the name received for $y'$ receives $\wtd w$, the breakdown of $w$.
This redundancy is perhaps more evident in~\Cref{pi:decomp:firstordertypes}, which gives the translation of types by composition: 
 the mimicked input action is five-level nested for 
 the original name. 
This resulting type is due to the composition of $\map{\cdot}^1_g$ and $\map{\cdot}^2$.   

\item[Redundant synchronizations on propagators]
Also, $\AmeliaMod{k}{\tilde x}{\binp{u_i}{w}{Q}}_g$ features redundant communications on propagators. 
For example, the bound name $y$ is locally propagated 
by $\prop_{k+1}$ and $\prop_{k+2}$. 
This is the result of breaking down sequential prefixes induced by 
 $\map{\cdot}^1_g$ (not present in the original process).  
%  and thus do not concern sequentiality in the original process.
 
\item[Trio structure is lost] Last but not least, the trio structure is lost as 
 subprocess $\widehat{Q}_{\tilde{x}}$ is guarded and nested, and 
 it 
 inductively invokes the function on continuation $Q$. 
 This results in an arbitrary level of process nesting, which is induced by the final application of encoding 
 $\map{\cdot}^2$ in the composition.  
\end{description}

The shortcomings of $\AmeliaMod{k}{\tilde x}{\cdot}_g$ are also evident in the treatment of recursive processes  and names. 
Because \HO does not feature recursion constructs, 
 $\map{\cdot}^1_g$ encodes them by  relying on abstraction passing 
 and shared abstractions. 
 Then, going back to \sessp via $\map{\cdot}^2_g$, 
 these representations are translated to a process involving 
 a replicated subprocess. 
However, going through this path the encoding of recursive process becomes convoluted. Moreover, all non-optimal features  already discussed for the case of input are also present in the decomposition of recursion. 

 %%% END MOTIVATION FOR OPTIMIZATION 

% \todo[inline]{continue}

\smallskip

Based on these observations, here we develop an optimized decomposition function, denoted  $\Do{\cdot}$ (\Cref{pi:def:decomp}),  
 that avoids the shortcomings described above. 
% it does not contain channels redirections, introduce propagators 
% only for mimicking sequentility of the original process, 
The optimized decomposition relies on a streamlined breakdown function, denoted $\Bopt{k}{\tilde x}{\cdot}$, which produces a composition of {trios processes}, with a fixed maximum number of nested prefixes. 
% and without
% parallel process nesting. 
The decomposed process avoids channel redirections and only introduces propagators that codify the sequentiality of the original process.

\paragraph{Roadmap.} To facilitate reading, we summarize notations and definitions related to $\F{\cdot}$ and their corresponding notions for the optimized decomposition $\Do{\cdot}$---see \Cref{pi:t:summ}.

\begin{table}[t!]
\begin{center}
  \resizebox{15.2cm}{!}{
\begin{tabular}{|l||c|c|}
\hline
                        & \Cref{pi:s:dbc} & This section \\ \hline \hline
Degree of  $P$    &   $\lenHO{P}$ (\Cref{pi:def:am-sizeproc})
& $\lenHOopt{P}$ (\Cref{pi:def:sizeproc})                                    \\ \hline
Decomposition of $S$ & $\mGt{S}$ and $\mRt{S}$ (\Cref{pi:f:decomp:firstordertypes})                                   & $\Gtopt{S}$ and $\Rtopt{S}$ (\Cref{pi:f:tdec})                                     \\ \hline
Breakdown of $P$     & $\AmeliaMod{k}{\tilde x}{P}_g$
(\Cref{pi:table:repo,pi:table:repo2})                                   & $\Bopt{k}{\tilde x}{P}$ and $\Brecpi{k}{\tilde x}{P}_g$ (\Cref{pi:t:bdowncorec,pi:t:bdowncorec-rec})                                \\ \hline
Decomposition of  $P$  & $\F{P}$ (\Cref{pi:def:am-decomp})                                  & $\Do{P}$  (\Cref{pi:def:decomp})                                  \\ \hline
  \end{tabular}
  }
\end{center}
\caption{Summary of notations used in our two decompositions. \label{pi:t:summ}}
\end{table}

\subsection{Preliminaries}
As before, we need to decompose a session type into a \emph{list} of minimal session types:
\begin{definition}[Decomposing Types] 
\label{pi:def:typesdecomp}
Let $S$ and $C$ be a session and a channel type, resp. 
(cf.~\Cref{top:f:sts}). 
 The \emph{type decomposition
function}  $\Gtopt{\cdot}$ is defined in~\Cref{pi:f:tdec}.
\end{definition}
As before, we need two auxiliary functions for decomposing recursive types, denoted
$\Rtopt{\cdot}$ and $\Rtsopt{}{}{\cdot}$.

A comparison between $\Gtopt{\cdot}$ and $\mGt{\cdot}$ (\Cref{pi:f:decomp:firstordertypes}) is already useful to understand the intent and scope of the optimized decomposition. Consider, e.g., the decompositions of the session type $\btout{C}{S}$ (with $S \neq \tinact$):
\begin{align*}
	     \mGt{\btout{C}{S}} &= \bbtout{ \lrangle{ \btinp{ \btinp{ \lrangle{ \btinp{ \mGt{C} }{\tinact} } }{\tinact} }{\tinact} } }{\tinact} , \mGt{S}
\\
\Gtopt{\btout{C}{S}} &= \btout{\Gtopt{C}}{\tinact}\, ,\Gtopt{S}
\end{align*}

These differences at the level of induced \msts will be useful in our formal comparison of the two decompositions, in \Cref{pi:ss:measure}.

\cralt{
 \todo[inline]{When giving examples about the optimized notions, it is useful to contrast with the results obtained via the analog notions for the non-optimized case.}}{}
\begin{example}[Decomposing a Recursive Type]
  \label{pi:ex:rtype}
  Let
  $S = \trec{t}S'$ be a recursive session type, with $S'=\btinp{\tint}\btinp{\bool}\btout{\bool}\vart{t}$.
  By~\Cref{pi:f:tdec},
   since $S$ is tail-recursive,  $\Gtopt{S} = \Rtopt{S'}$. 
  Further, 
  $$\Rtopt{S'} = \trec{t}\btinp{\mathsf{\Gtopt{\tint}}} \vart{t},
  \Rtopt{\btinp{\bool}\btout{\bool}\vart{t}}$$ By definition of
  $\Rtopt{\cdot}$, we obtain 
  $$\Gtopt{S} = \trec{t}\btinp{\tint} \vart{t},
  \trec{t}\btinp{\bool} \vart{t}, \trec{t}\btout{\bool}
  \vart{t}, \Rtopt{t}$$ 
  (using $\Gtopt{\tint} = \tint$ and
  $\Gtopt{\bool} = \bool$). Since 
  $\Rtopt{\vart{t}} = \epsilon$, we have 
  $$\Gtopt{S} = \trec{t}\btinp{\tint} \vart{t}, \trec{t}\btinp{\bool} \vart{t}, 
  \trec{t}\btout{\bool} \vart{t}$$  	
  %\hspace*{\fill} $\lhd$
  \end{example}

  \begin{example}[Decomposing an Unfolded Recursive Type]
   Let $T = \btinp{\bool}\btout{\bool}S$ be a derived unfolding of  $S$ from~\Cref{pi:ex:rtype}. Then, by~\Cref{pi:f:tdec}, 
    $\Rtsopt{}{s}{T}$ is the list of minimal recursive 
    types obtained as follows:  first, 
    $\Rtsopt{}{s}{T} = \Rtsopt{}{s}{\btout{\bool}\trec{t}S'}$ and after one more step, $\Rtsopt{}{s}{\btout{\bool}\trec{t}S'} = \Rtsopt{}{s}{\trec{t}S'}$. Finally, we have $\Rtsopt{}{s}{\trec{t}S'} = \Rtopt{S'}$. 
    We get the same list of minimal types as in~\Cref{pi:ex:rtype}: 
    $$\Rtsopt{}{s}{T} = \trec{t}{\btinp{\tint}\vart{t}}, \trec{t}{\btinp{\bool}\vart{t}}, \trec{t}{\btout{\bool}
    \vart{t}}$$ 
  %  \hspace*{\fill} $\lhd$
  \end{example}
 
\begin{figure}[!t]
\begin{mdframed}[style=alttight]
\begin{align*}
% \begin{tabular}{l l}
% \begin{tabular}{l}
% \end{tabular}	\\
% \begin{tabular}{c}
  \Gtopt{\btout{C}{S}} &=
  \begin{cases}
  \btout{\Gtopt{C}}{\tinact}  &  \text{if $S = \tinact$} \\
  \btout{\Gtopt{C}}{\tinact}\, ,\Gtopt{S} & \text{otherwise}
  \end{cases} \\
  \Gtopt{\btinp{C}{S}} &=
  \begin{cases}
  \btinp{\Gtopt{C}}{\tinact}  &  \text{if $S = \tinact$} \\
  \btinp{\Gtopt{C}}{\tinact}\, , \Gtopt{S} & \text{otherwise}
  \end{cases} \\
   \Gtopt{\tinact} &= \tinact \\
  \Gtopt{\chtype{S}} &= \chtype{\Gtopt{S}} \\
  \Gtopt{S_1,\ldots,S_n} &= \Gtopt{S_1},\ldots,\Gtopt{S_n} \\
  \Gtopt{\trec{t}{S'}} &= \Rtopt{S'} \\
  \Gtopt{S} &= \Rtsopt{}{}{S} \quad \text{where} \ S \neq \trec{t}S' \\
  \Rtopt{\vart{t}} &= \epsilon  \\
  \Rtopt{\btout{C}S} &= \trec{t}{\btout{\Gtopt{C}}} \tvar{t}, \Rtopt{S} \\
  \Rtopt{\btinp{C}S} &= \trec{t}{\btinp{\Gtopt{C}}} \tvar{t}, \Rtopt{S} \\ 
  \Rtsopt{}{}{\btinp{C}S} &=  
  \Rtsopt{}{}{\btout{C}S} = \Rtsopt{}{s}{S} \\ 
  \Rtsopt{}{}{\trec{t}{S}} &= \Rtopt{S}
% \end{tabular}	
% \end{tabular}
\end{align*}
\end{mdframed}
\caption[Optimized decomposition of session types $\Gtopt{\cdot}$]{Optimized decomposition of session types $\Gtopt{\cdot}$ (cf.~\Cref{pi:def:typesdecomp}) \label{pi:f:tdec}}
% \vspace{-3mm}
\end{figure}

\begin{definition}[Decomposing Environments] \label{pi:def:typesdecompenv}
  Given environments $\Gamma$ and $\Delta$, we  define $\Gtopt{\Gamma}$ and $\Gtopt{\Delta}$ inductively as $\Gtopt{\emptyset} =  \emptyset$ and
  \begin{align*}
        \Gtopt{\Delta \cat u_i:S} &= \Gtopt{\Delta},(u_i,\ldots,u_{i+\len{\Gtopt{S}}-1}) : \Gtopt{S} 
        \\ 
      \Gtopt{\Gamma \cat u_i:\chtype{S}} &= \Gtopt{\Gamma} \cat u_i : \Gtopt{\chtype{S}} 
&
  \end{align*}
\end{definition}

We now define the (optimized) degree of a process. A comparison with the previous definition (\Cref{pi:def:am-sizeproc}) provides further indication of the improvements induced by optimized decomposition.
\begin{definition}[Degree of a Process]
  \label{pi:def:sizeproc}
	The \emph{optimized degree} of a process $P$, denoted $\lenHOopt{P}$, is inductively defined as follows:
	$$
	\begin{cases}
	\lenHOopt{Q} + 1 & \text{if $P =\bout{u_i}{y}Q$ or $P=\binp{u_i}{y}Q$}
	\\
	\lenHOopt{Q} & \text{if $P = \news{s:S}Q$}
	\\
  \lenHOopt{Q} + 1 & \text{if $P = \news{r:S}Q$ and $\traux{S}$}
	\\
	\lenHOopt{Q} + \lenHOopt{R} + 1 & \text{if $P = Q \Par R$}
	\\
	1 & \text{if $P = \inact$ or $P = \rvar{X}$} \\
  \lenHOopt{Q} + 1 & \text{if } P = \recp{X}Q 
  % \\
  % 1 & \text{if } P = \rvar{X}
	\end{cases}
	$$
	\end{definition}

%\begin{definition}[Process Initialization]
%  \label{pi:d:counterinit}
As before, given a finite tuple of names
$\widetilde u = (a,b,s,s',\ldots)$, we write $\mathsf{init}(\widetilde u)$ to denote the tuple
$(a_1,b_1,s_1,s'_1,\ldots)$; recall that a process is {initialized} if all of its names are indexed.
%\end{definition}

Given two tuples of indexed names $\wtd u$ and $\wtd x$,  it is useful to 
collect those names in $\wtd x$ that appear 
in $\wtd u$.  
\begin{definition}[Free indexed names]
  \label{pi:d:fnb}
	Let $\wtd u$ and $\wtd x$ be two tuples 
	of names. We define the set  
	$\fnb{\wtd u}{\wtd x}$ as $\{ z_k : z_i \in \wtd u \wedge z_k \in \wtd x\}$.
\end{definition}
 
As usual, we treat sets of names as tuples (and vice-versa). By abusing
notation, given a process $P$, we shall write $\fnb{P}{\wtd y}$ to stand for
$\fnb{\fn{P}}{\wtd y}$. Then, we have that $\fnb{P}{\widetilde x} \subseteq
\widetilde x$. In the definition of the breakdown function, this notion allows
us to conveniently determine a \emph{context} for a subsequent trio.  

\begin{definition}\label{pi:d:fnv}
  Given a process $P$, we write $\recpx{P}$ 
  to denote that $P$ has a free recursive variable. 
\end{definition}

% \begin{definition}[Predicates on Names]
% 	Let $u$ be a name  with session type $C$ (i.e., $u:C$). We define the predicate $\lin{u}$ if $C = S$. 
% \end{definition}

% \begin{definition}[Subsequent index substitution]
%   \label{pi:d:nextn}
%   Let $n_i$ be an indexed name. We define $\nextn{n_i} = \linecondit{\lin{n_i}}{\incrname{n}{i}}{\{\}}$.
% %  follows 
% %  \begin{align*} 
% %    \nextn{n_i}= \begin{cases}
% %      \incrname{n}{i} & \text{if  } \ \lin{n_i} \\
% %      \{\} & \text{otherwise}
% %      \end{cases}
% %  \end{align*}
% \end{definition}

  \begin{remark}
    % \paragraph{Remark}
    \label{pi:r:prefix}
 Whenever 
$\apropinp{k}{\wtd y}$ (resp. $\apropout{k}{\wtd y}$) with $\wtd y = \epsilon$,
 we shall write $\apropinp{k}{}$ (resp. $\apropout{k}{}$) 
 to stand for $\apropinp{k}{y}$ (resp. $\apropout{k}{y}$) such that $\prop_{k} : \btinp{\chtype{\tinact}}\tinact$ (resp. $\dual {\prop_k}:\btout{\chtype{\tinact}}\tinact$).
\end{remark}

% \subsubsection{Core fragment} 
\begin{definition}[Index function]
  \label{pi:def:indexfunction}
  Let $S$ be an (unfolded) recursive session type. The function $\indf{S}$ is defined as follows:  
  \begin{align*}
   \indf{S} = \begin{cases}
     \indfaux{0}{S'\subst{S}{\vart{t}}} & \text{if $S =\trec{t}{S'}$} \\
     \indfaux{0}{S} & \text{otherwise}
   \end{cases}
\qquad 
\indfaux{l}{T} = 
\begin{cases}
	\indfaux{l+1}{S} & \text{if $T = \btout{U}S$ or $T = \btinp{U}S$} 
	\\
	\len{\Rtopt{S}} - l + 1 & \text{if $T =\trec{t}{S}$}
\end{cases}
    \end{align*}
  \end{definition}

\begin{definition}[Name breakdown]
\label{pi:d:name-breakdown}
Let $u_i:C$ be an indexed name with its session type. 
We write $\bname{u_i:C}$ to denote 
$$\bname{u_i:C}=(u_i,\ldots,u_{i + \len{\Gtopt{C}} - 1})$$
We extend $\bname{\cdot}$ to work on lists of assignments (name, type), as follows: 
$$\bname{(u^1_i, \ldots, u^n_j) : (C_1, \ldots, C_n)} =
\bname{u^1_i:C_1} \cdot \ldots \cdot \bname{u^n_j:C_n}$$
\end{definition}
 
%\begin{definition}[Recursive subprocess predicate]
%  \label{pi:d:recpx}
  % and only if there is $Q$ such that 
  % $Q.\rvar{X}$ is a subprocess of $P$.  
%\end{definition}

  % \begin{definition}[Depth of recursive variable]
  %   Let $P$ be a recursive process. 
  %   We define \rvardepth{P} to count sequential prefixes in $P$ 
  %   preceding recursive variable $X$ or a subprocess of shape $\recp{X}Q$. 
  %   Function \rvardepth{P} and \rvardepthaux{P} are mutually defined as follows: 
  %   \begin{align*}
  %     \rvardepth{P} &= 
  %     \begin{cases}
  %       \rvardepthaux{P} & \text{ if } \recpx{P} \\
  %         0 & \text{ otherwise } \\
  %     \end{cases}
  %     \\ 
  %     \rvardepthaux{P} &= 
  %     \begin{cases}
  %       \rvardepth{Q} + 1 & \text{ if } P = \alpha.Q \\
  %       \rvardepth{Q} + \rvardepth{R}  & \text{ if } P = Q \Par R \\
  %       0 & \text{ if } P = \recp{X}Q \text{ or } P = \rvar{X} 
  %       \\
  %       \rvardepth{Q} & \text{ if } P = \news{s}Q
  %     \end{cases}
  %     % \rvardepthaux{\alpha.Q} &= \rvardepth{Q} + 1\\ 
  %     % \rvardepthaux{(Q \Par R)} &= \rvardepth{Q} + \rvardepth{R} \\
  %     % \rvardepthaux{\recp{X}Q} &= 0 \\ 
  %     % \rvardepthaux{\rvar{X}} &= 0 \\
  %     % \rvardepthaux{\news{s}Q} & = \rvardepth{Q}
  %   \end{align*}
  % \end{definition}

\subsection{The Optimized Decomposition}
\begin{table}[t!]
\begin{center}
  % \begin{table*}
  \begin{tabular}{|l|l|l|}
    \rowcolor{gray!25}
    \hline
     & $P$ &
      \multicolumn{1}{l|}{
    \begin{tabular}{l}
      \noalign{\smallskip}
      $\Bopt{k}{\tilde x}{P}$
      \smallskip
    \end{tabular}
  }  \\
    \hline
    \multirow{2}{*}{1}
     & \multirow{2}{*}{$\binp{u_i}{y}Q$}
  & 
    \begin{tabular}{l}
        \noalign{\smallskip}
        $\propinp{k}{\wtd x}\binp{u_{\mstindex}}{\wtd y} \apropout{k+1}{\wtd z}
       \Par \Bopt{k+1}{\tilde z}{Q\sigma}$
        \smallskip
    \end{tabular}
      \\
    \cdashline{3-3}
     &  &     \begin{tabular}{ll}
      \noalign{\smallskip}
    $y_j:S$ 
    & 
    $\widetilde y = (y_1,\ldots,y_{\len{\Gtopt{S}}})$ 
    \\
    $\wtd w = \linecondit{\lin{u_i}}{\{ u_i \}}{\epsilon}$
    & 
      $\widetilde z = \fnb{Q}{\widetilde x \widetilde y \setminus \wtd w}$
      \\
      $\mstindex = \linecondit{\traux{u_i}}{\indf{S}}{i}$ 
      &
      $\sigma = \nextn{u_i} \cdot \subst{y_1}{y}$
      \smallskip
    \end{tabular} 
    \\ \hline 
    2 & $\bout{u_i}{y_j}{Q}$ &
      \begin{tabular}{l}
        \noalign{\smallskip}
        $\propinp{k}{\widetilde x}
        \bbout{u_{\mstindex}}{\widetilde y}
        \apropout{k+1}{\widetilde z}  \Par \Bopt{k+1}{\tilde z}{Q\sigma}$
       \smallskip
    \end{tabular}
    \\
    \cdashline{3-3}
     &  &     \begin{tabular}{ll}
      \noalign{\smallskip}
  $y_j : S$
  & 
   $\wtd y = (y_j, \ldots,y_{j+\len{\Gtopt{S}}-1})$ 
   \\
  $\wtd w = \linecondit{\lin{u_i}}{\{ u_i \}}{\epsilon}$ 
  &
       $\wtd z = \fnb{Q}{\wtd x \setminus \wtd w}$
       \\
       $\mstindex = \linecondit{\traux{u_i}}{\indf{S})}{i}$ 
       &
      $\sigma = \nextn{u_i}$ 
      \smallskip
    \end{tabular} 
    \\ \hline 
      3 &  $\news{s:C}{Q}$ &
    \begin{tabular}{l}
      \noalign{\smallskip}
     $\news{\widetilde{s}:\Gtopt{C}}{\,\Bopt{k}{\tilde x}
         {Q\sigma}}$
      \smallskip
    \end{tabular}
    %side-conditions
    \\
        \cdashline{3-3}
     &  &     \begin{tabular}{ll}
      \noalign{\smallskip}
      $\widetilde{s} = (s_1,\ldots,s_{\len{\Gtopt{C}}})$ 
      &
      $\sigma = \subst{s_1 \dual{s_1}}{s \dual{s}} $
      \smallskip
    \end{tabular} 
    \\
    \hline
        4 &    $\news{s:\trec{t}{S}}{Q}$
       &
        \begin{tabular}{l}
            \noalign{\smallskip}
             $\news{\widetilde{s}:\Rtopt{S}} \big(\propinp{k}{\wtd x}\propout{k+1}{\wtd z}\inact \Par \Bopt{k+1}{\tilde z}{Q}\big)$
        \smallskip
        \end{tabular} 
    \\
    \cdashline{3-3}
     &  &     \begin{tabular}{ll}
      \noalign{\smallskip}
          $\traux{\trec{t}{S}}$
         & 
          $\widetilde{s} = (s_1,\ldots,s_{\len{\Rtopt{S}}})$ 
          \\
            $\wtd z = \wtd x, \wtd s, \wtd {\dual s}$ 
            &
          $\widetilde {\dual{s}} = (\dual{s_1},\ldots,\dual{s_{\len{\Rtopt{S}}}})$
      \smallskip
    \end{tabular}  
    \\ \hline
         5 & $Q_1 \Par Q_2$ &
    \begin{tabular}{l}
      \noalign{\smallskip}
      $\propinp{k}{\widetilde x} \propout{k+1}{\widetilde y}
      \apropout{k+\degree+1}{\widetilde z}  
       \Par \Bopt{k+1}{\tilde y}{Q_1} \Par 
      %  \\ 
       \Bopt{k+\degree+1}{\tilde z}{Q_2}$
      \smallskip
    \end{tabular}
            \\
            \cdashline{3-3}
     &  &     \begin{tabular}{lll}
      \noalign{\smallskip}
            $\degree = \lenHOopt{Q_1}$ &
      $\widetilde y  = \fnb{Q_1}{\widetilde x}$
      & $\widetilde z = \fnb{Q_2}{\widetilde x}$
      \smallskip
    \end{tabular} 
    \\
    \hline
        6 & $\inact$ &
  \begin{tabular}{l}
    \noalign{\smallskip}
    $\propinp{k}{}\inact$
   \smallskip
  \end{tabular}
    \\
    \hline %\hline
      7 & $\recp{X}P$
      &
        \begin{tabular}{l} 
            \noalign{\smallskip}
            $\news{\prop^r_X}\big(\propinp{k}{\wtd x}
                \propoutrec{k+1}{\wtd z} \recp{X}\binp{\prop^r_{X}}{\wtd y}
        \propoutrec{k+1}{\wtd y}X
            \Par \Brecpi{k+1}{\tilde z}{P}_{g}\big)$
         \smallskip
        \end{tabular} 
        \\
                    \cdashline{3-3}
     &  &     \begin{tabular}{lll}
      \noalign{\smallskip}
          $\wtd n = \fs{P}$ & $\len{\wtd z} = \len{\wtd y}$ & $g = \{X \mapsto \wtd z\}$ 
          \\ 
          $\wtd n : \wtd C$ & $\wtd z = \bname{\wtd n : \wtd C}$ &
      \smallskip
    \end{tabular} 
    \\
    \hline 
\end{tabular}	
\end{center}
\caption[Optimized breakdown function  $\Bopt{k}{\tilde x}{\cdot}$ for processes.]{Optimized breakdown function  $\Bopt{k}{\tilde x}{\cdot}$.  
The auxiliary function $\Brecpi{k}{\tilde x}{\cdot}_g$ 
is given in~\Cref{pi:t:bdowncorec-rec}. \label{pi:t:bdowncorec}}
% \vspace{-3mm}
\end{table}

\begin{table}[t!]
  \thesisalt{
  
    \begin{tabular}{ |l|l|l|l|}
      \hline 
    \rowcolor{gray!25}
    %\hline
     & $P$ &
      \multicolumn{2}{l|}{
    \begin{tabular}{l}
      \noalign{\smallskip}
      $\Brecpi{k}{\tilde x}{P}_g$
      \smallskip
    \end{tabular}
  }  \\
    \hline
    8 & $\bout{u_i}{y_j}{Q}$ & 
      \begin{tabular}{ll}
      \noalign{\smallskip}
      $\recp{X}\binp{\prop^r_k}{\widetilde x}
        \bout{u_l}{\widetilde y} 
        \propoutrec{k+1}{\wtd z}\rvar{X} \Par$
%      		\bout{\prop^r_{k+1}}{\widetilde z}X
\\ 
       $\Brecpi{k+1}{\tilde z}{Q\sigma}_g$ 
       & $(\text{if } g \neq  \es)$  
      \smallskip
      \\
      \hdashline 
      \noalign{\smallskip}
      % \noalign{\hbox to \textwidth{\leaders\hbox to 3pt{\hss . \hss}\hfil}}
      $\recp{X}\propinprecsh{k}{\wtd x}
      \big(\bout{u_{\mstindex}}{\wtd y} 
        \apropoutrecsh{k+1}{\wtd z} \Par \rvar{X}\big) 
%      		\bout{\prop^r_{k+1}}{\widetilde z}X
       \Par$ 
       \\
      $\Brecpi{k+1}{\tilde z}{Q\sigma}_g 
      $ &  $(\text{if } g = \es)$
    \smallskip
  \end{tabular} & 
  \begin{tabular}{l}
         \noalign{\smallskip}
         $y_j : T$ 
        %  $\wedge$ 
        \\ 
         $\wtd y = (y_j,\ldots,y_{j+\len{\Gtopt{T}}-1})$ \\ 
      % $r:S \wedge \mathsf{tr}(S)$ \\
      $\wtd w = \linecondit{\lin{u_i}}{\{ u_i \}}{\epsilon}$ \\
      $\wtd z = g(X) \cup \fnb{Q}{\wtd x \setminus \wtd w}$  \\
      $l = \linecondit{\traux{u_i}}{\indf{S}}{i}$ \\
      $\sigma = \nextn{u_i}$
      \smallskip
  \end{tabular} 
  \\
\hline 
  9 & $\binp{u_i}{y}{Q}$
  &
  \begin{tabular}{ll}
      \noalign{\smallskip}
      $\recp{X}\binp{\prop^r_k}{\widetilde x}
        \binp{u_l}{\widetilde y} 
        \propoutrec{k+1}{\wtd z}
        \rvar{X}
       \Par$
      \\ 
       $\Brecpi{k+1}{\tilde z}{Q\sigma}_g$
       & $(\text{if } g \neq  \es)$  
      % \small
       \smallskip
      \\
      \hdashline 
      \noalign{\smallskip}
      % \noalign{\hbox to \textwidth{\leaders\hbox to 3pt{\hss . \hss}\hfil}}
      $\recp{X}\propinprecsh{k}{\wtd x}
      \big(\binp{u_{\mstindex}}{\wtd y} 
        \apropoutrecsh{k+1}{\wtd z} \Par \rvar{X}\big) 
%      		\bout{\prop^r_{k+1}}{\widetilde z}X
       \Par$
       \\  
       $\Brecpi{k+1}{\tilde z}{Q\sigma}_g$ 
       &
       $(\text{if } g = \es)$
      \smallskip 
  \end{tabular} & 
  \begin{tabular}{l}
         \noalign{\smallskip}
      % $r:S \wedge \mathsf{tr}(S)$ \\
      $y_j : T$ \\ 
      $\wtd y = (y_1,\ldots,y_{\len{\Gtopt{T}}})$ \\
      $\wtd w = \linecondit{\lin{u_i}}{\{ u_i \}}{\epsilon}$ \\
      $\wtd z = g(X) \cup \fnb{Q}{\wtd x \wtd y \setminus \wtd w}$  \\
      $l = \linecondit{\traux{u_i}}{\indf{S}}{i}$ \\
      $\sigma = \nextn{u_i} \cdot \subst{y_1}{y}$
      % $\wtd z = \fnb{Q}{\wtd x \wtd y}$  
      \smallskip
  \end{tabular} 
  \\
\hline
10 & $Q_1 \Par Q_2$
& 
\begin{tabular}{ll}
      \noalign{\smallskip}
      $\recp{X}\proprinpk{r}{k}{\wtd x}$
      \\ \quad 
      $\big(\proproutk{r}{k+1}{\wtd y_1}\rvar{X}
      \Par \apropoutrecsh{k+\degree+1}{\wtd y_2}\big)
       \Par$
      & 
       $(\text{if } g \neq  \es)$  
    \smallskip
    \\ \quad 
       $\Brecpi{k+1}{\tilde y_1}{Q_1}_{g} \Par 
      \Brecpi{k+\degree+1}{\tilde y_2}{Q_2}_{\es}$
      &   
    \smallskip
    \\
      \hdashline 
      \noalign{\smallskip}
      $\recp{X}\proprinpk{r}{k}{\wtd x}$
      \\ \quad 
      $\big(\apropoutrecsh{k+1}{\wtd y_1}
      \Par \apropoutrecsh{k+\degree+1}{\wtd y_2} \Par \rvar{X}\big)
       \Par$ 
       \smallskip
       \\ \quad 
        $\Brecpi{k+1}{\tilde y_1}{Q_1}_{\es} \Par 
      \Brecpi{k+\degree+1}{\tilde y_2}{Q_2}_{\es}$ &
      $(\text{if  $g = \es$})$
      \smallskip
  \end{tabular} & 
  \begin{tabular}{l}
         \noalign{\smallskip}
         $\recpx{Q_1}$ \\
         $\wtd y_1  = g(X) \cup \fnb{Q_1}{\wtd x}$ \\
         $\wtd y_2  =  \fnb{Q_2}{\wtd x}$ \\
        %  $i \in \{1,2\}.$ \\
        %  $\quad \wtd y_i' = \linecondit{\recpx{Q_i}}{g(X)}{\epsilon}$ \\
        %  $\quad g_i = \linecondit{\recpx{Q_i}}{g_i}{\epsilon}$ \\
        %  $\quad \wtd y_i  = \wtd y_i' \cup \fnb{Q_i}{\wtd x}$ \\
         $\degree = \lenHOopt{Q_1}$
      \smallskip
  \end{tabular} 
  \\
  \hline
  11 & ${\news{s}{Q}}$ & 
  \begin{tabular}{ll}
      \noalign{\smallskip}
      % $\news{\widetilde{s}:\Gt{C}}$ 
      $\recp{X}\news{\widetilde{s}:\Gtopt{C}}$
      \\ \qquad \quad 
      $\big(\proprinpk{r}{k}{\wtd x}\proproutk{r}{k+1}{\wtd z}X
      \Par$
      \\ \qquad \qquad 
      $\Brecpi{k+1}{\tilde z}{Q\sigma}_g \big)$
      & $(\text{if } g \neq \es)$
      \smallskip
      \\
      \hdashline 
      \noalign{\smallskip}
      $\recp{X}\news{\wtd{s}:\Gtopt{C}}$
      \\ \qquad \quad 
      $\big( \proprinpk{r}{k}{\wtd x}\big(\apropoutrecsh{k+1}{\wtd z} \Par X\big)
      \Par$
      \\ \qquad \qquad 
      $\Brecpi{k+1}{\tilde z}{Q\sigma}_g \big)$ 
      & $(\text{if } g = \es)$
   \smallskip
  \end{tabular} & 
  \begin{tabular}{l}
      \noalign{\smallskip}
      $s:C$
      \\ 
        %  $\mathsf{tr}(\trec{t}{S})$ \\
         $\widetilde{s} = (s_1,\ldots,s_{\len{\Gtopt{C}}})$ \\
         $\widetilde {\dual{s}}' = (\dual{s_1},\ldots,\dual{s_{\len{\Gtopt{C}}}})$ \\
         $\widetilde {\dual{s}} =
          \linecondit{\lin{C}}{\widetilde {\dual{s}}' }{\epsilon}$ \\
  %        $\widetilde {\dual{s}} = \begin{cases}
  %         (\dual{s_1},\ldots,\dual{s_{\len{\Gt{S}}}}) & \text{if} \ C = S \\
  %         \epsilon & \text{if} \ C = \chtype{U}
  % \end{cases}
  % $\\
        %  $\widetilde {\dual{s}} = (\dual{s_1},\ldots,\dual{s_{\len{\Rtopt{S}}}})$ \\
         $\wtd z = \wtd x, \wtd s, \wtd {\dual s}$ 
         \\ 
         $\sigma = \subst{s_1 \dual{s_1}}{s \dual{s}} $
      \smallskip
  \end{tabular} 
  \\
\hline
12 & $X$ 
 &  
 \begin{tabular}{ll}
  \noalign{\smallskip}
 $\recp{X}\proprinpk{r}{k}{\wtd x}\apropoutrecsh{\rvar{X}}{\wtd x}.\rvar{X}$
 & $(\text{if } g \not= \es)$
%  \qquad 
\smallskip
\\
\hdashline 
\noalign{\smallskip}
 $\recp{X}\proprinpk{r}{k}{}\big(\apropoutrecsh{\rvar{X}}{} \Par X\big)$
 & $(\text{if } g = \es)$
 \smallskip
 \end{tabular}
 & 
%  \begin{tabular}{l}
%         \noalign{\smallskip}
%         $\wtd x = g(X)$
%      \smallskip
%  \end{tabular} 
  \\
\hline
13 &  $\inact$ &
  \begin{tabular}{l}
    \noalign{\smallskip}
    $\proprinpk{r}{k}{}\inact$
   \smallskip
  \end{tabular}
  &
  \begin{tabular}{l}
    \noalign{\smallskip}
  \smallskip
  \end{tabular}
  \\
    \hline
\end{tabular}
  % \caption{Optimized breakdown function  $\Bopt{k}{\tilde x}{\cdot}$ for processes, and auxiliary function for recursive processes $\Brecpi{k}{\tilde x}{\cdot}_g$. \label{pi:t:bdowncorec}}
  % \vspace{-3mm}
  % \end{table*}
}
  {\resizebox{1.01\textwidth}{!}{}}
\caption{The auxiliary function $\Brecpi{k}{\tilde x}{\cdot}_g$. \label{pi:t:bdowncorec-rec}}
% \vspace{-3mm}
\end{table}
We  define the optimized decomposition $\Do{\cdot}$ by relying on the revised breakdown function 
$\Bopt{k}{\tilde x}{\cdot}$ (cf.~\Cref{pi:ss:bf}).
Given a context $\wtd x$ and a $k > 0$, $\Bopt{k}{\tilde x}{\cdot}$ is defined 
 on  initialized processes. 
\Cref{pi:t:bdowncorec} gives the definition: we use an auxiliary function for
recursive processes, denoted $\Brecpi{k}{\tilde x}{\cdot}_g$, where  
parameter $g$ maps recursive  variables 
 to a list of name variables (cf.~\Cref{pi:ss:brec}). 
%  We assume a recursive process does not have a nested recursive process: 
%  that is 

In the following, to keep presentation simple, we assume recursive processes $\recp{X}P$ in which $P$ does not contain a subprocess of shape $\recp{Y}P'$. The generalization of our decomposition without this assumption is not difficult, but is notationally heavy.

%   That is, function $\Brecpi{k}{\tilde x}{\cdot}_g$  assumes 
%   at most one recursive variable $X$  in a given process. 
%   We remark this is assumed to simplify the presentation
%   and the encoding can be generalized. We sketch it: 
%   $\Brecpi{k}{\tilde x}{\cdot}_g$ would need to treat $\recp{X}P'$ 
%   similarly to $\B{k}{\tilde x}{\cdot}_g$, but it 
%   would append an entry for $Y$ in $g$ 
%   and \emph{Parallel Composition} would need to be modified
%   (see below).}

\subsubsection{The Optimized Breakdown Function}
\label{pi:ss:bf}
We describe entries 1-7 in~\Cref{pi:t:bdowncorec}, assuming the side conditions given in the table.
\smallskip

\begin{description}
\item[1. Input]
%The breakdown of $\binp{u_i}{y}Q$ is 
Process $\Bopt{k}{\tilde x}{\binp{u_i}{y}Q}$
consists of a leading trio that 
mimics the input  and runs in parallel with the 
breakdown of~$Q$.
%\begin{align*}
%	  \B{k}{\tilde x}{\binp{u_i}{y}Q} = 
%    \propinp{k}{\wtd x}\binp{u_{\mstindex}}{\wtd y} \apropout{k+1}{\wtd z}
%     \Par \B{k+1}{\tilde z}{Q\sigma} 
%\end{align*}
In the trio, a context $\wtd x$ is expected along $\prop_k$. 
Then, an input on $u_{\mstindex}$ mimics the input action:
it expects the \textit{decomposition} of name $y$, denoted $\wtd y$. 
To decompose $y$ we use its type: if $y:S$ then 
$\wtd y = (y_1,\ldots,y_{\len{\Gtopt{S}}})$.
The index $\mstindex$ of $u_{\mstindex}$ depends on the type of $u_i$.
Intuitively, if $u_i$ is tail-recursive then $\mstindex = \indf{S}$ (\Cref{pi:def:indexfunction}) and we do not increment it, as the same decomposition of $u_i$ should be used to mimic a new instance   in the continuation.
Otherwise, if  $u_i$ is linear then we use the substitution 
$\sigma = \subst{u_{i+1}}{u_i}$ to increment the index in $Q$.
Next, the context $\wtd z = \fnb{Q}{\wtd x \wtd y \setminus \wtd w}$ is propagated, where $\wtd w = (u_i)$ or $\wtd w = \epsilon$.
%decomposed names from $\wtd x \wtd y$ without $u_i$ if $\lin{u_i}$ whose source names appear free in $Q$. 
% If  $u_i$ is linear (and not tail-recursive) then we use the substitution 
% $\sigma = \subst{u_{i+1}}{u_i}$ to increment it in $Q$.
% Intuitively, if a name is tail-recursive we use $f(S)$ for an index
% and do not increment it as the same decomposition of recursive name $u_i$ 
% should be used to mimic a new instance of $u_i$ in continuation $Q$. 

\item[2. Output]
% The breakdown of  $\bout{u_i}{y}Q$ is as follows: 
%\begin{align*}
% \B{k}{\tilde x}{\bout{u_{i}}{y}Q} = \propinp{k}{\widetilde x}
%			\bbout{u_{\mstindex}}{\widetilde y}
%			\apropout{k+1}{\widetilde z}  \Par
%      \B{k+1}{\tilde z}{Q\sigma}	
%\end{align*}
Process $\Bopt{k}{\tilde x}{\bout{u_{i}}{y_j}Q}$ sends the decomposition of  $y$ on $u_{\mstindex}$, with $\mstindex$ as in the {input} case. 
We decompose name $y_j$ based on its type $S$: 
$\wtd y = (y_j,\ldots,y_{j+\len{\Gtopt{S}}-1})$.
The context to be propagated is $\wtd z = \fnb{P}{\wtd x  \setminus \wtd w}$, 
where $\wtd w$ and $\sigma$ are as in the input case. 
%The substitution  is also   as in the input case. 

\item[3. Restriction (Non-recursive name)]
The breakdown of process $\news{s:C}{Q}$ is 
$$\news{\widetilde{s}:\Gtopt{C}}{\,\Bopt{k}{\tilde x}
 			{Q\sigma}}$$ where   $s$ is decomposed using $C$: 
$\wtd s = (s_1,\ldots,s_{\len{\Gtopt{C}}})$. 
%Since $\news{s}$ binds both $s$ and its dual $\dual s$, 
%the substitution $\sigma$ is simply $\subst{s_1\dual{s}_1}{s\dual{s}}$. 
Since $\news{s}$ binds $s$ and $\dual s$
(or only $s$ if $C$ is a shared type)
the substitution $\sigma$ is simply $\subst{s_1\dual{s}_1}{s\dual{s}}$
and initializes indexes in $Q$.

\item[4. Restriction (Recursive name)] As in the previous case, in the breakdown of $\news{s:\trec{t}{S}}{Q}$ the name $s$ is decomposed into $\wtd s$ by relying on $\trec{t}S$.
	Here the breakdown consists of the breakdown of $Q$  
	running in parallel with a control trio, which appends 
	restricted (recursive) names $\wtd s$ and $\wtd {\dual s}$ 
	to the context, i.e., $\wtd z = \wtd x, \wtd s, \wtd {\dual s}$.

\item[5. Composition]
The breakdown of process $Q_1 \Par Q_2$ uses
a control trio to trigger the breakdowns of $Q_1$ and $Q_2$, similarly as before. 

\item[6. Inaction]
The breakdown of $\inact$ is an input prefix that receives an empty context  ($\wtd x = \epsilon$). 
\smallskip

\item[7. Recursion] 
The breakdown of $\recp{X}P$ is as follows: 
\begin{align*}
  % &\B{k}{\tilde x}{\recp{X}P} = \\
  \news{\prop^r_X}(\propinp{k}{\wtd x}
  \propoutrec{k+1}{\wtd z}
\recp{X}\binp{\prop^r_{X}}{\wtd y}
\propoutrec{k+1}{\wtd y}X
\Par \Brecpi{k+1}{\tilde z}{P}_{g})
\end{align*}
We have a control trio and the breakdown of   $P$, obtained  using  the auxiliary function $\Brecpi{k}{\tilde x}{\cdot}_g$ (\Cref{pi:t:bdowncorec-rec}). The trio receives the context $\wtd x$ on $\prop_k$ and 
propagates it further. 
To ensure typability, we bind all session free names of $P$ using the context $\wtd z$, which contains the decomposition of those free names (cf. \Cref{pi:d:name-breakdown}). This context is needed to break down $P$, and so we record it as $g = \{ X \mapsto \wtd z\}$ in the definition of $\Brecpi{k+1}{\tilde x}{P}_g$. This way, $\wtd z$ will be propagated all the way until reaching $\rvar{X}$. 

Next, the recursive trio is enabled, and receives $\wtd y$ along $\prop^r_{X}$, with $\len{\wtd z}= \len{\wtd y}$ and $l = \len{P}$.
The tuple $\wtd y$ is propagated to the first trio of $\Brecpi{k+1}{\tilde x}{P}_g$. 
By definition of $\Brecpi{k+1}{\tilde x}{P}_g$, its propagator
$\prop^r_{X}$ will send the same context as received by the first 
trio. Hence, the recursive part of the control 
trio  keeps sending this context to the next instances of 
recursive trios of $\Brecpi{k+1}{\tilde x}{P}_g$.

Notice that the leading trio actually has four prefixes. This simplifies our presentation:  this trio can be broken down into two trios by introducing an extra propagator $\prop_{k+1}$ to send over $\prop^r_{k+2}$.
\end{description}

%\begin{align*}
%\propinp{k}{\widetilde x}
%  \propoutrec{k+2}{\wtd z}
%  \apropout{k+1}{\dual{\prop^r_{k+2}}} \Par 
%  \propinp{k+1}{x_\prop}
%  \recp{X}\binp{\prop^r_{k+l-1}}{\widetilde y}
%\bout{x_c}{\wtd y}
%        X
%\end{align*}

\subsubsection{Handling  $P$ in $\recp{X}P$}
\label{pi:ss:brec}
%We describe how decomposition handles \sessp processes 
%in which names can be typed with recursive session types

%$\trec{t}S$ and recursive processes $\recp{X}P$.
As already mentioned, we use the auxiliary function
   $\Brecpi{k}{\tilde x}{\cdot}_g$ in \Cref{pi:t:bdowncorec-rec} to generate \emph{recursive
   trios}. 
      We concentrate on discussing entries 8-11 in~\Cref{pi:t:bdowncorec-rec}; the other entries are similar as before.
A key observation is that parameter $g$ can be empty. 
To see this, consider a process like $P=\recp{X}(Q_1 \Par Q_2)$
where $\rvar{X}$ occurs free in $Q_1$ but not in $Q_2$. 
If $X$ occurs free in $Q_1$ then its decomposition will have a non-empty
$g$, whereas the $g$ for $Q_2$ will be empty. 
% by well-typedness, $\rvar{X}$ occurs free in $Q_1$ but not in $Q_2$ (or vice
% versa). If $X$ occurs free in $Q_1$ then its decomposition will have a non-empty
% $g$, whereas the $g$ for $Q_2$ will be empty. 
In the recursive trios of~\Cref{pi:t:bdowncorec-rec}, the difference between $g \not= \es$ and $g = \es$ is subtle: in the former case, $X$ appears guarded by a propagator; in the latter case, it appears unguarded in a parallel composition. 
% \acheck{so that these trios replicate themselves on a trigger.}
This way, trios in the breakdown of $Q_2$ replicate themselves on a trigger 
from the breakdown of $Q_1$.
  
%If $g \not= \es$ then we break down $P$ into a composition of recursive trios in which a trio generated for
%    $X$ enacts a new instance of the breakdown of recursive process by
%    communicating with a control trio generated by $\B{}{}{\mu X.P}$ via
%    $c^r_X$. 
%    
%    Otherwise, the case $g \not = \es$  denotes processes that are nested in
%    recursive parallel composition but do not invoke the recursive variable, which we treat differently. 
%     To illustrate
%   this, let   $P=\recp{X}(Q_1 \Par Q_2)$.
%   As $P$ is well-typed,  $\rvar{X}$ must appear in either $Q_1$ or
%   $Q_2$. Say $X$ appears in $Q_1$. So, an execution of an instance of $Q_1$
%    results in a copy of $Q_2$ in parallel. Thus, the trios in the breakdown
%   of $Q_2$ have to simulate the replication on a trigger from the breakdown of
%   $Q_1$. So, in this breakdown  each trio  
%    replicates itself on a trigger. 

Having discussed this difference, we only describe the cases when $g \not= \es$: 

\begin{description}
	\item[8 / 9. Output and Input]
The breakdown of $\bout{u_i}{y_j}{Q}$ consists of the breakdown of $Q$ in parallel with a leading trio, a recursive process whose body is defined as in  $\B{}{}{\cdot}$.
      As names $g(X)$ may not appear free in ${Q}$, 
       we must ensure that a context $\wtd z$ for 
      the recursive body is propagated. 
      %Hence,   $\wtd z = g(X) \cup \fnb{Q}{\wtd x \setminus \wtd w}$      where $\wtd w = \linecondit{\lin{u_i}}{\{ u_i \}}{\epsilon}$. 
      	 The breakdown of $\binp{r}{y}{Q}$ is defined similarly.
	\item[10. Parallel Composition]  We discuss the breakdown of $Q_1 \Par Q_2$ assuming $\recpx{Q_1}$. We take $\wtd y_1 = g(X) \cup \fnb{Q_1}{\wtd x}$ 
     to ensure that $g(X)$ is propagated to the breakdown of $X$. 
     The role of $\prop^r_{k+l+1}$ is to enact a new instance 
     of the breakdown of $Q_2$; it has a shared type to enable replication.
     In a running process, the number of these triggers in parallel 
      denotes the number of available instances of $Q_2$. 
%     \acheck{We remark that for a generalization to 
%     multiple recursive variables $X$, we would need to split 
%     context based on recursive variables appearing in parallel 
%     components (and look up for those in $g$) 
%     to ensure that each $X$  retrieves appropriate context. }

\item[11. Recursive Variable]  In this case, the breakdown is a control trio
     % $\recp{X}\proprinpk{r}{k}{\wtd x}\proproutk{r}{\rvar{X}}{\wtd x}X$, 
      that receives the context $\wtd x$ from a preceding trio 
    and  propagates it again to the first 
    control trio of the breakdown of 
    a recursive process along   $\prop^r_X$. 
    Notice that by construction we have  
    $\wtd x = g(X)$.   

\end{description}

  We may now define the optimized process decomposition $\Do{\cdot}$:
 
\begin{definition}[Decomposing Processes, Optimized]
  \label{pi:def:decomp}
    Let $P$ be a process with $\widetilde u = \fn{P}$ and $\wtd v = \rfn{P}$.
     Given the breakdown function $\Bopt{k}{\tilde x}{\cdot}$  in~\Cref{pi:t:bdowncorec}, the \emph{optimized decomposition} of $P$, denoted $\Do{P}$, is
  defined as
  $$
    \Do{P} = \news{\wtd \prop}\big(
    \propout{k}{\wtd r} \inact \Par \Bopt{k}{\tilde r}{P\sigma}\big)
    $$ where: $k >0$;
    $\widetilde \prop = (\prop_k,\ldots,\prop_{k+\lenHOopt{P}-1})$;
    $\wtd r$ such that for $v \in \wtd v$ and $v:S$
    $(v_1,\ldots,v_{\len{\Rtopt{S}}}) \subseteq \wtd r$; and
      $\sigma = \subst{\mathsf{init}(\widetilde u)}{\widetilde u}$.
  \end{definition}
  
  The definition of $\Do{\cdot}$ is similar to the definition of $\F{\cdot}$ given in \Cref{pi:def:am-decomp}. The optimizations are internal to the definition of the breakdown function; notice that the handling of recursive names needed in $\F{\cdot}$ is not needed in $\Do{\cdot}$, as they are now handled by the auxiliary function  $\Brecpi{k}{\tilde x}{\cdot}_g$. 
%  \cralt{
%  \todo[inline]{Please, contrast the optimized definition above with the definition !}}{}
   
 \subsection{Examples}
 Here we  illustrate
  $\Do{\cdot}$, $\Gtopt{\cdot}$, 
 $\Bopt{k}{\tilde x}{\cdot}$, and
 $\Brecpi{k}{\tilde x}{\cdot}_g$ by revisiting the two examples from \Cref{pi:s:dbc}.

\begin{example}[\Cref{pi:ex:decomp-cr}, Revisited]
  Consider again the process $P = \news{u}(A \Par B)$ as in~\Cref{pi:ex:decomp-cr}.
  Recall that $P$ implements session types $S = \btout{\dual T}\tinact$ 
  and $T = \btinp{\tint}\btout{\bool}\tinact$.
%  \begin{align*}
%    P &= \news{u:S}(\bout{u}{w} \binp{\dual w}{t} \bout{\dual w}{\unaryint{t}} 
%    \inact \Par 
%    \\ 
%    & \qquad \qquad \quad 
%      \binp{\dual u}{x} \bout{x}{\textsf{5}}\binp{x}{b} \inact) = 
%      \news{u}(A \Par B)	
%  \end{align*}

By~\Cref{pi:def:sizeproc},  $\lenHOopt{P} = 9$. 
  The optimized decomposition of $P$ is: 
  \begin{align*}
    \Do{P} &= \news{\wtd \prop}\big(\apropout{1}{}  
    \Par \news{u_1} \Bopt{1}{\epsilon}{(A \Par B)\sigma'} \big) \quad 
  \end{align*}

    \noindent where $ \sigma' = \mathsf{init}(\fn{P}) \cdot \subst{u_1 \dual{u_1}}{u\dual{u}}$ 
    and $\wtd \prop = (\prop_1, \ldots, \prop_9)$. 
  We have: 
  \begin{align*}
    \Bopt{1}{\epsilon}{(A \Par B)\sigma'} \big) &= 
    \propinp{1}{} \propout{2}{}  \apropout{6}{}  \Par 
    \Bopt{2}{\epsilon}{A \sigma'} \Par \Bopt{6}{\epsilon}{B \sigma'}
\end{align*}

\noindent 
The breakdowns of sub-processes $A$ and $B$ are as follows: 
\begin{align*}
  \Bopt{2}{\epsilon}{A \sigma'} &= \propinp{2}{}\bout{u_1}{w_1, w_2} 
  \apropout{3}{} \Par 
  \propinp{3}{}\binp{\dual w_1}{t}\apropout{4}{} \Par 
  \\
  &
  \qquad \propinp{4}{}\bout{\dual w_2}{\unaryint{t}}\apropout{5}{} \Par 
  \propinp{5}{}\inact  \\
  % B decomposition 
  \Bopt{6}{\epsilon}{B \sigma'} &= 
  \propinp{6}{}\binp{\dual u_1}{x_1,x_2}\apropout{7}{x_1, x_2} \!\Par\! 
  \propinp{7}{x_1,x_2}\bout{x_1}{\textsf{5}}\apropout{8}{x_2} \!\Par 
  \\
  &
  \qquad 
  \propinp{8}{x_2}\binp{x_2}{b_1}\apropout{9}{} \Par 
  \propinp{9}{}\inact
\end{align*}

\noindent 
Name $w$ is decomposed as indexed names
$\dual w_1, \dual w_2$; by using $\Gtopt{\cdot}$ (\Cref{pi:def:typesdecomp}) on $T$, their \msts are $M_1 = \btout{\tint}\tinact$ and $M_2 = \btinp{\bool}\tinact$, respectively. 
Name $u_1$ is the decomposition of name $u$ and it is typed with  
% $\btout{\btout{\tint}\tinact, \btinp{\bool}\tinact}\tinact$.
$\btout{\dual M_1, \dual M_2}\tinact$.

We discuss the reduction steps from $\Do{P}$.
    After a few administrative reductions on $\prop_1$, $\prop_2$, and $\prop_6$, $\Do{P}$ mimics the first source communication:
    \begin{align*}
      \Do{P} &\red^3
      \news{\wtd \prop_{*}}\big(
      \highlighta{\bout{u_1}{w_1, w_2}}
      \apropout{3}{} \Par 
       \Bopt{3}{\epsilon}{\binp{\dual w}{t} \bout{\dual w}{\unaryint{t}} \inact } \Par \\
      & \qquad \qquad  \quad 
      \highlighta{\binp{\dual u_1}{x_1,x_2}}\apropout{7}{x_1, x_2}\Par
      \Bopt{7}{x_1,x_2}{\bout{x}{\textsf{5}}\binp{x}{b} \inact}\big) \\
      &\red	
      \news{\wtd \prop_{*}}\big(
      \apropout{3}{} 
      \Par \Bopt{2}{\epsilon}{\binp{\dual w}{t} \bout{\dual w}{\unaryint{t}} \inact } 
      \Par  \apropout{7}{w_1,w_2} \Par
       \Bopt{7}{x_1,x_2}{\bout{x}{\textsf{5}}\binp{x}{b} \inact} \big) 
      % = Q_1
    \end{align*}

  %% MORE FORMAL ARGUMENT 
  %   \noindent 
  %   Notice that the decomposition of  $w$ (i.e., $w_1, w_2$)  
  %    is propagated as a 
  %   context to the first trio in $\B{7}{x_1,x_2}{\bout{x}{\textsf{5}}\binp{x}{b} \inact}$.
  % Now, we have $P \by{} P'$ where: 
  % % \begin{align*}
  % %   P \by{} P' 
  % % \end{align*}
  % \begin{align*}
  %   P' &= ( \binp{\dual w}{t} \bout{\dual w}{\unaryint{t}} \inact \Par
  %     \bout{x}{\textsf{5}}\binp{x}{b}\inact \subst{w}{x})	
  % \end{align*}

  % \noindent Now, we can see that 
  % \begin{align*}
  %   Q_1 \approx 
  %   \news{\wtd \prop}\big(
  %     \apropout{2}{w_1,w_2} \Par \propinp{2}{w_1,w_2}\propout{3}{}\apropout{7}{w_1,w_2}
  %     \Par \B{2}{\epsilon}{\binp{\dual w}{t} \bout{\dual w}{\unaryint{t}} \inact } 
  %      \Par
  %     \B{7}{x_1,x_2}{\bout{x}{\textsf{5}}\binp{x}{b} \inact} \big)
  %   \equiv \Do{P'}
  % \end{align*}

  % \noindent 
  % Here we can see that the transmission of name $w$ on name $u$ in $P$ 
% is directly mimicked by $\Do{P}$: it is not delegated to 
% some other channel through channel redirections, as in \exref{ex:decomp-cr}. 

%  This allows for less reductions to simulate 
% actions of the original process. 
% Further, propagators here only serve to 
% enforce sequentiality of prefixes present in the original process. 
% Finally, the trio structure is recovered. 
    \noindent Above, $\wtd{\prop}_*=(\prop_3,\prop_4,\prop_5,\prop_7,\prop_8,\prop_9)$.
      After reductions on $\prop_3$
      and $\prop_7$, name $w_1$ substitutes $x_1$ and the communication along 
      $w_1$ can be mimicked: 
      \begin{align*}
          \Do{P} &\red^6 
          \news{\wtd \prop_{**}}\big(
          \highlighta{\binp{\dual w_1}{t}}\apropout{4}{} \Par 
          \propinp{4}{}\bout{\dual w_2}{\unaryint{t}}\apropout{5}{} \Par  \propinp{5}{}\inact  \Par
          \\
          & \qquad \qquad \qquad 
           \highlighta{\bout{w_1}{\textsf{5}}}\apropout{8}{w_2} \Par 
  \propinp{8}{x_2}\binp{x_2}{b_1}\apropout{9}{} \Par 
  \propinp{9}{}\inact \big) \\
  &\red 
  \news{\wtd \prop_{**}} \big(
          \apropout{4}{\textsf{5}} \Par 
          \propinp{4}{t}\bout{\dual w_2}{\unaryint{t}}\apropout{5}{} \Par 
          \propinp{5}{}\inact  \Par 
          \\
          &
    \qquad \qquad \quad 
          \apropout{8}{w_2} \Par 
           \propinp{8}{x_2}\binp{x_2}{b_1}\apropout{9}{} \Par 
  \propinp{9}{}\inact \big) 
  % = Q_2 
      \end{align*} 
      Above, $\wtd \prop_{**} = (\prop_4,\prop_5,\prop_8,\prop_9)$.
Further reductions follow similarly. 
    %% MORE FORMAL ARGUMENT 
  %     \noindent 
  %     Further, we have $P' \by{} P''$ where: 
  % % \begin{align*}
  % %   P \by{} P' 
  % % \end{align*}
  % \begin{align*}
  %   P'' &=  \bout{\dual w}{\unaryint{\textsf{5}}} \inact \Par
  %     \binp{w}{b}\inact
  % \end{align*}
      
  %     \noindent We can see that 
  %     \begin{align*}
  %       Q_2 \approx 
  %       \news{\wtd \prop} \big(
  %         \apropout{3}{\textsf{5}, w_2} \Par 
  %         \propinp{3}{t, w_2}\propout{4}{t}\apropout{8}{w_2} \Par  
  %         \propinp{4}{t}\bout{\dual w_2}{\unaryint{t}}\apropout{5}{} \Par 
  %         \propinp{5}{}\inact  \Par 
  % \propinp{8}{x_2}\binp{x_2}{b_1}\apropout{9}{} \Par 
  % \propinp{9}{}\inact \big)
  %       \equiv \Do{P''}
  %     \end{align*}
\end{example}

 \begin{example}[\Cref{pi:ex:recproc}, Revisited]
 Consider again the tail-recursive session type $S = \trec{t}\btinp{\tint}\btout{\tint}\tvar{t}$. 
	Also, let $R$ be a process implementing a channel $r$ with type $S$: 
	 \begin{align*}
		 R=\recp{X}\binp{r}{z}\bout{r}{-z}X
  \end{align*}
  We decompose name $r$ using $S$ and obtain two channels 
  typed with \msts as in~\Cref{pi:f:tdec}. We have:
  $r_1:\trec{t}\btinp{\tint}\tvar{t}$ and
  $r_2:\trec{t}\btout{\tint}\tvar{t}$.

  The trios produced by $\Bopt{k}{\epsilon}{R}$ 
  satisfy two properties: they 
    (1) mimic the recursive behavior of $R$ 
  and (2) use the same decomposition of channel $r$ (i.e., $r_1$,$r_2$) in every instance. 
  
  To accomplish~(1), each trio of the breakdown of the recursion body
  is {a recursive trio}. 
  For~(2), we need two things.  
  First, we expect to receive all recursive names in the 
  context $\wtd x$ when entering the decomposition of the recursion body;  
  further,
  each trio should use one recursive name from the names 
  received and propagate 
  \emph{all} of them to subsequent trio. Second, we need an 
  extra control trio when breaking down prefix $\mu X$: this trio  
  (i)~receives recursive names from the 
  last trio in the breakdown of the recursion body and 
  (ii)~activates another instance with these recursive names.
  
  Using these ideas, process $\Bopt{1}{r_1,r_2}{R}$ is as follows (we    write $R'$ to stand for $\binp{r}{z}\bout{r}{-z}X$):
  \begin{align*}
  \propinp{1}{r_1,r_2}\propoutrec{2}{r_1,r_2}
  	\recp{X}\proprinpk{r}{X}{y_1,y_2}\propoutrec{2}{y_1,y_2}X
  	\Par 
     \Brecpi{2}{r_1,r_2}{R'}
  \end{align*}
  \noindent where $\Brecpi{2}{r_1,r_2}{R'}$ is the composition of three recursive trios:
  \begin{align*}	
  & \recp{X}\propinprec{2}{y_1,y_2}
  	\binp{r_1}{z_1}\propoutrec{3}{y_1,y_2,z_1}X 
   \Par 
   \\
   & 
  	 \qquad \recp{X}\propinprec{3}{y_1,y_2, z_1}
  	\binp{r_2}{-z_1}\propoutrec{4}{y_1,y_2}X \Par 
  	 \recp{X}\propinprec{4}{y_1,y_2}\proproutk{r}{\rvar{X}}{y_1,y_2}\rvar{X}
  \end{align*}
 $\prop^r_2$ will first
  activate the recursive trios 
  with context $(r_1,r_2)$.  
  Next, 
  each trio uses one of $r_1,r_2$ and propagate them both 
  mimicking the recursion body. 
  The last recursive trio   
  sends $r_1,r_2$ back to the top-level control trio, so it
 can enact another instance 
  of the  decomposition of the recursion body by activating the
  first recursive trio. 
 \end{example}

\subsection{Measuring the Optimization}
\label{pi:ss:measure}
Before discussing the static and dynamic correctness of $\Do{\cdot}$, here we measure the improvements   over $\F{\cdot}$.
A key metric for comparison is the number of prefixes/sychronizations induced by each decomposition. This includes (1)~the number of prefixes involved in \emph{channel redirections} and (2)~the \emph{number of propagators}; both can be counted by already defined notions: 
\begin{enumerate}
  \item {Channel redirections} can be  
  counted by the levels of nesting in the decompositions of types (cf.~\Cref{pi:f:decomp:firstordertypes,pi:f:tdec}). 
  \item The {number of propagators} is determined by  the degree of a process 
  (cf.~\Cref{pi:def:am-sizeproc,pi:def:sizeproc}).
    %  counts the number of 
    % propagators required by a respective decomposition 
\end{enumerate}

These two metrics are related; let us discuss them in detail.

\paragraph{Channel redirections.}
The decompositions of types for $\F{\cdot}$ and $\Do{\cdot}$ abstractly describe the respective channel redirections.  
The type decomposition for $\F{\cdot}$ (\Cref{pi:f:decomp:firstordertypes}) defines 5 levels of nesting for the translation of input/output types. Then, at the level of (decomposed) processes, channels with these types  implement redirections: the nesting levels correspond to 5 additional prefixes in the decomposed process that mimic a source input/output action. In contrast, the type decomposition for $\Do{\cdot}$ (\Cref{pi:f:tdec}) induces no nesting, and so at the level of processes there are no additional prefixes.

\paragraph{Number of propagators.}
We define auxiliary functions to count the number of propagators induced by
$\F{\cdot}$ and $\Do{\cdot}$. These functions, denoted $\numpropam{\cdot}$ and
$\numprop{\cdot}$, respectively, are defined using the degree functions
($\lenHO{\cdot}$ and $\lenHOopt{\cdot}$) given by~\Cref{pi:def:am-sizeproc,pi:def:sizeproc}, respectively.

Remarkably,  $\lenHO{\cdot}$ and
$\numpropam{\cdot}$ are not equal.
The difference lies in the
number of tail-recursive names in a process. In  $\F{\cdot}$
there are propagators $\prop_k$ but also propagators $\prop^r$  for recursive
names. 
\Cref{pi:def:am-sizeproc}, however, only counts the former kind of propagators. 
For any $P$,  the number of 
propagators $\prop^r$ in $\F{P}$ is the number of free and bound
tail-recursive names in $P$.  We remark that, by definition,  there may be more than one occurrence of a propagator $\prop^r$ in $\F{P}$: there is at least one prefix with subject $\prop^r$; further occurrences depend on the sequentiality structure of the (recursive)  type assigned to $r$.
      On the other hand, in $\Do{P}$ there are  propagators  
  $\prop_k$ and  
  propagators $\prop^r_X$,  
  whose number corresponds 
  to the number of recursive variables in the process. 
  % and function $\numprop{\cdot}$  corresponds directly to the  degree $\lenHOopt{\cdot}$. 
  To define 
  $\numpropam{\cdot}$ and
$\numprop{\cdot}$,
  we write $\lentr{P}$ to denote bound occurrences of recursive names 
  and $\lenrecpx{P}$ to denote the number of occurrences of recursive variables.

  \begin{definition}[Propagators in $\F{P}$ and $\Do{P}$]
  \label{d:propopt}
  Given a process $P$, the number of propagators in each decomposition is given by
 $$
  	\numpropam{P}  =\lenHO{P} + 2 \cdot \len{\lentr{P}} + \len{\rfn{P}}
  	\qquad 
  	\numprop{P}  = \lenHOopt{P} +  \lenrecpx{P}
$$
% \begin{align*}
%   \numpropam{P}  &=\lenHO{P} + 2 \cdot \len{\lentr{P}} + \len{\rfn{P}} \\
%   \numprop{P} = 
% \end{align*}
\end{definition}

% We can prove the following proposition:

% %%% OLD PROOF STARTS
%    \begin{proposition} 
%      Given a process $P$, we have $\numpropam{P} \geq \numprop{P}$.
%    \end{proposition} 
% %   \begin{proof}%(Sketch)
% %  Because $\numpropam{P} \geq
% %     \lenHO{P} + \lentr{P}$, it suffices to show that 
% %     $\lenHO{P} + \lentr{P} \geq \numprop{P}$.
% %       The proof is by induction on structure of $P$. 
% %       We show the base case and two inductive cases (input and restriction); other cases are similar:
% %       \begin{itemize}
% %       	\item Case $P = \inact$. Then $\lenHO{P} + \lentr{P} = \numprop{P} = 1$. 
% %       	\item Case $P = \binp{u}{x}Q$. Then $\lenHO{P} = \lenHO{Q}+5$ 
% %       	and $\lenHOopt{P} = \lenHOopt{Q} + 1$. Now, by IH, 
% %       	and as $\lentr{Q} = \lentr{P}$, we 
% %       	have $\lenHO{Q} + \lentr{Q} + 5 \geq \numprop{Q} + 1$.  
% %       	\item  Case $P = \news{r:S}Q$, with  $\tr(S)$. 
% %       Then $\lenHO{P} = \lenHO{Q}$ (by \defref{def:am-sizeproc})
% %       and  $\lentr{P} = \lentr{Q} + 1$. 
% %       Further, we have $\lenHOopt{\news{r:S}Q} = \lenHOopt{Q} +
% %       1$. 
% % %      (by \defref{def:sizeproc}).
% %       Now, by IH 
% % %      we have  $\lenHO{Q} + \lentr{Q} > \numprop{Q}$. So, 
% % 		we can
% %       conclude that $\lenHO{Q} + \lentr{Q} + 1 \geq \lenHOopt{Q} + 1$.
% %       \end{itemize}
% %         \end{proof} 

% %   We remark that when $P \not= \inact \Par \ldots \Par \inact$, 
% %   we have $\numpropam{P} > \numprop{P}$, as $\numpropam{P} = \numprop{P}$ holds only when $P = \inact$.
% %%% OLD PROOF ENDS

Notice that  
$\numprop{P}$ gives the exact number of actions induced by propagators in $\Do{P}$; in contrast, due to propagators $\prop^r$,  
$\numpropam{P}$ gives the \emph{least} number of such actions  in $\F{P}$.

In general, we have $\numpropam{P} \geq \numprop{P}$, but we  can be more precise for a broad class of processes. 
We say that a process $P \not\equiv \inact$ 
is in \emph{normal form} if $P  = \news{\wtd n}(Q_1 \Par \ldots \Par Q_n)$, where each $Q_i$ (with $i \in \{1,\ldots,n\}$) is not $\inact$ and does not contain restriction nor parallel composition at the top-level.
%\begin{remark}[Normal Form]
%  For every $P \not\equiv \inact$ there is $Q$ such that 
%  $P \equiv Q$ and $Q = \news{\wtd n}(Q_1 \Par \ldots \Par Q_n)$ 
%  where $Q_i$ is restriction-free and $Q_i \not\equiv Q'_i \Par Q''_i \not\equiv \inact$  
%  for $i \in \{1,\ldots,n\}$. 
%  We say $Q$ is a normal form of $P$. 
%\end{remark}
We have the following result; see  
  \longversion{\Cref{pi:app:propopt} for details.}{\cite{AAP21-full} for details.}

%%% NEW PROOF STARTS
\begin{restatable}[]{lemma}{propopt}
  \label{pi:p:propopt}
  If $P$ is in normal form 
   then $\numpropam{P} \geq
   \frac{5}{3} \cdot \numprop{P}$. 
\end{restatable}

This result implies that the number of (extra) synchronizations induced by propagators in $\F{P}$ is larger than in $\Do{P}$.

  \subsection{Static Correctness}
  \label{pi:ss:pi-results}

%\subsubsection{Static Correctness}
%We first state~\Cref{pi:t:typerecur}, which ensures the typability of $\Bopt{k}{\tilde x}{\cdot}$ under \msts. 
Here we establish the analog of
\Cref{pi:t:amtyprecdec}
(\Cref{ss:results}) but for $\Do{\cdot}$.
We rely on an auxiliary predicate: 

\begin{definition}[Indexed Names]
	\label{pi:def:indexed}
	Suppose some typing environments $\Gamma, \Delta$.
	Let $\wtd x$, $\wtd y$ be two tuples of indexed names. 
	We write $\indexed{\widetilde y}{\widetilde x}{\Gamma,\Delta}$ for the predicate
	$$
	\forall z_i.\left(z_i \in \widetilde x \Leftrightarrow
	 \exists m.((z_i,\ldots,z_{i+m-1}) \subseteq \widetilde y \wedge 
	 m = \len{\Gtopt{(\Gamma,\Delta)(z_i)}})\right)
%	 ~~
%	 \text{$m_i \geq 0$}
	$$
%	\noindent where $\wtd z = (z_n,\ldots,z_m)$.
\end{definition}

We now state our static correctness results (typability with respect to \msts) for the auxiliary function $\Brecpi{k}{\tilde y}{\cdot}_g$, the breakdown function $\Bopt{k}{\tilde y}{P}$, and the optimized decomposition $\Do{\cdot}$:

%In \longversion{\appref{app:proofbrec}}{\cite{AAP21-full}} we prove the  typability of $\B{k}{\tilde x}{\cdot}$:

\begin{restatable}[Typability of Auxiliary Breakdown: $\Brecpi{k}{\tilde y}{\cdot}_g$]{lemma}{pilemmbrec}
  \label{pi:lem:brec}
    Let $P$ be an initialized process. 
    If $\Gamma \cdot X : \envR;\Delta \proves P \hastype \Proc$ 
    then: 
    \begin{align*}
      \Gtopt{\Gamma \setminus \wtd x};\Theta 
      \proves \Brecpi{k}{\tilde y}{P}_g \hastype \Proc \quad (k > 0)
    \end{align*}
    \noindent where 
    \begin{itemize}
    	\item 
    % $g = \{X \mapsto \wtd m\}$,
    $\wtd x \subseteq \fn{P}$
    such that 
    $\Delta \setminus \wtd x = \emptyset$;
    \item $\wtd y = \wtd v \cdot \wtd m$, where 
    $\wtd m = \codom{g}$ and 
    $\wtd v$ is such that
   $\indexed{\wtd v}{\wtd x}{\Gamma, \Delta}$ 
    holds; 
    
    \item   
    $\Theta = \thetaR \cat \thetax(g)$ 
    where 
    \begin{itemize}
   \item 
    $\dom{\thetaR} = \{\prop^r_k,\prop^r_{k+1},\ldots,\prop^r_{k+\lenHOopt{P}-1} \} \cup 
    \{\dual{\prop^r_{k+1}},\ldots,\dual{\prop^r_{k+\lenHOopt{P}-1}}\}$ 
    
    \item Let $\wtd N = (\Gtopt{\Gamma},\Gtopt{\envR \cdot \Delta})(\wtd y)$. Then
    $$\thetaR (\prop^r_k)=
    \begin{cases}
    \trec{t}\btinp{\wtd N}\tvar{t}
    & \text{if $g \not= \es$}
    \\
    \chtype{\wtd N} & \text{otherwise}
       \end{cases} 
       $$ 
     \item 
    $\balan{\thetaR}$
    \item  $\thetax(g) = \bigcup_{X \in \dom{g}}\prop^r_X:\chtype{\wtd M_X}$
     where 
     $\wtd M_X = (\Gtopt{\Gamma},\Gtopt{\envR})(g(X))$. 
         \end{itemize}
\end{itemize}
  \end{restatable}

  \begin{proof} 
    By induction on the structure of $P$; see~\Cref{pi:app:lem:brec} for details. 
  \end{proof} 

 \begin{restatable}[Typability of Breakdown]{lemma}{thmtyperecur}
 \label{pi:t:typerecur}
 \label{pi:t:thmtyperecur}
    Let $P$ be an initialized process.  
      If $\Gamma;\Delta \proves P \hastype \Proc$ then
   $$\Gtopt{\Gamma \setminus \widetilde x};\Gtopt{\Delta \setminus \widetilde x} \cat
   \Theta \proves \Bopt{k}{\tilde y}{P} \hastype \Proc 
   \quad 
   (k > 0)
   $$ where
   \begin{itemize}
   	\item $\widetilde x \subseteq \fn{P}$ and $\wtd y$
 such that $\indexed{\widetilde y}{\widetilde x}{\Gamma,\Delta}$ holds.
    \item  $\dom{\Theta} = 
     \{\prop_k,\prop_{k+1},\ldots,\prop_{k+\lenHOopt{P}-1}\} 
     \cup \{\dual{\prop_{k+1}},\ldots,\dual{\prop_{k+\lenHOopt{P}-1}}\}$
     \item $\Theta(\prop_k)=
   \btinp{\widetilde M} \tinact$,
   where $\widetilde M = (\Gtopt{\Gamma}\cat\Gtopt{\Delta})(\widetilde y)$.
   \item 
    $\balan{\Theta}$ 
      \end{itemize}
 \end{restatable}
 \begin{proof}
   By induction of the structure of $P$ and~\Cref{pi:lem:brec}; 
   see~\Cref{pi:app:proofbrec} for details. 
 \end{proof}

We now (re)state the minimality result, now based on the decomposition $\Do{\cdot}$. 
% The proof follows from~\Cref{pi:t:typerecur};
%     \longversion{see~\Cref{pi:app:decompcore}.}{see \cite{AAP21-full}.}

 \begin{restatable}[Minimality Result for \sessp, Optimized]{theorem}{thmdecompcore}
  \label{pi:t:decompcore}
  \label{pi:t:thmdecompcore}
    Let $P$ be a process with $\widetilde u = \fn{P}$. 
    If $\Gamma;\Delta \proves P \hastype \Proc$ then
    $\Gtopt{\Gamma \sigma};\Gtopt{\Delta\sigma} \proves \Do{P} \hastype \Proc$, where
    $\sigma = \subst{\mathsf{init}(\widetilde u)}{\widetilde u}$.
  \end{restatable}
  \begin{proof} 
    Direct by~\Cref{pi:def:decomp} and~\Cref{pi:t:typerecur}; 
    see~\Cref{pi:app:decompcore} for details. 
  \end{proof} 
%  \begin{proof}
%    Direct by using \lemref{t:typerecur}
%    \longversion{(see \appref{app:decompcore})}{(see \cite{AAP21-full} for details)}.
%  \end{proof}

% \begin{definition}[Recursive subprocess predicate]
%   \label{pi:d:recpx}
%   Let $P$ be a process. We define predicate $\recpx{P}$ if 
%   and only if there is $Q$ such that 
%   $Q.\rvar{X}$ is a subprocess of $P$.  
% \end{definition}

\subsection{Dynamic Correctness}
\label{pi:ss:dyncorr}
As a complement to the minimality result just given, here we establish that $P$ and $\Do{P}$ are \emph{behaviorally equivalent} (\Cref{pi:t:mainbsthm}).
%We overview this result and its required notions. 
%\longversion{see \appref{app:ss:mst} and \appref{app:ss:oc} for details}{see \cite{AAP21-full} for details}.
The notion of behavioral equivalence that we consider is \emph{MST-bisimila\-ri\-ty} (cf.~\Cref{pi:d:fwb}, $\mstb$), a variant of \emph{characteristic bisimilarity}, one of the session-typed behavioral equivalences studied in~\cite{KouzapasPY17}.

\thesisalt{
\subsubsection{Preliminaries}
\begin{newaddenv}
  We require some auxiliary definitions and notations from \cite{KouzapasPY17}.

\paragraph{Typed Labeled Transition System.}
\label{top:ss:elts}
Our typed LTS is obtained by coupling the untyped LTS given in~\Cref{top:fig:untyped_LTS}
% \Cref{pi:s:prelim}  
with a labeled transition relation 
on typing environments, given in \Cref{top:fig:envLTS}. 
Building upon the reduction relation for session environments in \Cref{top:d:wtenvred},
such a relation
is defined on triples of environments by 
extending the LTSs
in \cite{KYHH2015,KY2015}; it is 
denoted
\[
	(\Gamma_1, \Lambda_1, \Delta_1) \by{\ell} (\Gamma_2, \Lambda_2, \Delta_2)
\]
\newc{Recall that  $\Gamma$ admits  weakening. 
Using this principle (not valid for $\Lambda$ and $\Delta$), we have  
$
	(\Gamma', \Lambda_1, \Delta_1) \hby{\ell} (\Gamma', \Lambda_2, \Delta_2)
$
whenever 
$
	(\Gamma, \Lambda_1, \Delta_1) \hby{\ell} (\Gamma', \Lambda_2, \Delta_2)
$.}
Some intuitions follow.
\begin{description}
	\item[Input Actions]
are defined by 
Rules~$\eltsrule{SRv}$ and $\eltsrule{ShRv}$.
In Rule~$\eltsrule{SRv}$
the type of value $V$
and the type of the object associated to the session type on $s$ 
should coincide. 
The resulting type tuple must contain the environments 
associated to $V$. 
The
dual endpoint $\dual{s}$ cannot be
present in the session environment: if it were present
the only possible communication would be the interaction
between the two endpoints (cf. Rule~$\eltsrule{Tau}$).
Following similar principles, 
Rule~$\eltsrule{ShRv}$ defines input actions for shared names.

\item[Output Actions] are defined by Rules~$\eltsrule{SSnd}$
and $\eltsrule{ShSnd}$.  
Rule~$\eltsrule{SSnd}$ states the conditions for observing action
$\news{\widetilde{m}} \bactout{s}{V}$ on a type tuple 
$(\Gamma, \Lambda, \Delta\cdot \AT{s}{S})$. 
The session environment $\Delta \,\cat\, \AT{s}{S}$ 
should include the session environment of the sent value $V$ \newc{(denoted $\Delta'$ in the rule)}, 
{\em excluding} the session environments of names $m_j$ 
in $\widetilde{m}$ which restrict the scope of value $V$ \newc{(denoted $\Delta_j$ in the rule)}. 
Analogously, the linear variable environment 
$\Lambda'$ of $V$ should be included in $\Lambda$. 
\newc{The rule defines the scope extrusion of session names in $\widetilde{m}$; consequently, 
environments associated to 
their dual endpoints  (denoted $\Delta'_j$ in the rule) appear in
the resulting session environment}. Similarly for shared 
names in $\widetilde{m}$ that are extruded.  
All free values used for typing $V$ \newc{(denoted $\Lambda'$ and $\Delta'$ in the rule)} are subtracted from the
resulting type tuple. The prefix of session $s$ is consumed
by the action.
Rule $\eltsrule{ShSnd}$ follows similar ideas for output actions on shared names:
the name must be typed with $\chtype{U}$; 
conditions on value $V$ are identical to those on Rule~$\eltsrule{SSnd}$.

\item[Other Actions]
%Rules $\eltsrule{Sel}$ and $\eltsrule{Bra}$ describe actions for select and branch.
%Both
%rules require the absence of the dual endpoint from the session
%environment.%, and the presence of the action labels in the type.
Rule~$\eltsrule{Tau}$ defines
internal transitions: 
it reduces the session environment (cf. \Cref{top:d:wtenvred}) or keeps it 
unchanged.
\end{description}

\smallskip

\newc{We illustrate Rule~$\eltsrule{SSnd}$ by means of an example:}

\begin{example}
	\newc{Consider the environment tuple
	$
		(\Gamma;\, \es;\, s: \btout{\lhot{(\btout{S} \tinact)}} \tinact \cat s': S)
	$
	and the typed value $V= \abs{x} \bout{x}{s'} \binp{m}{z} \inact$ with 
	\[
		\Gamma; \es; s': S \cat m: \btinp{\tinact} \tinact \proves V \, \hastype \, \lhot{(\btout{S} \tinact)}
	\]
%
%	\noi Let 
%	$
%		\Delta'_1=\set{\overline{m}: \btout{\tinact} \tinact}
%	$
%	and $U=\btout{\lhot{\btout{S} \tinact}} \tinact$.
	Then, by Rule $\eltsrule{SSnd}$, we can derive:
	\[
		(\Gamma; \es; s: \btout{\lhot{(\btout{S} \tinact)}} \tinact \cat s': S) \by{\news{m} \bactout{s}{V}} (\Gamma; \es; s: \tinact \cat \dual{m}: \btout{\tinact} \tinact)
	\]
%	\qed
Observe how the protocol along $s$ is partially consumed; also, the resulting session environment is extended with 
  $\dual{m}$, the dual endpoint of the extruded name $m$.}
\end{example}

\begin{notation}
Given a value $V$ of type $U$, we sometimes annotate the output action 
$\news{\widetilde{m}} \bactout{n}{V}$
with the type of $V$ 
as $\news{\widetilde{m}} \bactout{n}{\AT{V}{U}}$.
\end{notation}

%%%%%%%%%%%%%%%%%%%% Environment LTS Figure %%%%%%%%%%%%%%%%%%%%%%%%%
\begin{figure}[t!]
    \begin{mdframed}
        \begin{mathpar}
        \inferrule[\eltsrule{SRv}]{
            \dual{s} \notin \dom{\Delta}
            \and
            \Gamma; \Lambda'; \Delta' \proves V \hastype U
        }{
            (\Gamma; \Lambda; \Delta \cat s: \btinp{U} S) \by{\bactinp{s}{V}} (\Gamma; \Lambda\cat\Lambda'; \Delta\cat\Delta' \cat s: S)
        }
        \and
        \inferrule[\eltsrule{ShRv}]{
            \Gamma; \es; \es \proves a \hastype \chtype{U}
            \and
            \Gamma; \Lambda'; \Delta' \proves V \hastype U
        }{
            (\Gamma; \Lambda; \Delta) \by{\bactinp{a}{{V}}} (\Gamma; \Lambda\cat\Lambda'; \Delta\cat\Delta')
        }
        \and
        \inferrule[\eltsrule{SSnd}]{
            \begin{array}{l}
                \Gamma \cat \Gamma'; \Lambda'; \Delta' \proves V \hastype U
                \and
                \Gamma'; \es; \Delta_j \proves m_j  \hastype U_j
                \and
                \dual{s} \notin \dom{\Delta}
                \\
                \Delta'\backslash (\cup_j \Delta_j) \subseteq (\Delta \cat s: S)
                \and
                \Gamma'; \es; \Delta_j' \proves \dual{m}_j  \hastype U_j'
                \and
                \Lambda' \subseteq \Lambda
            \end{array}
        }{
            (\Gamma; \Lambda; \Delta \cat s: \btout{U} S)
            \by{\news{\widetilde{m}} \bactout{s}{V}}
            (\Gamma \cat \Gamma'; \Lambda\backslash\Lambda'; (\Delta \cat s: S \cat \cup_j \Delta_j') \backslash \Delta')
        }
        \and
        \inferrule[\eltsrule{ShSnd}]{
            \begin{array}{l}
                \Gamma \cat \Gamma' ; \Lambda'; \Delta' \proves V \hastype U
                \and
                \Gamma'; \es; \Delta_j \proves m_j \hastype U_j
                \and
                \Gamma ; \es ; \es \proves a \hastype \chtype{U}
                \\
                \Delta'\backslash (\cup_j \Delta_j) \subseteq \Delta
                \and
                \Gamma'; \es; \Delta_j' \proves \dual{m}_j\hastype U_j'
                \and
                \Lambda' \subseteq \Lambda
            \end{array}
        }{
            (\Gamma ; \Lambda; \Delta) \by{\news{\widetilde{m}}
            \bactout{a}{V}}
            (\Gamma \cat \Gamma'; \Lambda\backslash\Lambda'; (\Delta \cat \cup_j \Delta_j') \backslash \Delta')
        }
        \and
%        \inferrule[\eltsrule{Sel}]{
%            \dual{s} \notin \dom{\Delta}
%            \and
%            j \in I
%        }{
%            (\Gamma; \Lambda; \Delta \cat s: \btsel{l_i: S_i}_{i \in I}) \by{\bactsel{s}{l_j}} (\Gamma; \Lambda; \Delta \cat s:S_j)
%        }
%        \and
%        \inferrule[\eltsrule{Bra}]{
%            \dual{s} \notin \dom{\Delta} \quad j \in I
%        }{
%            (\Gamma; \Lambda; \Delta \cat s: \btbra{l_i: T_i}_{i \in I}) \by{\bactbra{s}{l_j}} (\Gamma; \Lambda; \Delta \cat s:S_j)
%        }
%        \and
        \inferrule[\eltsrule{Tau}]{
            \Delta_1 \red \Delta_2 \vee \Delta_1 = \Delta_2
        }{
            (\Gamma; \Lambda; \Delta_1) \by{\tau} (\Gamma; \Lambda; \Delta_2)
        }
    \end{mathpar}
\end{mdframed}

    \caption[Labeled Transition System for Typed Environments]{Labeled Transition System for Typed Environments. 
    \label{top:fig:envLTS}}
    \end{figure}

    %%%%%%%%%%%%%%%%%%%% End Environment LTS Figure %%%%%%%%%%%%%%%%%%%%%%

\noi
The typed LTS  combines
the LTSs in \Cref{top:fig:untyped_LTS,top:fig:envLTS}. 

\begin{definition}[Typed Labeled Transition System]
	\label{top:d:tlts}
	A {\em typed transition relation} is a typed relation
	$\horel{\Gamma}{\Delta_1}{P_1}{\by{\ell}}{\Delta_2}{P_2}$
	where:
	\begin{enumerate}
		\item
				$P_1 \by{\ell} P_2$ and 
		\item
				$(\Gamma, \emptyset, \Delta_1) \by{\ell} (\Gamma, \emptyset, \Delta_2)$ 
				with $\Gamma; \emptyset; \Delta_i \proves P_i \hastype \Proc$ ($i=1,2$).
%				\dk{We sometimes annotated the output action with
%				the type of value $V$ as in $\widetilde{m} \bactout{n}{V: U}$.}
	\end{enumerate}
	We 
	%extend to $\By{}$ and $\By{\hat{\ell}}$  where we 
	write  $\By{}$ for the reflexive and transitive closure of $\by{}$,
	$\By{\ell}$ for the transitions $\By{}\by{\ell}\By{}$, and $\By{\hat{\ell}}$
	for $\By{\ell}$ if $\ell\not = \tau$ otherwise $\By{}$.
\end{definition}

\newc{A typed transition relation requires type judgements with an empty $\Lambda$, i.e., an empty environment 
for linear higher-order types.
Notice that for open process terms (i.e., with free variables), 
we can always apply Rule~$\trule{EProm}$ (cf. \Cref{top:fig:typerulesmys}) and obtain an empty $\Lambda$. 
As it will be clear below (cf. \Cref{top:d:typedrel}), we will be working with closed process terms, i.e., processes without free  variables.
}

% \begin{newaddenv}
	\paragraph{Typed Relations.}
	% \todo[inline]{DO WE NEED TYPED RELATIONS DEFINITION?}

\noi We now define \emph{typed relations} and \emph{contextual equivalence} (i.e., barbed congruence).  
To define typed relations, we first define \emph{confluence}
over session environments $\Delta$.
\newc{Recall that 
$\Delta$ captures session communication, which is deterministic. 
The notion of confluence allows us to abstract away from alternative computation paths 
that may arise due to non-interfering reductions of session names.}

\begin{definition}[Session Environment Confluence]\label{top:d:conf}
	Two session environments $\Delta_1$ and $\Delta_2$
	are \emph{confluent}, denoted $\Delta_1 \bistyp \Delta_2$,
	if there exists a $\Delta$ such that:
	i)~$\Delta_1 \red^\ast \Delta$ and 
	ii)~$\Delta_2 \red^\ast \Delta$
	(here we write $\red^\ast$ for the multi-step reduction in \Cref{top:d:wtenvred}).
\end{definition}

We illustrate confluence by means of an example: 

\begin{example}[Session Environment Confluence]
	Consider the (balanced) session environments:
	\begin{eqnarray*}
	\Delta_1 & = & \set{s_1: T_1 \cat s_2: \btinp{U_2} \tinact \cat \dual{s_2}: \btout{U_2} \tinact} \\
	\Delta_2 & = & \set{s_1: T_1 \cat s_2: \btout{U_1} \btinp{U_2} \tinact \cat \dual{s_2}: \btinp{U_1} \btout{U_2} \tinact}
	\end{eqnarray*}
	Following \Cref{top:d:wtenvred}, we have that 
 $\Delta_1 \red \set{s_1: T_1 \cat s_2: \tinact \cat \dual{s_2}: \tinact}$
	and $\Delta_2 \red \red \set{s_1: T_1 \cat s_2: \tinact \cat \dual{s_2}: \tinact}$.
	Therefore,  $\Delta_1$ and $\Delta_2$ are confluent.
	\qed
\end{example}

Typed relations relate only closed processes whose
session environments are balanced  and confluent:

\begin{definition}[Typed Relation]\label{top:d:typedrel}
	We say that a binary relation over typing judgements
	\[
		\Gamma_1; \emptyset; \Delta_1 \proves P_1 \hastype \Proc\ \Re \ \Gamma_2; \emptyset; \Delta_2 \proves P_2 \hastype \Proc
	\]
	\noi
	is a {\em typed relation} whenever:
	\begin{enumerate}
		\item	$P_1$ and $P_2$ are closed;
		\item	$\Delta_1$ and $\Delta_2$ are balanced (cf. \Cref{top:d:wtenvred}); and
		\item	$\Delta_1 \bistyp \Delta_2$ (cf. \Cref{top:d:conf}).
	\end{enumerate}
	\end{definition}
	
	\begin{notation}[Typed Relations]
	\label{not:typedrel}
Given a typed relation $\Gamma_1; \emptyset; \Delta_1 \proves P_1 \hastype \Proc\ \Re \ \Gamma_2; \emptyset; \Delta_2 \proves P_2 \hastype \Proc$, to reduce eye strain 
%we shall adopt the following convenient notations. 
%\begin{enumerate}
%	\item When $\Gamma_1 = \Gamma_2 = \Gamma$, \fixed{to emphasize that $\Gamma$ applies to both $P_1$ and $P_2$,} we shall write: $$\horel{\Gamma}{\Delta_1}{P_1}{\ \Re \ }{\Delta_2}{P_2}$$
%	 \item When $\Gamma_1 \neq \Gamma_2$, to emphasize this difference, 
%	 \item 
we shall write: $$\horel{\Gamma_1}{\Delta_1}{P_1}{\ \Re \ }{\Gamma_2; \Delta_2}{P_2}$$
%\end{enumerate}
		\end{notation}	
%While notation (1) will be useful when defining characteristic bisimilarity (\Cref{top:d:fwb}), we shall use notation (2) to define MST-bisimilarity (\Cref{pi:d:fwb}) and associated results.

%\begin{notation}[Typed Relations]
%	To reduce eyestrain, we write
%	\[
%		\horel{\Gamma}{\Delta_1}{P_1}{\ \Re \ }{\Delta_2}{P_2}
%	\]
%	to denote the typed relation 
%		$\Gamma; \emptyset; \Delta_1 \proves P_1 \hastype \Proc\ \Re \ \Gamma; \emptyset; \Delta_2 \proves P_2 \hastype \Proc$. 
%		\fixed{Notice that $\Gamma$ applies to both $P_1$ and $P_2$.}
%\end{notation}	
%\end{definition}

% \end{newaddenv}

% \todo[inline]{lts end}
%%%%%%%%%%%%%%%%%%%%%% UNTYPED LTS END %%%%%%%%%%%%%%%%%%%%%%%%%%%%%%%%%%%%%%%%%%%%%%%%%%%%%
\end{newaddenv}

\begin{newaddenv}
  \paragraph{Characteristic  Bisimilarity}
Characteristic bisimilarity equates typed processes by relying on \emph{characteristic trigger processes}.
\fixed{Intuitively, characteristic processes arise in characterizations of contextual equivalence to (succintly) capture the arbitrary contexts in which an exchanged name can be used by a recipient.}
This notion, which we recall below, needs to be adjusted for our purposes. 

\thesisalt{
\begin{figure}
  \[
    \begin{array}{rclcrcl}
      \mapchar{\btinp{C} S}{u}
      &\defeq&
      \binp{u}{x} (\bout{t}{u} \inact \Par \mapchar{C}{x})
      &\qquad &
      \omapchar{S}  & \defeq &  s ~~ (s \textrm{ fresh})
      \\
      \mapchar{\btout{C} S}{u}
      &\defeq&
      \bout{u}{\omapchar{C}} \bout{t}{u} \inact
      &&
      \omapchar{\chtype{S}} &\defeq& a ~~ (a \textrm{ fresh})
      \\
%      \mapchar{\btsel{l : S}}{u}
%      & \defeq &
%      \bsel{u}{l} \bout{t}{u} \inact
%      &&
%      % \omapchar{\chtype{L}} &\defeq&a ~~ (a \textrm{ fresh})
%      \\
%      \mapchar{\btbra{l_i: S_i}_{i \in I}}{u}
%      & \defeq &
%      \bbra{u}{l_i: \bout{t_i}{u} \inact}_{i \in I}		 
%      &&
%      % \omapchar{\shot{U}} &\defeq& \abs{x}{\mapchar{U}{x}}
%      \\
       
  %		\\
  %		\mapchar{\tvar{t}}{u}
  %		&\defeq&
  %		\varp{X}_{\vart{t}}
  %		\\
  %
      \mapchar{ \trec{t}{S} }{u} &\defeq& \mapchar{S \subst{\tinact}{\vart{t}} }{u}
  
      &&
      % \omapchar{\lhot{U}} & \defeq &  \abs{x}{\mapchar{U}{x}}
      \\
      \mapchar{\tinact}{u}
      & \defeq &
      \inact
      \\
      \mapchar{\chtype{S}}{u} 
      &\defeq&
      \bout{u}{\omapchar{S}} \bout{t}{u} \inact		 
      \\
      % \mapchar{\chtype{L}}{u}
      % &\defeq&
      %  \bout{u}{\omapchar{L}} \bout{t}{u} \inact 
      % \\
      % \mapchar{\shot{U}}{u}
      % &\defeq& 
      % \appl{u}{\omapchar{U}}
      % \\
      % \mapchar{\lhot{U}}{u}
      % &\defeq &
      % \appl{u}{\omapchar{U}}
    \end{array}
    \]
  
  \caption[Characteristic processes and characteristic values]{Characteristic processes (left) and characteristic values (right), as introduced in   \Cref{pi:d:trigger}.\label{top:fig:char}}
  \end{figure}}{}

\begin{definition}[Characteristic trigger process~\cite{KouzapasPY17}]
  \label{pi:d:trigger}
The \emph{characteristic trigger process} for type $C$ is
  \begin{align*}
    \ftrigger{t}{v}{C}	& \defeq 	\fotrigger{t}{x}{s}{C}{v}
  \end{align*}
  \noindent where $\mapchar{C}{y}$ is the \emph{characteristic process} 
  for $C$ on name $y$ (cf.~\Cref{top:fig:char}). 
  % \cite{KouzapasPY17}.
\end{definition}

We may now state the definition of characteristic bisimilarity that applies in our (first-order) setting:

\begin{definition}[Characteristic Bisimilarity]
    \label{top:d:fwb}
        A typed relation $\Re$ is a {\em  characteristic bisimulation} if 
        for all $\horel{\Gamma_1}{\Delta_1}{P_1}{\ \Re \ }{\Gamma_2; \Delta_2}{Q_1}$, 
\begin{enumerate}[1)]
    \item 
            Whenever 
            $\horel{\Gamma_1}{\Delta_1}{P_1}{\by{\news{\widetilde{m_1}} \bactout{n}{V_1: U_1}}}{\Delta_1'}{P_2}$ 
            then there exist 
            $Q_2$, $V_2$,  $\Delta'_2$ such that 
            $\horel{\Gamma_2}{\Delta_2}{Q_1}{\by{\news{\widetilde{m_2}}\bactout{n}{V_2: U_2}}}{\Delta_2'}{Q_2}$
            and, for a fresh $t$,
            \[
                \Gamma_1; \Delta''_1  \proves  {\newsp{\widetilde{m_1}}{P_2 \Par \ftrigger{t}{V_1}{U_1}}}
                 \ \Re\ 
                \Gamma_2; \Delta''_2 \proves {\newsp{\widetilde{m_2}}{Q_2 \Par \ftrigger{t}{V_2}{U_2}}}
            \]

    \item	
            For all $\horel{\Gamma_1}{\Delta_1}{P_1}{\by{\ell}}{\Delta_1'}{P_2}$ such that 
            $\ell$ is not an output, there exist $Q_2$,   $\Delta'_2$ such that 
            $\horel{\Gamma_2}{\Delta_2}{Q_1}{\by{\hat{\ell}}}{\Delta_2'}{Q_2}$
            and
            $\horel{\Gamma_1}{\Delta_1'}{P_2}{\ \Re \ }{\Gamma_2; \Delta_2'}{Q_2}$; and 

    \item	The symmetric cases of 1 and 2.                
\end{enumerate}
The largest such bisimulation is called \emph{characteristic bisimilarity}, denoted by $\fwb$.
\end{definition}
\end{newaddenv}}{}

% We require some auxiliary notations and definitions.

\subsubsection{MST-bisimila\-ri\-ty and Main Result}
We introduce MST-bisimila\-ri\-ty (denoted $\mstb$), and discuss key differences with respect to characteristic bisimilarity. 
One of the differences is that we let an action along a name $n$ to be mimicked by an action on a possibly indexed name $n_i$, for some $i$. %We need the following notion: 

\begin{definition}[Indexed name]
  \label{pi:def:indexedname}
  Given a name $n$, we write $\iname{n}$ to either denote 
  $n$ or any indexed name $n_i$, with $i > 0$.
\end{definition}

\newcommand{\namesrelate}{\processrelate}
Suppose we wish to relate $P$ and $Q$ using $\mstb$, and that $P$ performs an
output action involving name $v$. In our setting, $Q$ should send a \emph{tuple}
of names:  the decomposition of $v$. Another difference is that output
objects should be related by the relation $\namesrelate$: 
\cralt{
\todo[inline]{This definition is repeated for clarity.}}{}
\begin{newaddenv}
  \begin{definition}[Relating names]
    \label{pi:def:namesrelate}
    We define the relation on names $\namesrelate$ as follows:
    $$
    \frac{}
    {\epsilon \namesrelate \epsilon} 
    \qquad 
    \frac{\Gamma;\Lambda;\Delta \proves n_i \hastype C}
    {n_i \namesrelate (n_i,\ldots,n_{i+\len{\Gtopt{C}}-1})} 
    \qquad 
     \frac{\tilde n \namesrelate \tilde m_1 \ 
    \quad  n_i \namesrelate \tilde m_2}
    {\tilde n,n_i \namesrelate \tilde m_1, \tilde m_2} 
    $$
    \noindent where $\epsilon$ denotes the empty list. 
  \end{definition}
\end{newaddenv}

Our variants of characteristic and trigger processes are defined as follows: 

\begin{definition}[Minimal characteristic processes]
\label{pi:d:mcp}
Given a type $C$, name $u$, and index $i$, we define the \emph{minimal} characteristic processes $\mapcharm{C}{u}_i$ and $ \omapcharm{C}$ as follows:
  \begin{align*}
    \mapcharm{\btinp{C} S}{u}_i
    &\defeq
    \binp{u_i}{x} (\bout{t_1}{u_{i+1}, \ldots,u_{i+\len{\Gtopt{S}}}} \inact \Par \mapcharm{C}{x}_i)
    \\
    \mapcharm{\btout{C} S}{u}_i
    &\defeq
    \bout{u_i}{\omapcharm{C}} \bout{t_1}{u_{i+1}, \ldots,u_{i+\len{\Gtopt{S}}}} \inact \\
    \mapcharm{\tinact}{u}_i & \defeq \inact \\
    \mapcharm{\chtype{C}}{u}_i
    &\defeq \bout{u_1}{\omapcharm{C}} \bout{t_1}{u_1} \inact \\
    \mapcharm{ \trec{t}{S} }{u}_i &\defeq \mapcharm{S \subst{\tinact}{\vart{t}} }{u}_i \\
    \omapcharm{S}  & \defeq   \wtd s ~~ (\len{\wtd s} = \len{\Gtopt{S}}, \wtd s \textrm{ fresh}) \\
    \omapcharm{\chtype{C}} &\defeq a_1 ~~ (a_1 \textrm{ fresh})
  \end{align*}
  \noindent where $t_1$ is a fresh (indexed) name. 
\end{definition}

Given this definition, we may now revise   \Cref{pi:d:trigger}.

\begin{definition}[Minimal characteristic trigger process]
  \label{pi:d:mintrigger}
Given a type $C$, the trigger process is
  \begin{align*}
    \ftriggerm{t}{v_i}{C}	& \defeq 	\fotriggerm{t_1}{x}{s_1}{C}{\tilde v}{i}
  \end{align*}
  \noindent where
  $v_i \namesrelate \wtd v$, $y_i \namesrelate \wtd y$, and
   $\mapcharm{C}{y}_i$  is a  
  minimal characteristic process for type $C$ on name $y$
  \longversion{(see~\Cref{pi:d:mcp})}{(see \cite{AAP21-full} for a definition)}. 
\end{definition}

%% $\horel{\Gamma}{\Delta_1}{P_1}{\ \Re \ }{\Delta_2}{Q_1}$

We are now ready to define MST-bisimilarity. 
In the following, we shall adopt \Cref{not:typedrel}(2) for typed relations in order to explicitly account for the effect of the decompositions in the types/assignments recorded in~$\Gamma$.
\begin{definition}[MST-Bisimilarity]
  \label{pi:d:fwb}
	A typed relation $\Re$ is an {\em  MST bisimulation} if 
	for all 
	%$\horelm{\Gamma_1;\Delta_1}{P_1}{\ \Re \ }{\Gamma_2;\Delta_2}{Q_1}$, 
	$\horelm{\Gamma_1;\Delta_1}{P_1}{\ \Re \ }{\Gamma_2;\Delta_2}{Q_1}$, 
	\begin{enumerate}
		\item 
				Whenever 
				$\horelm{\Gamma_1;\Delta_1}{P_1}
        {\by{\news{\widetilde{m_1}} \bactout{n}{v:C_1}}}
        {\Delta'_1}{P_2}$ 
        then there exist
				$Q_2$, $\Delta'_2$, and $\sigmav$ such that 
				$\horelm{\Gamma_2;\Delta_2}{Q_1}
        {\By{\news{\wtd{m_2}}\bactout{\iname{n}}{\tilde v : \Gtopt{C}}}}
        {\Delta_2'}{Q_2}$
        where 
        $v\sigmav \namesrelate \wtd v$
        and, for a fresh $t$,
%        \begin{align*}
%          \Gamma; \Delta''_1  &\proves  {\newsp{\wtd{m_1}}{P_2 \Par \ftrigger{t}{v}{C_1}}}
%             \Re \\
%          \Delta''_2 &\proves {\newsp{\widetilde{m_2}}{Q_2 \Par \ftriggerm{t}{v\sigma}{C_1}}}
%        \end{align*}
            \[
                \Gamma_1; \Delta''_1  \proves  {\newsp{\wtd{m_1}}{P_2 \Par \ftrigger{t}{v}{C_1}}}
                 \ \Re\ 
                \Gamma_2; \Delta''_2 \proves {\newsp{\widetilde{m_2}}{Q_2 \Par \ftriggerm{t}{v\sigma}{C_1}}}
            \]
        %
        % \[
        %   \Gamma; \Delta''_1  \proves  {\newsp{\wtd{m_1}}{P_2 \parallel \ftrigger{t}{v}{C_1}}}
        %      \Re\ 
        %   \Delta''_2 \proves {\newsp{\widetilde{m_2}}{Q_2 \parallel \ftriggerm{t}{v\sigma}{C_1}}}
        % \]
        
       \item Whenever $\horelm{\Gamma_1;\Delta_1}{P_1}
       {\by{\abinp{n}{v}}}{\Delta_1'}{P_2}$
       then there exist  $Q_2$, $\Delta_2'$, and $\sigmav$ such that 
       $\horelm{\Gamma_2;\Delta_2}{Q_1}{\By{\abinp{\iname{n}}{\tilde v}}}
       {\Delta_2'}{Q_2}$ 
       where 
       $v\sigmav \namesrelate \wtd v$
       and 
       $\horelm{\Gamma_1;\Delta_1'}{P_2}{\ \Re \ }
       {\Gamma_2;\Delta_2'}{Q_2}$, 
        
      \item	
				Whenever  
        $\horelm{\Gamma_1;\Delta_1}{P_1}{\by{\ell}}
        {\Delta_1'}{P_2}$, with  
				$\ell$ not an output or input, then there exist 
        $Q_2$ and $\Delta'_2$ such that 
				$\horelm{\Gamma_2;\Delta_2}{Q_1}
        {\By{\hat{\ell}}}{\Delta_2'}{Q_2}$
				and
        $\horelm{\Gamma_1;\Delta_1'}{P_2}{\ \Re \ }
        {\Gamma_2; \Delta_2'}{Q_2}$  
        and  
        $\subl{\ell} = n$ implies  
        $\subl{\hat \ell} = \iname{n}$.

		\item	The symmetric cases of 1, 2, and 3.                 
	\end{enumerate}
	The largest such bisimulation is called \emph{MST bisimilarity} ($\mstb$).
\end{definition}

\begin{newaddenv}
  
We can now state our dynamic correctness result:

\begin{restatable}[Operational Correspondence]{theorem}{mainbsthm}
\label{pi:t:mainbsthm}
  Let $P$ be a process such that 
  $\Gamma;\Delta \proves P$. We have 
  $$\horelm{\Gamma;\Delta}{P}{\ \mstb \ }
  {\Gtopt{\Gamma};\Gtopt{\Delta}}{\Do{P}}$$ 
\end{restatable}
%\begin{proof}
%   To prove that $\Do{P}$ is MST-bisimilar to $P$ we construct a relation
% $\relS$ that is an MST-bisimulation and  that contains $(P,\Do{P})$; see \defref{d:relation-s}.
The proof of this theorem is by coinduction: we exhibit a binary relation $\relS$  such that $(P,\Do{P}) \in \relS$
and is an MST bisimulation. 
%We define $\relS$
%\longversion{(\defref{d:relation-s})}{}
% using two auxiliary functions $\Cbpi{-}{-}{\cdot}$ and
%$\Dbpi{-}{-}{\cdot}$ 
%\longversion{(\tabref{t:tablecd} and \tabref{t:tablecd-rec}).}{.}
\longversion{The proof
that $\relS$ is an MST bisimulation is given by \lemref{pi:l:lemms} and
\Cref{pi:l:lemmsdir-typed} in the next sub-section.}
{See~\cite{AAP21-full} for the full account.}
%Finally, by the definition we have  \in \relS$.
% (see \appref{app:mainbsthm} for more details). 
  % The relation $\mathcal{S}$ is a MST bisimulation (\lemref{l:lemms} 
  % and \lemref{l:lemmsdir}) 
  % and $(P, \D{P}) \in \mathcal{S}$.
%\end{proof}
\end{newaddenv}

\begin{newaddenv}
  \subsection{Proof of \Cref{pi:t:mainbsthm}}
   \label{sec:t:mainbsthm}
\end{newaddenv}
 
\begin{newaddenv}
\subsubsection{Preliminaries}
In this section we define the relation $\relS$ used in the proof of \Cref{pi:t:mainbsthm}. 
We require some auxiliary notions. 

First, we define a relation $\processrelate$  on processes, which corresponds to the extension of the relation on indexed names given by \Cref{pi:def:namesrelate}: 
\end{newaddenv}
  % \todo[inline]{Check repetition of below Def}
\begin{definition}[Relation $\processrelate$ on processes]
Given the relation $\namesrelate$ on names (\Cref{pi:def:namesrelate}), we define the relation $\processrelate$ on processes as follows:
\label{pi:def:valuesprelation}
\begin{mathpar}
  \inferrule[\eltsrule{IPSnd}]
  {P\sigma \processrelate P' \and v\sigma \namesrelate \wtd v \and \sigma = \nextn{n_i}}
  {\bout{n_i}{v}P \processrelate \bout{n_i}{\wtd v}P'}
  \qquad 
  \inferrule[\eltsrule{IPInact}] {
  }{\inact \processrelate \inact} 
  \\
  \inferrule[\eltsrule{IPRcv}] {
    P\sigma \processrelate P' \and v\sigma \namesrelate \wtd v \and \sigma = \nextn{n_i}
  }{\binp{n_i}{y}P \processrelate \binp{n_i}{\wtd y}P'}  
 \quad 
  \inferrule[\eltsrule{IPNews}] 
  { P \processrelate P' \and \tilde m_1 \namesrelate \tilde m_2 
  }{\news{\tilde m_1}P \processrelate \news{\tilde m_2}P' } 
\end{mathpar}
\end{definition}

\begin{newaddenv}

We have the following properties for $\processrelate$:
  \begin{lemma}
    \label{pi:lemm:processrelate-trigger}
    We have: 
    $(\ftrigger{t}{v}{C}\subst{t_1,v_i}{t,v},\ \ftriggerm{t_1}{v_i}{C}) \in \processrelate$ 
    for $i > 0$. 
  \end{lemma} 
  \begin{proof}[Proof]
    % Directly by~\Cref{pi:d:trigger}, \Cref{pi:d:mintrigger}, and 
    % \Cref{pi:def:valuesprelation}. 
    Directly by \Cref{\thesisalt{pi:d:trigger}{top:d:trigger},pi:d:mintrigger,pi:def:valuesprelation}. 
  \end{proof}
\end{newaddenv}

\begin{newaddenv}
\begin{lemma}
  \label{pi:l:processrelate-mst}
  Relation $\processrelate$ is an MST bisimulation. 
  \end{lemma}
  \begin{proof}[Proof]
  Straightforward by transition induction. 
  \end{proof} 
\end{newaddenv}

As we have seen, the output clause of MST bisimilarity ``appends'' trigger processes in parallel to the processes under comparison.
The following definitions introduce notations which are useful for distinguishing the triggers included in a process. 

\begin{definition}[Trigger Collections]
\label{pi:d:triggerscollection}
%We define.
\hfill
\begin{itemize}
	\item 
We let $H, H'$ to range over \emph{trigger collections}: processes of the form
$P_1 \Par \cdots \Par P_n$ (with $n \geq 1$), where each $P_i$ is a trigger
process or a process that originates from a trigger process. 
%\begin{definition}[Process in parallel with a trigger or a characteristic process]
%\label{pi:def:parallel}
\item 
We write 
$P \parallel Q$ to stand for $P \Par Q$ 
where either $P$ or $Q$ is a trigger collection.  
\end{itemize}

\end{definition}

\begin{example}
Let $H_1 =  \ftrigger{t}{v}{C} \Par \mapchar{C}{u} \Par 
\bout{t'}{n}\inact$  
where $v, t,t',u,n$ are channel names, 
$C$ a channel type. 
We could see that $\mapchar{C}{u}$ and  $\bout{t'}{n}\inact$ originate from a trigger process. Thus, $H_1$ is a trigger collection.
\end{example}

\begin{definition}[Propagators of $P$]
 \label{pi:d:fpn}
  We define $\fpn{P}$ to denote the set of free propagator names in $P$. 
\end{definition}
 
The following definition is useful when constructing an MST bisimulation for \emph{recursive processes}, i.e., processes $P$ such that $\recpx{P}$ holds (cf. \Cref{pi:d:fnv}). 
The \emph{depth} of a recursive variable denotes the number of prefixes  
  preceding the occurrence of a recursive variable  $X$ or a subprocess of the form $\recp{X}Q$; in turn, this depth will be related to the current trio mimicking a given process.

\begin{definition}[Depth of recursive variable]
  Let $P$ be a recursive process.
%  We define \rvardepth{P} to count sequential prefixes in $P$   preceding recursive variable $X$ or a subprocess of shape $\recp{X}Q$. 
  The functions \rvardepth{P} and \rvardepthaux{P} are mutually defined as follows: 
  \begin{align*}
    \rvardepth{P} &= 
    \begin{cases}
      \rvardepthaux{P} & \text{ if } \recpx{P}   \\
        0 & \text{ otherwise } \\
    \end{cases}
    \qquad \qquad 
    \rvardepthaux{P} = 
    \begin{cases}
      0 & \text{ if } P = \recp{X}Q \text{ or } P = \rvar{X} 
      \\
      \rvardepth{Q} + 1 & \text{ if } P = \alpha.Q \\
      \rvardepth{Q} + \rvardepth{R}  & \text{ if } P = Q \Par R 
      \\
      \rvardepth{Q} & \text{ if } P = \news{s}Q
    \end{cases}
  \end{align*}
\end{definition}
\newcommand{\jdepth}[4]{\ensuremath{\mathscr{D}_#4(#1,#2,#3)}}
\begin{definition}\label{pi:d:jdepth}
The predicate \jdepth{P}{Q}{d}{X} holds whenever there exist $Q, Q'$ such that
(i)~$P \equiv Q' \subst{\recp{X}Q}{X}$
and 
(ii)~$\rvardepth{Q'} = d$.
%  if $Q \equiv Q'\subst{\recp{X}Q^*}{X}$ \\ where  $\alpha_d.\alpha_{d-1}. \cdots .\alpha_{0}.  (X \Par R) \equiv Q' \preceq \recp{X}Q^* $
\end{definition}

Finally, we introduce a notation on (indexed) names and substitutions:

% \begin{figure*}
\begin{figure}[t!]
  \begin{table}[H]
    \thesisalt{
    \begin{tabular}{ |l|l|l|}
      \rowcolor{gray!25}
  \hline
  $P$ &
    \multicolumn{2}{l|}{
  \begin{tabular}{l}
    \noalign{\smallskip}
    $\Cbpi{\tilde u}{\tilde x}{P}$
    \smallskip
  \end{tabular} 
  } 
  \\
  \hline
  $Q_1 \parallel Q_2$ 
  &
  \begin{tabular}{l}
  \noalign{\smallskip}
     $\{ R_1 \parallel R_2: 
     R_1 \in \Cbpi{\tilde u_1}{\tilde y}{Q_1}, 
     R_2 \in \Cbpi{\tilde u_2}{\tilde z}{Q_2}
     \}$
     \smallskip 
  \end{tabular}
  &
  %side-conditions
  \begin{tabular}{l}
    \noalign{\smallskip}
    $\wtd y = \fnb{Q_1}{\tilde x}$
    \\
    $\wtd z = \fnb{Q_2}{\tilde x}$ \\
    $\subst{\tilde u}{\tilde x} = \subst{\tilde u_1}{\tilde y} \cdot \subst{\tilde u_2}{\tilde z}$ 
    \smallskip
  \end{tabular}
     \\
     \hline
     $\news{s:C}{Q}$ &
     \begin{tabular}{l}
       \noalign{\smallskip}
        $\{\news{\widetilde{s}:\Gtopt{C}}{\,R} : \Cbpi{\tilde u}{\tilde x}{Q\sigma} \}$
       \smallskip
     \end{tabular}
     &
     %side-conditions
     \begin{tabular}{l}
       \noalign{\smallskip}
       $\widetilde{s} = (s_1,\ldots,s_{\len{\Gtopt{C}}})$ \\
       $\sigma =  \subst{s_1 \dual{s_1}}{s \dual{s}}$
      %  $\sigma = \begin{cases}
      %              \subst{s_1 \dual{s_1}}{s \dual{s}} & \text{if} \ C = S \\
      %              \subst{s_1}{s} & \text{if} \ C = \chtype{U}
  %  \end{cases}
  %  $
       \smallskip
     \end{tabular}
     \\
     \hline
  $Q$
  &
  \begin{tabular}{l}
  \noalign{\smallskip}
  if \jdepth{Q}{Q'}{d}{X}:
%  if $Q \equiv Q'\subst{\recp{X}Q^*}{X}$ \\
% where  $\alpha_d.\alpha_{d-1}. \cdots .\alpha_{0}.
%  (X \Par R) \equiv Q' \preceq \recp{X}Q^* $:
  \\
  \quad $\Cbrecpi{\tilde u}{\tilde x}{\recp{X}Q'}^d$
  % \\
  \qquad \qquad  
  \smallskip 
  \\
  \hdashline 
  otherwise:
  \\
  \noalign{\smallskip}
   \quad   $\{  \news{\wtd \prop} R \,:\,    R = \apropout{k}{\wtd u} \Par \Bopt{k}{\tilde x}{Q} \}$ 
  \\
  \quad $\cup$ 
  \\
\quad    $\{ \news{\wtd \prop}R \,:\, R \in \Dbpi{\tilde u}{\tilde x}{Q}\}$
  \smallskip 
  \end{tabular}
  &
  %side-conditions
  \begin{tabular}{l}
    \noalign{\smallskip}
    $\wtd \prop = \fpn{R}$
    \\
    $d \geq 1$
    \\
    (cf. \Cref{pi:d:jdepth} 
     for \jdepth{\cdot}{\cdot}{\cdot}{-})
    \smallskip
  \end{tabular} 
  \\
  \hline
  $H$
  &
  \begin{tabular}{l}
  \noalign{\smallskip}
   $\{ H' : H \subst{\tilde u}{\tilde x} \processrelate H'\}$ 
  \smallskip 
  \end{tabular}
  &
  %side-conditions
  \begin{tabular}{l}
    \noalign{\smallskip}
    % $\wtd \prop = \fpn{R}$
    \smallskip
  \end{tabular} 
  \\
  \hline 
  % \hline
  % $Q$
  % &
  % \begin{tabular}{l}
  % \noalign{\smallskip}
  % % $f$ \\
  % $\Cbrecpi{\tilde u}{\tilde x}{\recp{X}Q^*}^d$
  % \\
  % if $Q \equiv Q'\subst{\recp{X}Q^*}{X}$ \\
  % $\text{where  $Q' \equiv \alpha_d.\alpha_{d-1}. \ldots .\alpha_{0}.
  % (X \Par R)$ and $Q' \preceq \recp{X}Q^* $}$
  % % $\text{where  $Q' \equiv \alpha_d.\alpha_{d-1}. \ldots .\alpha_{d-p}.
  % % (P_X \Par R)$ and $\recp{X}Q^* \preceq P_X$}$
  % % $\text{if  $Q \equiv \alpha_d.\alpha_{d-1}. \ldots .\alpha_{d-p}.
  % % \recp{X}(P_X \Par R)$ and $\recp{X}Q^* \preceq P_X$}$
  % % \smallskip 
  % % \\
  % % \smallskip
  % \end{tabular}
  % &
  % %side-conditions
  % \begin{tabular}{l}
  %   \noalign{\smallskip}
  %   % $\wtd \prop = \fpn{R}$ \\
  %   % $\rho = \subst{\tilde u}{\tilde x}$
  %   \smallskip
  % \end{tabular} 
  % \\
  % \hline 
  %   $\inact$ &
  % \begin{tabular}{l}
  %   \noalign{\smallskip}
  %   $\{\inact\}$
  %  \smallskip
  % \end{tabular}
  % &
  % \begin{tabular}{l}
  %     \noalign{\smallskip}
  % %$\widetilde x = \es$ 
  % \smallskip
  % \end{tabular}
  % \\
  %  \hline 
  %   $\inact$ &
  % \begin{tabular}{l}
  %   \noalign{\smallskip}
  %   $\{\inact\}$
  %  \smallskip
  % \end{tabular}
  % &
  % \begin{tabular}{l}
  %     \noalign{\smallskip}
  % %$\widetilde x = \es$ 
  % \smallskip
  % \end{tabular}
  % \\ 
      \rowcolor{gray!25}
      \hline
      $P$ &
        \multicolumn{2}{l|}{
      \begin{tabular}{l}
        \noalign{\smallskip}
        $\Dbpi{\tilde u}{\tilde x}{P}$
        \smallskip
      \end{tabular}
    }  \\
      \hline
    $\binp{n_i}{y}Q$
    &
    
      \begin{tabular}{l}
          \noalign{\smallskip}
          $\{\binp{n_{\mstindex} \rho}{\wtd y} 
          \apropout{k+1}{\wtd z \rho}
         \Par \Bopt{k+1}{\tilde z}{Q\sigma}\}$ \\
          \smallskip
      \end{tabular}
      &
      %side-conditions
      \begin{tabular}{l}
        \noalign{\smallskip}
        $y:S$ 
        % $\wedge$
        \\ 
        $\wtd y = (y_1,\ldots,y_{\len{\Gtopt{S}}})$ \\
        $\wtd w = \linecondit{\lin{n_i}}{\{n_i\}}{\epsilon}$ \\
        $\wtd z = \fnb{Q}{\wtd x \wtd y \setminus \wtd w}$\\
        $\rho = \subst{\tilde u}{\tilde x}$ \\
        % $\rho= \rho
        % \cdot \subst{\tilde u_z}{\tilde z}$ \\
        $\sigma = \nextn{n_i}\cdot \subst{y_1}{y}$  \\
        $\mstindex = \linecondit{\traux{S}}{\indf{S}}{i}$
        \smallskip
      \end{tabular}
        \\
      \hline
      
      $\bout{n_i}{y_j}{Q}$ &
        \begin{tabular}{l}
          \noalign{\smallskip}
                $\{\bbout{n_{\mstindex} \rho}
                {\wtd y \rho}         
                \apropout{k+1}{\wtd z \rho}  \Par$ 
          $\Bopt{k+1}{\tilde z}{Q\sigma} \}$
         \smallskip
      \end{tabular}
      &
      %side-conditions
      \begin{tabular}{l}
        \noalign{\smallskip}
        $y_j : S$ 
        % $\wedge$ 
        \\ 
        $\wtd y = (y_j, \ldots,y_{j+\len{\Gtopt{S}}-1})$ \\
        $\wtd w = \linecondit{\lin{n_i}}{\{n_i\}}{\epsilon}$ \\
    % $\wtd w = \begin{cases}
    %  \{ n_i \} & \text{if} \ \lin{n_i} \\
    %  \epsilon & \text{otherwise} 	
    %  \end{cases}
    % $ \\ 
         $\wtd z = \fnb{Q}{\wtd x \setminus \wtd w}$\\
         $\rho = \subst{\tilde u}{\tilde x}$ \\
          % $\rho= \rho
          % \cdot \subst{\tilde u_z}{\tilde z}$ \\
         $\sigma = \nextn{n_i}$ \\
         ${\mstindex} = \linecondit{\traux{S}}{\indf{S}}{i}$
        \smallskip
      \end{tabular}
      \\
    
    \hline

      $Q_1 \Par Q_2$ &
      \begin{tabular}{l}
        \noalign{\smallskip}
        $\{ \propout{k}{\wtd y \rho}
        \apropout{k+\degree}{\wtd z \rho}   \Par$ 
        $\Bopt{k}{\tilde y}{Q_1} \Par \Bopt{k+\degree}{\tilde z}{Q_2}\}$ \\
        $\cup$
        \\
        $\{(R_1 \Par R_2) :
        R_1 \in \Cbpi{\tilde u_1}{\tilde y}{Q_1}, R_2 \in 
        \Cbpi{\tilde u_2}{\tilde z}{Q_2}\}$
        \smallskip
      \end{tabular}
      &
      %side-conditions
      \begin{tabular}{l}
        \noalign{\smallskip}
        $\wtd y  = \fnb{Q_1}{\wtd x}$ \\
        $\wtd z = \fnb{Q_2}{\wtd x}$ \\
        $\rho = \subst{\tilde u}{\tilde x}$ \\
      %  $\subst{\tilde u}{\tilde x} = \subst{\tilde u_1}{\tilde y}
      %  \cdot \subst{\tilde u_2}{\tilde z}$ \\
        $\degree = \lenHOopt{Q_1}$
        \smallskip
      \end{tabular}
          \\
      \hline
      $\recp{X}P$ &
    \begin{tabular}{l}
      \noalign{\smallskip}
      $\left\{ \propoutrec{k+1}{\wtd z\rho}
              \recp{X}\binp{\prop^r_{X}}{\wtd y}
        \propoutrec{k+1}{\wtd y}X
            \Par \Brecpi{k+1}{\tilde x}{P}_{g}
            \right\}$
     \smallskip
    \end{tabular}
    &
    \begin{tabular}{l}
        \noalign{\smallskip}
        $\wtd n = \fn{P}$ \\
        $\wtd n : \wtd C$ $\wedge$ $\wtd m = \bname{\wtd n : \wtd C}$ \\
        $\wtd z = \wtd x \cup \wtd m$, $\len{\wtd z} = \len{\wtd y}$ \\
        $g = \{X \mapsto \wtd m\}$
    \smallskip
    \end{tabular}
    \\
    \hline 
      $\inact$ &
    \begin{tabular}{l}
      \noalign{\smallskip}
      $\{\inact\}$
     \smallskip
    \end{tabular}
    &
    \begin{tabular}{l}
        \noalign{\smallskip}
    %$\widetilde x = \es$ 
    \smallskip
    \end{tabular}
  
      \\
      \hline 
    \end{tabular}}{\resizebox{1.01\textwidth}{!}{}}
    \caption{The sets $\Cbpi{\tilde u}{\tilde x}{P}$ (upper part) 
    and $\Dbpi{\tilde u}{\tilde x}{P}$ (lower part). \label{pi:t:tablecd}}
    % \vspace{-2mm}
    \end{table}
  \end{figure}

  \begin{figure}[t!]
    \begin{table}[H]

% \begin{figure*}

%     \begin{table}[H]
  \begin{tabular}{ |l|l|}
    \rowcolor{gray!25}
\hline
$P$ &
  \multicolumn{1}{l|}{
\begin{tabular}{l}
  \noalign{\smallskip}
  $\Cbrecpi{\tilde u}{\tilde x}{P}^d$
  \smallskip
\end{tabular} 
} 
\\
%    \hline
% $Q$
% &
% \begin{tabular}{l}
% \noalign{\smallskip}
% % $f$ \\
% $\Cbrecpi{\tilde u}{\tilde x}{\recp{X}Q^*}^d$
% \\
% if $Q \equiv Q'\subst{\recp{X}Q^*}{X}$ \\
% $\text{where  $Q' \equiv \alpha_d.\alpha_{d-1}. \ldots .\alpha_{0}.
% (X \Par R)$ and $Q' \preceq \recp{X}Q^* $}$
% % $\text{where  $Q' \equiv \alpha_d.\alpha_{d-1}. \ldots .\alpha_{d-p}.
% % (P_X \Par R)$ and $\recp{X}Q^* \preceq P_X$}$
% % $\text{if  $Q \equiv \alpha_d.\alpha_{d-1}. \ldots .\alpha_{d-p}.
% % \recp{X}(P_X \Par R)$ and $\recp{X}Q^* \preceq P_X$}$
% % \smallskip 
% % \\
% % \smallskip
% \end{tabular}
% &
% %side-conditions
% \begin{tabular}{l}
%   \noalign{\smallskip}
%   % $\wtd \prop = \fpn{R}$ \\
%   % $\rho = \subst{\tilde u}{\tilde x}$
%   \smallskip
% \end{tabular} 
% \\
\hline 
% $\Cbrecpi{\tilde u}{\tilde x}{Q}^d$
%\multirow{2}{*}{$\recp{X} Q$} 
\multirow{3}{*}{$\recp{X} Q$}
% $\recp{X} Q$
% $\recp{X}Q$
&
\begin{tabular}{l}
\noalign{\smallskip}
$ N \cup  \left\{\news{\wtd \prop} \big(
  \recp{X}\binp{\prop^r_{X}}{\wtd y}
\propoutrec{k+1}{\wtd y}X 
\Par
\proprinpk{r}{k}{\wtd x}\propoutrecsh{\rvar{X}}{\wtd x}\rvar{X} \Par R  \big)
\,:\,R \in \Dbrecpi{k}{\tilde x}{Q}^d_g \right\}$ 
\\
  {where:} 
  \\
     $N =
\begin{cases}
M  \cup  \left\{  \news{\wtd \prop} (\propoutrec{k}{\wtd z\rho}
\recp{X}\binp{\prop^r_{X}}{\wtd y}
\propoutrec{k}{\wtd y}X \Par   \Brecpi{k}{\tilde x}{Q}_g ) \right\}
% \\
 & \text{if } \recdep = 0 \\ 
\emptyset & \text{otherwise}
\end{cases}
$
\smallskip
\\
 $M = 
 \begin{cases}
  \{  \news{\wtd \prop} \big(
  \recp{X}\binp{\prop^r_{X}}{\wtd y}
\propoutrec{k+1}{\wtd y}X   \Par R \Par 
\\
\qquad \qquad \propoutrecsh{\rvar{X}}{\wtd x\rho}
\recp{X}\proprinpk{r}{k}{\wtd x}\propoutrecsh{\rvar{X}}{\wtd x}\rvar{X}  \big)
 \,:\, R \in \Dbrecpi{k}{\tilde x}{Q}^0_g \} & \text{ if } g \not= \es \\
\{  \news{\wtd \prop} \big(
  \recp{X}\binp{\prop^r_{X}}{}
\propoutrec{k+1}{}X \Par R  \Par  
\\
\qquad \qquad \apropoutrecsh{\rvar{X}}{} \Par 
\recp{X}\proprinpk{r}{k}{}(\apropoutrecsh{\rvar{X}}{} \Par \rvar{X})
\big)   \,:\, R \in \Dbrecpi{k}{\tilde x}{Q}^0_g \} & \text{ otherwise }
 \end{cases}$ 
 %\\
% $M_2 = \{  \news{\wtd \prop} (
%   \recp{X}\binp{\prop^r_{X}}{\wtd y}
% \propoutrec{k+1}{\wtd y}X \Par R$  \\
% $\qquad \Par 
% \proproutk{r}{\rvar{X}}{\wtd x\rho}
% \recp{X}\proprinpk{r}{k}{\wtd x}\propoutrecsh{\rvar{X}}{\wtd x}\rvar{X}
% )  \,:\, R \in \Dbrecpi{k}{\tilde x}{Q}^0_g \}$ \\
   \medskip 
\end{tabular}
%&
%%side-conditions
%\begin{tabular}{l}
%  \noalign{\smallskip}
%  % $\recdep = \rvardepth{Q}$
%  % $Q \equiv \recp{X} Q^*$ \\
%  $\wtd \prop = \fpn{R}$ \\
%  $\wtd x = \fs{Q}$ \\
%  $g = \{ X \mapsto \wtd x\}$ \\
%  $\rho = \subst{\tilde u}{\tilde x}$
%  \smallskip
%\end{tabular}
\\
% \hdashline 
% \chdas
% \\
\cdashline{2-2}
%\hline  
 & 
\begin{tabular}{l}
  \noalign{\smallskip}
   {Also:}
    \\
  % $\recdep = \rvardepth{Q}$
  % $Q \equiv \recp{X} Q^*$ \\
  $\wtd \prop = \fpn{R}$ \quad 
  $\wtd x = \fs{Q}$ \quad 
  $g = \{ X \mapsto \wtd x\}$ \quad 
  $\rho = \subst{\tilde u}{\tilde x}$
  \smallskip
\end{tabular}
\\
    % \rowcolor{gray!25}
    \hline
  \end{tabular}
         \caption[The definition of $\Cbrecpi{\tilde u}{\tilde x}{P}$ 
         (recursion extension). ]{The set $\Cbrecpi{\tilde u}{\tilde x}{P}$. 
      The set $\Dbrecpi{k}{\tilde x, \rho}{P}^d_g$ is defined in~\Cref{pi:t:tablecd-rec-j1,pi:t:tablecd-rec-j2}. 
      % Side conditions for 
      % last four rows ($g = \es$) are the same as for corresponding 
      % cases when $g \not = \es$. 
      \label{pi:t:tablecd-rec}}
      % \vspace{-2mm}
    \end{table}
  \end{figure}

  \begin{figure}[t!]
    \begin{table}[H]
      \newcommand{\auxp}{B}
% \begin{figure*}

%     \begin{table}[H]
      \begin{tabular}{ |l|l|}
        
        \hline
        \rowcolor{gray!25}
        % \noalign{\smallskip}
        $P$ 
        % \smallskip
        &
        % \noalign{\smallskip}
          % \multicolumn{1}{l|}{
        \begin{tabular}{ll}
          \noalign{\smallskip}
          $\Dbrecpi{k}{\tilde x, \rho}{P}^d_g$ $\text{ when } 
          g  \not= \es$ & 
          \smallskip
        \end{tabular}
      % } 
      % \smallskip
       \\
        \hline
               %%%%%%
      %% =======================
      %% =======================
      %%%%%%
      % $\Dbrecpi{k}{\tilde x, \rho}{\bout{u_i}{y_j}{Q}}^d_g$ & 
      % \multirow{3}{*}{
        $\bout{n_i}{y_j}{Q}$
        % } 
        &
      \begin{tabular}{ll}
      \noalign{\smallskip}	
      	$\left\{\bout{n_l}{\wtd y\rho} \propoutrec{k+1}{\wtd z\rho}
      \recp{X}\binp{\prop^r_k}{\wtd x}
      \bout{n_l}{\wtd y} 
      \propoutrec{k+1}{\wtd z}\rvar{X}
    \Par \Brecpi{k+1}{\tilde z}{Q\sigma}_g \right\}$
    & if $\rvardepth{Q} = \recdep$
    \\
    $N \cup \left\{\recp{X}\binp{\prop^r_k}{\wtd x}
      \bout{n_l}{\wtd y} 
      \propoutrec{k+1}{\wtd z}\rvar{X}
    \Par R \,:\, R \in \Dbrecpi{k+1}{\tilde z, \rho}{Q\sigma}^d_g \right\} $
    & \text{otherwise}
    \\
      \noalign{\smallskip}
      \end{tabular} 
    \\
     \cdashline{2-2}
      & \begin{tabular}{l}
        \noalign{\smallskip}
    $ N = 
    \begin{cases}
    \{ \auxp
    \Par 
    % \\ 
    \Brecpi{k+1}{\tilde z}{Q\sigma}_g \} & \text{if } \rvardepth{Q} = \recdep + 1 \\ 
    \es & \text{otherwise}
    \end{cases} 
    $
    \\   \noalign{\smallskip}
    $ \auxp = \propoutrec{k+1}{\wtd z\rho}
    \recp{X}\binp{\prop^r_k}{\wtd x}
    \bout{n_l}{\wtd y} 
    \propoutrec{k+1}{\wtd z}\rvar{X}$
    \smallskip
    \end{tabular} 
    \\ 
    \cdashline{2-2}
    % side conditions
    &    \begin{tabular}{lll}
          {\medskip}
         $y_j : T$ 
        &
         $\wtd y = (y_j,\ldots,y_{j+\len{\Gtopt{T}}-1})$ 
         & 
      $\wtd w = \linecondit{\lin{u_i}}{\{ u_i \}}{\epsilon}$ 
      \\
      $\wtd z = g(X) \cup \fnb{Q}{\wtd x \setminus \wtd w}$  
      &
      $l = \linecondit{\traux{u_i}}{\indf{S}}{i}$ 
      &
      $\sigma = \nextn{u_i}$
      \smallskip
    \end{tabular} 
    \\
    \hline %%%%=======================================
    % $\Dbrecpi{k}{\tilde x}{\binp{u_i}{y}{Q}}^d_g$
    $\binp{n_i}{y}{Q}$
    &
    \begin{tabular}{ll}
      \noalign{\smallskip}
       $\left\{\binp{n_l}{\wtd y} 
      \propoutrec{k+1}{\wtd z\rho}
      \recp{X}\binp{\prop^r_k}{\wtd x}
        \binp{n_l}{\wtd y} 
        \propoutrec{k+1}{\wtd z}
        \rvar{X}
       \Par \Brecpi{k+1}{\tilde z}{Q\sigma}_g \right\}$ 
       &
      if $\rvardepth{Q} = \recdep$
      \\
      \noalign{\smallskip}
      $N \cup \left\{\recp{X}\binp{\prop^r_k}{\widetilde x}
        \binp{n_l}{\widetilde y} 
        \propoutrec{k+1}{\wtd z}
        \rvar{X}
       \Par R : R \in \Dbrecpi{k+1}{\tilde z, \rho}{Q\sigma}^d_g \right\}$ 
      &
      otherwise
      \smallskip 
    \end{tabular}
    \\
       \cdashline{2-2} 
       &  \begin{tabular}{l}
      \noalign{\smallskip} 
      % \noalign{\smallskip}
      $N = 
      \begin{cases} 
        \{ \auxp 
       \Par
      %  \\
      \Brecpi{k+1}{\tilde z}{Q\sigma}_g \} & \text{ if } \rvardepth{Q} = \recdep + 1 \\
      \es & \text{ otherwise }
      \end{cases}
      $
      \\ 
      $\auxp =   \propoutrec{k+1}{\wtd z\rho}
      \recp{X}\binp{\prop^r_k}{\wtd x}
        \binp{n_l}{\wtd y} 
        \propoutrec{k+1}{\wtd z}
        \rvar{X}$
      \smallskip 
    \end{tabular} 
    \\
    \cdashline{2-2}
    &    \begin{tabular}{lll}
         \noalign{\smallskip}
      $y : T$ 
      &
      $\wtd y = (y_1,\ldots,y_{\len{\Gtopt{T}}})$ 
      &
      $\wtd w = \linecondit{\lin{u_i}}{\{ u_i \}}{\epsilon}$ 
      \\
      $\wtd z = g(X) \cup \fnb{Q}{\wtd x \wtd y \setminus \wtd w}$  
      &
      $l = \linecondit{\traux{u_i}}{\indf{S}}{i}$ 
      &
      $\sigma = \nextn{u_i} \cdot \subst{y_1}{y}$
      % $\wtd z = \fnb{Q}{\wtd x \wtd y}$  
      \smallskip
    \end{tabular} 
    \\
    \hline %%%%=======================================
    % $\Dbrecpi{k}{\tilde x,\rho}{Q_1 \Par Q_2}^d_g$
    $Q_1 \Par Q_2$
    & 
    \begin{tabular}{l}
      \noalign{\smallskip}
      $ N \cup \big\{\recp{X}\proprinpk{r}{k}{\wtd x}
      (\proproutk{r}{k+1}{\wtd y_1}\rvar{X}
      \Par \apropoutrecsh{k+\degree+1}{\wtd y_2}) 
       \Par R_1 \Par R_2 \,:\,$ 
     \smallskip  \\
     \qquad \qquad \qquad \qquad\qquad \qquad $ R_1 \in \Dbrecpi{k+1}{\tilde y_1, \rho}{Q_1}^d_{g_1}, 
      R_2 \in \Dbrecpi{k+\degree+1}{\tilde y_2, \rho}{Q_2}^d_{g_2}\big\}$ 
      % $\cup$ $N$
      \smallskip
        \end{tabular}
      \\
      %  \noalign{\smallskip}
      \cdashline{2-2}
       &  \begin{tabular}{l}
       \noalign{\smallskip}
      $ N = 
      \begin{cases} 
     \{ \auxp 
      % \proproutk{r}{k+1}{\wtd y_1\rho}
      % \recp{X}\proprinpk{r}{k}{\wtd x} (\proproutk{r}{k+1}{\wtd y_1}\rvar{X}
      % \Par 
      % \apropoutrecsh{k+\degree+1}{\wtd y_2})
       \Par
      %  \\
       \Brecpi{k+1}{\tilde y_1}{Q_1}_{g_1} \Par 
      \Brecpi{k+\degree+1}{\tilde y_2}{Q_2}_{g_2}\} & \text{ if } \rvardepth{Q_1} = \recdep \\
      \es & \text{ otherwise }
      \end{cases}
      $
      \smallskip
      \\ 
      $ \auxp = \proproutk{r}{k+1}{\wtd y_1\rho}
      \recp{X}\proprinpk{r}{k}{\wtd x} (\proproutk{r}{k+1}{\wtd y_1}\rvar{X}
      \Par \apropoutrecsh{k+\degree+1}{\wtd y_2})$
      \smallskip 
    \end{tabular} 
    \\
    \cdashline{2-2}
    % side conditions
    &    \begin{tabular}{llll}
         \noalign{\smallskip}
           $\recpx{Q_1}$ 
           &
           $\wtd y_1  = g(X) \cup \fnb{Q_1}{\wtd x}$ 
           &
           $\wtd y_2  =  \fnb{Q_2}{\wtd x}$ 
           &
           $\degree = \lenHOopt{Q_1}$
      \smallskip
    \end{tabular} 
    \\
    \hline  %%%%=======================================
    % $\Dbrecpi{k}{\tilde x, \rho}{\news{s:C}{Q}}^d_g$ & 
    $\news{s}{Q}$ &
    \begin{tabular}{l}
      \noalign{\smallskip}
      $N \cup \left\{\recp{X}\news{\widetilde{s}:\Gtopt{C}}\propinp{k}{\wtd x}\propout{k+1}{\wtd z}X \Par R : R \in \Dbrecpi{k+1}{\tilde z, \rho}{Q \sigma }^d_g \right\}$
      {\smallskip}
          \end{tabular} 
      \\
        \cdashline{2-2}
      &   \begin{tabular}{l}
           \noalign{\smallskip}
      $ N = 
      \begin{cases}
      \{\news{\wtd{s}:\Gtopt{C}} B 
      \Par 
      \Brecpi{k+1}{\tilde z}{Q\sigma}_g\} &  \text{ if } \rvardepth{Q_1} = \recdep \\
      \es & \text{ otherwise }
      \end{cases}
      $
      \\ 
       $B = \propout{k+1}{\wtd z\rho}
      \recp{X}\news{\wtd{s}:\Gtopt{C}}\propinp{k}{\wtd x}\propout{k+1}{\wtd z}X$ 
    \smallskip
    \end{tabular} 
    \\
    \cdashline{2-2}
    % side conditions
    &    \begin{tabular}{lll}
         \noalign{\smallskip}
        $s : S$ 
        & 
         $\widetilde{s} = (s_1,\ldots,s_{\len{\Gtopt{S}}})$ 
         &
         $\widetilde {\dual{s}} =  (\dual{s_1},\ldots,\dual{s_{\len{\Gtopt{S}}}})$ 
         \\
         $\wtd n =
          \linecondit{\lin{s}}{\wtd n}{\epsilon}$ 
          &
         $\wtd z = \wtd x, \wtd s, \wtd n$ 
         &
         $\sigma = \subst{s_1 \dual{s_1}}{s \dual{s}} $
      \smallskip
    \end{tabular} 
    \\
    \hline  %%%%=======================================
    % $\Brecpi{k}{\tilde x}{X}_g$ 
    $X$
    &  
    \begin{tabular}{l}
    \noalign{\smallskip}
    $\{ \inact \}$
    \smallskip
    \end{tabular}
    \\ 
        % \rowcolor{gray!25}
        \hline
      \end{tabular}
         \caption{
          % The extension of $\Cbpi{\tilde u}{\tilde x}{P}$ 
      The set $\Dbrecpi{k}{\tilde x, \rho}{P}^d_g$ 
      when  $g \not = \es$. 
      (\Cref{pi:t:tablecd-rec-j2} covers the case $g = \es$.)
      % Side conditions for 
      % last four rows ($g = \es$) are the same as for corresponding 
      % cases when $g \not = \es$.
       \label{pi:t:tablecd-rec-j1}}
      % \vspace{-2mm}
    \end{table}
  \end{figure}

  \begin{figure}[t!]
    \begin{table}[H]
      \newcommand{\auxp}{B}
% \begin{figure*}

%     \begin{table}[H]
      \begin{tabular}{ |l|l|}
        \rowcolor{gray!25}
        \hline
        $P$ &
          % \multicolumn{2}{l|}{
        \begin{tabular}{l}
          \noalign{\smallskip}
          $\Dbrecpi{k}{\tilde x, \rho}{P}^d_g$ $\text{ when } g = \es$
          \smallskip
        \end{tabular}
      % } 
       \\
        \hline
               %%%%%%
      %% =======================
      %% =======================
      %%%%%%
      $\bout{n_i}{y_j}{Q}$
      &  
      \begin{tabular}{l}
      \noalign{\smallskip}
      $ \{
      \bout{n_{\mstindex}}{\wtd y\rho} \apropoutrecsh{k+1}{\wtd z\rho} \Par  
      \recp{X}\propinprecsh{k}{\wtd x}
      (\bout{n_{\mstindex}}{\wtd y} \apropoutrecsh{k+1}{\wtd z}
      \Par \rvar{X}) 
     \Par$
    %  \\ 
    $\Brecpi{k+1}{\tilde z}{Q\sigma}_g \}$
      \smallskip
      \end{tabular}
      %  %  side conditions 
      % & 
      % \begin{tabular}{l}
      %        \noalign{\smallskip}
      %       %  $\wtd x = g(X)$
      %     \smallskip
      % \end{tabular} 
    \\ \hline     
    $\binp{n_i}{y}{Q}$
    &  
    \begin{tabular}{l}
    \noalign{\smallskip}
    $
    \{\binp{n_{\mstindex}}{\wtd y} \apropoutrecsh{k+1}{\wtd z\rho}  \Par 
    \recp{X}\propinprecsh{k}{\wtd x}
    (\binp{n_{\mstindex}}{\wtd y} 
    \apropoutrecsh{k+1}{\wtd z} \Par \rvar{X}) 
    \Par$ 
    %  \\
    $\Brecpi{k+1}{\tilde z}{Q\sigma}_g\}$ 
    \smallskip
    \end{tabular}
    % % side conditions
    % & 
    % \begin{tabular}{l}
    %      \noalign{\smallskip}
    %     %  $\wtd x = g(X)$
    %   \smallskip
    % \end{tabular} 
    \\ \hline    
    $Q_1 \Par Q_2$
    &  
    \begin{tabular}{l}
    \noalign{\smallskip}
    $
    \big\{\apropoutrecsh{k+1}{\wtd y_1\rho}
    \Par 
    \apropoutrecsh{k+\degree+1}{\wtd y_2\rho} \Par$
    \smallskip \\ 
    \qquad $\recp{X}\proprinpk{r}{k}{\wtd x}(\apropoutrecsh{k+1}{\wtd y_1}
    \Par \apropoutrecsh{k+\degree+1}{\wtd y_2} \Par \rvar{X}) \Par $
    \smallskip \\ 
    \qquad \qquad $\Brecpi{k+1}{\tilde y_1}{Q_1}_{g_1} \Par 
    \Brecpi{k+\degree+1}{\tilde y_2}{Q_2}_{g_2}\big\}$ 
    \smallskip
    \end{tabular}
    % %  side 
    % & 
    % \begin{tabular}{l}
    %      \noalign{\smallskip}
    %     %  $\wtd x = g(X)$
    %   \smallskip
    % \end{tabular} 
    \\ \hline    
    $\news{s:C}{Q}$ 
    &  
    \begin{tabular}{l}
    \noalign{\smallskip}
    $
    \{\apropoutrecsh{k+1}{\wtd z\rho} \Par 
    \recp{X}\news{\wtd{s}:\Gtopt{C}}
    \propinp{k}{\wtd x}(\apropoutrecsh{k+1}{\wtd z} \Par X)$
    $\Par$
    % \\
    $\Brecpi{k+1}{\tilde z}{Q}_g\}$
    \smallskip
    \end{tabular}
    % %  side
    % & 
    % \begin{tabular}{l}
    %      \noalign{\smallskip}
    %     %  $\wtd x = g(X)$
    %   \smallskip
    % \end{tabular} 
    \\ \hline    
      \end{tabular}
    %   \caption{The extension of $\Cbpi{\tilde u}{\tilde x}{P}$ 
    %   and definition of $\Dbrecpi{k}{\tilde x, \rho}{P}^d_g$
    %   (recursion extension). Side conditions for 
    %   last four rows ($g = \es$) are the same as for corresponding 
    %   cases when $g \not = \es$. \label{pi:t:tablecd-rec}}
    %   \vspace{-2mm}
    %   \end{table}
    % \end{figure*}
         \caption[The definition of $\Dbrecpi{k}{\tilde x, \rho}{P}^d_g$ when $g = \es$.]{
          % The extension of $\Cbpi{\tilde u}{\tilde x}{P}$ 
      The set $\Dbrecpi{k}{\tilde x, \rho}{P}^d_g$ when $g = \es$. Side conditions  are the same as for corresponding 
      cases from~\Cref{pi:t:tablecd-rec-j1} (when $g \not = \es$). 
      \label{pi:t:tablecd-rec-j2}
      }
      % \vspace{-2mm}
    \end{table}
  \end{figure}

  % \end{figure}
  % \end{figure*}

  \begin{definition}[Indexed names substitutions]
    \label{pi:d:indexedsubstitution}
    Let 
    $\widetilde u = (a,b,s, \dual s, \dual{s}', s',r, r', \ldots)$
    be a finite tuple of names.
    We write $\indices{\wtd u}$ to denote 
    $$\indices{\wtd u} = 
    \left\{\subst{a_{1}, b_{1}, s_{i}, \dual s_{i},  
    s'_{j}, \dual s'_{j}, r_{1}, r'_{1}, \ldots}{a,b,s,s',r,r',\ldots} : 
    i,j,\ldots > 0 \right\}$$
   \end{definition}

\begin{newaddenv}
\subsubsection{The relation $\relS$: Ingredients and Properties}
  \noindent Having all auxiliary notions in place, we are ready to define 
 $\relS$: 
\end{newaddenv} 

% \thesisalt{
       \begin{definition}[Relation \relS]
  \label{pi:d:relation-s}
  Let $P\subst{\tilde u}{\tilde x}$ be a well-typed process, with $\wtd u : \wtd C$.  
  We define the relation $\relS$ as follows: 
  \begin{align*}
    \relS &= \{(P\subst{\tilde u}{\tilde x}, R) \,:\,
        R \in \Cbpi{\tilde{w}}{\tilde{y}}{P\sigma},
        \\ & \qquad \qquad 
 \sigma \in \indices{\wtd u \cup \wtd x \cup  \fn{P} },~ 
  \wtd{w} = \bname{\wtd u \sigma : \wtd C},~ 
    \wtd{y} = \bname{\wtd x \sigma : \wtd C} 
    \}
  \end{align*}
  \noindent where the set $\Cbpi{-}{-}{\cdot}$ as in~\Cref{pi:t:tablecd}.
 \end{definition}
 
The definition of $\relS$ follows Parrow's proof of dynamic correctness for the trios in the untyped $\pi$-calculus. 
 Intuitively, processes in the set $\Cbpi{\wtd{w}}{\wtd{y}}{P}$ represent processes that are ``correlated'' to $P$, up to synchronizations induced by propagators. 
 In our case, the presence of trigger processes and recursive processes induces significant differences with respect to Parrow's definition. 
Given a process $P$, we have two mutually defined sets: $\Cbpi{\tilde u}{\tilde x}{P}$  and $\Dbpi{\tilde u}{\tilde x}{P}$, which are both given in \Cref{pi:t:tablecd}. The idea is that $\Cbpi{\tilde u}{\tilde x}{P}$ deals with trigger processes and top-level activation of propagators, whereas $\Dbpi{\tilde u}{\tilde x}{P}$ collects processes without triggers that are involved in the overall activation of propagators. 
Handling recursive processes requires dedicated treatment; this is formalized by the auxiliary sets $\Cbrecpi{\tilde u}{\tilde x}{P}$  (defined in  \Cref{pi:t:tablecd-rec}) and $\Dbrecpi{k}{\tilde x, \rho}{P}^d_g$ (defined in
       \Cref{pi:t:tablecd-rec-j1,pi:t:tablecd-rec-j2}).

% \\
% \newadd{\textbf{TODO}}
% \\
%  \todo[inline]{TODO}
\cralt{
\todo[inline]{Below paragraph is repeated from HO part for clarity}}{}
\begin{newaddenv}
\paragraph{Properties.}
We prove a series of lemmas that establish a form of \emph{operational correspondence}, divided in completeness and soundness properties.
We first  need the following result. Following Parrow~\cite{DBLP:conf/birthday/Parrow00},  we refer to prefixes 
  that do not correspond to prefixes of the original process (i.e. prefixes on propagators $\prop_i$),
  as \emph{non-essential prefixes}. 
  The relation $\relS$ is closed under reductions that involve non-essential prefixes.
\end{newaddenv} 
  
\begin{newaddenv}
  \begin{lemma} 
  \label{pi:l:c-prop-closed}
  Given an indexed process $P_1\subst{\tilde u}{\tilde x}$, 
  the set $\Cbpi{\tilde u}{\tilde x}{P_1}$ is closed under $\tau$ transitions on
  non-essential prefixes. That is, 
   if $R_1 \in \Cbpi{\tilde u}{\tilde x}{P_1}$ 
  and $R_1 \by{\tau} R_2$ is inferred from the actions on non-essential prefixes, 
  then $R_2 \in \Cbpi{\tilde u}{\tilde x}{P_1}$.
  \end{lemma}
  \begin{proof}[Proof (Sketch)]
  By the induction on the structure of $P_1$ and 
  the inspection of definition of $\Cbpi{-}{-}{\cdot}$ 
  and $\Dbpi{-}{-}{\cdot}$
  in \Cref{pi:t:tablecd} and \Cref{pi:t:tablecd-rec}. 
  \end{proof}
\end{newaddenv}

%\jcheck{Add a sentence describing the definition.}

% We construct relation $\mathcal{S}$ (\defref{d:relation-s}) using auxiliary functions $\Cbpi{-}{-}{\cdot}$
% and $\Dbpi{-}{-}{\cdot}$ (\tabref{t:tablecd} and \tabref{t:tablecd-rec}). 
% Then, we prove that $\mathcal{S}$ is a MST bisimulation 
% by \lemref{l:lemms} and \lemref{l:lemmsdir}. 
% Finally, we show that $(P,\Do{P}) \in \relS$.

% and prove that it is a MST Bisimulation...

% \textcolor{blue}{Lemmas are now here}
% \todo{check lemmas}
\begin{newaddenv}
  \paragraph{Operational Completeness.} 
  We first consider transitions using the untyped LTS; in~\Cref{pi:l:lemms-typed} we will 
  consider transitions with the typed LTS. 
\end{newaddenv}

\begin{restatable}[]{lemma}{lemms}
  \label{pi:l:lemms}
  Assume $P_1 \subst{\tilde u}{\tilde x}$ is a 
  process and $P_1\subst{\tilde u}{\tilde x} \relS Q_1$.
  \begin{enumerate}
    \item 
				Whenever 
        ${P_1\subst{\tilde u}{\tilde x}}
        {\by{\news{\widetilde{m_1}} \bactout{n}{v:C_1}}}{P_2}$,
        \newadd{such that $\dual n \not\in P_1\subst{\tilde u}{\tilde x}$},
				then there exist
				$Q_2$ and $\sigmav$ such that 
        ${Q_1}{\By{\news{\widetilde{m_2}}\bactout{\iname n}{\wtd v: \Gtopt{C_1}}}}{Q_2}$
        where $v\sigmav \namesrelate \wtd v$
        % where $V_1 \vrelate V_2$
        % and ${P_2}{\ \relS \ }{Q_2}$, 
        and, for a fresh $t$,
        \[
           {\newsp{\wtd{m_1}}{P_2 \parallel \ftrigger{t}{v}{C_1}}}
            \ \relS \ 
          {\newsp{\widetilde{m_2}}{Q_2 \parallel \ftriggerm{t_1}{v\sigmav}{C_1}}}
        \]

    \item	
      Whenever ${P_1\subst{\tilde u}{\tilde x}}
      {\by{\abinp{n}{v}}}{P_2}$, 
      \newadd{such that $\dual n \not\in P_1\subst{\tilde u}{\tilde x}$}, 
      then there exist $Q_2$ and $\sigmav$ such that 
      ${Q_1}{\By{\abinp{\iname n}{\tilde v}}}{Q_2}$
		where $v \sigmav \namesrelate \wtd v$
      and
      ${P_2}{\ \relS \ }{Q_2}$, 

      \item	
       Whenever ${P_1}{\by{\tau}}{P_2}$ 
       then there exists $Q_2$ such that 
				${Q_1}{\By{\tau}}{Q_2}$
				and
        ${P_2}{\ \relS \ }{Q_2}$. 
  \end{enumerate} 
\end{restatable}
%\vrelat
\begin{proof}
  By transition induction. See~\Cref{pi:app:lemms} for details. 
\end{proof} 

The following statement builds upon the previous one to the case of 
typed LTS (\Cref{top:d:tlts}): 
\begin{restatable}[]{lemma}{lemms-typed}
  \label{pi:l:lemms-typed}
  Assume $P_1 \subst{\tilde u}{\tilde x}$ is a  
  process and $P_1\subst{\tilde u}{\tilde x} \ \relS \ Q_1$.
  \begin{enumerate}
		\item 
				Whenever 
				$\horelm{\Gamma_1;\Delta_1}{P_1}
        {\by{\news{\widetilde{m_1}} \bactout{n}{v:C_1}}}
        {\Delta'_1}{P_2}$ 
        then there exist
				$Q_2$, $\Delta'_2$, and $\sigmav$ such that 
				$\horelm{\Gamma_2;\Delta_2}{Q_1}
        {\By{\news{\wtd{m_2}}\bactout{\iname{n}}{\tilde v : \Gtopt{C}}}}
        {\Delta_2'}{Q_2}$
        where 
        $v\sigmav \namesrelate \wtd v$
        and, for a fresh $t$,
        \[
          {\newsp{\wtd{m_1}}{P_2 \parallel \ftrigger{t}{v}{C_1}}}
           \ \relS \ 
         {\newsp{\widetilde{m_2}}{Q_2 \parallel \ftriggerm{t}{v\sigmav}{C_1}}}
       \]
        
       \item Whenever $\horelm{\Gamma_1;\Delta_1}{P_1}
       {\by{\abinp{n}{v}}}{\Delta_1'}{P_2}$
       then there exist  $Q_2$, $\Delta_2'$, and $\sigmav$ such that 
       $\horelm{\Gamma_2;\Delta_2}{Q_1}{\By{\abinp{\iname{n}}{\tilde v}}}
       {\Delta_2'}{Q_2}$ 
       where 
       $v\sigmav \namesrelate \wtd v$
       and 
       $P_2 \ \relS \ Q_2$, 
        
      \item	
				Whenever  
        $\horelm{\Gamma_1;\Delta_1}{P_1}{\by{\tau}}
        {\Delta_1'}{P_2}$  
				then there exist 
        $Q_2$ and $\Delta'_2$ such that 
				$\horelm{\Gamma_2;\Delta_2}{Q_1}
        {\By{{\tau}}}{\Delta_2'}{Q_2}$
				and
        $P_2 \ \relS \ Q_2$. 
        % $\horelm{\Gamma_1;\Delta_1'}{P_2}{\ \relS \ }
        % {\Gamma_2;\Delta_2'}{Q_2}$.  
        % and  
        % $\subl{\ell} = n$ implies  
        % $\subl{\hat \ell} = \iname{n}$.

		% \item	The symmetric cases of 1, 2, and 3.                 
	\end{enumerate}
\end{restatable}

\begin{proof} 
  \begin{newaddenv}
    The proof uses results of~\Cref{pi:l:lemms}. 
    We consider the first case, as the other two are similar. 
    By the definition of the typed LTS (\Cref{top:d:tlts}) we have: 
    \begin{align}
      &\Gamma_1; \Lambda_1; \Delta_1 \proves P_1\subst{\tilde u}{\tilde x}  \hastype \Proc 
      \label{pi:pt:cr-typed-1}
      \\
      &(\Gamma_1; \es; \Delta_1)  
      % \by{\news{\widetilde{m}} \bactout{n}{V}}
      \by{\news{\widetilde{m_1}} \bactout{n}{v:C_1}}
      (\Gamma_1; \es; \Delta_2)
      \label{pi:pt:cr-typed-2}
      % \\
      % & \horelm{\Gamma_1;\Delta_1}{P_1}
      % {\by{\news{\widetilde{m_1}} \bactout{n}{v:C_1}}}
      % {\Delta'_1;\Lambda'_1}{P_2} 
    \end{align}
  By \eqref{pi:pt:cr-typed-2} we further have 
  \begin{align*}
    \AxiomC{
      \begin{tabular}{c}
        $\Gamma \cat \Gamma'; \Lambda'; \Delta' \proves V \hastype U$ \\
    $\Delta'\backslash (\cup_j \Delta_j) \subseteq (\Delta \cat n: S)$
      \end{tabular}} 
    \AxiomC{\begin{tabular}{c}
      $\Gamma'; \es; \Delta_j \proves m_j  \hastype U_j$ \\ 
      $\Gamma'; \es; \Delta_j' \proves \dual{m}_j  \hastype U_j'$
    \end{tabular}} 
    \AxiomC{
      \begin{tabular}{c}
        $\dual{n} \notin \dom{\Delta}$ \\
        $\Lambda' \subseteq \Lambda$
      \end{tabular}
    } 
    \LeftLabel{\scriptsize \trule{SSnd}}
    \TrinaryInfC{
      $(\Gamma; \Lambda; \Delta \cat n: \btout{C_1} S)
      % \by{\news{\widetilde{m}} \bactout{n}{V}}
      \by{\news{\widetilde{m_1}} \bactout{n}{v:C_1}}
      (\Gamma \cat \Gamma'; \Lambda\backslash\Lambda';
       (\Delta \cat n: S \cat \cup_j \Delta_j') \backslash \Delta')$
    }
    \DisplayProof 
  \end{align*}
  \noindent By 
  \eqref{pi:pt:cr-typed-1} 
  and the condition $\dual{n} \notin \dom{\Delta}$ 
  we have $\dual n \not\in \fn{P_1\subst{\tilde u}{\tilde x}}$. 
  Therefore, we can apply Item 1 of~\Cref{pi:l:lemms}. 
\end{newaddenv}
\end{proof}

\paragraph{Operational Soundness.}
\begin{newaddenv}
  For the proof of operational soundness we follow the same strategy  as before: we first establish  a lemma for the untyped LTS, then we extend it to the case of the typed LTS. 
\end{newaddenv}

\begin{restatable}[]{lemma}{lemmsdir}
  \label{pi:l:lemmsdir}
  Assume $P_1 \subst{\tilde u}{\tilde x}$ is a  
  process and $P_1\subst{\tilde u}{\tilde x} \ \relS \ Q_1$.
  \begin{enumerate}
    \item 
                Whenever 
        ${Q_1}{\by{\news{\widetilde{m_2}}\bactout{n_i}{\wtd v: \Gtopt{C_1}}}}{Q_2}$
        , such that $\dual n_i \not\in \fn{Q_1}$, 
        then there exist
        $P_2$ and $\sigmav$ such that 
${P_1\subst{\tilde u}{\tilde x}}{\by{\news{\widetilde{m_1}} \bactout{n}{v:C_1}}}{P_2}$
where $v\sigmav \namesrelate \wtd v$
        where $v\sigmav \namesrelate \wtd v$
        % where $V_1 \vrelate V_2$
        % and ${P_2}{\ \relS \ }{Q_2}$, 
        and, for a fresh $t$,
        \[
           {\newsp{\wtd{m_1}}{P_2 \parallel \ftrigger{t}{v}{C_1}}}
            \ \relS \ 
          {\newsp{\widetilde{m_2}}{Q_2 \parallel \ftriggerm{t_1}{v\sigmav}{C_1}}}
        \]
  
        \item	
      Whenever 
      ${Q_1}{\by{\abinp{\iname n}{\tilde v}}}{Q_2}$
      then there exist $P_2$ and $\sigmav$ such that 
      ${P_1\subst{\tilde u}{\tilde x}}
      {\by{\abinp{n}{v}}}{P_2}$ 
        where $v \sigmav \namesrelate \wtd v$
      and
      ${P_2}{\ \relS \ }{Q_2}$, 
  
      \item	
       Whenever 
       ${Q_1}{\by{\tau}}{Q_2}$
       either \rom{1} ${P_1\subst{\tilde u}{\tilde x}}{\ \relS \ }{Q_2}$ 
       or 
       \rom{2}
        there exists $P_2$ such that 
       ${P_1}{\by{\tau}}{P_2}$ 
                and
        ${P_2}{\ \relS \ }{Q_2}$. 
    
  \end{enumerate} 
  \end{restatable}
%  \begin{proof}[Proof (Sketch)]
%    \begin{newaddenv}
%      See~\Cref{pi:app:lemmsdir}. 
%    \end{newaddenv}
%  \end{proof}

\begin{proof}[Proof (Sketch)]
\label{pi:app:lemmsdir}
% By transition induction. 
Following Parrow's approach,  we refer to prefixes 
corresponding to prefixes of the original process 
as \emph{essential prefixes}. 
We remark that a prefix in 
$\Do{P}$ is non-essential if and only if is a prefix on a propagator name. 
First, we discuss the case when a transition is inferred without any 
actions from essential prefixes. In this case 
we know that an action can only involve propagator prefixes and  
by inspection of definition of  $\Cbpi{\tilde u}{\tilde x}{P_1}$ that 
$\ell = \tau$. 
This concerns the sub-case \rom{1} of Part 3
and it directly follows by  \Cref{pi:l:c-prop-closed}. 

% First, we analyze case of $\tau$ transitions 
% on propagator prefixes.  This concerns the sub-case \rom{1} of Part 3. 
% This follows directly by the \Cref{pi:l:c-prop-closed}. 
Now, assume $Q_1 \by{\ell} Q_2$ when $\ell$ involves essential prefixes. 
This concerns Part 1, Part 2, and 
sub-case \rom{2} of Part 3. This case is mainly the inverse of the proof 
of \Cref{pi:l:lemms}. As Parrow, here we note that 
 an essential prefix is unguarded in $Q_1$ if and only if it is 
 unguarded in $P_1$. That is, by inspection of the definition, we see that 
 function $\Cbpi{\tilde u}{\tilde x}{P_1}$ 
does not unguard essential prefixes of $P_1$ that its members mimic
(the propagators serve as guards). 
% noting that only additional prefixes that 
% $\Cbpi{\tilde u}{\tilde x}{P_1}$ introduce are propagator prefixes. 
\end{proof} 

\begin{restatable}[]{lemma}{lemmsdirtyped}
  \label{pi:l:lemmsdir-typed}
  Assume $P_1 \subst{\tilde u}{\tilde x}$ is a 
  process and $P_1\subst{\tilde u}{\tilde x} \ \relS \ Q_1$.
  \begin{enumerate}
    \item 
                Whenever 
                $\horelm{\Gamma_2;\Delta_2}
                {Q_1}
                {\by{\news{\widetilde{m_2}}\bactout{\iname n}{\wtd v: \Gtopt{C_1}}}}
                {\Delta'_2}{Q_2}$
        % ${Q_1}{\by{\news{\widetilde{m_2}}\bactout{\iname n}{\wtd v: \Gt{C_1}}}}{Q_2}$
        then there exist
        $P_2$, $\sigmav$, and $\Delta_1'$ such that 
        $\horelm{\Gamma_1; \Delta_1}{P_1\subst{\tilde u}{\tilde x}}
        {\By{\news{\widetilde{m_1}}\bactout{n}{v:C_1}}}
        {\Delta_1'}{P_2}$
        % where $V_1 \vrelate V_2$
        % and ${P_2}{\ \relS \ }{Q_2}$, 
        and, for a fresh $t$,
        \[
           {\newsp{\wtd{m_1}}{P_2 \parallel \ftrigger{t}{v}{C_1}}}
            \ \relS \ 
          {\newsp{\widetilde{m_2}}{Q_2 \parallel \ftriggerm{t_1}{v\sigmav}{C_1}}}
        \]

        \item	
      Whenever 
      % $\horelm{}{}{}{}{}$
      $\horelm{\Gamma_2;  \Delta_2}{Q_1}
      {\by{\abinp{\iname{n}}{\tilde v}}}{\Delta_2'}{Q_2}$
      % ${Q_1}{\by{\abinp{\iname n}{\tilde v}}}{Q_2}$
      then there exist 
      $P_2$, $\sigmav$, and $\Delta_1'$ such that 
      $\horelm{\Gamma_1; \Delta_1}{P_1\subst{\tilde u}{\tilde x}}{\By{\abinp{{n}}{v}}}
      {\Delta'_1}{P_2}$
% 
      % ${P_1\subst{\tilde u}{\tilde x}}
      % {\by{\abinp{n}{v}}}{P_2}$ 
        where $v \sigmav \namesrelate \wtd v$
      and
      ${P_2}{\ \relS \ }{Q_2}$, 
  
      \item	
       Whenever 
       $\horelm{\Gamma_2; \Delta_2}{Q_1}{\by{\tau}}
       {\Delta_2'}{Q_2}$
% 
      %  ${Q_1}{\by{\tau}}{Q_2}$
       either \rom{1} 
       ${P_1\subst{\tilde u}{\tilde x}}{\ \relS \ }{Q_2}$ 
       or 
       \rom{2}
       there exist $P_2$  and $\Delta_1'$ such that 
       $\horelm{\Gamma_1; \Delta_1}{P_1\subst{\tilde u}{\tilde x}}
       {\By{\tau}}{\Delta_1'}{P_2}$
       and
       ${P_2}{\ \relS \ }{Q_2}$. 
      %   there exists $P_2$ such that 
      %  ${P_1}{\by{\tau}}{P_2}$ 
      %           and
      %   ${P_2}{\ \relS \ }{Q_2}$. 

  \end{enumerate} 
  \end{restatable}
  \begin{proof}[Proof (Sketch)]
    \begin{newaddenv}
    Analogous to the proof of~\Cref{pi:l:lemms-typed}, using 
\Cref{pi:l:lemmsdir}. 
    \end{newaddenv}
  \end{proof}

  We close this section by stating again \Cref{pi:t:mainbsthm} and giving its proof.
  
  \mainbsthm* 

\begin{proof}
\label{pi:app:mainbsthm}
%   To prove that $\Do{P}$ is MST-bisimilar to $P$ we construct a relation
% $\relS$ that is an MST-bisimulation and  that contains $(P,\Do{P})$; see \Cref{pi:d:relation-s}.

% To prove that $\Do{P}$ is MST-bisimilar to $P$
% Then, we prove that $\mathcal{S}$ is a MST bisimulation 
By \Cref{pi:l:lemmsdir,pi:l:lemmsdir-typed}   we know $\mathcal{S}$ is an MST bisimulation. 
So, we need to show  $(P,\Do{P}) \in \relS$.
Let $P_1$ be such that $P_1\subst{\tilde r}{\tilde x}=P$
where $\wtd r = \rfn{P}$. Further, let $\sigma = \bigcup_{v \in \tilde r} \subst{v_1}{v}$, 
so we have $\sigma \in \indices{\wtd r}$. Then, let $\wbns r = \bname{\wtd r : \wtd S}$
and $\wbns x = \bname{\wtd x : \wtd S}$ where $\wtd r : \wtd S$. 
Therefore, by \Cref{pi:def:decomp}  and \Cref{pi:t:tablecd}  
we have $\Do{P} \in \Cbpi{\bns r}{\bns x}{P_1}$. 
Finally, by \Cref{pi:d:relation-s} we have  $(P,\Do{P}) \in \relS$. 
% The relation $\mathcal{S}$ is a MST bisimulation ( \Cref{pi:l:lemms} 
% and \Cref{pi:l:lemmsdir}) 
% and $(P, \D{P}) \in \mathcal{S}$.
\end{proof}

  \begin{comment} 
We can now state our dynamic correctness result:

\begin{restatable}[Operational Correspondence]{thm}{mainbsthm}
\label{pi:t:mainbsthm}
  Let $P$ be a \sessp process such that 
  $\Gamma_1;\Delta_1 \proves P_1$. We have 
  $$\horelm{\Gamma;\Delta}{P}{\ \mstb \ }
  {\Gtopt{\Gamma};\Gtopt{\Delta}}{\Do{P}}$$ 
\end{restatable}
\begin{proof}
%   To prove that $\Do{P}$ is MST-bisimilar to $P$ we construct a relation
% $\relS$ that is an MST-bisimulation and  that contains $(P,\Do{P})$; see \defref{d:relation-s}.
By coinduction: we exhibit a binary relation $\relS$  that contains $(P,\Do{P})$
and prove that it is an MST bisimulation. 
%We define $\relS$
%\longversion{(\defref{d:relation-s})}{}
% using two auxiliary functions $\Cbpi{-}{-}{\cdot}$ and
%$\Dbpi{-}{-}{\cdot}$ 
%\longversion{(\tabref{t:tablecd} and \tabref{t:tablecd-rec}).}{.}
\longversion{The proof
that $\relS$ is an MST bisimilarity is given by \lemref{pi:l:lemms} and
\Cref{pi:l:lemmsdir-typed} (see~\Cref{pi:app:mainbsthm} for details).}
{See~\cite{AAP21-full} for the full account.}
%Finally, by the definition we have  \in \relS$.
% (see \appref{app:mainbsthm} for more details). 
  % The relation $\mathcal{S}$ is a MST bisimulation (\lemref{l:lemms} 
  % and \lemref{l:lemmsdir}) 
  % and $(P, \D{P}) \in \mathcal{S}$.
\end{proof}
\end{comment} 

\thesisalt{

\section{Extension with Labeled Choices}
\label{pi:sec:sb}
% \subsection{Minimality Result}

% \begin{align*}
%   \tau := \lto{\wtd \tau} \bnfbar \lti{\wtd \tau} 
%   \bnfbar \ltab{\wtd \tau} \bnfbar \ltend 
%   \bnfbar \ltc{\wtd \tau}
%   \bnfbar \ltbra  \bnfbar \texttt{Unit}
% \end{align*}

% \begin{align*}
%   P,Q, &:= \bout{x}{\tilde v}P \bnfbar 
%   \binp{x}{\tilde y}P \bnfbar \news{v}P
%   \bnfbar \inact \bnfbar P \Par Q 
%   \bnfbar \ltcase{v}{l_i}{x_i}{P_i}{i \in I} 
%   \\
%     v &:= x  \bnfbar \star \bnfbar l\_v
% \end{align*}

% \input{type_system_lt.tex}

% \begin{newaddenv}

%In this section, we address three intertwined aspects: (i)~the extension of our approach to support constructs for \emph{labeled choice} (branching and selection); (ii)~a detailed comparison with the CPS approach (Kobayashi, Dardha et al.) to deconstruct session-typed processes into linearly-typed processes; (iii)~a discussion on the notion of minimality enforced by \msts. 

%\subsection{Labelled Choices}
\begin{newaddenv}
%\Boptsb
In this section, we briefly discuss the extension of $\Do{\cdot}$ with \emph{labeled choices}. 
In session-based concurrency, labeled choices are implemented via branching and selection constructs, denoted $\bbra{u_i}{l_j : P_j}_{j \in I}$ (for some finite set $I$) and $\bsel{u_i}{l_j}Q$, respectively. Intuitively, a branching construct specifies the offer of a finite number of alternative behaviors ($P_1, P_2, \ldots$), whereas a selection construct signals the choice of one of them. 
Put differently, branching and selection implement deterministic choices via the input and output of labels; the reduction rule %(cf. \Cref{fig:reduction}) 
is as follows:
$$
	\bsel{n}{l_j} Q \Par \bbra{\dual{n}}{l_i : P_i}_{i \in I}  \red  Q \Par P_j \quad (j \in I)
$$

Branching and selection come with dedicated session types, denoted 
$\btbra{l_i : S_i}_{i \in I}$  and $\btsel{l_i : S_i}_{i \in I}$, respectively, for some finite index set $I$.
The corresponding typing rules are:
\[
				\inferrule[(Bra)]{
			 \forall i \in I \quad \Gamma; \Lambda; \Delta \cat u:S_i \proves P_i \hastype \Proc
		}{
			\Gamma; \Lambda; \Delta \cat u: \btbra{l_i:S_i}_{i \in I} \proves \bbra{u}{l_i:P_i}_{i \in I}\hastype \Proc
		}
		\qquad
	 	\inferrule[(Sel)]{
			\Gamma; \Lambda; \Delta \cat u: S_j  \proves P \hastype \Proc \quad j \in I
		}{
			\Gamma; \Lambda; \Delta \cat u:\btsel{l_i:S_i}_{i \in I} \proves \bsel{u}{l_j} P \hastype \Proc
		}
		\]
    
Our decomposition strategy can be extended to account for labeled choice (and their types).
This entails the extension of the type decomposition function $\Gtopt{\cdot}$ (\Cref{pi:def:typesdecomp}) and of the breakdown function $\Bopt{k}{\tilde x}{P}$ (\Cref{pi:t:bdowncorec}); this latter extension will be denoted  $\Boptsb{k}{\tilde x}{\cdot}$.

For simplicity, we focus on \emph{finite} processes,  without recursion / recursive session types. %; this will also facilitate comparisons with the CPS approach. 
\fixed{One of the challenges involved in extending our approach to account for both labeled choices and recursion concerns \emph{non-uniform} termination: in a branching type $\btbra{l_i:S_i}_{i \in I}$ not only each $S_i$ can be different, but some of them may denote recursive (non terminating) behaviors while the rest may not. Handling this potential non-uniformity requires care. }
We base our presentation on the corresponding breakdown for the case of the higher-order calculus \HO, as presented in~\cite{DBLP:journals/corr/abs-2301-05301}.

   To uniformly handle the potential differences in a branching type  $\btbra{l_i:S_i}_{i \in I}$, we break down branching and selection types  as follows: 
\end{newaddenv}
\begin{definition}[Decomposing Types: Labeled Choices]
  \label{pi:d:typessb}
  We extend the type decomposition function $\Gtopt{\cdot}$ (\Cref{pi:def:typesdecomp}) as follows:
  \begin{align*}
    \Gtopt{\btbra{l_i : S_i}_{i \in I}} &= 
    \btbra{l_i :\, \btinp{\Gtopt{S_i}}\tinact}_{i \in I}  
    \\
    \Gtopt{\btsel{l_i : S_i}_{i \in I}} &=  
    \btsel{l_i :\, \btout{\Gtopt{S_i}}\tinact}_{i \in I} 
  \end{align*} 
\end{definition}
\begin{newaddenv}
  This decomposition follows the intuition that branching and selection correspond to the input and output of labels, respectively. 

%The decomposition of branching and selection processes is defined accordingly: 
\end{newaddenv}

\begin{definition}[Decomposing Processes: Labeled Choice] 
  \label{pi:d:processsb}
  % \begin{align*}
  %   \Bopt{k}{\tilde x}{
  %     \propinp{k}{\tilde x}\bbra{u_i}{l_j : P_j}_{j \in I}}  
  %   &= \bbra{u_i}{l_j : \binp{n_i}{\tilde y}\Bopt{}{}{P_i\subst{y_1}{x_i}}}_{j \in I} \\
  %   \Bopt{}{}{\bsel{x_i}{l_i}Q}  
  %   &= \bsel{x_i}{}
  % \end{align*}
  Given a $k \geq 0$ and a tuple of names $\tilde x$, the decomposition function $\Boptsb{k}{\tilde x}{\cdot}$ is defined as 
    \begin{align*}
      \Boptsb{k}{\tilde x}{
       \bbra{u_i}{l_j : P_j}_{j \in I}}  
      &= 
      \propinp{k}{\tilde x}
      \bbra{u_i}{l_j : \news{\tilde \prop_{j}}\binp{u_i}{\tilde y}\apropout{k+1}{\tilde x \tilde y} \Par 
      \Boptsb{k+1}{\tilde x \tilde y}{P_j\subst{y_1}{u_i}}}_{j \in I} 
      \\
      \Boptsb{k}{\tilde x}{\bsel{u_i}{l_j}Q}  
      &= 
      \propinp{k}{\tilde x}\bsel{u_i}{l_j}
      \news{\tilde u : \Gtopt{S_j}}\bout{u_i}{
        {\tilde{\dual u}}}
        \apropout{k+1}{\tilde x} \Par 
        \Boptsb{k+1}{\tilde x}{Q\subst{s_1}{u_i}}
      % \apropout{k+1}{\tilde x} \Par 
    \end{align*} 
    \noindent where $\widetilde u=(u_{i+1},\ldots,u_{i+\len{\Gtopt{S_j}}})$ 
    and $\widetilde {\dual {u}}=(\dual {u_{i+1}},\ldots,\dual 
    {u_{i+\len{\Gtopt{S_j}}}})$. 
    \begin{newaddenv}
    \\ 
      For the remaining constructs, $\Boptsb{k}{\tilde x}{\cdot}$ corresponds 
      to %$\Boptsb{k}{\tilde x}{\cdot}$ 
      $\Bopt{k}{\tilde x}{\cdot}$ (\Cref{pi:t:bdowncorec}). 
    \end{newaddenv} 
  \end{definition}
 % \Boptsb
 
\fixed{Observe how, in the case of branching, once a particular branch $l_i$ has been selected, we receive the names on which to provide sessions from the branch $\Gtopt{S_i}$; the corresponding selection process communicates these names. 
To account for the different session types of the branches $S_i$, the names involved in the decomposition are bound (i.e., hidden).}

\smallskip

  \begin{newaddenv}
    We illustrate the workings of $\Boptsb{k}{\tilde x}{\cdot}$ by means of 
    the following example, adapted from~\cite{DBLP:journals/corr/abs-2301-05301}: 
  \end{newaddenv}

\begin{example}
\label{pi:ex:server}
  Consider a mathematical server $Q$ that offers clients two operations: addition and negation of integers. 
  The server uses name $u$ to implement the following session type: 
    \begin{align*}
      S = \btbra{\textsf{add}: 
      \underbrace{\btinp{\textsf{int}} \btinp{\textsf{int}} 
      \btout{\textsf{int}} \tinact}_{S_{\textsf{add}}}
      ~,\ 
      \textsf{neg} : 
      \underbrace{\btinp{\textsf{int}} \btout{\textsf{int}} \tinact}_{S_{\textsf{neg}}} 
      }
    \end{align*}
    This way, the $\textsf{add}$ branch receives two integers and sends
    over their sum;  the $\textsf{neg}$ branch has a single input of an integer followed by an
    output of its negation. 
    
    Let us consider a possible implementation for the server $Q$ and for a client $R$ that {selects} the first branch to add integers 16 and 26:
    \begin{align*}
       Q & \defas \bbra{u}{\textsf{add} : 
      \underbrace{\binp{u}{a}\binp{u}{b}\boutt{u}{a+b} }_{Q_{\textsf{add}}},\ 
      \textsf{neg} : 
      \underbrace{\binp{u}{a}\boutt{u}{-a}}_{Q_{\textsf{neg}}}} 
      \\
      R & \defas \bsel{\dual{u}}{\textsf{add}} 
      \bout{\dual {u}}{\mathbf{16}} \bout{\dual u}{\mathbf{26}} \binpt{\dual u}{r} 
      \end{align*}

      \noindent 
      The composed process $P \defas \news{u}(Q \Par R)$ can reduce as follows: 
  \begin{align*}
    P & ~\red~ 
    \news{u} 
    (\binp{u}{a}\binp{u}{b}\about{u}{a+b} \Par 
    \bout{\dual {u}}{\mathbf{16}} \bout{\dual u}{\mathbf{26}} \binpt{\dual u}{r} ) \\
    & ~\red^2~ 
    \news{u} (\about{u}{\mathbf{16+26}}  \Par \binpt{\dual u}{r} )
    = P'
  \end{align*}
First, \newadd{following~\Cref{pi:d:typessb}}, the decomposition  of $S$ is the minimal session type $M$, defined as follows:
\begin{align*}
  M = \Gtopt{S} =
      \btbra{
        &\textsf{add}:
      \btinpt{{\big(\btinpt{\textsf{int}}, \btinpt{\textsf{int}}, \btoutt{\textsf{int}}  \big)}},
      \\ 
      &\textsf{neg}:
      \btinpt{{\big(\btinpt{\textsf{int}}, \btoutt{\textsf{int}} \big)}}} 
\end{align*}

Let us now discuss the decomposition of $P$. 
\newadd{Using~\Cref{pi:d:processsb}} we have: 
  \begin{align}
    D = \news{\prop_1, \ldots, \prop_7} (\propinp{1}{} \propout{2}{} \propoutt{3}{} 
    \Par 
    \Boptsb{2}{\epsilon}{Q\sigma_2} \Par 
    \Boptsb{3}{\epsilon}{R\sigma_2} )
    \label{pi:ex:dec}
  \end{align}
  \noindent where $\sigma_2=\subst{u_1\dual{u_1}}{u\dual u}$. 
  The breakdown of process $Q$  is as follows: 
\begin{align*}
  \Boptsb{2}{\epsilon}{Q\sigma_2} = 
  \propinp{2}{}
  \bbra{u_1}{
    &\textsf{add} : \news{\tilde \prop_{j}}
    \binp{u_1}{y_1, y_2, y_3}\apropout{1}{y_1,y_2,y_3} \Par 
  \Boptsb{1}{y_1, y_2, y_3}{Q_{\mathsf{add}}\subst{y_1}{u_1}}, 
  \\ 
  &\textsf{neg}: 
  \news{\tilde \prop_{j}}\binp{u_1}{\tilde y}\apropout{1}{y_1, y_2} \Par 
  \Boptsb{1}{\y_1, y_2}{Q_{\mathsf{neg}}\subst{y_1}{u_1}}
  }
\end{align*}
\noindent where 
\begin{align*}
  \Boptsb{1}{y_1, y_2, y_3}{Q_{\mathsf{add}}\subst{y_1}{u_1}} &= 
  \apropinp{1}{y_1, y_2, y_3} \Par \propinp{1}{y_1, y_2, y_3}
  \binp{y_1}{a}\apropout{2}{y_2,y_3,a} \Par 
  \\
  & \qquad 
  \propinp{2}{y_2,y_3,a}\binp{y_2}{b}\apropout{3}{y_3,a,b} \Par 
  \propinp{3}{y_3,a,b}\bout{y_3}{a+b}\apropinp{4}{} \Par 
  \apropinp{4}{} 
  \\ 
  \Boptsb{1}{y_1, y_2}{Q_{\mathsf{neg}}\subst{y_1}{u_1}} &= 
  \apropinp{1}{y_1, y_2} \Par \propinp{1}{y_1, y_2}
  \binp{y_1}{a}\apropout{2}{y_2,a} \Par 
  \\
  & \qquad 
  \propinp{2}{y_2,a}\bout{y_2}{-a}\apropout{3}{} \Par 
  % \propinp{3}{y_3,a,b}\bout{y_3}{a+b}\apropinp{4}{} \Par 
  \apropinp{3}{} 
\end{align*}
In process $\Boptsb{1}{y_1, y_2, y_3}{Q_{\mathsf{add}}\subst{y_1}{u_1}}$, name $u_1$ implements $M$. 
Following the common trio structure, the first prefix awaits activation on 
$\prop_2$. The next prefix mimics the
branching action of $Q$ on $u_1$. 
Then, each branch consists of the input of 
the breakdown of the continuation of channel $u$, that is 
$y_1,y_2,y_3$. This input does not have a counterpart in $Q$; 
it is meant to synchronize with process $\Boptsb{3}{\epsilon}{R\sigma_2}$, 
the breakdown of the corresponding selection process.

In the bodies of the abstractions we break down $Q_{\textsf{add}}$ and $Q_{\textsf{neg}}$, but not before adjusting the names on which the broken down processes provide the sessions.
For this,  we substitute $u$ with $y_1$ in both processes, ensuring that the broken down names are bound by the input.
By binding decomposed names in the input we account for different 
session types of the original name in branches, while preserving typability: 
this way the decomposition of different branches can use (i)~the same names
but typed with different minimal types and (ii)~a different number of names, as it is the
case in this example.

The decomposition of the client process $R$, which implements the selection, 
is as follows: 
\begin{align*}
  \Boptsb{3}{\epsilon}{R\sigma_2} &= 
  \propinp{3}{\epsilon}\bsel{\dual u_1}{\mathsf{add}}
  \news{u_2,u_3, u_4}\bout{\dual u_1}{ u_2,  u_3,  u_4}
  \apropout{4}{} \Par  \Boptsb{4}{\epsilon}{\bout{\dual u_2}{\mathbf{16}}\bout{\dual u_2}{\mathbf{26}} \binpt{\dual u_2}{r}}
  % \\
  % & 
  % \propinp{4}{} ... 
\end{align*}
\noindent where:
\begin{align*}
  \Boptsb{4}{\epsilon}{\bout{\dual u_2}{\mathbf{16}}\bout{\dual u_2}{\mathbf{26}} \binpt{\dual u_2}{r}} = 
  \propinp{4}{}\bout{\dual u_2}{\mathbf{16}}\apropout{5}{} \Par 
  \propinp{5}{}\bout{\dual u_3}{\mathbf{26}}\apropout{6}{} \Par 
  \propinp{6}{}\binp{\dual u_4}{r}\apropout{7}{} \Par 
  \apropinp{7}{} 
\end{align*}

After receiving the context on $\prop_3$ (empty in this case), the selection action on $u_1$ is mimicked; then, the breakdown of 
channel continuation, $ u_2,  u_3,  u_4$, that are locally bound, 
are sent along name $u_1$. 
The intention is to use these names to connect 
the selected branch and the continuation of a selection process: the subprocess encapsulated in the branch  will use
$(u_2,  u_3,  u_4)$, while the dual names $(\dual u_2, \dual u_3, \dual
u_4)$ are present in the breakdown of the continuation.

Let us briefly examine the reductions of the decomposed process $D$ in (\ref{pi:ex:dec}).
First, $\prop_1$, $\prop_2$, and $\prop_3$ will synchronize. 
We have $D \red^4 D_1$, where
\begin{align*}
  D_1 = \news{\prop_4 \ldots \prop_7} 
  &\news{u_1} \big(
    \bbra{u_1}{
      \textsf{add} : \news{\tilde \prop_{j}}
      \binp{u_1}{y_1, y_2, y_3}\apropout{1}{\tilde y} \Par 
    \Boptsb{1}{y_1, y_2, y_3}{Q_{\mathsf{add}}\subst{y_1}{u_1}}, 
    \\ 
    & \qquad \qquad \qquad 
    \textsf{neg}: 
    \news{\tilde \prop_{j}}\binp{u_1}{\tilde y}\apropout{1}{y_1, y_2} \Par 
    \Boptsb{1}{\y_1, y_2}{Q_{\mathsf{neg}}\subst{y_1}{u_1}}
    }
  \\ 
    & \qquad \quad   \Par
    \bsel{\dual u_1}{\mathsf{add}}
    \news{u_2,u_3, u_4}\bout{\dual u_1}{ u_2,  u_3,  u_4}
    \apropout{4}{} \Par  \Boptsb{4}{\epsilon}{\bout{\dual u_2}{\mathbf{16}}\bout{\dual u_2}{\mathbf{26}} \binpt{\dual u_2}{r}}
    \big )
    \end{align*} 
 
    \noindent 
    Now, the processes chooses the label \textsf{add} 
    on $u_1$. The process $D_1$ will reduce further as 
    $D_1 \red D_2 \red^2 D_3$, where: 
    \begin{align*}
      D_2 =~ & \news{\tilde \prop_{j}}
      \binp{u_1}{y_1, y_2, y_3}\apropout{1}{\tilde y} \Par 
    \Boptsb{1}{y_1, y_2, y_3}{Q_{\mathsf{add}}\subst{y_1}{u_1}}
    \Par 
    \\ 
    & 
    \qquad  \news{u_2,u_3, u_4}\bout{\dual u_1}{ u_2,  u_3,  u_4}
    \apropout{4}{} \Par  \Boptsb{4}{\epsilon}{\bout{\dual u_2}{\mathbf{16}}\bout{\dual u_2}{\mathbf{26}} \binpt{\dual u_2}{r}} 
    \\
    D_3 =~ &
    \binp{\dual u_2}{a}\apropout{2}{ u_3, u_4,a} \Par  \propinp{2}{y_2, y_3,a}\binp{ y_2}{b}\apropout{3}{y_3,a,b} \Par 
    \\
    & \qquad  
    \propinp{3}{y_3,a,b}\bout{ y_3}{a+b}\apropout{4}{} \Par 
    \\ 
    & \qquad  
  \propinp{4}{} \bout{\dual u_2}{\mathbf{16}}\apropout{5}{} \Par 
  \propinp{5}{}\bout{\dual u_3}{\mathbf{26}}\apropout{6}{} \Par 
  \propinp{6}{}\binp{\dual u_4}{r}\apropout{7}{} \Par 
  \apropinp{7}{} 
    \end{align*}
Now, process $D_3$ can mimic the original 
transmission of the integer $16$ on channel $u_2$ as follows: 
\begin{align*}
  D_3 &\red 
  \apropout{2}{\dual u_3, \dual u_4, \mathbf{16}} \Par \propinp{2}{y_2, y_3,a}\binp{y_2}{b}\apropout{3}{y_3,a,b} \Par 
    \\
    & \qquad 
    \propinp{3}{y_3,a,b}\bout{y_3}{a+b}\apropout{4}{} \Par 
    \apropinp{4}{} 
    \Par 
    \\ 
    & \qquad 
  \apropout{5}{} \Par 
  \propinp{5}{}\bout{\dual u_3}{\mathbf{26}}\apropout{6}{} \Par 
  \propinp{6}{}\binp{\dual u_4}{r}\apropout{7}{} \Par 
  \apropinp{7}{} = D_4 
\end{align*}

Finally, $D_4$ reduces to $D_5$ as follows: 
\begin{align*}
  D_4 &\red^5 
  \bout{\dual u_4}{a+b}\apropout{4}{} \Par 
    \apropinp{4}{} 
    \Par 
    % \\ 
    % & \quad 
 \binp{ u_4}{r}\apropout{7}{} \Par 
  \apropinp{7}{} = D_5 
\end{align*}
This way, the steps from $D$ to $D_5$ attest to the fact that our extended decomposition correctly simulates the steps from $P$ to $P'$. 
\end{example}

%  
%\begin{newaddenv}
%  The example illustrating the decomposition of 
%  labelled choices constructs following 
%  the compositional approach (presented in~\Cref{pi:s:dbc}) 
%  can be found in~\Cref{pi:app:ss:exsb}.
%\end{newaddenv}
}{}

\section{Related Work}
\label{pi:s:rw}

Closely related work has been already discussed throughout the paper. In this section, we comment on other related literature.

A source of inspiration for our developments is  the trios decomposition by Parrow~\cite{DBLP:conf/birthday/Parrow00}, which he studied for an untyped $\pi$-calculus with replication; in contrast, \sessp processes are typed and feature recursion. 
We stress that our goal is to  clarify the role of sequentiality in session types by using processes  with \msts, which lack sequentiality. 
While Parrow's approach elegantly induces processes typable with \msts (and suggests a clean approach to establish dynamic correctness), defining trios decompositions for \sessp is just one   path towards our goal. %Indeed, our decomposition for choices uses quintets, which are still minimal and typable using \msts. 

	\thesisalt{The present work differs significantly with respect to our previous work~\cite{APV19,DBLP:journals/corr/abs-2301-05301}, which used  \HO as 
	 source language. Session communication in \HO is based on abstraction-passing, whereas here we focus on the name-passing calculus \sessp. 
	 This difference has several ramifications. 
		While in \HO propagators carry abstractions, 
in our case propagators 
are binding and carry names. Also,  names must be decomposed and propagating them requires care. 
Further novelties appear when decomposing  processes with recursion, which require a dedicated collection of \emph{recursive trios}; in contrast, an explicit construct for recursion is not present in \HO.
The proof of dynamic correctness for \sessp shares some similarities with the same proof for \HO, given in~\cite{DBLP:journals/corr/abs-2301-05301}; however, the technical details of moving from higher-order concurrency to first-order concurrency are substantial---from the behavioral equivalences used to the construction of the required bisimilarities (cf.    \Cref{sec:t:mainbsthm}).}{}
	
	Our work aims to understand session types in terms of themselves, by considering to the sub-class of session types without sequentiality, as defined by \msts. 
	Prior works have related session types with \emph{different} type systems---see, e.g.,~\cite{DBLP:conf/unu/Kobayashi02,DBLP:conf/ppdp/DardhaGS12,DBLP:journals/iandc/DardhaGS17,DBLP:conf/concur/DemangeonH11,DBLP:journals/corr/GayGR14}. 
	Kobayashi~\cite{DBLP:conf/unu/Kobayashi02} was the first to define a formal relationship between session types and \emph{usage types}, expressed as typed encodings of processes; this relationship was thoroughly studied by Dardha et al.~\cite{DBLP:conf/ppdp/DardhaGS12,DBLP:journals/iandc/DardhaGS17,DBLP:journals/corr/Dardha14} (see below).
	Demangeon and Honda~\cite{DBLP:conf/concur/DemangeonH11} connect a linearly-typed $\pi$-calculus with subtyping and a session-typed calculus via a full abstraction result. 
	Both works rely on convenient constructs in the target language (e.g., case constructors and variant types) to achieve correct encodability. The work of Gay et al.~\cite{DBLP:journals/corr/GayGR14} addresses a similar problem but by adopting a $\pi$-calculus without such additional features: they consider the correct encodability of session types into a generic type system for a simple $\pi$-calculus. Their work demonstrates that the translation of session types for branching and selection in the presence of subtyping is a challenging endeavor.
	
The work by Dardha et al.~\cite{DBLP:conf/ppdp/DardhaGS12,DBLP:journals/iandc/DardhaGS17,DBLP:journals/corr/Dardha14} develops further the translation first suggested by Kobayashi~\cite{DBLP:conf/unu/Kobayashi02}.
They compile a session $\pi$-calculus down into a   $\pi$-calculus  with the linear type system of~\cite{LinearPi} extended with variant types.
They represent sequentiality  using a continuation-passing style (CPS): a session type is interpreted as a linear type carrying a pair, consisting of 
 the original payload type and  a new linear channel type, to be used for ensuing interactions.
The differences between this CPS approach and our work are also technical: the approach in~\cite{DBLP:conf/ppdp/DardhaGS12} thus involves translations connecting \emph{two} different $\pi$-calculi and \emph{two} different type systems. In contrast, our approach based on \msts  justifies sequentiality using a single typed process framework. 
Another difference concerns process recursion and recursive session types: while the works~\cite{DBLP:conf/ppdp/DardhaGS12,DBLP:journals/iandc/DardhaGS17} consider only finite processes (no recursion nor recursive types), the work~\cite{DBLP:journals/corr/Dardha14} considers recursive session types as a way of supporting replicated processes (a specific class of unrestricted behaviors). In contrast, the decompositions we have considered here support full process recursion. 

%A detailed comparison between our approach and~\cite{DBLP:conf/ppdp/DardhaGS12} is interesting follow-up work.

Despite these concrete differences, it is interesting to compare our approach and the CPS approach. 
To substantiate our comparisons, we introduce some selected notions for the linearly-typed $\pi$-calculus considered by Dardha et al.~(the reader is referred to~\cite{DBLP:journals/iandc/DardhaGS17} for a thorough description and technical results).
    \begin{definition}[Linearly-Typed $\pi$-calculus Processes~\cite{DBLP:journals/iandc/DardhaGS17}] 
    The syntax of processes $P, Q, \ldots$, values $v, v', \ldots$,  and of linear types $\tau, \tau', \ldots$ is as follows:
\begin{align*}
 \text{Processes~~}   P,Q, &:= \bout{x}{\tilde v}P \bnfbar 
  \binp{x}{\tilde y}P \bnfbar \news{v}P
  \bnfbar \inact \bnfbar P \Par Q 
  \bnfbar \ltcase{v}{l_i}{x_i}{P_i}{i \in I} 
  \\
   \text{Values~~}   v, v' &:= x  \bnfbar \star \bnfbar l\_v
   \\
    \text{Types~~}   \tau, \tau' & := \lto{\wtd \tau} \bnfbar \lti{\wtd \tau} 
  \bnfbar \ltab{\wtd \tau} \bnfbar \ltend 
  \bnfbar \ltc{\wtd \tau}
  \bnfbar \ltbra  \bnfbar \texttt{Unit}
\end{align*}
\end{definition} 
 
At the level of processes, the main difference with respect to the syntax of \HOp in \Cref{top:fig:syntax} is the case construct `$\ltcase{v}{l_i}{x_i}{P_i}{i \in I}$', which together with the variant value `$l\_v$' implement a form of deterministic choice. This choice is decoupled from a synchronization: indeed, the process $\ltcase{l_j\_v}{l_i}{x_i}{P_i}{i \in I}$ (with $j \in I$) autonomously reduces to $P_j\subst{v}{x_j}$.
	At the level of types, the grammar above first defines  types 
	$\lto{\wtd \tau}$, $ \lti{\wtd \tau}$, and  
$\ltab{\wtd \tau}$ 
	for the linear exchange messages of type $\widetilde{\tau}$: output, input, and both output and input, respectively. Then, the types $\ltend$ and $\ltc{\wtd \tau}$ are assigned to channels without any capabilities and with arbitrary capabilities, respectively. Finally, we have the variant type $\ltbra$ associated to the case construct  and the unit type $\texttt{Unit}$.
	
%The typing rules for processes, given in~\Cref{top:fig:typeslinear}, should be self-explanatory; we omit definitions for contexts and their operations, which follow expected intuitions. The type system ensures the type preservation property (subject reduction).

%\input{type_system_lt.tex}
	
As already mentioned, the key idea of the CPS approach is to represent  one message exchange using \emph{two} channels: one is the message itself, the other is a continuation for the next sequential action in the session. We illustrate this approach  by example.  
	
    \newcommand{\contc}{c}
  \begin{example}[The CPS Approach, By Example]
  Consider the session type $S$ and the processes $Q$ and $R$ from \Cref{pi:ex:server}. Under the CPS approach, the session type $S$ is represented as linear types as follows:
%    \begin{align*}
%      S = \btbra{\textsf{add}: 
%      \underbrace{\btinp{\textsf{int}} \btinp{\textsf{int}} 
%      \btout{\textsf{int}} \tinact}_{S_{\textsf{add}}}
%      ~,\ 
%      \textsf{neg} : 
%      \underbrace{\btinp{\textsf{int}} \btout{\textsf{int}} \tinact}_{S_{\textsf{neg}}} 
%      }
%    \end{align*}
    \begin{align*}
      \ltenc{S} &=
      \lti{\ltsel{\textsf{add}}{\ltenc{S_{\textsf{add}}},\ 
      \textsf{neg}: {\ltenc{S_{\textsf{neg}}}}}{}} 
      & 
      \ltenc{S_{\textsf{add}}} &= \lti{\textsf{int}, \lti{\textsf{int}, 
      \lto{\textsf{int}, \ltend}}}
      \\
      & &
      \ltenc{S_{\textsf{neg}}} &= \lti{\textsf{int}, \lto{\textsf{int},  \ltend}}
    \end{align*}
 This way, sequentiality in $S$ arises in  $\ltenc{S}$ as nesting of pairs---the later that an action appears in a session type, the deeper it will appear in the corresponding linear type.
 Accordingly, the encoding of session-typed processes as linear processes, denoted $\ltenc{\cdot}_{f}$,  is illustrated below; the parameter $f$ records continuations:
   \begin{align*}
    \ltenc{P}_{\emptyset} &= \news{u}\ltenc{Q \Par R}_{\emptyset} = 
    \news{u}\ltenc{Q}_{\emptyset} \Par \ltenc{R}_{\emptyset} \\
   \ltenc{Q}_{\emptyset} &=  
  %  \binp{u}{y}\ltcase{y}{l_i}{c}{\ltenc{P_i}}{}  
   \binp{u}{y}\ltcasep{y}{\textsf{add}\_\contc_1~\triangleright~\ltenc{Q_{\textsf{add}}}_{u \mapsto \contc_1},\ 
   \textsf{neg}\_\contc_1~\triangleright~\ltenc{Q_{\textsf{neg}}}_{u \mapsto \contc_1}} 
   \\
   \ltenc{Q_{\textsf{add}}}_{u \mapsto \contc_1} &= 
   \binp{\contc_1}{a, \contc_2}\binp{\contc_2}{b,\contc_3}\news{\contc_4}
   \bout{\contc_3}{a+b, \contc_4}\inact
   \\
   \ltenc{Q_{\textsf{neg}}}_{u \mapsto \contc_1} &= 
   \binp{\contc_1}{a, \contc_2}\news{\contc_3}
   \bout{\contc_2}{-a, \contc_3}\inact 
   \\ 
   \ltenc{R}_{\emptyset} &= 
   \news{\contc_1}\bout{u}{\textsf{add}\_\contc_1} 
   \news{\contc_2}\bout{\contc_1}{\textbf{16}, \contc_2}
   \news{\contc_3}\bout{\contc_2}{\textbf{26}, \contc_3}
   \binp{\contc_3}{r}\inact
   \end{align*}
   Observe how the branching construct in $Q$ is encoded using two constructs: an input prefix immediately followed by a case construct. 
   Also, notice the use of restrictions $\news{\contc_i}$ (with $i \in \{1,2,3\}$) in the encoding of output prefixes: this is crucial to avoid interferences and ensure that session communications are mimicked in the intended order.
  \end{example} 
  
Based on the above example, we may identify two sources of comparison between our decompositions into \msts and the CPS approach.
	 At the level of types, there is clear resemblance between \msts and the class of linear types needed to encode finite session types in the CPS approach. 
  At the level of processes, both approaches use a similar principle, namely to generate fresh names to encode the sequential structure of sessions. While the CPS approach follows a \emph{dynamic} discipline to generate such names (i.e., they are generated by the encoding of output-like actions), the decomposition approach follows a \emph{static discipline}, i.e., fresh names are generated based on the length of the source session types. 
 
 \begin{figure}[!t]
    \begin{mdframed}%[style=alttight]
    \centering
		\begin{tikzpicture}[scale=1.5]
		\tikzstyle{ann} = [draw=none,fill=none,right]
		%\node[draw=none] at (0,-0.575) {$text$};
			%\node[draw](A) at (0, 0) {$Q_{\mathtt{Pay}}$};
			%\node[draw](A) at (0, 0) {$Q_{\mathtt{Pay}} = \binp{u}{a}\binp{u}{b}\bout{u}{a < 42} \inact$};
			\node[draw] (B) at (0, 0) {$\HOp$};
			%\draw[->, thick] (0,-0.85) -- (0,-1.2);
			\node[draw] (T1) at (-1.5,-1.25) {$\sessp$};
			%\node at (-1.75,-3) {$\parallel$};
			\node[draw] (T2) at (1.5,-1.25) {$\HO$};
			%\node at (1.5,-3) {$a\parallel$};
			\node[draw] (T3) at (1.5,-2.75) {$\mHO~{ (\HO+\text{MSTs})}$};
			\node[draw] (T4) at (-1.5,-2.75) {$\msessp~{(\sessp+\text{MSTs})}$};
			\draw[->, dotted, thick,>=stealth] (T1) edge  (B); 
			\draw[->, dotted, thick,>=stealth] (T2) edge (B);
			\draw[->, dotted, thick,>=stealth] (T3) edge (T2); 
			\draw[->, dotted, thick,>=stealth] (T4) edge (T1); 
			\draw[<->, color=gray,  thick,>=stealth] (T2) edge [bend left=15] (T1);
			\draw[->, color=black, thick,>=stealth](B) edge[bend right] node [left] {$\map{\cdot}^2$ {\small\cite{DBLP:conf/esop/KouzapasPY16,DBLP:journals/iandc/KouzapasPY19}}~\textcolor{white}{.}} (T1);
			\draw[->, color=black, thick,>=stealth](B) edge[bend left] node [right] {\textcolor{white}{.}~$\map{\cdot}^1_{g}$ {\small\cite{DBLP:conf/esop/KouzapasPY16,DBLP:journals/iandc/KouzapasPY19}}} (T2);
			\draw[->, color=black, thick,>=stealth](T2) edge[bend left] node [right] {$\D{\cdot}$ {\small \cite{APV19,DBLP:journals/corr/abs-2301-05301}}} (T3);
			\draw[->, color=\deccolor, thick,>=stealth](T1) edge[bend right] node [left] {$\F{\cdot}$} (T4);
			\draw[->, color=\decoptcolor, thick,>=stealth](T1) edge[bend left] node [right] {$\Do{\cdot}$} (T4);
		\end{tikzpicture}
	\end{mdframed}
		\caption[Summary of expressiveness results for \HOp]{Summary of expressiveness results for \HOp. Solid lines indicate correct encodings and  decompositions; dotted lines indicate sub-languages.\label{pi:f:taxonomy}}
\end{figure}
\section{Concluding Remarks}
\label{pi:s:concl}

We studied minimal session types (\msts) for \sessp---the sub-language of \HOp~\cite{DBLP:conf/esop/KouzapasPY16,DBLP:journals/iandc/KouzapasPY19} with first-order communication, recursion, and recursive types---and obtained new minimality results based on two different decompositions of processes. 
Introduced in \cite{APV19}, \msts express a specific form of minimality, which eschews from sequentiality in types;  
hence, our minimality results  for \sessp mean that sequentiality in types is a convenient and yet not indispensable feature, as it can be represented by name-passing processes while remaining in a session-typed setting.  

Following the approach for \HO~\thesisalt{\cite{APV19,DBLP:journals/corr/abs-2301-05301}}{\Cref{mst:ch:mst}},  which is inspired by Parrow's trios processes~\cite{DBLP:conf/birthday/Parrow00},
we %introduced minimal session types  for \sessp and 
defined two process decompositions for \sessp. They use type information to transform processes typable with standard session types into processes typable with \msts.
The first decomposition, denoted $\F{\cdot}$, is obtained by composing existing encodability results and the decomposition / minimality result for \HO; the second decomposition, denoted $\Do{\cdot}$, optimizes the first one by (i)~removing redundant synchronizations and (ii)~using the native support of recursion in \sessp. 
For both decompositions 
we establish the minimality result (cf. \Cref{pi:t:amtyprecdec,pi:t:decompcore}), which is actually a result of static correctness (i.e., preservation of typability). 
The gains in moving from $\F{\cdot}$ to $\Do{\cdot}$ can be accounted for in very precise terms, as attested by \Cref{pi:p:propopt}.
For the optimized decomposition $\Do{\cdot}$ we proved also a dynamic correctness result (\Cref{pi:t:mainbsthm}), which formally attests that a process and its decomposition are behaviorally equivalent. This result thus encompasses  a result of operational correspondence; its proof leverages  existing characterizations of contextual equivalence for \sessp~\cite{KouzapasPY17}, and adapts them to the case of \msts.
This way, our technical results together confirm the  significance of \msts; they also indicate that a minimality result is independent from the kind of communicated objects, either abstractions (functions from names to processes, as studied in~\cite{APV19,DBLP:journals/corr/abs-2301-05301}) or names (as studied here).

%We have also remarked that 
%decomposition on process-level is not essential to encode 
%a session typed process 
%into a minimal session typed process. 
%To argue for this, 
%we isolate features of decomposition that only 
%concerns minimality of session types 
%sketching such an encoding by the example. 

%As claiming redundancy in session types may seem a bit 
Sequentiality is the key distinguishing feature in the specification of message-passing programs using session types.
Our minimality results for \sessp and \HO should not be interpreted as meaning that sequentiality in session types is redundant in \emph{modeling and specifying} processes; rather, we claim that it is not an essential notion to \emph{verifying} them.  
Because we can type-check {session typed}
processes using type systems that do not directly support sequentiality in types, our decompositions suggest a technique for implementing session types into languages whose type systems do not support sequentiality. 

%We claim that \msts are ``minimal'' from the perspective of the type constructs available to express and verifying (correct) processes. 

%For the sake of space, we have not considered choice constructs  (selection and
%branching). There is no fundamental obstacle in treating them, apart from a very
%minor caveat: the decomposition in~\thesisalt{\cite{APV19}}{\Cref{mst:ss:exti} (\Cref{mst:ch:mst})} assumes typed processes in which
%every selection construct comes with a corresponding branching 
%\longversion{(see~\Cref{pi:app:ss:exsb} for an example)}{(see \cite{AAP21-full} for an example)}.

All in all, besides settling a question left open in~\thesisalt{\cite{APV19,DBLP:journals/corr/abs-2301-05301}}{\Cref{mst:ch:mst}},
our work deepens our understanding about the essential mechanisms in session-based concurrency and about the connection between the first-order and higher-order paradigms in the typed setting.
\Cref{pi:f:taxonomy} depicts the formal connections between \HOp and its sub-languages, based on:  \rom{1}~the mutual encodability results between \sessp and \HO~\cite{DBLP:conf/esop/KouzapasPY16,DBLP:journals/iandc/KouzapasPY19}; \rom{2}~the minimality result for \HO~\cite{APV19,DBLP:journals/corr/abs-2301-05301}; and \rom{3}~the decompositions and minimality results for \sessp obtained here.

There are several interesting items for future work. 
First, it would be interesting to consider asynchronous communication and subtyping in the context of our decompositions, both first-order and higher-order. 
Second, following the discussion in \Cref{pi:s:rw}, 
it would be insightful to formulate a ``hybrid'' approach that combines and unifies the CPS approach~\cite{DBLP:conf/unu/Kobayashi02,DBLP:conf/ppdp/DardhaGS12,DBLP:journals/iandc/DardhaGS17} and our approach based on \msts. Strengths of our approach include direct support for recursion and recursive types, and dynamic correctness guarantees given in terms of well-studied behavioral equivalences; on the other hand, the CPS approach offers a simple alternative to represent non-uniform session structures, such as those present in labeled choices. 

\paragraph{Acknowledgments.} We are grateful to the anonymous reviewers and to Dan Frumin for useful comments and suggestions, which helped us to improve our paper. 
This research has been supported by the Dutch Research Council (NWO) under project No. 016.Vidi.189.046 (`Unifying Correctness for Communicating Software').

%            \draw[->, dotted] (HO) -- node[above, small] {$\D{\cdot}$} (MHO);
%            \draw[->, dotted] (PI) -- node [right, small] {$\map{\cdot}^1_{g}$} (HO);
%            \draw[->, dotted] (MHO) -- node [right, small] {$\map{\cdot}^2$} (MPI);
%            \draw[->] (PI) -- node[below, scale=0.85] {$\F{\cdot}$} (MPI);

% \input{sources/conclusionrelwork.tex}

%%
%% The acknowledgments section is defined using the "acks" environment
%% (and NOT an unnumbered section). This ensures the proper
%% identification of the section in the article metadata, and the
%% consistent spelling of the heading.

% \begin{acks}
% This work has been partially supported by the Dutch Research Council (NWO) under project No. 016.Vidi.189.046 (Unifying Correctness for Communicating Software).

% We are grateful to the anonymous reviewers for their careful reading and constructive remarks.
% \end{acks}

%%
%% The next two lines define the bibliography style to be used, and
%% the bibliography file.
% \bibliographystyle{ACM-Reference-Format}
\bibliographystyle{abbrv}
\bibliography{session.bib}

%%
%% If your work has an appendix, this is the place to put it.

\newpage

\appendix
\tableofcontents
% \subimport{sources/}{appendix-ppdp2021.tex}

% \newcommand{\thesisalt}[2]{\ifx\thesis\undefined#1\else#2\fi}

\providecommand{\thesisalt}[2]{\ifx\thesis\undefined#1\else#2\fi}

 \newpage
%\section{Additional Example}

% \subsection{First Decomposition: Labelled Choice}
%\label{pi:app:ss:exsb}
%\input{example-am-selbra}

% \newpage
% \subimport{sources/}{appendix-mst-prelim.tex} 

% \subimport{sources/}{appendix-mst-pi-examples.tex}

% \subimport{sources/}{appendix-mst-pi-aux.tex}
% \section{Proofs}

% \newcommand{\thesisalt}[2]{\ifx\thesis\undefined#1\else#2\fi}

\providecommand{\thesisalt}[2]{\ifx\thesis\undefined#1\else#2\fi}

\ifx\thesisalt\undefined 
{\newcommand{\thesisalt}[2]{\ifx\thesis\undefined#1\else#2\fi}}
\else \fi

\thesisalt{
% \subsection{Decompose by Composition}
% \subsection{Appendix to \Cref{pi:s:dbc}}
  \section{Proofs}
}
{\section{Appendix to \Cref{pi:s:dbc}}}

\thesisalt{
  
 \subsection{Auxiliary Results}

We rely on the following properties of the type system in \Cref{pi:ss:types}.

\begin{restatable}[Substitution Lemma]{lemm}{lemsubst}
  \label{top:lem:subst}
  $\Gamma;\Delta \cat x:S \proves P \hastype \Proc$ and 
  $u \notin \dom{\Gamma,\Delta}$ implies 
  $\Gamma; \Delta \cat u:S \proves P\subst{u}{x} \hastype \Proc$.
  \end{restatable}
  
%  \begin{comment} 
%  The following property follows immediately from 
%  \defref{def:typesdecomp} and 
%  \defref{def:typesdecompenv}:
%  
%
%  \begin{restatable}[Typing Broken-down Variables]{lemm}{lemvarbreak}
%  \label{lem:varbreak}
%    If $\Gamma;\Delta \proves z_i \hastype C$ then 
%    $\Gt{\Gamma};\Gt{\Delta} \proves \wtd z \hastype \Gt{C}$ 
%    where $\wtd z = (z_i,\ldots,z_{i + \len{\Gt{C}} - 1})$.
%  \end{restatable}
%\end{comment} 
%  
  \begin{restatable}[Shared environment weakening]{lemm}{lemweaken}
    \label{top:lem:weaken}
    If $\Gamma;\Delta \proves P \hastype \Proc$
    then
    $\Gamma \cat u:\chtype{C}; \Delta \proves P \hastype \Proc$.
  \end{restatable}

  % \begin{comment} 
  \begin{restatable}[Shared environment strengthening]{lemm}{lemstrength}
    \label{top:lem:strength}
    If $\Gamma;\Delta \proves P \hastype \Proc$ and
    $u \notin \fn{P}$ then $\Gamma \setminus u;\Delta \proves P \hastype \Proc$.
  \end{restatable}
% \end{comment} 

% \thesisalt{\input{rec-aux.tex}}{}
  
}{}

% \begin{restatable}[]{lemm}{}
% 	\label{pi:lem:indexcor}
% 	Let $\widetilde r$ be tuple of channel names and $S$ a recursive 	session type.
% 	If $\widetilde r : \Rtsopt{}{s}{\btout{C}S}$ and $k = f(\btout{C}S)$
% 	then $r_k:\trec{t}\btout{\Gt{C}}\vart{t}$.
% \end{restatable}

\thesisalt{
  \subsection{Proof of \Cref{pi:t:amtyprecdec}}
}{
  \subsection{Proof of \Cref{pi:t:amtyprecdec}}
}
\thmamtyprec* 

\begin{proof}
  \label{pi:app:thmamtyprec}
  \begin{align}
      \Gamma; \Delta, \envR &\proves P \hastype \Proc & &( \text{Assumption}) \\
      \tmap{\Gamma}{1}; \tmap{\Delta}{1}, \tmap{\envR}{1} 
      &\proves \map{P}^1_g \hastype \Proc 
      & &(\text{Theorem 5.1 \cite{DBLP:journals/iandc/KouzapasPY19}, (10)}) \\
      \Gt{\tmap{\Gamma}{1}}, \Phi; \ 
      \Gt{\tmap{\Delta}{1}}, \Theta &\proves \B{k}{\epsilon}{ \map{P}^1_g } 
      \hastype \Proc & &( 
        \text{\thesisalt{\Cref{mst:t:typecore}}{\Cref{mst:t:typecore}}, (11)}) \\
      \map{ \Gt{\tmap{\Gamma}{1}} }^2, \map{\Phi}^2 ; 
      \map{ \Gt{\tmap{\Delta}{1}} }^2, \ \map{\Theta}^2 &\proves 
      \map{\B{k}{\epsilon}{\map{P}^1_g}}^2 \hastype \Proc 
      & &(\text{Theorem 5.2 \cite{DBLP:journals/iandc/KouzapasPY19}, (12)}) \\
      \mGt{\Gamma}, \Phi'; \mGt{\Delta}, \Theta' 
      &\proves \AmeliaMod{k}{\epsilon}{P}_g \hastype \Proc 
      & &(\text{Definition of } \AmeliaMod{k}{\epsilon}{\cdot}_g,\\
      & & &\text{\Cref{pi:decomp:firstordertypes,pi:ThetaPrimeDef}, (4)}) \nonumber 
  \end{align}
\end{proof}

\thesisalt{}{
  \subsection{Proof of \Cref{pi:t:amtyprecdec}}
}
\thmamtyprecdec*

\begin{proof}
\label{pi:app:thmamtyprecdec}
\begin{align}
     \Gamma; \Delta &\proves P \hastype \Proc & &( \text{Assumption}) \\
     \Gamma \sigma; \Delta \sigma &\proves P \sigma \hastype \Proc 
     & &( \text{\Cref{top:lem:subst}, (6)}) \\
     \mGt{\Gamma \sigma} \cat \Phi'; \mGt{\Delta \sigma}, \Theta' 
     &\proves \AmeliaMod{k}{\epsilon}{P \sigma}_g \hastype \Proc 
     & &( \text{$P$ is initialized, \Cref{pi:t:am-typ-rec}})
     \label{pi:eq:am-dec-rec1}
\end{align}

% By \thmref{t:decompcore} we have that: 

% We shall show the following: 

% $$\mGt{\Gamma \sigma}; \mGt{\Delta \sigma}
%       \cat \mGt{\envR\sigma} \proves 
%       \news{\widetilde \prop}
%     \news{\widetilde \prop_r}
%     \big(
%       \prod_{r \in \tilde{v} } P^r \Par   
%     \propout{k}{} \inact 
%     \Par \AmeliaMod{k}{\epsilon}{P\sigma}_g \big)$$

To complete the proof, let us construct a well-formed derivation tree
\begin{comment} 
\thesisalt{Note that we apply one of the polyadic rules, i.e. PolyResS, used for typing \HOp, 
derived by Arslanagi\'{c} et al.~\cite{APV19}}{}. 
\end{comment}

\begin{align}
  \AxiomC{for $r \in \widetilde v$} 
  \AxiomC{$
          \mGt{\Gamma \sigma} \cat \Phi'; \wtd r : \mGt{S} 
         \proves  \proprinp{r}{b}\news{s}(\bout{b}{s}\about{\dual s}{\wtd r}) 
         \hastype \Proc$} 
  \LeftLabel{\scriptsize (\textsc{Par}, $\len{\widetilde v}-1$ times)}
  \BinaryInfC{$\mGt{\Gamma \sigma} \cat \Phi'; 
  \mGt{\Delta_\mu \sigma}
   \proves
   \prod_{r \in \tilde{v} } P^r$} 
  \DisplayProof
  \label{pi:pt:am-dec-rec2}
\end{align}
\noindent where $S = \envR(r)$. 
By the definition of $\Phi'$ we have 
$c^r:\chtype{\chtype{\btinp{\mRts{}{}{\envR(r)}}\tinact}} \in \Phi'$ 
and by  \Cref{pi:decomp:firstordertypes} we have $\mRts{}{}{\envR(r)} = \mGt{S}$, 
 so the right-hand of \eqref{pi:pt:am-dec-rec2} is well-typed. 
\begin{align}
  \AxiomC{} 
  \LeftLabel{\scriptsize (\textsc{Nil})}
  \UnaryInfC{$
  \mGt{\Gamma \sigma} \cat \Phi'; 
    \mGt{\Delta \sigma} 
     \proves \inact \hastype \Proc$}
  \LeftLabel{\scriptsize (\textsc{Send})}
  \UnaryInfC{$
    \mGt{\Gamma \sigma} \cat \Phi'; 
    \mGt{\Delta \sigma}, \apropout{k}{\cdot};
    \tinact \proves\apropout{k}{\cdot}
  $} 
  \AxiomC{\eqref{pi:pt:am-dec-rec2}} 
  \LeftLabel{\scriptsize (\textsc{Par})}
  \BinaryInfC{ $\mGt{\Gamma \sigma} \cat \Phi'; 
  \mGt{\Delta \sigma} \cat \mGt{\Delta_\mu \sigma} 
  \cat \apropout{k}{\cdot};
   \tinact \proves
   \propout{k}{\cdot}
    \inact \Par 
    \prod_{r \in \tilde{v} } P^r$}
    \DisplayProof
    \label{pi:pt:am-dec-rec1}
\end{align}
\begin{align*}
  \AxiomC{
   \eqref{pi:pt:am-dec-rec1}
  } 
  \AxiomC{\eqref{pi:eq:am-dec-rec1}}
  \LeftLabel{\scriptsize (\textsc{Par})}
  \BinaryInfC{
    $\mGt{\Gamma \sigma} \cat \Phi'; 
    \mGt{\Delta \sigma} \cat  \mGt{\envR\sigma} \cat \apropout{k}{\cdot};
     \tinact, \Theta' \proves \propout{k}{\cdot}
      \inact \Par 
      \prod_{r \in \tilde{v} } P^r \Par  
      \AmeliaMod{k}{\epsilon}{P \sigma}_g \hastype \Proc$
  } 
  \LeftLabel{\scriptsize (\textsc{PolyResS})}
  \UnaryInfC{
    $\mGt{\Gamma \sigma}; \mGt{\Delta \sigma}
    \cat \mGt{\envR\sigma} \proves 
    \news{\widetilde \prop}
  \news{\widetilde \prop_r}
  \big(
    \prod_{r \in \tilde{v} } P^r \Par   
  \propout{k}{} \inact 
  \Par \AmeliaMod{k}{\epsilon}{P\sigma}_g \big)
    $
  }
  \DisplayProof 
\end{align*}
\end{proof}

\thesisalt{}{
  \section{Appendix to \Cref{pi:s:pi-optimizations}}
}

\thesisalt{
  % \subsection{Measuring the Optimization}
  \subsection{Proof of \Cref{pi:p:propopt}}
}
{
  \subsection{Proof of \Cref{pi:p:propopt}}
}

\propopt* 

\begin{proof}%(Sketch)
  % TODO: check if this label p:propt is called 
% \label{pi:p:propopt}
\label{pi:app:propopt}
\label{pi:proof:propopt}
Because $\numpropam{P} \geq
   \lenHO{P} + 2 \cdot \len{\lentr{P}}$ (\Cref{d:propopt}), it suffices to show that 
   $$\lenHO{P} + 2 \cdot \len{\lentr{P}} \geq \frac{5}{3} \cdot \numprop{P}$$
   The proof is by induction on structure of $P$. We show one base case (output prefix followed by 
   inaction) and
   three inductive cases (input, restriction, and parallel composition); other
   cases are similar:
     \begin{itemize}
       \item Case $P = \bout{u}{x}\inact$. Then 
        $\lenHO{P} = 4$, $\len{\lentr{P}} = 0$, and $\lenHOopt{P} = 2$. 
      So, we have $\lenHO{P} + 2 \cdot \len{\lentr{P}} \geq \frac{5}{3} \cdot \numprop{P}$.

       \item Case $P = \bout{u}{x}P'$ with $P' \not\equiv \inact$.
       By IH we know $\lenHO{P'} + \lentr{P'} \geq \frac{5}{3} \cdot \numprop{P'}$.
       We know $\lenHO{P} = \lenHO{P'} + 3$, $\lenHOopt{P} = \lenHOopt{P'} + 1$ and $\lenrecpx{P} = \lenrecpx{P'}$. 
       So,  
       $\lenHO{P'} + 2 \cdot \len{\lentr{P'}} + 3 \geq \frac{5}{3} \cdot \numprop{P} + 1 
       > \frac{5}{3}\cdot(\numprop{P} + 1)$. 
 
       \item  Case $P = \news{r:S}P'$ with  $\traux{S}$. 
     Then $\lenHO{P} = \lenHO{P'}$ (by \Cref{pi:def:am-sizeproc})
     and  $\len{\lentr{P}} = \len{\lentr{P'}} + 1$. 
     Further, we have $\lenrecpx{P} = \lenrecpx{P'}$ and $\lenHOopt{P} = \lenHOopt{P'} +
     1$. 
%      (by \Cref{pi:def:sizeproc}).
     Now, by IH 
%      we have  $\lenHO{Q} + \lentr{Q} > \numprop{Q}$. So, 
    we can
     conclude that $\lenHO{P} + 2 \cdot (\len{\lentr{P}} + 1)  \geq 
      \frac{5}{3} \cdot \numprop{P} + 1
     > \frac{5}{3} \cdot (\numprop{P} + 1)$.

     \item Case $P = P_1 \Par \ldots \Par P_n$. 
     By IH we know 
     \begin{align}
        \lenHO{P_i} + 2 \cdot \len{\lentr{P_i}} \geq \frac{5}{3} \cdot \numprop{P_i}
        \label{pi:eq:optproof-1}
     \end{align}
    %  $$\lenHO{Q_i} + 2 \cdot \lentr{Q_i} \geq \frac{5}{3} \cdot \numprop{Q_i}$$ 
     \noindent for $i \in \{1,\ldots,n\}$. 
     We know 
     $\lenHO{P} = \sum_{i=1}^{n} \lenHO{P_i} + (n-1)$, 
     $\len{\lentr{P}} = \sum_{i=1}^{n} \len{\lentr{P_i}}$,  
     and $\lenHOopt{P} = \sum_{i=1}^{n}\numprop{P_i} + (n-1)$.
     So, we should show 
     \begin{align}
       \sum_{i=1}^{n} \lenHO{P_i} + 2 \cdot \sum_{i=1}^{n} \len{\lentr{P_i}}
       + (n-1) \geq 
       \frac{5}{3} \cdot \big(\sum_{i=1}^{n}\numprop{P_i} + (n-1)\big)
       \label{pi:eq:measure-prop1}
     \end{align}
     % $$\sum_{i=1}^{n} \lenHO{Q_i} + (n-1) \geq k \cdot \big(\sum_{i=1}^{n}\lenHOopt{Q_i} + (n-1)\big)$$

     That is, 
     $$\sum_{i=1}^{n} (\lenHO{P_i} + 2 \cdot \len{\lentr{P_i}} + 1) \geq 
     \frac{5}{3} \cdot \sum_{i=1}^{n} \big(\numprop{P_i} + 1\big)$$

     Equivalent to 
     $$\sum_{i=1}^{n} (\lenHO{P_i} + 2 \cdot \len{\lentr{P_i}} + 1) \geq 
     \sum_{i=1}^{n} \big(\frac{5}{3} \cdot (\numprop{P_i} + 1)\big)$$

     We show that for each $i = \{1, \ldots, n\}$ the following holds:  
     $$\lenHO{P_i} + 2 \cdot \len{\lentr{P_i}} + 1 \geq  \frac{5}{3} \cdot (\numprop{P_i} + 1)$$ 

     That is 
     $$A = \frac{\lenHO{P_i} + 2 \cdot \len{\lentr{P_i}} + 1}{\numprop{P_i} + 1} \geq  \frac{5}{3}$$ 

    %  That is we need to detect $Q_i$ for which $\frac{\lenHO{Q_i} + 2 \cdot \lentr{Q_i}}{\lenHOopt{Q_i}}$
    %  is the least.
      As $P$ is in normal form, 
     we know that $P_i \equiv \alpha.P_i'$
     where $\alpha$ is some prefix. So,
     by \Cref{pi:def:am-sizeproc} and  \Cref{pi:def:sizeproc} we know that 
    for some $p^*$ and $p$ we have 
     $\lenHOopt{P_i} = p^* + \lenHOopt{P_i'}$, $\lenrecpx{P_i} = \lenrecpx{P_i'}$, 
     and $\lenHO{P_i} = p + \lenHO{P_i'}$. 
     We can distinguish two sub-cases: \rom{1} $P'_i \not\equiv \inact$ 
     and \rom{2} otherwise. 
     We consider sub-case \rom{1}. 
     By \eqref{pi:eq:optproof-1} we have: 
     $$A \geq \frac{\frac{5}{3}\cdot \numprop{P'_i} + p + 1}{\numprop{P'_i} + p^* + 1} \geq  \frac{5}{3}$$ 

     We need to find a prefix $\alpha$ such that $p$ and $\frac{p}{p^*}$ are the least. 
     We notice that for 
     the output prefix we have $p = 3$ and $\frac{p}{p^*} = \frac{3}{1}$. 
     So, the following holds 
     $$\frac{\frac{5}{3}\cdot \numprop{P'_i} + 4}{\numprop{P'_i} + 2} = 
     \frac{5}{3} \cdot \frac{\numprop{P'_i} + \frac{12}{5}}{\numprop{P'_i} + 2} \geq  \frac{5}{3}$$ 

     Now, we consider sub-case \rom{2} when $P'_i = \inact$. 
    %  We may notice that this term $A$ is the least when $Q'_i  = \inact$. 
     In this case we have $\lenHO{P'_i} + 2 \cdot \lentr{P_i'}= \numprop{P'_i} = 1$. 
     We pick $p$ and $p^*$ as in the previous sub-case. 
     So, we have 
     $$\frac{\lenHO{P_i} + 2 \cdot \len{\lentr{P_i}} + 1}{\numprop{P_i} + 1} =     \frac{1 + 3 + 1}{2 + 1} = \frac{5}{3}$$ 

    %  Now, we need to find $Q_i'$ for which $\lenHOopt{Q'_i}$ is the least, 
    %  and prefix $\alpha$ such that $p$ and $\frac{p}{p^*}$ are the least. We 
    %  can see that $Q_i' = \inact$ we have $\lenHOopt{Q'_i} = 1$ 
    %  and for the output prefix we have $p = 3$ and $\frac{p}{p^*} = \frac{3}{1}$.  
    %  Now, by substituting those values we have: 
    % %  $$\frac{\frac{5}{3}\cdot 1 + 3 + 1}{1 + 1 + 1} =  \frac{5}{3}$$ 

    %  Now, by IH we know that when
    %  $Q_i' = \inact$ then $\frac{\lenHO{Q_i'} + 2 \cdot \lentr{Q_i}}{\lenHOopt{Q_i'}} = 1$, 
    %  (otherwise $\frac{\lenHO{Q_i} + 2 \cdot \lentr{Q_i}}{\lenHOopt{Q_i}} \geq k$). 
    %  Also, for 
    %  $\alpha$ we take output. 
    %  % and $Q_i' = \inact$ then $\frac{\lenHO{Q_i}}{\lenHOopt{Q_i}}$ is the least. 
    %  That is, we have $\lenHO{Q_i} + 2 \cdot \lentr{Q_i}= 4$ and $\lenHOopt{Q_i} = 2$. 
     We can then conclude that 
    inequality \eqref{pi:eq:measure-prop1} holds. 
     \end{itemize}
       \end{proof}

\thesisalt{
\subsection{Proof of \Cref{pi:lem:brec}}}
{ 
  \subsection{Proof of \Cref{pi:lem:brec}}
}
       \begin{restatable}[]{lemm}{}
        \label{pi:lem:indexcor}
        Let $\widetilde r$ and $S$ be tuple of channel names and  a recursive 	session type, respectively.
        If $\widetilde r : \Rtsopt{}{s}{\btout{C}S}$ and $k = \indf{\btout{C}S}$
        then $r_k:\trec{t}\btout{\Gtopt{C}}\vart{t}$.
      \end{restatable}
       \begin{restatable}[Typing Broken-down Variables]{lemm}{lemvarbreak}
        \label{pi:lem:varbreak}
          If $\Gamma;\Delta \proves z_i \hastype C$ then 
          $\Gtopt{\Gamma};\Gtopt{\Delta} \proves \wtd z \hastype \Gtopt{C}$ 
          where $\wtd z = (z_i,\ldots,z_{i + \len{\Gtopt{C}} - 1})$.
        \end{restatable}

%\indf

\label{pi:app:lem:brec}
\pilemmbrec* 

% \todo[inline]{TODO: refer to lemma here}

% pi:lem:brec

\begin{proof}
By induction on the structure of $P$. 
  We consider five cases, taking $g \not= \es$; the analysis when $g = \es$ is similar. 
  % We have five cases, depending on shape of $P$: 
%   We 
   % consider following cases: 
   \begin{enumerate}
  % \item Case $P = \inact$. Directly by taking $g = \es$
  % and as $\wtd x = \es$
  %  thus $\wtd y = \es$ and $\Theta = \prop^r_k : \btinp{}\tinact$. 
   \item Case $P = X$. The only rule that can be applied here is 
           \textsc{RVar}: 
           \begin{align}
           \AxiomC{} 
           \LeftLabel{\scriptsize \textsc{(RVar)}}
               \UnaryInfC{$\Gamma \cdot X:\Delta;\Delta \proves X \hastype \Proc$}
               \DisplayProof
           \end{align}
   
Let $\wtd x = \emptyset$ as $\fn{P}=\emptyset$. 
So, $\wtd y = \wtd m$ where $\wtd m = g(X)$. 
Since $\lenHOopt{X}=1$ we have
   $\thetaR = \{\prop^r_k:\trec{t}\btinp{\wtd N}\tvar{t}\}$ where $\wtd N =
   (\Gtopt{\Gamma},\Gtopt{\Delta})(\wtd y)$. 
 In this case $\envR = \Delta$, thus $\wtd N = \wtd M$. 
           We shall then prove the following judgment: 
           \begin{align}
               \Gtopt{\Gamma \setminus x}; 
       \Theta
       \proves \Brecpi{k}{\tilde y}{X}_g	\hastype \Proc 
           \end{align}
           By \Cref{pi:t:bdowncorec-rec} we have:  
           $$\Brecpi{k}{\tilde y}{X}_g = 
     \recp{X}\proprinpk{r}{k}{\wtd y}\propoutrecsh{\rvar{X}}{\wtd y}\rvar{X}$$
                The following tree proves this case: 
     \begin{align}
      \AxiomC{}
      \LeftLabel{\scriptsize \textsc{(RVar)}}
      \UnaryInfC{$\Gtopt{\Gamma \setminus x}\cat X : \Theta; 
     \Theta \proves
    \rvar{X}$}
    \AxiomC{}
    \LeftLabel{\scriptsize \textsc{(PolySend)}}
      \UnaryInfC{$\Gtopt{\Gamma \setminus x}\cat X : \Theta; 
      \wtd y : \wtd M \proves
      \wtd y \hastype \wtd M$}
      \LeftLabel{\scriptsize \textsc{(Req)}}
      \BinaryInfC{$\Gtopt{\Gamma \setminus x}\cat X : \Theta; 
      {\prop^r_X}:\chtype{\wtd M} \cat \wtd y : \wtd M \proves
      \propoutrecsh{\rvar{X}}{\wtd y}\rvar{X}$}
       \DisplayProof
       \label{pi:pt:recvar-send1}
     \end{align}
           \begin{align}
      \AxiomC{\eqref{pi:pt:recvar-send1}} 
      \AxiomC{} 
      \LeftLabel{\scriptsize \textsc{(PolySend)}}
      \UnaryInfC{$\Gtopt{\Gamma \setminus x}\cat X : \Theta;\wtd y : \wtd M \proves \wtd y \hastype \wtd M$}
      \LeftLabel{\scriptsize \textsc{(Rcv)}}
           \BinaryInfC{$\Gtopt{\Gamma \setminus x}\cat X: \Theta;\Theta
     \proves \proprinpk{r}{k}{\wtd y}
     \propoutrecsh{\rvar{X}}{\wtd y}\rvar{X}	\hastype \Proc$} 
           \LeftLabel{\scriptsize \textsc{(Rec)}}
               \UnaryInfC{$\Gtopt{\Gamma \setminus x};\Theta
       \proves \recp{X}\proprinpk{r}{k}{\wtd y}
       \propoutrecsh{\rvar{X}}{\wtd y}\rvar{X}	\hastype \Proc$}
               \DisplayProof
           \end{align}
     \noindent where $\Theta=\thetaR \cat \dual{\prop^r_X}:\trec{t}\btout{\wtd M}\tvar{t}$.

       \item Case $P = \bout{u_i}{z_j}P'$. 
       Let $u_i :C$. 
   We distinguish three sub-cases: \rom{1} 
   $C = S = \trec{t}\btout{C_z}S'$, 
   \rom{2} 
   $C = S = \btout{C_z}S'$, and 
   \rom{3} 
   $C = \chtype{C_z}$. 
   We consider first two sub-cases, as \rom{3} is shown similarly. 
   The only rule that can be applied here is 
       \textsc{Send}: 
       \begin{align}
       \label{pi:pt:brec-out-inv}
           \AxiomC{$\Gamma \cdot X:\envR; \Delta \cdot u_i:{S'}
           \proves P' \hastype \Proc 
           $} 
           \AxiomC{$\Gamma \cdot X:\envR; \Delta_z \proves z_j \hastype C$} 
           \LeftLabel{\scriptsize \textsc{(Send)}}
           \BinaryInfC{$\Gamma \cdot X:\envR;\Delta \cdot u_i:S
           \cdot \Delta_z \proves \bout{u_i}{z_j}P' \hastype \Proc $} 
           \DisplayProof	
       \end{align}
     Then, by IH on the right assumption of \eqref{pi:pt:brec-out-inv} we have: 
   \begin{align}
       \label{pi:pt:brec-out-ih}
       \Gtopt{\Gamma \setminus \wtd x'}; \Theta' 
  %  \cdot \Gtopt{\Delta \cat r : S'}
           \proves \Brecpi{k+1}{\tilde y'}{P'}_g \hastype \Proc 
   \end{align}
   \noindent where $\wtd x' \subseteq \fn{P}$ such that 
      $(\Delta \cat r:{S'}) \setminus \wtd x' = \emptyset$, 
  and $\wtd y' = \wtd v' \cup \wtd m$ where  
       $\indexed{\wtd v'}{\wtd x'}{\Gamma,\Delta \cat r:{S'}}$. 
      Also, $\Theta' = \thetaR'  \cat \thetax$
  where $\thetaR'$ such that 
   $\balan{\thetaR'}$ with 
   $$\dom{\thetaR'} = \{\prop^r_{k+1},\prop^r_{k+2},\ldots,\prop^r_{k+\lenHOopt{P'}+1} \} \cup 
   \{\dual{\prop^r_{k+2}},\ldots,\dual{\prop^r_{k+\lenHOopt{P'}-1}},\dual{\prop^r_{k+\lenHOopt{P'}+1}} \}$$ 
   and 
   $\thetaR' (\prop^r_{k+1})=\trec{t}\btinp{\wtd N'}\tvar{t}$
   where 
  $\wtd N' = (\Gtopt{\Gamma},\Gtopt{\envR \cdot 
  \Delta \cdot u_i:S})(\wtd y')$.
      By applying \Cref{pi:lem:varbreak} on the second assumption of \eqref{pi:pt:brec-out-inv} we have: 
       \begin{align}
       \label{pi:eq:rec-out-ih2}
       \Gtopt{\Gamma \cdot X : \envR} ; \Gtopt{\Delta_z} \proves \wtd z \hastype \Gtopt{C_z}
       \hastype \Proc 
       \end{align}

       Let $\sigma = \nextn{u_i}$ 
   and in sub-case \rom{1} $\sigma_1 = \subst{\tilde n}{\tilde u}$ 
   where $\wtd n = (u_{i+1}, \ldots, u_{i + {\Gtopt{S}}})$ 
   and $\wtd u = (u_{i}, \ldots, u_{i + {\Gtopt{S}}-1})$, 
   otherwise \rom{2} $\sigma_1 = \epsilon$. 
       We define $\wtd x = \wtd x' \cat z$ and 
       $\wtd y = \wtd y'\sigma_1 \cat \wtd z \cdot u_i$. 

  By construction $\wtd x \subseteq P$ and 
  $(\Delta \cdot u_i:S \cdot \Delta_z) \setminus \wtd x = \emptyset$. 
  Further, we may notice that $\wtd y = \wtd v \cdot \wtd m$ such that 
                $\indexed{\wtd v}{\wtd x}{\Gamma, \Delta \cat u_i:{S} \cat \Delta_z}$ and 
       $\wtd y' = \wtd m \cup \fnb{P'}{\wtd y}$. 
       % Let $\thetaR = \thetaR' \cat \Theta'$ where: 
   Let $\Theta = \thetaR  \cat \thetax$ where 
       $$\thetaR = \thetaR' \cat 
   \prop^r_{k}:\trec{t}\btinp{\wtd N}\tvar{t} 
   \cat \dual{\prop^r_{k+1}}:\trec{t}\btout{\wtd N'}\tvar{t}$$
       
  \noindent where 
$\wtd N = (\Gtopt{\Gamma},\Gtopt{\envR \cdot \Delta \cdot u_i:{S} \cdot \Delta_z})(\wtd y)$. 
By construction and since $\lenHOopt{P}=\lenHOopt{P'}+1$ we have 
  $$\dom{\thetaR} = \{\prop^r_k,\prop^r_{k+1},\ldots,\prop^r_{k+\lenHOopt{P}-1} \} \cup 
   \{\dual{\prop^r_{k+1}},\ldots,\dual{\prop^r_{k+\lenHOopt{P}-1}} \}$$
   
   and $\balan{\thetaR}$. 

 By \Cref{pi:t:bdowncorec-rec} we have: 
       \begin{align}
           \Brecpi{k}{\tilde y}{P}_g = \recp{X}\binp{\prop^r_k}{\wtd y}
            \bout{u_l}{\wtd z} 
            \bout{\prop^r_{k+1}}{\wtd y'\sigma_1}X
         \Par
      \Brecpi{k+1}{\tilde y'\sigma_1}{P'\sigma}_g
       \end{align}
   \noindent where in sub-case \rom{1} $l = i$ and 
   in sub-case \rom{2} $l = \indf{S}$. 
  We shall prove the following judgment: 
  \begin{align}
      \Gtopt{\Gamma \setminus \wtd x}; \Theta \proves \Brecpi{k}{\tilde y}{P}_g \hastype \Proc 
  \end{align}
  
  Let $\Delta_1 = \Delta \cat u_i:{S} \cat \Delta_z$ 
and $\wtd u_i = \bname{u_i : S}$. 
% Let $\sigma = \nextn{u_i}$. 
We use some auxiliary
           sub-trees: 
  \begin{align}
  \label{pi:pt:brec-out4}
      \AxiomC{} 
      \LeftLabel{\scriptsize \textsc{(RVar)}}
      \UnaryInfC{$\Gtopt{\Gamma \setminus \wtd x} \cdot X : \Theta;\Theta 	\proves  
            X \hastype \Proc $}	
      \DisplayProof
  \end{align}
  \begin{align}
  \label{pi:pt:brec-out2}
      \AxiomC{\eqref{pi:pt:brec-out4}} 
      \AxiomC{} 
      \LeftLabel{\scriptsize \textsc{(PolyVar)}}
      \UnaryInfC{$\Gtopt{\Gamma \setminus \wtd x} \cdot X : \Theta;
      \Gtopt{\Delta \cat u_i\sigma:{S'}}  \proves 
      \wtd y'\sigma_1 \hastype \wtd N'$} 
      \LeftLabel{\scriptsize \textsc{(PolySend)}}
      \BinaryInfC{$\Gtopt{\Gamma \setminus \wtd x} \cdot X : \Theta;\Theta \cat 
  \Gtopt{\Delta \cat u_i\sigma:{S'}}
  \proves  
\proproutk{r}{k+1}{\wtd y'\sigma_1}X \hastype \Proc $}	
      \DisplayProof
  \end{align}

  Here, in typing the right-hand assumption, we may notice that in 
  sub-case \rom{1} 
  $\Gtopt{u_i:S} = \wtd u_i : \btout{\Gtopt{C_z}}\tinact \cat \Gtopt{S'} $ and 
  $\Gtopt{u_{i}\sigma:S'} =\wtd u_{i+1}:\Gtopt{S'}$
  where $\wtd u_{i+1} = (u_{i+1}, \ldots, u_{i + \len{\Gtopt{S'}}})$. 
  So the right-hand side follows by  \Cref{pi:def:indexed} and \Cref{top:lem:subst}.
  % and by definition $\wtd u_{i+1} \subseteq \wtd y'\sigma_1$
  Otherwise, in sub-case~\rom{2} 
  by \Cref{pi:def:typesdecompenv} and \Cref{pi:f:tdec} we know 
  $\Gtopt{u_i:S} = \Gtopt{u_i\sigma:S'}$ and 
  by definition $\wtd u_i \subseteq \wtd m \subseteq \wtd y'\sigma_1$. 
  \begin{align}
  \label{pi:pt:brec-out1}
  \AxiomC{\eqref{pi:pt:brec-out2}}
  \AxiomC{\eqref{pi:eq:rec-out-ih2}} 
  \LeftLabel{\scriptsize \textsc{(PolySend)}} 
  \BinaryInfC{$\Gtopt{\Gamma \setminus \wtd x} \cdot X : 
\Theta;\Theta \cat \Gtopt{\Delta_1}\proves  
            \bout{u_l}{\wtd z} 
        \proproutk{r}{k+1}{\wtd y'\sigma_1}X \hastype \Proc $}
  \DisplayProof	
  \end{align}
  \begin{align}
  \label{pi:pt:brec-out0}
  \AxiomC{\eqref{pi:pt:brec-out1}} 
  \AxiomC{}
  \LeftLabel{\scriptsize \textsc{(PolyVar)}} 
  \UnaryInfC{$\Gtopt{\Gamma \setminus \wtd x} \cdot X : \Theta;\Gtopt{\Delta_1}
  \proves 
  \wtd y \hastype \wtd N$} 
  \LeftLabel{\scriptsize \textsc{(PolyRcv)}}
      \BinaryInfC{$\Gtopt{\Gamma \setminus \wtd x} \cdot X : \Theta; \Theta \proves 
            \binp{\prop^r_k}{\wtd y}
            \bout{u_l}{\wtd z} 
        \proproutk{r}{k+1}{\wtd y'\sigma_1}X \hastype \Proc $}
      \LeftLabel{\scriptsize \textsc{(Rec)}}
      \UnaryInfC{$\Gtopt{\Gamma \setminus \wtd x};\Theta \proves 
      \recp{X}\binp{\prop^r_k}{\wtd y}
            \bout{u_l}{\wtd z} 
        \proproutk{r}{k+1}{\wtd y'\sigma_1}X \hastype \Proc $}
      \DisplayProof
  \end{align}
We may notice that 
by \Cref{pi:def:typesdecompenv} and \Cref{pi:f:tdec} we have 
$\Gtopt{\Delta_1}=\Gtopt{\Delta} \cat \wtd u_i : \Gtopt{S} \cat \Gtopt{\Delta_z}$.
% where $\wtd u_i = \bname{u_i : S}$. 
So, in sub-case \rom{1} as $u_i = u_l$ we have $\Gtopt{\Delta_1}(u_l) = \btout{\Gtopt{C_z}}\tinact$. 
           % $\Gtopt{\Delta_1}=\Gtopt{\Delta} \cat \wtd r : \Rtsopt{}{}{S} \cat 
           % \Gtopt{\Delta_z}$
     In sub-case \rom{2}, by \Cref{pi:lem:indexcor}
     and as $u_l = u_{\indf{S}}$
           we know $\Gtopt{\Delta_1}(u_l) = \trec{t}\btout{\Gtopt{C_z}}\vart{t}$. 
    %  Also, by \Cref{pi:def:typesdecompenv} and \Cref{pi:f:tdec} we know 
    %   $\Gtopt{u_i:S} = \Gtopt{u_i\sigma:S'}$ and 
    %   by definition. 
The following tree proves this case: 
  \begin{align}
      \AxiomC{\eqref{pi:pt:brec-out0}}
      \AxiomC{\eqref{pi:pt:brec-out-ih}} 
      \LeftLabel{\scriptsize \textsc{(Par)}}
      \BinaryInfC{$\Gtopt{\Gamma \setminus \wtd x}; \Theta \proves 
      \recp{X}\binp{\prop^r_k}{\wtd y}
            \bout{u_l}{\wtd z} 
            \bout{\prop^r_{k+1}}{\wtd y'\sigma_1}X 
        \Par \Brecpi{k+1}{\tilde y'\sigma_1}{P'\sigma}_g \hastype \Proc $} 
      \DisplayProof	
  \label{pi:pt:brec-out-ptree}
  \end{align}

Note that we have used the following for the right assumption
of \eqref{pi:pt:brec-out-ptree}:  
$$\Brecpi{k+1}{\tilde y'}{P'} \equiv_{\alpha} \Brecpi{k+1}{\tilde y'\sigma_1}{P'\sigma}$$

      \item Case $P = \binp{u_i}{z}P'$. 
  We distinguish three sub-cases: \rom{1} 
  $C = S = \trec{t}\btinp{C_z}S'$, 
  \rom{2} 
  $C = S = \btinp{C_z}S'$, and 
  \rom{3} 
  $C = \chtype{C_z}$. 
  The only rule that can be applied here is 
       \textsc{Rcv}: 
       \begin{align}
       \label{pi:pt:brec-inp-inv}
           \AxiomC{$\Gamma \cdot X:\envR; \Delta \cat u_i:{S'}
           \cat 
           \Delta_z 
           \proves P' \hastype \Proc 
           $} 
           \AxiomC{$\Gamma \cdot X:\envR; \Delta_z \proves z \hastype C$} 
           \LeftLabel{\scriptsize \textsc{(Rcv)}}
           \BinaryInfC{$(\Gamma \setminus z)\cdot X:\envR;\Delta \cat u_i:S
        \proves \binp{u_i}{z}P' \hastype \Proc $} 
           \DisplayProof	
       \end{align}
       
      Let $\wtd x' \subseteq \fn{P'}$ such that 
      $(\Delta \cdot u_i:{S'} \cdot \Delta_z) \setminus \wtd x' = \emptyset$ 
      and $\wtd y' = \wtd v \cup \wtd m$ such that
   $\indexed{\wtd v'}{\wtd x'}{\Gamma,\Delta \cat u_i:{S'}}$. 
   Also, $\Theta' = \thetaR' \cat \thetax(g)$ where 
   $\balan{\thetaR'}$ with 
   $$\dom{\thetaR'} = \{\prop^r_{k+1},\prop^r_{k+2},\ldots,\prop^r_{k+\lenHOopt{P'}+1} \} \cup 
   \{\dual{\prop^r_{k+2}},\ldots,\dual{\prop^r_{k+\lenHOopt{P'}+1}} \}$$ 
   and 
   $\thetaR' (\prop^r_{k+1})=\trec{t}\btinp{\wtd N'}\tvar{t}$, 
   where  $\wtd N' = (\Gtopt{\Gamma} \cat \Gtopt{\envR \cdot \Delta \cdot u_i:{S} })(\wtd y')$.
   Then, by IH on the right assumption of \eqref{pi:pt:brec-out-inv} we have: 
   \begin{align}
       \label{pi:pt:brec-inp-ih}
       \Gtopt{\Gamma \setminus \wtd x'}; \thetaR' 
           \proves \Brecpi{k+1}{\tilde y'}{P'}
           \hastype \Proc 
   \end{align}
      
        By applying \Cref{pi:lem:varbreak}  to the second 
        assumption of \eqref{pi:pt:brec-out-inv}, we have: 
        \begin{align}
        \label{pi:pt:brecinp-ih2}
            \Gtopt{\Gamma} \cdot X : \thetaR; \Gtopt{\Delta_z} \proves \wtd z \hastype \Gtopt{C}
        \end{align}

       Let $\sigma = \nextn{u_i}$ 
   and in sub-case \rom{1} $\sigma_1 = \subst{\tilde n}{\tilde u}$ 
   where $\wtd n = (u_{i+1}, \ldots, u_{i + {\Gtopt{S}}})$ 
   and $\wtd u = (u_{i}, \ldots, u_{i + {\Gtopt{S}}-1})$, 
   otherwise in sub-case \rom{2} $\sigma_1 = \epsilon$. 
       We define $\wtd x = \wtd x'\sigma \setminus z$ and 
       $\wtd y = \wtd y'\sigma_1 \setminus \wtd z$ with 
       $\len{\wtd z} = \len{\Gtopt{C}}$. By construction 
   $\wtd x \subseteq \fn{P}$ and $(\Delta \cat r:S) \setminus \wtd x = \es$. 
       Further, we may notice that $\wtd y=\wtd v \cdot \wtd m$, where $\wtd v$ is 
   such that 
   $\indexed{\wtd v}{\wtd x}{\Gamma, \Delta \cat r:{S} \cat \Delta_z}$
       and  
       $\wtd y = \wtd m \cup \fnb{P'}{\wtd y \wtd z}$. 
   Let $\Theta = \thetaR  \cat \thetax$ where 
       $$\thetaR = \thetaR' \cat \prop^r_{k}:\trec{t}\btinp{\wtd N}\tvar{t} \cat 
   \dual{\prop^r_{k+1}}:\trec{t}\btout{\wtd N'}\tvar{t}$$
       
  \noindent where 
$\wtd N = (\Gtopt{\Gamma},\Gtopt{\envR \cdot \Delta \cdot u_i:{S} \cat \Delta_z})(\wtd y)$. 
By construction and since $\lenHOopt{P}=\lenHOopt{P'}+1$ we have 
  $$\dom{\thetaR} = \{\prop^r_k,\prop^r_{k+1},\ldots,\prop^r_{k+\lenHOopt{P}-1} \} \cup 
   \{\dual{\prop^r_{k+1}},\ldots,\dual{\prop^r_{k+\lenHOopt{P}-1}} \}$$
   
   and $\balan{\thetaR}$. 
% 	Since $\wtd x = \wt$
% 	Since $\mathsf{tr}(S)$ we know
% 	$\fn{P} = \fn{P'} \cat z$ and by this 
% 	$(\Delta \cat r:S' \cat \Delta_z) \setminus \wtd x = \emptyset$.

 By \Cref{pi:t:bdowncorec-rec} we have:  
       \begin{align}
       \Brecpi{k}{\tilde y}{P}_g = 
           \recp{X}\binp{\prop^r_k}{\wtd y}
            \binp{u_l}{\wtd z} 
            \bout{\prop^r_{k+1}}{\wtd y'\sigma_1}X
         \Par
      \Brecpi{k+1}{\tilde y}{P'\sigma}_g
       \end{align}
   \noindent where in sub-case \rom{1} $l = i$ and 
   in sub-case \rom{2} $l = \indf{S}$. 
Let $\Gamma_1 = \Gamma \setminus \wtd x$.
  We shall prove the following judgment: 
  \begin{align}
      \Gtopt{\Gamma_1 \setminus z}; \Theta \proves \Brecpi{k}{\tilde y}{P}_g \hastype \Proc 
  \end{align}
  
  %  Let $\Delta_1 = \Delta \cat u_i:{S} 
   % 		\cat \Delta_z$. 
  We use some auxiliary sub-trees: 
  \begin{align}
  \label{pi:pt:brec-inp4}
      \AxiomC{} 
      \LeftLabel{\scriptsize \textsc{(RVar)}}
      \UnaryInfC{$\Gtopt{\Gamma} \cdot X : \Theta;\Theta 	\proves  
            X \hastype \Proc$}	
      \DisplayProof
  \end{align}
  \begin{align}
  \label{pi:pt:brec-inp2}
      \AxiomC{\eqref{pi:pt:brec-inp4}} 
      \AxiomC{} 
      \LeftLabel{\scriptsize \textsc{(PolyVar)}}
      \UnaryInfC{$\Gtopt{\Gamma} \cdot X : \Theta;
      \Gtopt{\Delta \cat u_l:{S'} 
  \cat \Delta_z} \proves 
      \wtd y'\sigma_1 \hastype \wtd N'$} 
      \LeftLabel{\scriptsize \textsc{(PolySend)}}
      \BinaryInfC{$\Gtopt{\Gamma} \cdot X : \Theta;\Theta 
      \cat 
  \Gtopt{\Delta_1}
  \proves  
            \bout{\prop^r_{k+1}}{\wtd y'\sigma_1}X
            \hastype \Proc$}	
      \DisplayProof
  \end{align}
Here, in typing the right-hand assumption, we may notice that in 
sub-case \rom{1} 
$\Gtopt{u_i:S} = \wtd u_i : \btinp{\Gtopt{C_z}}\tinact \cat \Gtopt{S'} $ and 
$\Gtopt{u_{i}\sigma:S'} =\wtd u_{i+1}:\Gtopt{S'}$
where $\wtd u_{i+1} = (u_{i+1}, \ldots, u_{i + \len{\Gtopt{S'}}})$
and by definition $\wtd u_{i+1} \subseteq \wtd y'\sigma_1$. 
So the right-hand side follows by  \Cref{pi:def:indexed} and \Cref{top:lem:subst}.
Otherwise, in sub-case \rom{2} 
by \Cref{pi:def:typesdecompenv} and \Cref{pi:f:tdec} we know 
$\Gtopt{u_i:S} = \Gtopt{u_i\sigma:S'}$. 
  \begin{align}
  \label{pi:pt:brec-inp1}
  \AxiomC{\eqref{pi:pt:brec-inp2}}
  \AxiomC{\eqref{pi:pt:brecinp-ih2}} 
  \LeftLabel{\scriptsize \textsc{(PolyRcv)}} 
  \BinaryInfC{$\Gtopt{\Gamma \setminus z} \cdot X : \Theta;\Theta \cat 
  \Gtopt{\Delta \cat u_i : S}
  \proves  
            \binp{u_l}{\wtd z} 
            \bout{\prop^r_{k+1}}{\wtd y'\sigma_1}X
            \hastype \Proc$}
  \DisplayProof	
  \end{align}
  \begin{align}
  \label{pi:pt:brec-inp0}
  \AxiomC{\eqref{pi:pt:brec-inp1}} 
  \AxiomC{}
  \LeftLabel{\scriptsize \textsc{(PolyVar)}} 
  \UnaryInfC{$\Gtopt{\Gamma \setminus z} \cdot X : \Theta;
  \Gtopt{\Delta \cat r : S}
  \proves 
  \wtd y \hastype \wtd N$} 
  \LeftLabel{\scriptsize \textsc{(PolyRcv)}}
      \BinaryInfC{$\Gtopt{\Gamma_1 \setminus z} \cdot X : \Theta;\Theta \proves 
            \binp{\prop^r_k}{\wtd y}
            \binp{u_l}{\wtd z} 
            \bout{\prop^r_{k+1}}{\wtd y'\sigma_1}X \hastype \Proc $}
      \LeftLabel{\scriptsize \textsc{(Rec)}}
      \UnaryInfC{$\Gtopt{\Gamma_1 \setminus z};\Theta \proves 
      \recp{X}\binp{\prop^r_k}{\wtd y}
            \binp{u_l}{\wtd z} 
            \bout{\prop^r_{k+1}}{\wtd y'\sigma_1}X
            \hastype \Proc $}
      \DisplayProof
  \end{align}
% We may notice that 
% by \Cref{pi:def:typesdecompenv} and \Cref{pi:f:tdec} we have 
%  			$\Gtopt{\Delta \cat r:S}=\Gtopt{\Delta} \cat \wtd r : \Rtsopt{}{}{S}$.
%  			 Further, by \Cref{pi:lem:indexcor}
%  			we know $\Gtopt{\Delta \cat r:S}(r_{f(S)}) = \trec{t}\btinp{\Gtopt{C}}\vart{t}$.
We may notice that 
by \Cref{pi:def:typesdecompenv} and \Cref{pi:f:tdec} we have 
$$\Gtopt{\Delta_1}=\Gtopt{\Delta} \cat \wtd u_i : \Gtopt{S} \cat \Gtopt{\Delta_z}$$
% where $\wtd u_i = \bname{u_i : S}$. 
So, in sub-case \rom{1} as $u_i = u_l$ we have
$\Gtopt{\Delta_1}(u_l) = \btinp{\Gtopt{C_z}}\tinact$. 
     In sub-case \rom{2}, by \Cref{pi:lem:indexcor}
     and as $u_l = u_{\indf{S}}$
           we know $\Gtopt{\Delta_1}(u_l) = \trec{t}\btinp{\Gtopt{C_z}}\vart{t}$. 
The following tree proves this case: 
  \begin{align}
      \AxiomC{\eqref{pi:pt:brec-inp0}}
      \AxiomC{\eqref{pi:pt:brec-inp-ih}} 
      \LeftLabel{\scriptsize \textsc{(Par)}}
      \BinaryInfC{$\Gtopt{\Gamma_1 \setminus z};\Theta \proves 
      \recp{X}\binp{\prop^r_k}{\wtd y}
            \binp{u_l}{\wtd z} 
            \bout{\prop^r_{k+1}}{\wtd y'\sigma_1}X
         \Par
      \Brecpi{k+1}{\tilde y'\sigma_1}{P'\sigma}_g$} 
      \DisplayProof	
  \label{pi:pt:brec-inp-ptree}
  \end{align}	
Note that we have used the following for the right assumption
of \eqref{pi:pt:brec-inp-ptree}:  
$$\Brecpi{k+1}{\tilde y'}{P'} \equiv_{\alpha} \Brecpi{k+1}{\tilde y'\sigma_1}{P'\sigma}$$

\item Case $P = Q_1 \Par Q_2$. The only rule that can be applied here 
is \textsc{Par}: 
\begin{align}
  \label{pi:pt:brec-par-inv}
  \AxiomC{$\Gamma \cdot X : \envR;\Delta_1 \proves Q_1 \hastype \Proc$}
  \AxiomC{$\Gamma \cdot X : \envR;\Delta_2 \proves Q_2 \hastype \Proc$}
  \LeftLabel{\scriptsize \textsc{(Par)}}
  \BinaryInfC{$\Gamma \cdot X : \envR;\Delta_1 \cat \Delta_2
                       \proves Q_1 \Par Q_2 \hastype \Proc$}
  \DisplayProof
\end{align}

Here we assume $\recpx{Q_1}$. 
So, by IH on the first and second assumption of \eqref{pi:pt:brec-par-inv} we have: 
\begin{align}
\label{pi:pt:brec-par-ih1}
\Gtopt{\Gamma \setminus x_1};\Theta_1 &\proves \Brecpi{k}{\tilde y_1}{Q_1}_{g} \hastype \Proc \\
\label{pi:pt:brec-par-ih2}  
\Gtopt{\Gamma \setminus x_1};\Theta_2 &\proves \Brecpi{k+\degree +1}{\tilde y_2}{Q_2}_{\es} \hastype \Proc
\end{align}
\noindent where for $i \in \{1,2\}$ we have $\wtd x_i \subseteq \fn{Q_i}$ such that 
$\Delta_i \setminus \wtd x = \es$,  
and $\wtd y_i = \wtd v_i \cdot \wtd m_i$, where $\wtd v_i$ is such that 
$\indexed{\wtd v_1}{\wtd x_i}{\Gamma, \Delta_i}$ and 
$\wtd m_i = \codom{g_i}$.
Further, $\Theta_i = \thetaR^i \cdot \thetax(g_i)$ where 
\begin{align*}
\dom{\thetaR^1} &= \{\prop^r_{k+1},\prop^r_{k+2},\ldots,\prop^r_{k+\lenHOopt{Q_1}} \} \cup 
   \{\dual{\prop^r_{k+2}},\ldots,\dual{\prop^r_{k+\lenHOopt{Q_1}}} \} \\
 \dom{\thetaR^2} &= \{\prop^r_{k+\degree+1},\prop^r_{k+\degree+2},\ldots,
 \prop^r_{k+\degree+\lenHOopt{Q_2}} \} 
%  \cup 
   % \{\dual{\prop^r_{k+2}},\ldots,\dual{\prop^r_{k+\degree+\len{Q_2}}} \}
\end{align*}
   and 
   $\thetaR^1 (\prop^r_{k+1})=\trec{t}\btinp{\wtd N^1}\tvar{t}$
 and 
 $\thetaR^2 (\prop^r_{k+\degree+1})=\chtype{\wtd N^2}$
 with $\wtd N^i = (\Gtopt{\Gamma},\Gtopt{\envR \cdot \Delta_i})(\wtd y_i)$. 
 
Let $\Delta = \Delta_1 \cdot \Delta_2$. 
 We define $\wtd x = \wtd x_1 \cdot \wtd x_2$ 
 and $\wtd y = \wtd y_1 \cdot \wtd y_2$. 
%  Let $g = g_1 \cdot g_2$.
 By construction $\wtd x \subseteq \fn{P}$ and $(\Delta_1\cdot \Delta_2) \setminus \wtd x = \es$. 
Further, we may notice that $\wtd y = \wtd v \cdot \wtd m$, where 
$\wtd m = \codom{g}$ and 
$\indexed{\wtd v}{\wtd x}{\Gamma, \Delta}$.
 We shall prove the following judgment: 
 \begin{align}
   \Gtopt{\Gamma \setminus \wtd x}; \Theta \proves \Brecpi{k}{\tilde y}{P}_g \hastype \Proc 
 \end{align}
 \noindent where 
 $\Theta = \Theta_1 \cdot \Theta_2 \cdot \Theta'$ 
  with 
  $$\Theta'=\prop^r_k:  \trec{t}\btinp{\wtd N}\tvar{t}
  \cdot \dual{\prop^r_{k+1}}: \trec{t}\btout{\wtd N_1}\tvar{t}$$
  %  \cdot \dual{\prop^r_{k+\degree+1}}: \trec{t}\btout{\wtd N_2}\tvar{t}$$
 By \Cref{pi:t:bdowncorec-rec} we have: 
 \begin{align*}
  \Brecpi{k}{\tilde y}{P}_g = 
    \recp{X}\proprinpk{r}{k}{\wtd y} 
    (\proproutk{r}{k+1}{\wtd y_1}X \Par 
    \apropoutrecsh{k+\degree+1}{\wtd y_2})
    \Par
    \Brecpi{k+1}{\tilde y_1}{Q_1}_{g}\Par 
    \Brecpi{k+\degree+1}{\tilde y_2}{Q_2}_{\es}
 \end{align*}

 We use some auxiliary sub-trees: 
 \begin{align}
  \AxiomC{} 
  \LeftLabel{\scriptsize \textsc{(Nil)}}
  \UnaryInfC{$\Gtopt{\Gamma \setminus \wtd x} 
  \cdot X : \Theta' \proves \inact \hastype \Proc$} 
  \AxiomC{}
    \LeftLabel{\scriptsize \textsc{(PolyVar)}}
    \UnaryInfC{$\Gtopt{\Gamma \setminus \wtd x} 
             \cdot X : \Theta'
              ; \wtd y_2 : \wtd N_2 
              \proves \wtd y_2 \hastype \wtd N_2$} 
  \LeftLabel{\scriptsize \textsc{(Snd)}}
  \BinaryInfC{$\Gtopt{\Gamma \setminus \wtd x} 
  \cdot X : \Theta'
  ;  \cdot  \wtd y_2 : \wtd N_2
  \proves
        % \aproproutk{k+\degree+1}{\wtd y_2}\rvar{X}
        \apropoutrecsh{k+\degree+1}{\wtd y_2}
  $} 
   \DisplayProof
   \label{pi:pt:brec-par-2}
 \end{align}
 \begin{align}
  \AxiomC{} 
  \LeftLabel{\scriptsize \textsc{(RVar)}}
  \UnaryInfC{$\Gtopt{\Gamma \setminus \wtd x} 
  \cdot X : \Theta'; \Theta' \proves X \hastype \Proc$} 
  \AxiomC{}
    \LeftLabel{\scriptsize \textsc{(PolyVar)}}
    \UnaryInfC{$\Gtopt{\Gamma \setminus \wtd x} 
             \cdot X : \Theta'
              ; \wtd y_1 : \wtd N_1
              \proves \wtd y_1 \hastype \wtd N$} 
  \LeftLabel{\scriptsize \textsc{(Snd)}}
  \BinaryInfC{$\Gtopt{\Gamma \setminus \wtd x} 
  \cdot X : \Theta'
  ; \Theta' \cdot  \wtd y_1 : \wtd N_1
  \proves
      \proproutk{r}{k+1}{\wtd y_1}X 
  \hastype \Proc$} 
  \AxiomC{\eqref{pi:pt:brec-par-2}}
  \LeftLabel{\scriptsize \textsc{(Par)}}
  \BinaryInfC{
    $\Gtopt{\Gamma \setminus \wtd x} 
    \cdot X : \Theta'
    ; \Theta' \cdot  \wtd y: \wtd N 
    \proves
        \proproutk{r}{k+1}{\wtd y_1}X \Par 
          \apropoutrecsh{k+\degree+1}{\wtd y_2}
    \hastype \Proc$
  }
   \DisplayProof
   \label{pi:pt:brec-par-1}
 \end{align}
  \begin{align}
    \AxiomC{\eqref{pi:pt:brec-par-1}} 
    \AxiomC{}
    \LeftLabel{\scriptsize \textsc{(PolyVar)}}
    \UnaryInfC{$\Gtopt{\Gamma \setminus \wtd x} 
             \cdot X : \Theta'
              ; \wtd y : \wtd N 
              \proves \wtd y \hastype \wtd N$} 
    \LeftLabel{\scriptsize \textsc{(Rcv)}}
    \BinaryInfC{$\Gtopt{\Gamma \setminus \wtd x} 
                 \cdot X : \Theta'
                    ; \Theta' \proves 
                \proprinpk{r}{k}{\wtd y} 
                (\proproutk{r}{k+1}{\wtd y_1}X \Par 
                  \apropoutrecsh{k+\degree+1}{\wtd y_2})
                  \hastype \Proc$} 
    \LeftLabel{\scriptsize \textsc{(Rec)}}
    \UnaryInfC{$
        \Gtopt{\Gamma \setminus \wtd x}; \Theta' \proves 
        \recp{X}\proprinpk{r}{k}{\wtd y} 
        (\proproutk{r}{k+1}{\wtd y_1}X \Par 
        \apropoutrecsh{k+\degree+1}{\wtd y_2})
      \hastype \Proc
    $}
    \DisplayProof
    \label{pi:pt:brec-par-0}
  \end{align}

 The following tree proves this case: 
 \begin{align*}
  \AxiomC{\eqref{pi:pt:brec-par-0}} 
  \AxiomC{\eqref{pi:pt:brec-par-ih1}} 
  \AxiomC{\eqref{pi:pt:brec-par-ih2}}
  \LeftLabel{\scriptsize \textsc{(Par)}}
  \BinaryInfC{$\Gtopt{\Gamma \setminus \wtd x}; \Theta \proves
  \Brecpi{k+1}{\tilde y_1}{Q_1}_{g_1} \Par 
  \Brecpi{k+\degree+1}{\tilde y_2}{Q_2}_{g_2} \hastype \Proc$}
  \LeftLabel{\scriptsize \textsc{(Par)}}
  \BinaryInfC{$
      \Gtopt{\Gamma \setminus \wtd x}; \Theta_1 \cdot \Theta_2 \proves 
      \recp{X}\proprinpk{r}{k}{\wtd y} 
      (\proproutk{r}{k+1}{\wtd y_1}X \Par 
      \apropoutrecsh{k+\degree+1}{\wtd y_2})
      \Par
      \Brecpi{k+1}{\tilde y_1}{Q_1}_{g}
      \Par 
      \Brecpi{k+\degree+1}{\tilde y_2}{Q_2}_{\es} \hastype \Proc
  $}
   \DisplayProof
 \end{align*}

 \item Case $P = \news{s:C}P'$. 
 We distinguish two sub-cases: \rom{1} $C = S$ and \rom{2}
 $C = \chtype{C'}$. 
 We only consider sub-case \rom{1} as the other  is similar. 
 First, we $\alpha$-convert $P$ as follows: 
 \begin{align*}
  P \equiv_{\alpha} \news{s_1:C}P'\subst{s_1 \dual{s_1}}{s \dual{s}}
\end{align*}
The only rule can that can be applied is \textsc{ResS}: 
\begin{align}
  \label{pi:pt:brec-news-inv}
  \AxiomC{$\Gamma; \envR \cdot \Delta \cdot s_1:S \cdot \dual {s_1}:\dual S
        \proves P'\subst{s_1 \dual{s_1}}{s \dual{s}} \hastype \Proc$}
  \LeftLabel{\scriptsize \textsc{(ResS)}}
  \UnaryInfC{$\Gamma; \envR \cdot \Delta \proves
              \news{s_1:S}P'\subst{s_1 \dual{s_1}}{s \dual{s}} \hastype \Proc$}
  \DisplayProof
\end{align}

By IH on the assumption of \eqref{pi:pt:brec-news-inv} we have:
\begin{align}
  \label{pi:eq:brec-news-ih}
  \Gtopt{\Gamma \setminus \wtd x};
  % \Gtopt{s_1:S \cat \dual {s_1}:\dual S}
        \Theta' \proves
        \Brecpi{k}{\tilde y'}{P'\subst{s_1 \dual{s_1}}{s \dual{s}}}_g \hastype \Proc
\end{align}

\noindent where $\wtd x' \subseteq \fn{P}$ such that 
      $(\Delta \cdot s_1:S \cdot \dual{s_1}:\dual S) \setminus \wtd x' = \emptyset$, 
  and $\wtd y' = \wtd v' \cup \wtd m$ where  
       $\indexed{\wtd v'}{\wtd x'}{\Gamma,\Delta  \cdot s_1:S \cdot \dual{s_1}:\dual S}$. 
      Also, $\Theta' = \thetaR'  \cat \thetax$, 
  where $\thetaR'$ such that 
   $\balan{\thetaR'}$ with 
   $$\dom{\thetaR'} = \{\prop^r_{k+1},\prop^r_{k+2},\ldots,\prop^r_{k+\lenHOopt{P'}+1} \} \cup 
   \{\dual{\prop^r_{k+2}},\ldots,\dual{\prop^r_{k+\lenHOopt{P'}+1}} \}$$ 
   and 
   $\thetaR' (\prop^r_{k+1})=\trec{t}\btinp{\wtd N'}\tvar{t}$
   where 
  $\wtd N' = (\Gtopt{\Gamma},\Gtopt{\envR \cdot 
  \Delta  \cdot s_1:S \cdot \dual{s_1}:\dual S})(\wtd y')$.

  We define $\wtd x = \wtd x' \setminus (s_1, \dual{s_1})$ 
  and $\wtd y = \wtd y' \setminus (\wtd s \cdot \wtd{\dual{s}})$ 
  where $\wtd s = \bname{s_1:S}$ and $\wtd{\dual{s}} = \bname{\dual{s_1}:\dual S}$. 
  By construction $\wtd x \subseteq \fn{P}$ and $\Delta \setminus \wtd x =\es$. 
  Further, $\wtd y = \wtd v \cdot \wtd m$ where
  $\indexed{\wtd v}{\wtd x}{\Gamma,\Delta}$.  
  
  Let $\Theta = \Theta' \cat \thetaR''$ where 
  % $$\thetaR = \thetaR' \cat 
  % \prop^r_{k}:\trec{t}\btinp{\wtd N}\tvar{t} 
  % \cat \dual{\prop^r_{k+1}}:\trec{t}\btout{\wtd N'}\tvar{t}$$
  $$\thetaR'' = 
  \prop^r_{k}:\trec{t}\btinp{\wtd N}\tvar{t} 
  \cat \dual{\prop^r_{k+1}}:\trec{t}\btout{\wtd N'}\tvar{t}$$

  \noindent where 
$\wtd N = (\Gtopt{\Gamma},\Gtopt{\envR \cdot \Delta \cdot u_i:{S}})(\wtd y)$. 
By construction and since $\lenHOopt{P}=\lenHOopt{P'}+1$ we have 
  $$\dom{\thetaR} = \{\prop^r_k,\prop^r_{k+1},\ldots,\prop^r_{k+\lenHOopt{P}-1} \} \cup 
   \{\dual{\prop^r_{k+1}},\ldots,\dual{\prop^r_{k+\lenHOopt{P}-1}} \}$$

 By \Cref{pi:t:bdowncorec-rec} we have: 
\begin{align*}
  \Brecpi{k}{\tilde y}{P}_g = 
    \recp{X}\news{\widetilde{s}:\Gtopt{C}}\propinp{k}{\wtd y}\propout{k+1}{\wtd y'}X
    \Par
    \Brecpi{k}{\tilde y'}{P'\subst{s_1 \dual{s_1}}{s \dual{s}}}_g
\end{align*}

 We shall prove the following judgment: 
 \begin{align}
   \Gtopt{\Gamma \setminus \wtd x}; \Theta \proves \Brecpi{k}{\tilde y}{P}_g \hastype \Proc 
 \end{align}

 We use the  auxiliary sub-trees: 
 \begin{align} 
  \label{pi:pt:brec-news-pt1-1}
  \AxiomC{} 
  \LeftLabel{\scriptsize \textsc{(PolyVar)}}
  \UnaryInfC{$\Gtopt{\Gamma} \cdot X: \thetaR''; \Gtopt{\Delta}
  \cat 
  \wtd s : \Gtopt{S} \cat \wtd {\dual s}: \Gtopt{\dual S} \proves 
  \wtd y' \hastype \wtd N'$} 
  \DisplayProof 
 \end{align} 
\begin{align}
\label{pi:pt:brec-news-pt1}
\AxiomC{} 
\LeftLabel{\scriptsize \textsc{(RVar)}}
\UnaryInfC{$\Gtopt{\Gamma} \cdot X: \thetaR; \thetaR'' 
 \proves X 
\hastype \Proc$} 
% \AxiomC{} 
% \LeftLabel{\scriptsize PolyVar}
% \UnaryInfC{$\Gtopt{\Gamma} \cdot X: \thetaR''; \Gtopt{\Delta}
% \cat 
% \wtd s : \Gtopt{S} \cat \wtd {\dual s}: \Gtopt{\dual S} \proves 
% \wtd y' \hastype \wtd N'$} 
\AxiomC{\eqref{pi:pt:brec-news-pt1-1}} 
\LeftLabel{\scriptsize \textsc{(PolySend)}}
\BinaryInfC{$\Gtopt{\Gamma}\cdot X:\thetaR;\thetaR \cdot \Gtopt{\Delta}
\cat 
\wtd s : \Gtopt{S} \cat \wtd {\dual s}: \Gtopt{\dual S}
\proves \propout{k+1}{\wtd y'}X
\hastype \Proc$}
\DisplayProof
\end{align}
 The following tree proves this case: 
 \begin{align*}
  \AxiomC{\eqref{pi:pt:brec-news-pt1}} 
  \AxiomC{} 
 \LeftLabel{\scriptsize \textsc{(PolyVar)}}
 \UnaryInfC{$\Gtopt{\Gamma};\Gtopt{\Delta} \proves \wtd y \hastype \wtd N$} 
  \LeftLabel{\scriptsize \textsc{(PolyRcv)}}
  \BinaryInfC{$\Gtopt{\Gamma \setminus \wtd x} \cdot X : \thetaR''; \thetaR'' 
  \cdot \wtd s : \Gtopt{S} \cat \wtd {\dual s}: \Gtopt{\dual S}
  \proves 
  \propinp{k}{\wtd y}\propout{k+1}{\wtd y'}X
      \hastype \Proc$}
  \LeftLabel{\scriptsize \textsc{(PolyResS)}}
  \UnaryInfC{$
  \Gtopt{\Gamma \setminus \wtd x} \cdot X : \thetaR''; \thetaR'' \proves 
  \news{\wtd{s}:\Gtopt{S}}
  \propinp{k}{\wtd y}\propout{k+1}{\wtd y'}X
      \hastype \Proc$} 
  \LeftLabel{\scriptsize \textsc{(Rec)}}
  \UnaryInfC{$	\Gtopt{\Gamma \setminus \wtd x}; \thetaR'' \proves 
  \recp{X}
  \news{\wtd s:\Gtopt{S}}
  \propinp{k}{\wtd y}\propout{k+1}{\wtd y'}X
      \hastype \Proc$} 
  \AxiomC{\eqref{pi:eq:brec-news-ih}}
  \LeftLabel{\scriptsize \textsc{(Par)}}
  \BinaryInfC{$	\Gtopt{\Gamma \setminus \wtd x}; \Theta \proves 
        \recp{X}
        \news{\wtd s:\Gtopt{S}}
        \propinp{k}{\wtd y}\propout{k+1}{\wtd y'}X
        \Par
        \Brecpi{k}{\tilde y'}{P'\subst{s_1 \dual{s_1}}{s \dual{s}}}_g
   \hastype \Proc$}
   \DisplayProof
 \end{align*}
   \end{enumerate}
% \qed
 This concludes the proof of \Cref{pi:lem:brec}.
\end{proof}

\thesisalt{}{
  \subsection{Proof of \Cref{pi:t:thmtyperecur}}
}

\thmtyperecur*
\label{pi:app:proofbrec}
\label{pi:app:thmtyperecur}

\begin{proof} 
% \label{pi:app:proofbrec}
By induction on the structure of $P$. By assumption $\Gamma;\Delta \proves P
\hastype \Proc~$.
We consider nine cases. We separately treat {input} and {output}  cases 
depending on whether the subject name of the prefix is recursive or not. 
%  First, we consider six cases not involving recursive names 
%  in a prefix: 
%  Similarly, \texttt{Restriction} case depending if 
%  restricted name is recursive. 
% We consider six cases, depending on the shape of $P$, when
% subject names are not recursive (where applicable). Then, we separately treat
% \texttt{Input} and \texttt{Output} cases when subject name is recursive, and
% \texttt{Restriction} when restricted name is recursive.  
\begin{enumerate}
  \item Case $P = \inact$. The only rule that can be applied here is \textsc{Nil}.
   By inversion of this rule, we have:
  $\Gamma;\es \proves \inact$.
  We shall then prove the following judgment:
  \begin{align}
      \Gtopt{\Gamma};\Theta \proves \Bopt{k}{\tilde y}{\inact} \hastype \Proc
  \end{align}

  \noindent where $\widetilde x \subseteq \fn{\inact}=\epsilon$ and $\Theta =
  \{c_k:\btinp{\chtype{\tinact}}\tinact\}$.  Since by \Cref{pi:r:prefix} we know that
  $\propinp{k}{} \inact$ stands for $\propinp{k}{y} \inact$ with
  $\prop_k : \btinp{\chtype{\tinact}}\tinact$.
  % \ref{t:}

  By \Cref{pi:t:bdowncorec}: 
  $\Bopt{k}{\epsilon}{\inact} =
  \propinp{k}{} \inact$. 
  The following tree proves this case:
  \def\proofSkipAmount{\vskip 1.2ex plus.8ex minus.4ex}
  \begin{prooftree}
  \AxiomC{}
  \LeftLabel{\scriptsize \textsc{(Nil)}}
    \UnaryInfC{$\Gamma';\es;\es \proves \inact \hastype \Proc$}
    \AxiomC{$\prop_k \notin \dom{\Gamma}$}
    \LeftLabel{\scriptsize \textsc{(End)}}
    \BinaryInfC{$\Gamma';\es;\prop_k : \tinact \proves \inact \hastype \Proc$}
    \AxiomC{}
    \LeftLabel{\scriptsize \textsc{(Sh)}}
    \UnaryInfC{$\Gamma';\es \proves y \hastype \chtype{\tinact}$}
    \LeftLabel{\scriptsize \textsc{(Rcv)}}
    \BinaryInfC{$\Gtopt{\Gamma}\cat y : \chtype{\tinact};\Theta \proves \propinp{k}{} \inact \hastype \Proc$}
%      \LeftLabel{\scriptsize \Cref{pi:lem:strength}
%      with $y$}
%      \UnaryInfC{$\Gtopt{\Gamma};\Theta \proves \propinp{k}{} \inact \hastype \Proc$}
  \end{prooftree}

  \noindent where $\Gamma' = \Gtopt{\Gamma} \cat y : \chtype{\tinact}$. 
  We know $\prop_k \notin \dom{\Gamma}$ since we use reserved names for 	propagator channels. %In the continuation of the proof 
  %we omit typing terminated processes. 
  
  \item Case $P = \binp{u_i}{z}P'$. We distinguish two sub-cases, depending on whether $u_i$ is linear or not: \rom{1}
$u_i \in \dom{\Delta}$ and \rom{2} $u_i \in \dom{\Gamma}$.
We consider
sub-case \rom{1} first. For this case Rule \textsc{Rcv} can be applied:
  \begin{align}
    \label{pi:pt:inputInv}
    \AxiomC{$\Gamma; \Delta \cat u_i : S  \cat \Delta_z \proves P' \hastype \Proc$}
    \AxiomC{$\Gamma; \Delta_z \proves z \hastype C$}
    \LeftLabel{\scriptsize \textsc{(Rcv)}}
    \BinaryInfC{$\Gamma \setminus z;
                    \Delta \cat u_i:
                  \btinp{C}S \proves \binp{u_i}{z}P' \hastype \Proc$}
    \DisplayProof
  \end{align}

By IH on the first assumption of \eqref{pi:pt:inputInv} we know:
\begin{align}
  \label{pi:eq:input-ih}
  \Gtopt{\Gamma'_1};\Gtopt{\Delta'_1} \cat \Theta_1 \proves
  \Bopt{k+1}{\tilde y'}{P'} \hastype \Proc
\end{align}
\noindent where  $\li x' \subseteq \fn{P'}$ 
and $\li y'$ such that $\indexed{\wtd y'}{\wtd x'}{\Gamma,\Delta}$. 
Also, 
 $\Gamma_1'=\Gamma \setminus \widetilde x'$, 
 $\Delta_1' = \Delta \setminus \widetilde x'$, and $\balan{\Theta_1}$ 
 with 
 $$
 \dom{\Theta_1}=\{\prop_{k+1},\ldots,\prop_{k+\lenHOopt{P'}}\}
 \cup \{\dual{\prop_{k+2}},\ldots,\dual{\prop_{k+\lenHOopt{P'}}}\}
 $$
 and ${\Theta_1(\prop_{k+1})}=\btinp{\widetilde M'}\tinact$
 where $\widetilde M' = (\Gtopt{\Gamma},\Gtopt{\Delta,u_i:S,\Delta_z})(\widetilde y')$.

By applying \Cref{pi:lem:varbreak} to the second
assumption of \eqref{pi:pt:inputInv} we have:
\begin{align}
    \label{pi:eq:input-ih2}
    \Gtopt{\Gamma};\Gtopt{\Delta_z} \proves \wtd z \hastype \Gtopt{C}
\end{align}

Let $\wtd x = \wtd x',u \setminus z$ 
and $\wtd y = \wtd y' \sigma ,u_i \setminus \wtd z$ such that 
$\len{\wtd z}=\Gtopt{C}$, where $\sigma = \subst{\wtd n}{\wtd u}$ with $\widetilde n =
    (u_{i+1},\ldots,u_{i+\len{\Gtopt{S}}})$ and $\widetilde u =
    (u_i,\ldots,u_{i+\len{\Gtopt{S}}-1})$. We may notice that 
by \Cref{pi:def:indexed} 
 $\indexed{\wtd y}{\wtd x}{\Gamma,\Delta}$ holds. 
We define $\Theta = \Theta_1 \cat \Theta'$, where
\begin{align*}
\Theta' = \prop_k:\btinp{\widetilde M} \tinact \cat \dual {\prop_{k+1}}:\btout{\widetilde M'} \tinact
\end{align*}
with $\widetilde M = (\Gtopt{\Gamma},\Gtopt{\Delta \cat \cat u_i:
                  \btinp{C}S})(\widetilde y)$.
By \Cref{pi:def:sizeproc},
$\lenHOopt{P} = \lenHOopt{P'} + 1$ so
$$\dom{\Theta} = \{\prop_k,\ldots,\prop_{k+\lenHOopt{P}-1}\}
    \cup \{\dual{\prop_{k+1}},\ldots,\dual{\prop_{k+\lenHOopt{P}-1}}\}$$
    and $\Theta$ is balanced since $\Theta(\prop_{k+1}) \dualof \Theta(\dual{\prop_{k+1}})$ and
     $\Theta_1$ is balanced.
By \Cref{pi:t:bdowncorec}:
\begin{align*}
 \Bopt{k}{\tilde y}{\binp{u_i}{z}P'} = \propinp{k}{\widetilde y}
 \binp{u_i}{\widetilde z}
 \propout{k+1}
 {\widetilde y'\sigma} \inact \Par \Bopt{k+1}{\tilde y'\sigma}{P'\incrname{u}{i}}
\end{align*}

%   Let $\wtd x = \wtd x',u \setminus z$ 
%   and $\wtd y = \wtd y',u_i \setminus \wtd z$ such that 
%   $\len{\wtd z}=\Gtopt{C}$. We may notice that 
%   by \Cref{pi:def:indexed} 
%   $\indexed{\wtd y}{\wtd x}{\Gamma,\Delta}$ holds. 
Also, let $\Gamma_1 = \Gamma \setminus \widetilde x$ and 
$\Delta_1 = \Delta \setminus \widetilde x$. We may notice that 
$\Delta_1 = \Delta'_1$.
We shall prove the following judgment:
\begin{align*}
\Gtopt{\Gamma_1 \setminus z};\Gtopt{\Delta_1}
  \cat \Theta
  \proves
      \Bopt{k}{\tilde y}{\binp{u_i}{z}P'}
\end{align*}

%\noindent where $\Gamma_1 = \Gamma \setminus \widetilde x$.
We type sub-process $ \propinp{k}{\widetilde y}
 \binp{u_i}{\widetilde z}
 \propout{k+1}
 {\widetilde y'} \inact$ 
with some auxiliary derivations:
 
\begin{align}
   \label{pi:pt:input2}
  \AxiomC{}
  \LeftLabel{\scriptsize \textsc{(Nil)}}
  \UnaryInfC{$\Gtopt{\Gamma};\es \proves \inact \hastype \Proc$}
  \LeftLabel{\scriptsize \textsc{(End)}}
  \UnaryInfC{$\Gtopt{\Gamma};\dual{\prop_{k+1}}:\tinact \proves \inact \hastype \Proc$}
  \DisplayProof
\end{align}
\begin{align}
  %\gamma
  \label{pi:st:input1}
%    \AxiomC{}
%    \LeftLabel{\scriptsize Nil}
%    \UnaryInfC{$\Gtopt{\Gamma};\es \proves \inact \hastype \Proc$}
%    \LeftLabel{\scriptsize End}
%    \UnaryInfC{$\Gtopt{\Gamma};\dual{\prop_{k+1}}:\tinact \proves \inact \hastype \Proc$}
\AxiomC{\eqref{pi:pt:input2}} 
  \AxiomC{}
  \LeftLabel{\scriptsize \textsc{(PolyVar)}}
  \UnaryInfC{$\Gtopt{\Gamma};\Gtopt{\Delta\setminus\Delta_1\cat u_{i+1}:S} \cat \Gtopt{\Delta_z}
              \proves \widetilde y'\sigma \hastype \widetilde M'$}
  \LeftLabel{\scriptsize \textsc{(PolySend)}}
  \BinaryInfC{%\begin{tabular}{c}
  $\Gtopt{\Gamma};
                \dual {\prop_{k+1}}:\btout{\widetilde M'} \tinact
                \cat \Gtopt{\Delta\setminus\Delta_1\cat u_{i+1}:S} \cat \Gtopt{\Delta_z}
                \proves 
                \propout{k+1}{\widetilde y'\sigma} \inact \hastype \Proc$
                %\end{tabular}
                }
  \LeftLabel{\scriptsize \textsc{(End)}}
  \UnaryInfC{%\begin{tabular}{c}
  $
  \Gtopt{\Gamma};
                \dual {\prop_{k+1}}:\btout{\widetilde M'} \tinact
                \cat u_i:\tinact \cat 
                  \Gtopt{\Delta \setminus \Delta_1 \cat u_{i+1}:S}\cat \Gtopt{\Delta_z}
                \proves %$ \\ $
                \propout{k+1}{\widetilde y'\sigma} \inact \hastype \Proc$
                %\end{tabular}
                }
  \DisplayProof
\end{align}
\begin{align}
  %beta
  \label{pi:st:input-rcv}
  \AxiomC{\eqref{pi:st:input1}}
    \AxiomC{\eqref{pi:eq:input-ih2}}
  \LeftLabel{\scriptsize \textsc{(PolyRcv)}}
  \BinaryInfC{
  %\begin{tabular}{c}
      $\Gtopt{\Gamma \setminus z}; u_i:\btinp{\Gtopt{U}} \tinact \cat
              \dual {\prop_{k+1}}:\btout{\widetilde M'} \tinact \cat 
%                \widetilde y : \widetilde M
              \Gtopt{\Delta_2}
              \proves %$ \\
              \binp{u_i}{\widetilde z}\propout{k+1}{\widetilde y'\sigma} \inact \hastype \Proc$
  %\end{tabular}
}
   \LeftLabel{\scriptsize \textsc{(End)}}
  \UnaryInfC{ 
  %\begin{tabular}{c}
  $\Gtopt{\Gamma \setminus z}; 
              \dual {\prop_{k+1}}:\btout{\widetilde M'} \tinact
              \cat \prop_k: \tinact \cat
              \Gtopt{\Delta_2}
              \proves %$ \\ 
              \binp{u_i}{\widetilde z}\propout{k+1}{\widetilde y'\sigma} \inact \hastype \Proc$
   %\end{tabular}
   }
  \DisplayProof
\end{align}
\begin{align}
  \label{pi:st:input3}
  \AxiomC{\eqref{pi:st:input-rcv}}
  \AxiomC{}
  \LeftLabel{\scriptsize \textsc{(PolyVar)}}
  \UnaryInfC{$\Gtopt{\Gamma \setminus z};
  \Gtopt{\Delta_2}
  \proves \widetilde y \hastype \widetilde M$}
  \LeftLabel{\scriptsize \textsc{(PolyRcv)}}
  \BinaryInfC{$\Gtopt{\Gamma_1 \setminus z};
   \Theta' \proves \propinp{k}{\widetilde
  y}\binp{u_i}{z}\propout{k+1}{\widetilde y'\sigma} \inact \hastype \Proc$}
\DisplayProof
\end{align}
  \noindent 
  where $\Delta_2 = \Delta \cat u_i:\btinp{C}S \setminus \Delta_1$. 
Using \eqref{pi:st:input3}, 
the following tree
proves this case:
\begin{align}
    \label{pi:pt:input}
  \AxiomC{\eqref{pi:st:input3}}
  \AxiomC{\eqref{pi:eq:input-ih}}
   \LeftLabel{\scriptsize}
  \UnaryInfC{
  %\begin{tabular}{c}
  $\Gtopt{\Gamma'_1 \setminus z}; \Gtopt{\Delta_1} \cat \Theta_1 \proves \Bopt{k+1}{\tilde y'\sigma}{P'\incrname{u}{i}} \hastype \Proc$
  %\end{tabular}
  }
  \LeftLabel{\scriptsize \textsc{(Par)}}
  \BinaryInfC{%\begin{tabular}{c}
              $\Gtopt{\Gamma_1 \backslash z};\Gtopt{\Delta_1}
                \cat \Theta \proves \propinp{k}{\widetilde y}
                \binp{u_i}{\widetilde z}\propout{k+1}{\widetilde y'\sigma} \inact \Par \Bopt{k+1}
                {\tilde y'\sigma}{P'\incrname{u}{i}} \hastype \Proc$
                %\end{tabular}
                }
  \DisplayProof
\end{align}
    \noindent

    Note that we have used the following for the right assumption of \eqref{pi:pt:input}:
    \begin{align*}
        % \Gtopt{\Delta'_1 \cat u_i:S}\subst{\tilde n}{\tilde u} &=
        % \Gtopt{\Delta'_1 \cat u_{i+1}:S} \\
        \Bopt{k+1}{\tilde y'}{P'} &\equiv_{\alpha}
        \Bopt{k+1}{\tilde y'\sigma}{P'\incrname{u}{i}}
    \end{align*}

    Next, we comment the case when $u_i \notin \wtd x$. 
    In this case $\Delta_1 = \Delta \setminus \widetilde x \cat 
    u_i:\btinp{C}S$. Hence, in the right hand-side of 
  \eqref{pi:pt:input} instead of $\Gtopt{\Delta_1}$ we would have 
  $\Gtopt{\Delta \setminus \widetilde x \cat 
    u_{i+1}:S}$ and in the left-hand side we have 
    $u_i : \btinp{\Gtopt{C}}\tinact$ as a linear environment. 
    Then, we would need to apply \Cref{top:lem:subst}  with 
    $\subst{u_{i}}{u_{i+1}}$ to the right-hand side before 
    invoking \eqref{pi:eq:input-ih}. 
    We remark that similar provisos apply  
    to the following cases when the assumption is $u_i \notin \wtd x$.

 This concludes sub-case \rom{1}. We now consider sub-case \rom{2}, i.e.,
 $u_i \in \dom{\Gamma}$.
 Here Rule~\textsc{Acc} can be applied:
 \begin{align}
     \label{pi:pt:inputptr-subcase2}
     \AxiomC{$\Gamma;\es \proves u_i \hastype \chtype{C}$}
     \AxiomC{$\Gamma;\Delta \cat z : C \proves P' \hastype \Proc$}
    \AxiomC{$\Gamma;z:C \proves z \hastype C$}
    \LeftLabel{\scriptsize \textsc{(Acc)}}
     \TrinaryInfC{$\Gamma;\Delta
                     \proves \binp{u_i}{z}P' \hastype \Proc$}
     \DisplayProof
 \end{align}

% Let $\widetilde x, \widetilde x', \widetilde y, \widetilde y'$,
%   $\Theta_1$, $\Theta$, $\Gamma_1$, and
%   $\Delta_1$  be defined as in sub-case
%   \rom{1}.
 By IH on the second assumption of \eqref{pi:pt:inputptr-subcase2} we have:
 \begin{align}
 \label{pi:eq:input-ih-2}
  \Gtopt{\Gamma'_1};\Gtopt{\Delta'_1} \cat \Theta_1 \proves \Bopt{k+1}{\tilde y'}{P'} \hastype \Proc
 \end{align}
 
\noindent where $\wtd x'$ and $\wtd y'$ are as in sub-case \rom{1}. 
Also, $\Gamma'_1 = \Gamma \setminus \wtd x'$
and $\Delta'_1 = \Delta  \setminus \wtd y'$ 
and $\balan{\Theta_1}$ with
 $$
 \dom{\Theta_1}=\{\prop_{k+1},\ldots,\prop_{k+\lenHOopt{P'}}\}
 \cup \{\dual{\prop_{k+2}},\ldots,\dual{\prop_{k+\lenHOopt{P'}}}\}
 $$
 and ${\Theta_1(\prop_{k+1})}=\btinp{\widetilde M'}\tinact$
 where $\widetilde M' = (\Gtopt{\Gamma},\Gtopt{\Delta,z:C})(\widetilde y')$.

  By applying \Cref{pi:lem:varbreak} to the first and third
     assumptions of \eqref{pi:pt:inputptr-subcase2} we have:
    \begin{align}
        \label{pi:eq:input-ih2-2}
        \Gtopt{\Gamma};\es \proves u_i \hastype \chtype{\Gtopt{C}}
\\
        \label{pi:eq:input-ih3-2}
        \Gtopt{\Gamma};\Gtopt{z:C} \proves \wtd z \hastype \Gtopt{C}
    \end{align}

  We define $\wtd x = \wtd x' \cup u \setminus z$ and $\wtd y = \wtd y' \cup u_i 
  \setminus \wtd z$ where $\len{\wtd z} = \len{\Gtopt{C}}$. 
  Notice that $\indexed{\wtd y}{\wtd x}{\Gamma,\Delta}$.
  Let $\Gamma_1 = \Gamma \setminus x$. 
  We define $\Theta = \Theta_1 \cat \Theta'$, where
\begin{align*}
\Theta' = \prop_k:\btinp{\widetilde M} \tinact \cat \dual {\prop_{k+1}}:\btout{\widetilde M'} \tinact
\end{align*}
with $\widetilde M = (\Gtopt{\Gamma},\Gtopt{\Delta})(\widetilde y)$.
By \Cref{pi:def:sizeproc},
$\lenHOopt{P} = \lenHOopt{P'} + 1$ so
$$\dom{\Theta} = \{\prop_k,\ldots,\prop_{k+\lenHOopt{P}-1}\}
    \cup \{\dual{\prop_{k+1}},\ldots,\dual{\prop_{k+\lenHOopt{P}-1}}\}$$
    and $\Theta$ is balanced since $\Theta(\prop_{k+1}) \dualof \Theta(\dual{\prop_{k+1}})$ and
     $\Theta_1$ is balanced.

   By \Cref{pi:t:bdowncorec}, we have:
   \begin{align}
        \Bopt{k}{\tilde y}{\binp{u_i}{z}P'} = \propinp{k}{\widetilde y}
        \binp{u_i}{z}\propout{k+1}{\widetilde y'} \inact \Par \Bopt{k+1}{\tilde y'}
        {P'}
   \end{align}

   We shall prove the following judgment:
   \begin{align}
       \Gtopt{\Gamma_1};\Gtopt{\Delta_1} \cat \Theta \proves
       \Bopt{k}{\tilde y}{\binp{u_i}{z}P'} \hastype \Proc
   \end{align}

  To this end, we use some auxiliary derivations: 
  \begin{align}
      \label{pi:st:input1-3}
  \AxiomC{}
  \LeftLabel{\scriptsize \textsc{(Nil)}}
  \UnaryInfC{$\Gtopt{\Gamma};\es;\es \proves \inact \hastype \Proc$}
  \LeftLabel{\scriptsize \textsc{(End)}}
  \UnaryInfC{$\Gtopt{\Gamma};\es;\dual{\prop_{k+1}}:\tinact \proves \inact \hastype \Proc$}
  \DisplayProof
  \end{align}

\begin{align}
  %\gamma
  \label{pi:st:input1-2}
%    \AxiomC{}
%    \LeftLabel{\scriptsize Nil}
%    \UnaryInfC{$\Gtopt{\Gamma};\es;\es \proves \inact \hastype \Proc$}
%    \LeftLabel{\scriptsize End}
%    \UnaryInfC{$\Gtopt{\Gamma};\es;\dual{\prop_{k+1}}:\tinact \proves \inact \hastype \Proc$}
\AxiomC{\eqref{pi:st:input1-3}} 
  \AxiomC{}
  \LeftLabel{\scriptsize \textsc{(PolyVar)}}
  \UnaryInfC{$\Gtopt{\Gamma};\Gtopt{\Delta_2} \cat z : \Gtopt{C}
              \proves \widetilde y' \hastype \widetilde M'$}
  \LeftLabel{\scriptsize \textsc{(PolySend)}}
  \BinaryInfC{$\Gtopt{\Gamma};
                \dual {\prop_{k+1}}:\btout{\widetilde M'} \tinact
                \cat \Gtopt{\Delta_2}
                \cat z : \Gtopt{C}
                \proves \propout{k+1}{\widetilde y'} \inact \hastype \Proc$}
  \DisplayProof
\end{align}
\begin{align}
  %beta
  \label{pi:st:input2-2}
  \AxiomC{\eqref{pi:eq:input-ih2-2}}
  \AxiomC{\eqref{pi:st:input1-2}}
  \AxiomC{\eqref{pi:eq:input-ih3-2}}
  \LeftLabel{\scriptsize \textsc{(PolyAcc)}}
  \TrinaryInfC{$\Gtopt{\Gamma};
              \dual{\prop_{k+1}}:\btout{\widetilde M'} 
               \tinact \cat \Gtopt{\Delta_2}
               \proves \binp{u_i}{z}\propout{k+1}{\widetilde y'} \inact \hastype \Proc$}
   \LeftLabel{\scriptsize \textsc{(End)}}
  \UnaryInfC{$\Gtopt{\Gamma};
              \dual{\prop_{k+1}}:\btout{\widetilde M'}
               \tinact \cat \prop_k:\tinact 
                   \cat \Gtopt{\Delta_2}
               \proves \binp{u_i}{z}\propout{k+1}{\widetilde y'} \inact \hastype \Proc$}
  \DisplayProof
\end{align}
\begin{align}
  \label{pi:st:input3-2}
  \AxiomC{\eqref{pi:st:input2-2}}
  \AxiomC{}
  \LeftLabel{\scriptsize \textsc{(PolyVar)}}
  \UnaryInfC{$\Gtopt{\Gamma};
  \Gtopt{\Delta_2}
  \proves
              \widetilde y \hastype \widetilde M$}
  \LeftLabel{\scriptsize \textsc{(PolyRcv)}}
  \BinaryInfC{$\Gtopt{\Gamma_1};\Theta' \proves
                  \propinp{k}{\widetilde y}
                    \binp{u_i}{z}\propout{k+1}{\widetilde y'} \inact
                    \hastype \Proc$}
  \DisplayProof
\end{align}
\noindent where $\Delta_2 = \Delta \setminus \Delta_1$.

Using \eqref{pi:eq:input-ih-2} and \eqref{pi:st:input3-2}, 
the following tree
proves this sub-case:
\begin{align}
    \label{pi:pt:input-2}
  \AxiomC{\eqref{pi:st:input3-2}}
  \AxiomC{\eqref{pi:eq:input-ih-2}}
  \LeftLabel{\scriptsize \textsc{(Par)}}
  \BinaryInfC{$\Gtopt{\Gamma_1};\Gtopt{\Delta_1}
  \cat \Theta \proves
  \propinp{k}{\widetilde y}\binp{u_i}{z}
  \propout{k+1}{\widetilde y'}
  \inact \Par \Bopt{k+1}{\tilde y'}{P'} \hastype \Proc$}
  \DisplayProof
\end{align}
%   As in sub-case \rom{1}, we may notice that if $y \in \fv{P'}$ then $\Gamma'_1 = \Gamma_1 \setminus y$.
%   On the other hand, if $y \notin \fv{P'}$ then $\Gamma'_1 = \Gamma_1$ so we need to apply
%   \Cref{pi:lem:strength} with $y$ to \eqref{pi:eq:input-ih-2} in \eqref{pi:pt:input-2}.
Note that if $u \notin \fn{P'}$ we need to apply 
 \Cref{top:lem:strength}  with $u_i$ to the 
right assumption of \eqref{pi:pt:input-2} before applying \eqref{pi:eq:input-ih-2}. 
This concludes the analysis of the input case.
%  $P = \binp{u_i}{z}P'$.

\item Case $P = \bout{u_i}{z_j}P'$. 
We distinguish two
sub-cases: \rom{1} $u_i \in \dom{\Delta}$ and \rom{2} $u_i \in \dom{\Gamma}$.
We consider sub-case \rom{1} first. For this case Rule~\textsc{Send} can be applied:
\begin{align}
  \label{pi:pt:outputitr}
  \AxiomC{$\Gamma;\Delta_1 \cat  u_i:S \proves P' \hastype \Proc$}
  \AxiomC{$\Gamma;\Delta_z \proves z_j \hastype C$}
  \AxiomC{$u_i:S \in \Delta$}
  \LeftLabel{\scriptsize \textsc{(Send)}}
  \TrinaryInfC{$\Gamma;\Delta
                \proves \bout{u_i}{z_j}P' \hastype \Proc$}
  \DisplayProof
\end{align}
  \noindent where $\Delta = \Delta_1 \cat\Delta_z
                \cat u_i:\btout{C}S$.

    By IH on the first assumption of \eqref{pi:pt:outputitr} we have:
    \begin{align}
        \label{pi:eq:output-ih-1}
      \Gtopt{\Gamma'_1};\Gtopt{\Delta'_1 } \cat \Theta_1 \proves
      \Bopt{k+1}{\tilde y'}{P'} \hastype \Proc
    \end{align}
  \noindent where $\widetilde x' \subseteq \fv{P'}$ 
  and $\widetilde y'$ such that 
  $\indexed{\widetilde y'}{\widetilde x'}{\Gamma,\Delta_1 \cat u_i:S}$. 
   Also, $\Gamma'_1 =
  \Gamma \setminus \widetilde y'$, 
  $\Delta'_1 = \Delta_1 \setminus \widetilde y'$, and $\balan{\Theta_1}$ 
  with
  $$
  \dom{\Theta_1}=\{\prop_{k+1},\ldots,\prop_{k+\lenHOopt{P'}}\}\cup
  \{\dual{\prop_{k+2}},\ldots,\dual{\prop_{k+\lenHOopt{P'}}}\}
  $$
  and $\Theta_1(\prop_{k+1})=\btinp{\widetilde M_1}\tinact$ where
  $\widetilde M_1 =
  (\Gtopt{\Gamma},\Gtopt{\Delta_1, u_i:S})(\widetilde y')$.
    
    By \Cref{pi:lem:varbreak} and the first
     assumption of \eqref{pi:pt:outputitr} we have:
     \begin{align}
         \label{pi:eq:output-ih2}
         \Gtopt{\Gamma};\Gtopt{\Delta_z} \proves \widetilde z \hastype \Gtopt{C}
     \end{align}

  \noindent where $\wtd z = (z_j,\ldots,z_{j+\len{\Gtopt{C}}-1})$.
  We assume $\wtd x = \wtd x',u, z$. Since 
  $\wtd x' \subseteq \fn{P'}$ follows that  
 $\widetilde x \subseteq \fn{P}$.
 Let $\wtd y = \wtd y'\sigma, u_i, \wtd z$ where 
 $\sigma = \subst{\wtd n}{\wtd u}$ with $\widetilde n =
    (u_{i+1},\ldots,u_{i+\len{\Gtopt{S}}})$ and $\widetilde u =
    (u_i,\ldots,u_{i+\len{\Gtopt{S}}-1})$. We have
    $\wtd z = \fnb{z}{\wtd y}$. 
%   $\len{\wtd z} = \len{\Gtopt{C}}$. 
 By \Cref{pi:def:indexed} it follows that  
 $\indexed{\wtd y}{\wtd x}{\Gamma, \Delta}$. 
    We define $\Theta = \Theta_1 \cat \Theta'$, where:
    \begin{align*}
      \Theta' = \prop_{k}: \btinp{\widetilde M} \tinact \cat
      \dual{\prop_{k+1}}:\btout{\widetilde M_1} \tinact
    \end{align*}

\noindent with $\widetilde M = (\Gtopt{\Gamma}\cat \Gtopt{\Delta})(\widetilde y)$. By \Cref{pi:def:sizeproc}, we know $\lenHOopt{P} =\lenHOopt{P'} + 1$, so
$$
\dom{\Theta}=\{\prop_k,\ldots,\prop_{k+\lenHOopt{P}-1}\} \cup
\{\dual{\prop_{k+1}},\ldots,\dual{\prop_{k+\lenHOopt{P}-1}}\}
$$
and $\Theta$ is balanced since $\Theta(\prop_{k+1}) 
\dualof \Theta(\dual{\prop_{k+1}})$ and $\Theta_1$ is 
balanced. 

By \Cref{pi:t:bdowncorec}, we have:
\begin{align*}
  \Bopt{k}{\tilde y}{\bout{u_i}{z}P'} =
  \propinp{k}{\widetilde y}
  \bbout{u_i}{\wtd z} \propout{k+1}{\widetilde y'\sigma}
  \inact \Par
  \Bopt{k+1}{\tilde y'\sigma}{P'\incrname{u}{i}}
\end{align*}
% \noindent where $\len{\wtd z}=\len{\Gtopt{C}}$. 
Let $\Gamma_1 = \Gamma \setminus \widetilde x = \Gamma'_1$ 
and $\Delta \setminus \wtd x = \Delta'_1$.
We shall prove the following judgment:
\begin{align}
  \AxiomC{$\Gtopt{\Gamma_1}; \Gtopt{\Delta'_1} \cat \Theta \proves
  \Bopt{k}{\tilde y}{\bout{u_i}{z}P'} \hastype \Proc$}
  \DisplayProof
\end{align}
%   \noindent where we may notice that following holds: 

%   $$\Delta'_1 = \Delta \setminus \wtd y' = (\Delta_1 \cat u_i:\btout{C}S \cat \Delta_z) \setminus \widetilde y $$

To type the sub-process $ \propinp{k}{\widetilde y}
  \bbout{u_i}{\wtd z} \propout{k+1}{\widetilde y'}
  \inact$ of
$\Bopt{k}{\tilde y}{\bout{u_i}{z}P'}$
we use the following auxiliary derivations:
\begin{align}
    \label{pi:pt:send2}
    \AxiomC{}
    \LeftLabel{\scriptsize \textsc{(Nil)}}
    \UnaryInfC{$\Gtopt{\Gamma};\es \proves \inact \hastype \Proc$}
    \LeftLabel{\scriptsize \textsc{(End)}}
    \UnaryInfC{$\Gtopt{\Gamma};\dual{\prop_{k+1}} : \tinact \proves \inact \hastype \Proc$}
    \DisplayProof
\end{align}
\begin{align}
    \label{pi:pt:zsend}
%  	\AxiomC{}
%  	\LeftLabel{\scriptsize Nil}
%  	\UnaryInfC{$\Gtopt{\Gamma};\es \proves \inact \hastype \Proc$}
%  	\LeftLabel{\scriptsize End}
%  	\UnaryInfC{$\Gtopt{\Gamma};\dual{\prop_{k+1}} : \tinact \proves \inact \hastype \Proc$}
\AxiomC{\eqref{pi:pt:send2}} 
    \AxiomC{}
    \LeftLabel{\scriptsize \textsc{(PolyVar)}}
    \UnaryInfC{$\Gtopt{\Gamma};\Gtopt{\Delta_1 \cat u_{i+1}:S \setminus \Delta'_1}
                \proves \widetilde y'\sigma \hastype \widetilde M_1$}
    \LeftLabel{\scriptsize \textsc{(PolySend)}}
    \BinaryInfC{
    \begin{tabular}{c}
    $\Gtopt{\Gamma};
                \dual{\prop_{k+1}}:\btout{\widetilde M_1} \tinact
                    \cat \Gtopt{\Delta_1 \cat u_{i+1}:S \setminus \Delta'_1}
                \proves \bout{\dual {\prop_{k+1}}}{\widetilde y'\sigma} 
                \hastype \Proc$
                \end{tabular}}
    \LeftLabel{\scriptsize \textsc{(End)}}
    \UnaryInfC{
    \begin{tabular}{c}$\Gtopt{\Gamma};
                \dual{\prop_{k+1}}:\btout{\widetilde M_1} 
                 \tinact
                \cat u_i:\tinact
                \cat \Gtopt{\Delta_1 \cat u_{i+1}:S \setminus \Delta'_1}
                \proves \bout{\dual {\prop_{k+1}}}{\widetilde y'\sigma} \inact
                \hastype \Proc$
                \end{tabular}}
    \DisplayProof
\end{align}
\begin{align}
    \label{pi:pt:uisend}
    \AxiomC{\eqref{pi:pt:zsend}}
    \AxiomC{\eqref{pi:eq:output-ih2}}
%  	\LeftLabel{\scriptsize \Cref{pi:lem:subst} with $\subst{\tilde z_1}{\tilde z}$}
%  	\UnaryInfC{$\Gtopt{\Gamma};\Gtopt{\Delta_z \subst{\tilde z}{\tilde z_1}} \proves \widetilde z \hastype \Gtopt{C}$}
    \LeftLabel{\scriptsize \textsc{(Send)}}
    \BinaryInfC{\begin{tabular}{c}
                $\Gtopt{\Gamma};
                \Gtopt{\Delta \setminus \Delta'_1}
                      \cat
                       \dual{\prop_{k+1}}:\btout{\widetilde M_1}\tinact
                       \proves \bbout{u_i}{\widetilde z}
                   \propout{k+1}{\widetilde y'} \inact \hastype \Proc$
                   \end{tabular}}
      \LeftLabel{\scriptsize \textsc{(End)}}
    \UnaryInfC{\begin{tabular}{c}
                $\Gtopt{\Gamma};
                \Gtopt{\Delta \setminus \Delta'_1}
                       \cat
                       \dual{\prop_{k+1}}:\btout{\widetilde M_1}\tinact \cat \prop_k:\tinact 
                       \proves \bbout{u_i}{\wtd z}
                   \propout{k+1}{\wtd y'} \inact \hastype \Proc$
                   \end{tabular}}
    \DisplayProof
\end{align}
% TODO continue updating proof here 
\begin{align}
\label{pi:pt:stvalue}
\AxiomC{\eqref{pi:pt:uisend}}
\AxiomC{}
\LeftLabel{\scriptsize \textsc{(PolyVar)}}
\UnaryInfC{$\Gtopt{\Gamma};\Gtopt{\Delta \setminus \Delta_1'} \proves
                \widetilde y : \widetilde M$}
\LeftLabel{\scriptsize \textsc{(PolyRcv)}}
\BinaryInfC{\begin{tabular}{c}
                $\Gtopt{\Gamma_1};
               \Theta'
               \proves \propinp{k}{\widetilde y}
                \bbout{u_i}{\wtd z}
              \propout{k+1}{\wtd y'} \inact
              \hastype \Proc$
              \end{tabular}}
\DisplayProof
\end{align}
%    \noindent where $\Delta_3 = \Delta \setminus \Delta_2$. 
Using \eqref{pi:eq:output-ih-1} and \eqref{pi:pt:stvalue}, the following tree proves this case:
\begin{align}
  \label{pi:pt:output1}
  \AxiomC{\eqref{pi:pt:stvalue}}
  \AxiomC{\eqref{pi:eq:output-ih-1}}
  \UnaryInfC{$\Gtopt{\Gamma_1};\Gtopt{\Delta'_1} \cat
              \Theta_1 \proves
              \Bopt{k+1}{\tilde y'\sigma}{P'\incrname{u}{i}} \hastype \Proc$}
   \LeftLabel{\scriptsize \textsc{(Par)}}
  \BinaryInfC{$\Gtopt{\Gamma_1}; \Gtopt{\Delta'_1} \cat \Theta \proves
                \Bopt{k}{\tilde y}{\bout{u_i}{z}P'} \hastype \Proc$}
  \DisplayProof
\end{align}

      Note that we have used the following for the right assumption of \eqref{pi:pt:input}:
    \begin{align*}
        % \Gtopt{\Delta'_1 \cat u_i:S}\subst{\tilde n}{\tilde u} &=
        % \Gtopt{\Delta'_1 \cat u_{i+1}:S} \\
        \Bopt{k+1}{\tilde y'}{P'} &\equiv_{\alpha}
        \Bopt{k+1}{\tilde y'\sigma}{P'\incrname{u}{i}}
    \end{align*}

%  In this case $u_i \notin \wtd x$ we would have to 
%%  apply substitution $\subst{u_{i+1}}{u_{i}}$ to $\Gamma_1$
%apply \Cref{pi:lem:subst} with 
%$\subst{u_{i+1}}{u_{i}}$

%   \noindent where $\widetilde n =
%   (u_{i+1},\ldots,u_{i+\len{\Gtopt{S}}})$ and $\widetilde u =
%   (u_i,\ldots,u_{i+\len{\Gtopt{S}}-1})$.

We now consider sub-case \rom{2}. For this sub-case Rule~\textsc{Req}
      can be applied:
 \begin{align}
     \label{pi:pt:outputitr-2}
     \AxiomC{$\Gamma;\es \proves u \hastype \chtype{C}$}
     \AxiomC{$\Gamma;\Delta_1 \hastype P' \hastype \Proc$}
    \AxiomC{$\Gamma;\Delta_z \proves z \hastype C$}
    \LeftLabel{\scriptsize \textsc{(Req)}}
     \TrinaryInfC{$\Gamma;\Delta_1 \cat \Delta_z \proves \bout{u_i}{z}P' \hastype \Proc$}
     \DisplayProof
 \end{align}

 Let $\widetilde x' \subseteq \fn{P'}$ and 
 $\wtd y$ such that $\indexed{\wtd y'}{\wtd x'}{\Gamma, \Delta_1}$
 . Further, let $\Gamma'_1 =
 \Gamma \setminus \widetilde x'$ and
 $\Delta'_1 = \Delta_1 \setminus \wtd x'$.
  Also, let $\Theta_1$ be
   environment defined as in sub-case \rom{1}.
%TODO dfdfdok

By IH on the second assumption of \eqref{pi:pt:outputitr-2} we have:
\begin{align}
    \label{pi:eq:output-ih-1-2}
  \Gtopt{\Gamma_1'};\Gtopt{\Delta'_1} \cat \Theta_1 \proves
  \Bopt{k+1}{\tilde y'}{P'} \hastype \Proc
\end{align}

  By \Cref{pi:lem:varbreak} and the first and third
     assumptions of \eqref{pi:pt:outputitr-2} we have:
     \begin{align}
         \label{pi:eq:output-ih-2-1}
         \Gtopt{\Gamma};\es \proves u_i \hastype \Gtopt{\chtype{C}}
\\
         \label{pi:eq:output-ih-2}
         \Gtopt{\Gamma};\Gtopt{\Delta_z} \proves \widetilde z \hastype \Gtopt{C}
     \end{align}

%  % need to construct \Theta
%  Let $\widetilde x = \fv{P}$ and $\Gamma_1 = \Gamma \setminus \widetilde x$.
%  We define $\Theta = \Theta_1 \cat \Theta_2 \cat \Theta'$, where:
%  \begin{align*}
%    \Theta' = \prop_{k}: \btinp{\widetilde M} \tinact \cat
%    \dual{\prop_{k+r+1}}:\btout{\widetilde M_2} \tinact
%  \end{align*}
%    \noindent with
%  $\widetilde M = (\Gtopt{\Gamma}\cat \Gtopt{\Lambda})(\widetilde x)$.
%  By \Cref{pi:def:sizeproc}, we know $\len{P} =
%  \len{V} + \len{P'} + 1$, so
%  $$\dom{\Theta}=(\prop_k,\ldots,\prop_{k+\len{P}-1}) \cup
%  (\dual{\prop_{k+1}},\ldots,\dual{\prop_{k+\len{P}-1}})$$
%  By construction $\Theta$ is balanced since $\Theta(\prop_{k+\degree+1}) \dualof \Theta(\dual{\prop_{k+\degree+1}})$ and
%  	 $\Theta_1$ and $\Theta_2$ are balanced.

We define $\wtd x = \wtd x' \cup z \cup u$ and 
$\wtd y = \wtd y' \cup \wtd z \cup u_i$ where $\len{\wtd z} = \len{\Gtopt{C}}$. 
  Notice that $\indexed{\wtd y}{\wtd x}{\Gamma,\Delta}$.
  Let $\Gamma_1 = \Gamma \setminus \wtd x = \Gamma'_1 \setminus u$
  and $\Delta \setminus \wtd x = \Delta'_1$.
% We define $\widetilde x$, $\wtd y$, and $\Theta$ as in sub-case
% \rom{1}. 

By \Cref{pi:t:bdowncorec}, we have:
\begin{align*}
\Bopt{k}{\tilde y}{\bout{u_i}{z}P'} =
\propinp{k}{\widetilde y}
\bbout{u_i}{\wtd z} \propout{k+1}{\widetilde y'} \inact \Par
\Bopt{k+1}{\tilde y'}{P'}
\end{align*}

We shall prove the following judgment: 
$$\Gtopt{\Gamma_1};\Gtopt{\Delta'_1} \cat \Theta
                \proves \Bopt{k}{\tilde y}{\bout{u_i}{z}P'}$$
We use some auxiliary derivations:
\begin{align}
    \label{pi:pt:zsend-2}
    \AxiomC{}
    \LeftLabel{\scriptsize \textsc{(Nil)}}
    \UnaryInfC{$\Gtopt{\Gamma};\es;\es \proves \inact \hastype \Proc$}
    \LeftLabel{\scriptsize \textsc{(End)}}
    \UnaryInfC{$\Gtopt{\Gamma};\es;\dual{\prop_{k+1}}:\tinact \proves \inact \hastype \Proc$}
    \AxiomC{}
    \LeftLabel{\scriptsize \textsc{(PolyVar)}}
    \UnaryInfC{$\Gtopt{\Gamma}; \Gtopt{\Delta_1\setminus \Delta'_1}
                \proves \widetilde y' : \wtd M_1$}
    \LeftLabel{\scriptsize \textsc{(PolySend)}}
    \BinaryInfC{$\Gtopt{\Gamma}; \Gtopt{\Delta_1 \setminus \Delta'_1} \cat
                \dual{\prop_{k+1}}:\btout{\widetilde M_1} \tinact
                \proves \bout{\dual {\prop_{k+1}}}{\widetilde y'} \inact \hastype \Proc$}
    \DisplayProof
\end{align}

  % \begin{align}
  %     \AxiomC{} 
  %     \UnaryInfC{}
  % \end{align}

\begin{align}
    \label{pi:pt:uisend-2}
    \AxiomC{\eqref{pi:eq:output-ih-2-1}} 
    \AxiomC{\eqref{pi:pt:zsend-2}}
%  	\AxiomC{$\Gamma;\wtd z:\Gtopt{C} \proves \wtd z \hastype \Gtopt{C}$}
\AxiomC{\eqref{pi:eq:output-ih-2}} 
    \LeftLabel{\scriptsize \textsc{(PolyReq)}}
    \TrinaryInfC{$\Gtopt{\Gamma};
    \Gtopt{\Delta_1 \setminus \Delta'_1}
            \cat 
            \dual{\prop_{k+1}}:\btout{\widetilde M_1} \tinact
            \cat \wtd z : \Gtopt{C}
            \proves \bbout{u_i}{\wtd z}
            \propout{k+1}{\widetilde y'} \inact \hastype \Proc$}
    \DisplayProof
\end{align}
\begin{align}
\label{pi:pt:stvalue-2}
\AxiomC{\eqref{pi:pt:uisend-2}}
\AxiomC{}
\LeftLabel{\scriptsize \textsc{(PolyVar)}}
\UnaryInfC{$\Gtopt{\Gamma};
\Delta_1 \setminus \Delta_1' \cat \wtd z:\Gtopt{C}
          \proves
            \widetilde y \hastype \widetilde M$}
\LeftLabel{\scriptsize \textsc{(PolyRcv)}}
\BinaryInfC{$\Gtopt{\Gamma_1};
        \Theta'
       \proves \propinp{k}{\widetilde y}
        \bbout{u_i}{\wtd z} \propout{k+1}{\widetilde y'} \inact \hastype \Proc$}
\DisplayProof
\end{align}

The following tree proves this case:
\begin{align}
  \label{pi:pt:output1-2}
  \AxiomC{\eqref{pi:pt:stvalue-2}}
  \AxiomC{\eqref{pi:eq:output-ih-1-2}}
  \LeftLabel{\scriptsize}
  \UnaryInfC{$\Gtopt{\Gamma_1};\Gtopt{\Delta'_1} \cat \Theta_1 \proves
  \Bopt{k+1}{\tilde y'}{P'} \hastype \Proc$}
   \LeftLabel{\scriptsize \textsc{(Par)}}
  \BinaryInfC{$\Gtopt{\Gamma_1};\Gtopt{\Delta'_1} \cat \Theta
                \proves \Bopt{k}{\tilde y}{\bout{u_i}{z}P'} \hastype \Proc$}
  \DisplayProof
\end{align}
We remark that if $u_i \notin \fn{P'}$ we need to apply 
 \Cref{top:lem:strength}  with $u_i$ to the 
right assumption of \eqref{pi:pt:output1-2} before applying \eqref{pi:eq:output-ih-1-2}. 
This concludes the analysis for the output case $P = \bout{u_i}{z}P'$.

\item Case $P = \news{s:C}P'$. We distinguish two sub-cases:
\rom{1} $C = S$ and \rom{2} $C = \chtype{C}$. 
First, we $\alpha$-convert $P$ as follows:
\begin{align*}
  P \equiv_{\alpha} \news{s_1:C}P'\subst{s_1 \dual{s_1}}{s \dual{s}}
\end{align*}

We consider sub-case \rom{1} first. For this case Rule~\textsc{ResS} can be applied:
\begin{align}
  \label{pi:pt:restritr}
  \AxiomC{$\Gamma;\Delta \cat s_1:S \cat \dual {s_1}:\dual S
              \proves P'\subst{s_1 \dual{s_1}}{s \dual{s}} \hastype \Proc$}
  \LeftLabel{\scriptsize \textsc{(ResS)}}
  \UnaryInfC{$\Gamma;\Delta \proves
            \news{s_1:S}P'\subst{s_1 \dual{s_1}}{s \dual{s}} \hastype \Proc$}
  \DisplayProof
\end{align}

By IH on the assumption of \eqref{pi:pt:restritr} we have:
\begin{align}
  \label{pi:eq:restr-ih}
  \Gtopt{\Gamma \setminus \wtd x};\Gtopt{\Delta \setminus \wtd x \cat s_1:S \cat \dual {s_1}:\dual S}
              \cat \Theta_1 \proves
              \Bopt{k}{\tilde y}{P'\subst{s_1 \dual{s_1}}{s \dual{s}}} \hastype \Proc
\end{align}
\noindent where 
$\widetilde x \subseteq \fn{P'}$
such that $s_1,\dual{s_1} \notin \wtd x$ and 
$\wtd y$ such that $\indexed{\wtd y}{\wtd x}{\Gamma,\Delta}$.
%where $\Delta' = \Delta \cat s_1:S \cat \dual {s_1}:\dual S$.
Also, $\balan{\Theta_1}$ with
$$
\dom{\Theta_1}=\{\prop_k,\ldots,\prop_{k+\lenHOopt{P'}-1}\}
\cup \{\dual{\prop_{k+1}},\ldots,\dual{\prop_{k+\lenHOopt{P'}-1}}\}
$$
and $\Theta_1(\prop_k) =
\btinp{\widetilde M} \tinact$ with $\widetilde M =
(\Gtopt{\Gamma}\cat\Gtopt{\Delta})(\widetilde y)$.

Note that we take $\Theta = \Theta_1$ since $\lenHOopt{P}=\lenHOopt{P'}$. By \Cref{pi:def:typesdecomp,pi:def:typesdecompenv} and~\eqref{pi:eq:restr-ih}, we know that:
\begin{align}
  \label{pi:eq:restr-ih1}
  \Gtopt{\Gamma \setminus \wtd x};\Gtopt{\Delta \setminus \wtd x} \cat \widetilde s:\Gtopt{S} \cat	\dual{\widetilde s}:
\Gtopt{\dual S} \proves \Bopt{k}{\tilde y}{P'\subst{s_1 \dual{s_1}}{s \dual{s}}}
  \hastype \Proc
\end{align}
\noindent where $\widetilde s = (s_1,\ldots,s_{\len{\Gtopt{S}}})$ and
$\dual{\widetilde s} = (\dual{s_1},\ldots,\dual{s_{\len{\Gtopt{S}}}})$.
By \Cref{pi:t:bdowncorec}, we have:
\begin{align*}
  \Bopt{k}{\tilde y}{\news{s}P'} = \news{\widetilde s : \Gtopt{S}}
                              \Bopt{k}{\tilde y}{P'\subst{s_1 \dual{s_1}}{s \dual{s}}}
\end{align*}

The following tree proves this sub-case:
\begin{align}
  \label{pi:pt:restr1}
  \AxiomC{\eqref{pi:eq:restr-ih1}}
  \LeftLabel{\scriptsize \textsc{(PolyResS)}}
  \UnaryInfC{$\Gtopt{\Gamma \setminus \wtd x};\Gtopt{\Delta \setminus \wtd x} \proves
              \news{\widetilde s : \Gtopt{S}}
              \Bopt{k}{\tilde y}{P'\subst{s_1 \dual{s_1}}{s \dual{s}}} \hastype \Proc$}
  \DisplayProof
\end{align}

We now consider sub-case \rom{2}. Similarly to sub-case \rom{1} we first $\alpha$-convert $P$ as
follows:
\begin{align*}
  P \equiv_\alpha \news{s_1}{P'\subst{s_1}{s}}
\end{align*}

For this sub-case Rule~\textsc{Res} can be applied:
\begin{align}
  \label{pi:eq:restritr2}
  \AxiomC{$\Gamma \cat s_1:\chtype{C};\Delta \proves
          P'\subst{s_1}{s} \hastype \Proc$}
  \LeftLabel{\scriptsize \textsc{(Res)}}
  \UnaryInfC{$\Gamma;\Delta \proves \news{s_1}P'\subst{s_1}{s}
                  \hastype \Proc$}
  \DisplayProof
\end{align}

By IH on the
first assumption of \eqref{pi:eq:restritr2} we have:
\begin{align}
  \label{pi:eq:restr-ih2}
  \Gtopt{\Gamma \setminus \wtd x \cat s_1:\chtype{C}};\Gtopt{\Delta \setminus \wtd x}
      \cat \Theta_1
      \proves \Bopt{k}{\tilde y}{P'\subst{s_1}{s}} \hastype \Proc
\end{align}

\noindent where 
$\widetilde x \subseteq \fn{P'}$
such that $s_1 \notin \wtd x$ and 
$\wtd y$ such that $\indexed{\wtd y}{\wtd x}{\Gamma,\Delta}$.
Also, $\balan{\Theta_1}$ with
$$
\dom{\Theta_1}=\{\prop_k,\ldots,\prop_{k+\lenHOopt{P'}-1}\}
\cup \{\dual{\prop_{k+1}},\ldots,\dual{\prop_{k+\lenHOopt{P'}-1}}\}
$$
and $\Theta_1(\prop_k) =
\btinp{\widetilde M} \tinact$ with $\widetilde M =
(\Gtopt{\Gamma}\cat\Gtopt{\Delta})(\widetilde y)$.

Here we also take $\Theta = \Theta_1$ since  $\lenHOopt{P}=\lenHOopt{P'}$.
We notice that by \Cref{pi:def:typesdecomp,pi:def:typesdecompenv} and
\eqref{pi:eq:restr-ih2}:
\begin{align}
  \label{pi:eq:restr-ih3}
  \Gtopt{\Gamma \setminus \wtd x} \cat s_1:\Gtopt{\chtype{C}};\Gtopt{\Delta \setminus \wtd x}
      \cat \Theta_1
      \proves \Bopt{k}{\tilde y}{P'\subst{s_1}{s}} \hastype \Proc
\end{align}

By \Cref{pi:t:bdowncorec}, we have:
\begin{align*}
  \Bopt{k}{\tilde y}{\news{s}P'} = \news{s_1 : \Gtopt{\chtype{C}}}
                              \Bopt{k}{\tilde y}{P'\subst{s_1}{s}}
\end{align*}
%\noindent where $\widetilde s = s_1$ since $\len{\Gtopt{\chtype{U}}}=1$.
The following tree proves this sub-case:
\begin{align}
  \AxiomC{\eqref{pi:eq:restr-ih3}}
  \LeftLabel{\scriptsize \textsc{(PolyRes)}}
  \UnaryInfC{$\Gtopt{\Gamma \setminus \wtd x};\Gtopt{\Delta \setminus \wtd x}\cat \Theta
                  \proves \news{s_1 : \Gtopt{\chtype{C}}}
                              \Bopt{k}{\tilde y}{P'\subst{s_1}{s}} \hastype \Proc$}
  \DisplayProof
\end{align}

\item Case $P = Q \Par R$. For this case only Rule~\textsc{Par} can be applied:
\begin{align}
  \label{pi:pt:compitr}
  \AxiomC{$\Gamma;\Delta_1 \proves Q \hastype \Proc$}
  \AxiomC{$\Gamma;\Delta_2 \proves R \hastype \Proc$}
  \LeftLabel{\scriptsize \textsc{(Par)}}
  \BinaryInfC{$\Gamma;\Delta_1 \cat \Delta_2
                       \proves Q \Par R \hastype \Proc$}
  \DisplayProof
\end{align}

By IH on the first assumption of \eqref{pi:pt:compitr} we have:
\begin{align}
  \label{pi:eq:comp-ih-1}
  \Gtopt{\Gamma'_1};\Gtopt{\Delta'_1} \cat \Theta_1 \proves
\Bopt{k+1}{\tilde y_1}{Q} \hastype \Proc
\end{align}

\noindent where 
$\widetilde x_1 \subseteq \fn{Q}$ and 
$\widetilde y_1$ such that $\indexed{\widetilde y_1}{\wtd x_1}{\Gamma,\Delta_1}$. 
Also, $\Gamma'_1 = \Gamma \setminus \wtd x_1$, 
$\Delta'_1 = \Delta_1 \setminus \wtd x_1$, 
and $\balan{\Theta_1}$ with
$$\dom{\Theta_1}=\{\prop_{k+1},\ldots,\prop_{k+\lenHOopt{Q}}\}
\cup \{\dual{\prop_{k+2}},\ldots,\dual{\prop_{k+\lenHOopt{Q}}}\}
$$
and
$\Theta_1(\prop_{k+1}) = \btinp{\widetilde M_1} \tinact$ with $\widetilde M_1 =
(\Gtopt{\Gamma}\cat\Gtopt{\Delta_1})(\widetilde y_1)$.

By IH on the second assumption of \eqref{pi:pt:compitr} we have:
\begin{align}
  \label{pi:eq:comp-ih-2}
  \Gtopt{\Gamma'_2};\Gtopt{\Delta_2'} \cat \Theta_2 \proves
\Bopt{k+\degree+1}{\tilde y_2}{R} \hastype \Proc
\end{align}

\noindent where 
$\widetilde x_2 \subseteq \fn{R}$ and 
$\widetilde y_2$ such that $\indexed{\widetilde y_2}{\wtd x_2}{\Gamma,\Delta_2}$
and  
$\degree = |Q|$. Also, 
$\Gamma'_2 = \Gamma \setminus \wtd x_2$, 
$\Delta'_2 = \Delta_2 \setminus \wtd x_2$
and $\balan{\Theta_2}$ with
$$\dom{\Theta_2}=\{\prop_{k+\degree+1},\ldots,\prop_{k+\degree+\lenHOopt{R}}\}
\cup \{\dual{\prop_{k+\degree+2}},\ldots,\dual{\prop_{k+\degree+\lenHOopt{R}}}\}$$ and
$\Theta_2(\prop_{k+\degree+1}) = \btinp{\widetilde M_2} \tinact$ with $\widetilde M_2 =
(\Gtopt{\Gamma}\cat\Gtopt{\Delta_2})(\widetilde y_2)$.

We define $\wtd x = \wtd x_1 \cup \wtd x_2$. 
We may notice that $\wtd x \subseteq \fn{P}$ since 
$\fn{P} = \fn{Q} \cup \fn{R}$. 
Accordingly, we define $\wtd y = \wtd y_1, \wtd y_2$. 
By definition, $\indexed{\wtd y}{\wtd x}{\Gamma, \Delta_1 \cat \Delta_2}$ holds. 
Further, let $\widetilde M = (\Gtopt{\Gamma},\Gtopt{\Delta_1\cat\Delta_2})(\widetilde y)$. 
%$\wtd x_1 \subseteq \fn{Q}$ and $\wtd x_2 \subseteq \fn{R}$. 
%
%Let $\widetilde x = \fv{P}$ and $\widetilde M = (\Gtopt{\Gamma},\Gtopt{\Lambda_1\cat\Lambda_2})(\widetilde x)$.
%We may notice that $\widetilde M_i \subseteq \widetilde M$ for $i \in \{1,2\}$.
We define $\Theta = \Theta_1 \cat \Theta_2 \cat \Theta'$ where:
\begin{align*}
  \Theta' = \prop_k:\btinp{\widetilde M} \tinact \cat
              \dual{\prop_{k+1}}:\btout{\widetilde M_1} \tinact
              \cat
              \dual{\prop_{k+\degree+1}}:\btout{\widetilde M_2} \tinact
\end{align*}

By construction $\Theta$ is balanced since $\Theta(\prop_{k+1}) \dualof \Theta(\dual{\prop_{k+1}})$, $\Theta(\prop_{k+\degree+1}) \dualof \Theta(\dual{\prop_{k+\degree+1}})$, and
     $\Theta_1$ and $\Theta_2$ are balanced.

By \Cref{pi:t:bdowncorec} we have:
\begin{align*}
    \Bopt{k}{\tilde y}{Q \Par R} =\propinp{k}{\widetilde y} \propout{k+1}{\widetilde y_1}
  \propout{k+\degree+1}{\widetilde y_2} \inact \Par
  \Bopt{k+1}{\tilde y_1}{Q} \Par \Bopt{k+\degree+1}{\tilde y_2}{R}
\end{align*}

We may notice that $\wtd y_1 = \fnb{Q}{\widetilde y}$
and $\wtd y_2 = \fnb{R}{\widetilde y}$ hold
by the construction of $\wtd y$. 
Let $\Gamma_1 = \Gamma \setminus \widetilde x$. We shall prove the following judgment:
\begin{align*}
\Gtopt{\Gamma_1};\Gtopt{\Delta'_1 \cat \Delta'_2}
              \cat \Theta \proves
              \propinp{k}{\widetilde y} \bout{\dual {\prop_{k+1}}}{\widetilde y_1}
                           \bout{\dual {\prop_{k+\degree+1}}}{\widetilde y_2} \inact \Par
                        \Bopt{k+1}{\tilde y_1}{Q} \Par \Bopt{k+\degree+1}{\tilde y_2}{R} \hastype \Proc
\end{align*}

To type  $\propinp{k}{\widetilde y} \propout{k+1}{\widetilde y_1}
  \propout{k+\degree+1}{\widetilde y_2} \inact$, we use some auxiliary derivations:
\begin{align}
  \label{pi:pt:compsend2}
  \AxiomC{}
\LeftLabel{\scriptsize \textsc{(Nil)}}
  \UnaryInfC{$\Gtopt{\Gamma};\es \proves \inact$}
  \LeftLabel{\scriptsize \textsc{(End)}}
  \UnaryInfC{$\Gtopt{\Gamma};\dual{\prop_{k+\degree+1}}:\tinact \proves \inact$}
  \AxiomC{}
  \LeftLabel{\scriptsize \textsc{(PolyVar)}}
  \UnaryInfC{$\Gtopt{\Gamma};\Gtopt{\Delta_2} \cat 
              \proves \widetilde z \hastype \widetilde M_2$}
  \LeftLabel{\scriptsize \textsc{(PolySend)}}
  \BinaryInfC{$\Gtopt{\Gamma};\Gtopt{\Delta'_2} \cat 
              \dual{\prop_{k+\degree+1}}:\btout{\widetilde M_2}\tinact
              \proves \propout{k+\degree+1}{\widetilde z} \inact \hastype \Proc$}
  \LeftLabel{\scriptsize \textsc{(End)}}
  \UnaryInfC{$\Gtopt{\Gamma};\Gtopt{\Delta'_2} \cat 
              \dual{\prop_{k+\degree+1}}:\btout{\widetilde M_2}\tinact
              \cat \dual{\prop_{k+1}}:\tinact
              \proves \propout{k+\degree+1}{\widetilde z} \inact \hastype \Proc$}
  \DisplayProof
\end{align}

\begin{align}
  \label{pi:pt:compsend1}
  \AxiomC{\eqref{pi:pt:compsend2}}
  \AxiomC{}
  \LeftLabel{\scriptsize \textsc{(PolyVar)}}
  \UnaryInfC{$\Gtopt{\Gamma};\Gtopt{\Delta_1}
              \proves \widetilde y_1 \hastype \widetilde M_1$}
  \LeftLabel{\scriptsize \textsc{(PolySend)}}
  \BinaryInfC{\begin{tabular}{c}
  $\Gtopt{\Gamma};\Gtopt{\Delta'_1\cat\Delta'_2} \cat             \dual{\prop_{k+1}}:\btout{\widetilde M_1}\tinact \cat
               \dual{\prop_{k+\degree+1}}:\btout{\widetilde M_2}\tinact
               \proves \propout{k+1}{\widetilde y_1}
                 \propout{k+\degree+1}{\widetilde y_2} \inact \hastype \Proc$
                 \end{tabular}}
     \LeftLabel{\scriptsize \textsc{(End)}}
 \UnaryInfC{\begin{tabular}{c}
  $\Gtopt{\Gamma};\Gtopt{\Delta'_1\cat\Delta'_2} \cat
              \dual{\prop_{k+1}}:\btout{\widetilde M_1}\tinact \cat
               \dual{\prop_{k+\degree+1}}:\btout{\widetilde M_2}\tinact
               \cat \prop_k:\tinact
               \proves \propout{k+1}{\widetilde y_1}
                 \propout{k+\degree+1}{\widetilde y_2} \inact \hastype \Proc$
                 \end{tabular}}
  \DisplayProof
\end{align}

\begin{align}
  \label{pi:pt:compinput1}
  \AxiomC{\eqref{pi:pt:compsend1}}
  \AxiomC{}
  \LeftLabel{\scriptsize \textsc{(PolyVar)}}
  \UnaryInfC{$\Gtopt{\Gamma};\Gtopt{\Delta'_1\cat\Delta'_2}
                    \proves \widetilde x \hastype \widetilde M$}
  \LeftLabel{\scriptsize \textsc{(PolyRcv)}}
  \BinaryInfC{$\Gtopt{\Gamma_1};\Theta' \proves
                       \propinp{k}{\widetilde y}
                       \propout{k+1}{\widetilde y_1}
                          \propout{k+\degree+1}{\widetilde y_2} \inact \hastype \Proc$}
  \DisplayProof
\end{align}

\begin{align}
  \label{pi:pt:comp-par-1}
  \AxiomC{\eqref{pi:eq:comp-ih-1}}
  \LeftLabel{\scriptsize (\Cref{top:lem:weaken}) with $\Gamma_1' \setminus \Gamma_1$}
  \UnaryInfC{$\Gtopt{\Gamma_1};\Gtopt{\Delta_1} \cat \Theta_1 \proves
                    \Bopt{k+1}{\tilde y_1}{Q} \hastype \Proc$}
    \DisplayProof
\end{align}

\begin{align}
  \label{pi:pt:comp-par-2}
  \AxiomC{\eqref{pi:eq:comp-ih-2}}
  \LeftLabel{\scriptsize (\Cref{top:lem:weaken}) with $\Gamma_2' \setminus \Gamma_1$}
  \UnaryInfC{$\Gtopt{\Gamma_1};\Gtopt{\Delta'_2} \cat \Theta_2 \proves
                    \Bopt{k+\degree+1}{\tilde y_2}{R} \hastype \Proc$}
    \DisplayProof
\end{align}

\begin{align}
  \label{pi:pt:par}
  \AxiomC{\eqref{pi:pt:comp-par-1}}
  \AxiomC{\eqref{pi:pt:comp-par-2}}
  \LeftLabel{\scriptsize \textsc{(Par)}}
  \BinaryInfC{$\Gtopt{\Gamma_1};\Gtopt{\Delta'_1 \cat \Delta'_2} \cat
              \Theta_1 \cat \Theta_2 \proves
              \Bopt{k+1}{\tilde y}{Q} \Par \Bopt{k+\degree+1}{\tilde z}{R} \hastype \Proc$}
   \DisplayProof
\end{align}

The following tree proves this case:
\begin{align*}
  \AxiomC{\eqref{pi:pt:compinput1}}
  \AxiomC{\eqref{pi:pt:par}}
  \LeftLabel{\scriptsize \textsc{(Par)}}
  \BinaryInfC{
  \begin{tabular}{c}
  $\Gtopt{\Gamma_1};\Gtopt{\Delta'_1 \cat \Delta'_2}
              \cat \Theta \proves \propinp{k}{\widetilde y}
              \bout{\dual {\prop_{k+1}}}{\widetilde y_1}
                           \bout{\dual {\prop_{k+\degree+1}}}{\widetilde y_2} \inact \Par
                        \Bopt{k+1}{\tilde y_1}{Q} \Par \Bopt{k+\degree+1}{\tilde y_2}{R} \hastype \Proc$
                        \end{tabular}
                        }
  \DisplayProof
\end{align*}

\item Case $P = \recp{X}P'$. The only rule that can be 
applied here is \textsc{Rec}:
\begin{align}
\label{pi:pt:recInv}
  \AxiomC{$\Gamma \cat X : \Delta;\Delta \proves P' \hastype \Proc$} 
  \UnaryInfC{$\Gamma;\Delta \proves \recp{X}P'$} 
  \DisplayProof
\end{align}

We remark that by \eqref{pi:pt:recInv} we know 
$x:S \in \Delta \implies \traux{S}$.
Then, by applying \Cref{pi:lem:brec} on the assumption of \eqref{pi:pt:recInv} we have: 
\begin{align}
\label{pi:pt:recIH}
\Gtopt{\Gamma \setminus \wtd x};\Theta 
\proves \Brecpi{k+1}{\tilde y'}{P'}_g \hastype \Proc	
\end{align}

% \noindent where $g = \{X \mapsto \wtd m\}$
% with $\wtd n = \fn{P'}$ and $\wtd m = \bname{\wtd n : \wtd S}$.w
We take  $\wtd x' = \fn{P'}$, so we have 
$\Delta \setminus x' = \emptyset$. Further, 
$\wtd y'$ is such that 
$\indexed{\wtd y'}{\wtd x'}{\Gamma,\Delta}$. 
Let  $\Theta' = \thetaR' \cat \thetax(g)$ with 
$\thetax(g) =  {\prop^r_{X}} : \chtype{\wtd M'}$ 
with $\wtd M' = (\Gtopt{\Gamma},\Gtopt{\Delta})(\wtd y')$.
Further,  $\wtd x' \subseteq P'$ and $\wtd y'$ such that 
$\Delta \setminus x' = \emptyset$ and 
$\indexed{\wtd y'}{\wtd x'}{\Gamma,\Delta}$. 
Also, $\balan{\thetaR'}$ with 
$$\dom{\thetaR'} = \{\prop^r_{k+1},\ldots,\prop^r_{k+\lenHOopt{P'}} \} \cup 
\{\dual{\prop^r_{k+1}},\ldots,\dual{\prop^r_{k+\lenHOopt{P'}}} \}$$ where
$\thetaR' (\prop_{k+1})=\prop_k:\btinp{\wtd M'}\tinact$. 

Let $l = \lenHOopt{P'}$ and $\wtd z$ 
such that $\len{\wtd y'} = \len{\wtd z}$.
% We take $\wtd x = \wtd x'$ since $\fn{P} = \fn{P'}$.
% As $\Delta \setminus \wtd x = \emptyset$ we 
% have that $\wtd x' \supseteq \fn{P} \setminus \fs{P}$. 
% so we have $\wtd y'$ is properly defined. 
Further, let $\wtd x \subseteq \fn{P}$  
and $\wtd y = \bname{\wtd x : \wtd S}$. 
By definition we have $\wtd x \subseteq \wtd x'$. 
% We remark that by $\wtd x' = \fn{P}$ that 
% $\wtd y'$ is properly defined for this case. 
By \Cref{pi:t:bdowncorec}, we have: 
\begin{align*}
\Bopt{k}{\tilde y}{P} &=
      \news{\prop^r_X}
      (\propinp{k}{\wtd y}
      \proproutk{r}{k+1}{\wtd y'}
      \recp{X}
      \proprinpk{r}{X}{\wtd z}
      \proproutk{r}{k+1}{\wtd z}X
    \Par 
     \Brecpi{k+1}{\tilde y'}{P'}_g)
\end{align*}
Let $\thetaR = \thetaR' \cdot \thetaR''$ where: 
$$\thetaR'' =  \dual{\prop^r_{k}} : \trec{t}\btinp{\wtd M}\tvar{t}
\cdot \dual{\prop^r_{k+1}} : \trec{t}\btout{\wtd M'}\tvar{t}$$

% Let $\Theta = \Theta' \cdot \thetaR''$ where: 
% $$\thetaR'' = 
% \prop^r_{k} \cdot
% \prop^r_{X} : \trec{t}\btinp{\wtd M}\tvar{t} 
% $$ 
% \noindent where $\wtd M = (\Gtopt{\Gamma}\cat\Gtopt{\Delta})(\widetilde y)$. 
%  By construction $\balan{\thetaR}$. 

% We know $\wtd x \subseteq \fn{P}$ since $\fn{P} = \fn{P'}$.
% We take $\wtd x = \wtd x'$ since $\fn{P} = \fn{P'}$.
% As $\Delta \setminus \wtd x = \emptyset$ we 
% have that $\wtd x \supseteq \fn{P} \setminus \fs{P}$. 
We shall prove the following judgment: 
\begin{align} 
\Gtopt{\Gamma \setminus \wtd x}; \Gtopt{\Delta \setminus \wtd x}
\cdot \envR \proves 	\Bopt{k}{\tilde y}{P}
\hastype \Proc
\end{align}

Let $\thetax' = \prop^r_{X} : \trec{t}\btinp{\wtd M'}\tvar{t}
\cdot \dual{\prop^r_{X}} : \trec{t}\btout{\wtd M'}\tvar{t}$. 
We use some auxiliary sub-trees: 

%	\begin{align}
%		\AxiomC{} 
%		\UnaryInfC{}
%		\DisplayProof	
%	\end{align}
%	
%	\begin{align}
%	\label{pi:pt:rec4}
%		\AxiomC{} 
%		\LeftLabel{\scriptsize RVar}
%		\UnaryInfC{$\Gtopt{\Gamma \setminus \wtd x} \cat X : \Theta''; 
%		\Theta''
%		\proves 
%        	X
%		$}
%		\DisplayProof	
%	\end{align}

\begin{align}
\label{pi:pt:rec4}
\AxiomC{} 
\LeftLabel{\scriptsize \textsc{(RVar)}}
\UnaryInfC{$\Gtopt{\Gamma \setminus \wtd x} \cat X : \Theta''; 
\Theta''
\proves 
      X
      \hastype \Proc
$}
\AxiomC{} 
\LeftLabel{\scriptsize \textsc{(PolyVar)}}
\UnaryInfC{$\Gtopt{\Gamma \setminus \wtd x} \cat X : \Theta'';
\Gtopt{\Delta} \proves 
\wtd z \hastype \wtd M$} 
\LeftLabel{\scriptsize \textsc{(Send)}}
\BinaryInfC{$\Gtopt{\Gamma \setminus \wtd x} \cat X : \Theta'';
\Gtopt{\Delta} \cat 
\Theta''
\proves 
\proproutk{r}{k+1}{\wtd z}X
      \hastype \Proc
$}	
\DisplayProof
\end{align}
\begin{align}
\label{pi:pt:rec3}
\AxiomC{\eqref{pi:pt:rec4}}
\AxiomC{} 
\LeftLabel{\scriptsize \textsc{(PolyVar)}}
\UnaryInfC{$\Gtopt{\Gamma \setminus \wtd x}; \Gtopt{\Delta} 
\proves \wtd z \hastype \wtd M$}
\LeftLabel{\scriptsize \textsc{(PolyRcv)}}
\BinaryInfC{$\Gtopt{\Gamma \setminus \wtd x} \cat X : \Theta'';
\Theta'' \proves 
\binp{\prop^r_{X}}{\wtd z}
\proproutk{r}{k+1}{\wtd z}X
      \hastype \Proc
$} 
\LeftLabel{\scriptsize \textsc{(Rec)}}
\UnaryInfC{\begin{tabular}{c}
$
\Gtopt{\Gamma \setminus \wtd x}; 
\Theta''
\proves$ 
% \\ 
$
      \recp{X}\binp{\prop^r_{X}}{\wtd z}
      \proproutk{r}{k+1}{\wtd z}X
      \hastype \Proc$
      \end{tabular}}
\DisplayProof	
\end{align}
\noindent where $\Theta'' = \dual{\prop^r_{k+1}} : \trec{t}\btout{\wtd M}\tvar{t}
\cat \prop^r_{X} : \trec{t}\btinp{\wtd M}\tvar{t}$.
\begin{align}
\label{pi:pt:rec2}
\AxiomC{\eqref{pi:pt:rec3}} 
\AxiomC{} 
\LeftLabel{\scriptsize \textsc{(PolyVar)}}
\UnaryInfC{$\Gtopt{\Gamma \setminus \wtd x}; 
\Gtopt{\Delta} \proves 
\wtd y' \hastype \wtd M'$} 
\LeftLabel{\scriptsize \textsc{(PolySend)}}
\BinaryInfC{
\begin{tabular}{c}
$
\Gtopt{\Gamma \setminus \wtd x}; 
\Gtopt{\Delta} \cat 
\dual{\prop^r_{k+1}} : \trec{t}\btout{\wtd M}\tvar{t}
\cat \prop^r_{X} : \trec{t}\btinp{\wtd M}\tvar{t} 
\proves 
      \bout{\prop^r_{k+1}}{\wtd y}
      \recp{X}\binp{\prop^r_{X}}{\wtd z}
      \proproutk{r}{k+1}{\wtd z}X
      \hastype \Proc$
      \end{tabular}}
\DisplayProof
\end{align}
\begin{align}
\label{pi:pt:rec1}
\AxiomC{\eqref{pi:pt:rec2}} 
\AxiomC{} 
\LeftLabel{\scriptsize \textsc{(PolyVar)}}
\UnaryInfC{$\Gtopt{\Gamma \setminus \wtd x};
\wtd y : \wtd M 
\proves \wtd y \hastype \wtd M$} 
\LeftLabel{\scriptsize \textsc{(PolyRcv)}}
\BinaryInfC{
\begin{tabular}{c}
$\Gtopt{\Gamma \setminus \wtd x}; 
\Gtopt{\Delta \setminus \wtd x} \cdot 
\thetaR'' \cdot \prop^r_{X} : \trec{t}\btinp{\wtd M}\tvar{t}
\proves \qquad \qquad \propinp{k}{\wtd y}
\proproutk{r}{k+1}{\wtd y'}
\recp{X}
\proprinpk{r}{X}{\wtd z}
\proproutk{r}{k+1}{\wtd z}X
      \hastype \Proc$
      \end{tabular}
} 
\DisplayProof
\end{align}

The following tree proves this case: 
\begin{align}
\AxiomC{\eqref{pi:pt:rec1}} 
%		\AxiomC{$\Gtopt{\Gamma \setminus \wtd x};  
%		\thetaR 
%		\proves
%		\Brecpi{k+1}{\tilde y'}{P'}$} 
\AxiomC{\eqref{pi:pt:recIH}} 
\LeftLabel{\scriptsize \textsc{(Par)}}
\BinaryInfC{
\begin{tabular}{c}
$
\Gtopt{\Gamma \setminus \wtd x}; 
\Gtopt{\Delta \setminus \wtd x} \cdot \thetaR \cdot \thetax'
\proves 
\propinp{k}{\wtd y}
      \proproutk{r}{k+1}{\wtd y'} 
      \recp{X}\binp{\prop^r_{X}}{\wtd z}
      \proproutk{r}{k+1}{\wtd z}X
    \Par 
    \Brecpi{k+1}{\tilde y}{P'}
\hastype \Proc$
\end{tabular}
} 
\LeftLabel{\scriptsize \textsc{(ResS)}}
\UnaryInfC{
\begin{tabular}{c}
$
\Gtopt{\Gamma \setminus \wtd x}; 
\Gtopt{\Delta \setminus \wtd x} \cdot \thetaR
\proves 
 \news{\prop^r_X}(\propinp{k}{\wtd y}
\proproutk{r}{k+1}{\wtd y'}
    \recp{X}\binp{\prop^r_{X}}{\wtd z}
    \proproutk{r}{k+1}{\wtd z}X
  \Par 
  \Brecpi{k+1}{\tilde y'}{P'})
\hastype \Proc$
\end{tabular}
}
\DisplayProof	
\end{align}

\item Case $P = \bout{r}{z}P'$ when $\traux{r}$. 
For this case Rule~\textsc{Send} can be applied:
\begin{align}
\label{pi:pt:rec-outputitr}
\AxiomC{$\Gamma;\Delta \cat  r:S' \proves P' \hastype \Proc$}
\AxiomC{$\Gamma;\Delta_z \proves z \hastype C$}
\AxiomC{}
\LeftLabel{\scriptsize \textsc{(Send)}}
\TrinaryInfC{$\Gamma;\Delta \cat 
r : S \cat 
\Delta_z
              \proves \bout{r}{z}P' \hastype \Proc$}
\DisplayProof
\end{align}
\noindent where $S = \btout{C}S'$.

Then, by IH on the first assumption of \eqref{pi:pt:rec-outputitr} we have:
\begin{align}
  \label{pi:eq:rec-output-ih-1}
  \Gtopt{\Gamma'_1};\Gtopt{\Delta'_1 } \cat \Theta_1 \proves
  \Bopt{k+1}{\tilde y'}{P'} \hastype \Proc
\end{align}
\noindent where $\widetilde x' \subseteq \fn{P'}$ 
such that $r \in \wtd x'$ 
and $\widetilde y'$ such that 
$\indexed{\widetilde y'}{\widetilde x'}{\Gamma,\Delta \cat r:S}$. 
By this follows that $(r_1,\ldots,r_{\len{\Gtopt{S}}}) \subseteq \wtd y'$. 
Also, $\Gamma'_1 =
\Gamma \setminus \widetilde x'$, 
$\Delta'_1 = \Delta \setminus \widetilde y'$, and $\balan{\Theta_1}$
with
$$
\dom{\Theta_1}=\{\prop_{k+1},\ldots,\prop_{k+\lenHOopt{P'}}\}\cup
\{\dual{\prop_{k+2}},\ldots,\dual{\prop_{k+\lenHOopt{P'}}}\}
$$
and $\Theta_1(\prop_{k+1})=\btinp{\widetilde M_1}\tinact$ where
$\widetilde M_1 =
(\Gtopt{\Gamma},\Gtopt{\Delta, r:S})(\wtd y')$.

By applying \Cref{pi:lem:varbreak} on the
 assumption of \eqref{pi:pt:outputitr} we have:
 \begin{align}
   \label{pi:eq:rec-outputih2}
   \Gtopt{\Gamma};\Gtopt{\Delta_z} \proves \widetilde z \hastype \Gtopt{C}
 \end{align}

We assume $\wtd x = \wtd x', z$. Since 
$\wtd x' \subseteq \fn{P'}$ follows that  
$\widetilde x \subseteq \fn{P}$.
Let $\wtd y = \wtd y', \wtd z$ where 
$\len{\wtd z} = \len{\Gtopt{C}}$. 
By \Cref{pi:def:indexed} it follows that  
$\indexed{\wtd y}{\wtd x}{\Gamma, \Delta \cat r : S \cat \Delta_z}$. 
We define $\Theta = \Theta_1 \cat \Theta'$, where:
\begin{align*}
  \Theta' = \prop_{k}: \btinp{\widetilde M} \tinact \cat
  \dual{\prop_{k+1}}:\btout{\widetilde M_1} \tinact
\end{align*}

\noindent with $\widetilde M = (\Gtopt{\Gamma}\cat \Gtopt{\Delta \cat r:S \cat \Delta_z})(\widetilde y)$. By \Cref{pi:def:sizeproc}, we know $\lenHOopt{P} =\lenHOopt{P'} + 1$, so
$$
\dom{\Theta}=\{\prop_k,\ldots,\prop_{k+\lenHOopt{P}-1}\} \cup
\{\dual{\prop_{k+1}},\ldots,\dual{\prop_{k+\lenHOopt{P}-1}}\}
$$
and $\Theta$ is balanced since $\Theta(\prop_{k+1}) 
\dualof \Theta(\dual{\prop_{k+1}})$ and $\Theta_1$ is 
balanced. 

By \Cref{pi:t:bdowncorec}, we have:
\begin{align*}
\Bopt{k}{\tilde y}{\bout{r}{z}P'} =
\propinp{k}{\widetilde y}
\bbout{r_{\indf{S}}}{\wtd z} \propout{k+1}{\widetilde y'}
\inact \Par
\Bopt{k+1}{\tilde y'}{P'}
\end{align*}
% \noindent where $\len{\wtd z}=\len{\Gtopt{C}}$. 
Let $\Gamma_1 = \Gamma \setminus \widetilde x = \Gamma'_1$ 
and $\Delta \setminus \wtd x = \Delta'_1$.
We shall prove the following judgment:
\begin{align}
\AxiomC{$\Gtopt{\Gamma_1}; \Gtopt{\Delta'_1} \cat \Theta \proves
\Bopt{k}{\tilde y}{\bout{r}{z}P'} \hastype \Proc$}
\DisplayProof
\end{align}
%   \noindent where we may notice that following holds: 

%   $$\Delta'_1 = \Delta \setminus \wtd y' = (\Delta_1 \cat u_i:\btout{C}S \cat \Delta_z) \setminus \widetilde y $$

Let $\Delta_1 =\Delta \cat r : S \cat \Delta_z$.
To type the left-hand side component of
$\Bopt{k}{\tilde y}{\bout{r}{z}P'}$
we use some auxiliary derivations:
\begin{align}
\label{pi:pt:rec-output3}
\AxiomC{}
\LeftLabel{\scriptsize \textsc{(Nil)}}
\UnaryInfC{$\Gtopt{\Gamma};\es \proves \inact \hastype \Proc$}
\LeftLabel{\scriptsize \textsc{(End)}}
\UnaryInfC{$\Gtopt{\Gamma};\dual{\prop_{k+1}} : \tinact \proves \inact \hastype \Proc$}
\AxiomC{}
\LeftLabel{\scriptsize \textsc{(PolyVar)}}
\UnaryInfC{$\Gtopt{\Gamma};\Gtopt{\Delta \cat r:S \setminus \Delta'_1}
      \proves \widetilde y' \hastype \widetilde M_1$}
\LeftLabel{\scriptsize \textsc{(PolySend)}}
\BinaryInfC{$\Gtopt{\Gamma};
      \dual{\prop_{k+1}}:\btout{\widetilde M_1} \tinact
        \cat \Gtopt{\Delta \cat r:S \setminus \Delta'_1}
      \proves \bout{\dual {\prop_{k+1}}}{\widetilde y'} 
      \hastype \Proc$}
\DisplayProof
\end{align}

\begin{align}
\label{pi:pt:rec-output2}
\AxiomC{\eqref{pi:pt:rec-output3}}
\AxiomC{\eqref{pi:eq:rec-outputih2}}
%  	\LeftLabel{\scriptsize \Cref{pi:lem:subst} with 
%  	$\subst{\tilde z_1}{\tilde z}$}
%  	\UnaryInfC{$\Gtopt{\Gamma};\Gtopt{\Delta_z \subst{\tilde z}{\tilde z_1}} \proves \widetilde z\hastype \Gtopt{C}$}
\LeftLabel{\scriptsize \textsc{(PolySend)}}
\BinaryInfC{\begin{tabular}{c}
      $\Gtopt{\Gamma};
      \Gtopt{\Delta_1 \setminus \Delta_1'}
            \cat
             \dual{\prop_{k+1}}:\btout{\widetilde M_1}\tinact
             \proves$ 
             $\bbout{r_{\indf{S}}}{\widetilde z}
               \propout{k+1}{\widetilde y'} \inact \hastype \Proc$
               \end{tabular}}
  \LeftLabel{\scriptsize \textsc{(End)}}
\UnaryInfC{\begin{tabular}{c}
      $\Gtopt{\Gamma};
      \Gtopt{\Delta_1 \setminus \Delta'_1}
             \cat 
             \dual{\prop_{k+1}}:\btout{\widetilde M_1}\tinact \cat \prop_k:\tinact 
             \proves$ 
             $\bbout{r_{\indf{S}}}{\wtd z}
               \propout{k+1}{\wtd y'} \inact \hastype \Proc$
               \end{tabular}}
\DisplayProof
\end{align}
% TODO continue updating proof here 
\begin{align}
\label{pi:pt:rec-output1}
\AxiomC{\eqref{pi:pt:rec-output2}}
\AxiomC{}
\LeftLabel{\scriptsize \textsc{(PolyVar)}}
\UnaryInfC{$\Gtopt{\Gamma};\Gtopt{\Delta_1 \setminus \Delta_1'} \proves
      \widetilde y \hastype \widetilde M$}
\LeftLabel{\scriptsize \textsc{(PolyRcv)}}
\BinaryInfC{\begin{tabular}{c}
      $\Gtopt{\Gamma_1}; 
             \Theta'
           \proves$ 
           $\propinp{k}{\widetilde y}
          \bbout{r_{\indf{S}}}{\wtd z}
            \propout{k+1}{\wtd y'} \inact
            \hastype \Proc$
            \end{tabular}}
\DisplayProof
\end{align}

We may notice that 
by \Cref{pi:def:typesdecompenv} and \Cref{pi:f:tdec} we have 
   $\Gtopt{\Delta_1 \setminus \Delta_1'}=\Gtopt{\Delta} \cat \wtd r : \Rtsopt{}{}{S} \cat 
   \Gtopt{\Delta_z} \setminus \Gtopt{\Delta_1'}$. Further, by \Cref{pi:lem:indexcor}
   we know $\Gtopt{\Delta_1}(r_{\indf{S}}) = \trec{t}\btout{\Gtopt{C}}\vart{t}$.

The following tree proves this case:
\begin{align}
\label{pi:pt:output1}
\AxiomC{\eqref{pi:pt:rec-output1}}
\AxiomC{\eqref{pi:eq:rec-output-ih-1}}
\UnaryInfC{$\Gtopt{\Gamma_1};\Gtopt{\Delta'_1} \cat
      \Theta_1 \proves
            \Bopt{k+1}{\tilde y'}{P'} \hastype \Proc$}
 \LeftLabel{\scriptsize \textsc{(Par)}}
\BinaryInfC{$\Gtopt{\Gamma_1}; \Gtopt{\Delta'_1} \cat \Theta \proves
              \Bopt{k}{\tilde y}{\bout{r}{z}P'} \hastype \Proc$}
\DisplayProof
\end{align}

\item Case $P = \binp{r}{z}P'$ when $\traux{r}$.
For this case Rule \textsc{Rcv} can be applied:
\begin{align}
  \label{pi:pt:rec-inputInv}
  \AxiomC{$\Gamma; \Delta \cat r : S'  \cat \Delta_z \proves P' \hastype \Proc$}
  \AxiomC{$\Gamma; \Delta_z \proves z \hastype C$}
  \LeftLabel{\scriptsize Rcv}
  \BinaryInfC{$\Gamma \setminus z;
          \Delta \cat r:
                S \proves \binp{r}{z}P' \hastype \Proc$}
  \DisplayProof
\end{align}
\noindent where $S = \btinp{C}S'$. 

Then, by IH on the first assumption of \eqref{pi:pt:inputInv} we know:
\begin{align}
\label{pi:eq:rec-input-ih}
\Gtopt{\Gamma'_1};\Gtopt{\Delta'_1} \cat \Theta_1 \proves
\Bopt{k+1}{\tilde y'}{P'} \hastype \Proc
\end{align}

\noindent where 
$\widetilde x' \subseteq \fn{P'}$ 
such that $r \in \wtd x'$ 
and $\widetilde y'$ such that 
$\indexed{\widetilde y'}{\widetilde x'}{\Gamma,\Delta \cat r:S}$. 
By this it follows that $(r_1,\ldots,r_{\len{\Gtopt{S}}}) \subseteq \wtd y'$. 

Also,
$\Gamma_1'=\Gamma \setminus \widetilde x'$, 
$\Delta_1' = \Delta \setminus \widetilde x'$, and $\balan{\Theta_1}$ with
$$
\dom{\Theta_1}=\{\prop_{k+1},\ldots,\prop_{k+\lenHOopt{P'}}\}
\cup \{\dual{\prop_{k+2}},\ldots,\dual{\prop_{k+\lenHOopt{P'}}}\}
$$
and ${\Theta_1(\prop_{k+1})}=\btinp{\widetilde M'}\tinact$
where $\widetilde M' = (\Gtopt{\Gamma},\Gtopt{\Delta,r:S,\Delta_z})(\widetilde y')$.

By applying \Cref{pi:lem:varbreak} to the second
assumption of \eqref{pi:pt:inputInv} we have:
\begin{align}
\label{pi:eq:rec-input-ih2}
\Gtopt{\Gamma};\Gtopt{\Delta_z} \proves z \hastype \Gtopt{C}
\end{align}

Let $\wtd x = \wtd x' \setminus z$ 
and $\wtd y = \wtd y' \setminus \wtd z$ such that 
$\len{\wtd z}=\Gtopt{C}$ where. We may notice that 
by \Cref{pi:def:indexed} 
$\indexed{\wtd y}{\wtd x}{\Gamma,\Delta \cat r:S}$ holds. 
We define $\Theta = \Theta_1 \cat \Theta'$, where
\begin{align*}
\Theta' = \prop_k:\btinp{\widetilde M} \tinact \cat \dual {\prop_{k+1}}:\btout{\widetilde M'} \tinact
\end{align*}
with $\widetilde M = (\Gtopt{\Gamma},\Gtopt{\Delta \cat \cat u_i:
                \btinp{C}S})(\widetilde y)$.
By \Cref{pi:def:sizeproc},
$\lenHOopt{P} = \lenHOopt{P'} + 1$ so
$$\dom{\Theta} = \{\prop_k,\ldots,\prop_{k+\lenHOopt{P}-1}\}
\cup \{\dual{\prop_{k+1}},\ldots,\dual{\prop_{k+\lenHOopt{P}-1}}\}$$
and $\Theta$ is balanced since $\Theta(\prop_{k+1}) \dualof \Theta(\dual{\prop_{k+1}})$ and
 $\Theta_1$ is balanced.
By \Cref{pi:t:bdowncorec}:
\begin{align*}
\Bopt{k}{\tilde y}{\binp{r}{z}P'} = \propinp{k}{\wtd y}
\binp{r_{\indf{S}}}{\wtd z}
\propout{k+1}
{\wtd y'} \inact \Par \Bopt{k+1}{\tilde y'}{P'}
\end{align*}

Also, let $\Gamma_1 = \Gamma \setminus \widetilde x$ and 
$\Delta_1 = \Delta \setminus \widetilde x$. We may notice that 
$\Delta_1 = \Delta'_1$.
We shall prove the following judgment:
\begin{align*}
\Gtopt{\Gamma_1 \setminus z};\Gtopt{\Delta_1}
\cat \Theta
\proves
  \Bopt{k}{\tilde y}{\binp{r}{z}P'}
\end{align*}

%\noindent where $\Gamma_1 = \Gamma \setminus \widetilde x$.
Let $\Delta_1 = \Delta \cat r:S$.
The left-hand side component of $\Bopt{k}{\tilde y}{\binp{r}{z}P'}$ is typed using
some auxiliary derivations:
\begin{align}
%\gamma
\label{pi:pt:rec-input3}
\AxiomC{}
\LeftLabel{\scriptsize \textsc{(Nil)}}
\UnaryInfC{$\Gtopt{\Gamma};\es \proves \inact \hastype \Proc$}
\LeftLabel{\scriptsize \textsc{(End)}}
\UnaryInfC{$\Gtopt{\Gamma};\dual{\prop_{k+1}}:\tinact \proves \inact \hastype \Proc$}
\AxiomC{}
\LeftLabel{\scriptsize \textsc{(PolyVar)}}
\UnaryInfC{$\Gtopt{\Gamma};
\Gtopt{\Delta \setminus \Delta_1}
            \proves \widetilde y' \hastype \widetilde M'$}
\LeftLabel{\scriptsize \textsc{(PolySend)}}
\BinaryInfC{$\Gtopt{\Gamma};
              \dual {\prop_{k+1}}:\btout{\widetilde M'} \tinact
              \cat 
              \Gtopt{\Delta \setminus \Delta_1}
              \proves \propout{k+1}{\widetilde y'} \inact \hastype \Proc$}
\DisplayProof
\end{align}
\begin{align}
%beta
\label{pi:pt:rec-input2}
\AxiomC{\eqref{pi:pt:rec-input3}}
\AxiomC{\eqref{pi:eq:rec-input-ih2}}
\LeftLabel{\scriptsize \textsc{(Rcv)}}
\BinaryInfC{
\begin{tabular}{c}
    $\Gtopt{\Gamma \setminus z}; 
            \dual {\prop_{k+1}}:\btout{\widetilde M'} \tinact \cat 
    \Gtopt{\Delta_2}
            \proves$ 
            $\binp{r_{\indf{S}}}{\widetilde z}\propout{k+1}{\widetilde y'} \inact \hastype \Proc$
\end{tabular}
}
 \LeftLabel{\scriptsize \textsc{(End)}}
\UnaryInfC{ 
\begin{tabular}{c}
$\Gtopt{\Gamma \setminus z}; 
            \dual {\prop_{k+1}}:\btout{\widetilde M'} \tinact
            \cat \prop_k: \tinact \cat
            \Gtopt{\Delta_2}
            \proves$ 
            $\binp{r_{\indf{S}}}{\wtd z}\propout{k+1}{\widetilde y'} \inact \hastype \Proc$
 \end{tabular}}
\DisplayProof
\end{align}
\begin{align}
\label{pi:pt:rec-input1}
\AxiomC{\eqref{pi:pt:rec-input2}}
\AxiomC{}
\LeftLabel{\scriptsize \textsc{(PolyVar)}}
\UnaryInfC{$\Gtopt{\Gamma \setminus z};
\Gtopt{\Delta_2}
\proves \widetilde y \hastype \widetilde M$}
\LeftLabel{\scriptsize \textsc{(PolyRcv)}}
\BinaryInfC{$\Gtopt{\Gamma_1 \setminus z};
 \Theta' \proves \propinp{k}{\widetilde
y}\binp{r_{\indf{S}}}{z}\propout{k+1}{\widetilde y'} \inact \hastype \Proc$}
\DisplayProof
\end{align}
\noindent 
where $\Delta_2 = \Delta \cat r:S \setminus \Delta_1$. 
We may notice that 
by \Cref{pi:def:typesdecompenv} and \Cref{pi:f:tdec} we have 
   $\Gtopt{\Delta_2}=\Gtopt{\Delta} \cat \wtd r : \Rtsopt{}{}{S} \setminus  
   \Gtopt{\Delta_1}$. Further, by \Cref{pi:lem:indexcor}
   we know $\Gtopt{\Delta_2}(r_{\indf{S}}) = \trec{t}\btinp{\Gtopt{C}}\vart{t}$.

The following tree
proves this case:
\begin{align}
\label{pi:pt:rec-input}
\AxiomC{\eqref{pi:pt:rec-input1}}
\AxiomC{\eqref{pi:eq:rec-input-ih}}
 \LeftLabel{\scriptsize}
\UnaryInfC{
\begin{tabular}{c}
$\Gtopt{\Gamma'_1 \setminus z}; \Gtopt{\Delta_1} \cat \Theta_1 \proves$ 
$\Bopt{k+1}{\tilde y'}{P'} \hastype \Proc$
\end{tabular}}
\LeftLabel{\scriptsize \textsc{(Par)}}
\BinaryInfC{\begin{tabular}{c}
      $\Gtopt{\Gamma_1 \backslash z};\Gtopt{\Delta_1}
              \cat \Theta \proves$ 
              $\propinp{k}{\widetilde y}
              \binp{r_{\indf{S}}}{\widetilde z}\propout{k+1}{\widetilde y'} \inact \Par \Bopt{k+1}
              {\tilde y'}{P'} \hastype \Proc$
              \end{tabular}}
\DisplayProof
\end{align}
\noindent

\item Case $P = \news{s:\trec{t}{S}}{P'}$. For this case 
Rule~\textsc{ResS} can be applied: 
\begin{align}
\label{pi:pt:resrecInv}
\AxiomC{$\Gamma;\Delta \cat s:\trec{t}S \cat 
\dual{s}:\dual{\trec{t}S} \proves P'$} 
\LeftLabel{\scriptsize \textsc{(ResS)}}
\UnaryInfC{$\Gamma;\Delta \proves 
\news{s:\trec{t}{S}}P'$}
\DisplayProof
\end{align}

By IH on the assumption of \eqref{pi:pt:resrecInv} we have: 
\begin{align}
\label{pi:eq:recresIh}
\Gtopt{\Gamma \setminus \wtd x'};\Gtopt{\Delta \setminus \wtd x'}
\cat \Theta_1 \proves \Bopt{k+1}{\tilde y'}{P'}	
\end{align}

where we take $\wtd y'$ such that $\wtd s, \wtd {\dual s} \subseteq \wtd y'$ with $\wtd s = (s_1,\ldots,s_{\len{\Rtopt{S}}})$
and $\wtd {\dual s} = (\dual s_1,\ldots,\dual s_{\len{\Rtopt{S}}})$.
. Accordingly, $\wtd x'$ is such that
$s,\dual s \subseteq \wtd x'$. Since $\lin{s}$ and 
$\lin{\dual s}$ we know $\wtd x' \subseteq \fn{P'}$.  
Also, $\indexed{\wtd y'}{\wtd x'}{\Gamma,\Delta_1}$ 
where $\Delta_1 = \Delta \cat s:\trec{t}S \cat 
\dual{s}:\dual{\trec{t}S} $. 
Also, $\balan{\Theta_1}$ with 
$$
\dom{\Theta_1}=\{\prop_{k+1},\ldots,\prop_{k+\lenHOopt{P'}}\}
\cup \{\dual{\prop_{k+2}},\ldots,\dual{\prop_{k+\lenHOopt{P'}}}\}
$$
\noindent and $\Theta_1(\prop_{k+1}) = \btinp{\wtd M'}\tinact$
where $\wtd M = (\Gtopt{\Gamma}, \Gtopt{\Delta})(\wtd y')$. 

Let $\wtd y = \wtd y' \setminus (\wtd s, \wtd {\dual s})$ 
and $\wtd x = \wtd x' \setminus (s,\dual s)$. 
Since $s,\dual s \notin \fn{P}$ we know $\wtd x \subseteq \fn{P}$
and $\indexed{\wtd y}{\wtd x}{\Gamma,\Delta}$.

We define $\Theta = \Theta_1\cat \Theta'$ where 
$$\Theta' = \prop_k:\btinp{\wtd M}\tinact \cat 
\dual{\prop_{k+1}} : \btout{\wtd M'}\tinact$$

By \Cref{pi:t:bdowncorec}, we have: 
\begin{align*}
 \Bopt{k}{\tilde y}{P} = \news{\widetilde{s}:\Rtopt{S}}
 (\propinp{k}{\wtd y}\propout{k+1}{\wtd y'}\inact \Par \Bopt{k}{\tilde y'}{P'})
\end{align*}

We should prove the following judgment: 
\begin{align*}
 \Gtopt{\Gamma \setminus \wtd x};\Gtopt{\Delta \setminus \wtd x} \cat \Theta \proves \Bopt{k}{\tilde y}{P} \hastype \Proc
\end{align*}

We use an auxiliary sub-tree: 
\begin{align}
\label{pi:pt:resrec1}
\AxiomC{} 
\LeftLabel{\scriptsize \textsc{(Nil)}}
\UnaryInfC{$\Gtopt{\Gamma};\es \proves \inact 
\hastype \Proc$} 
\AxiomC{} 
\LeftLabel{\scriptsize \textsc{(PolyVar)}}
\UnaryInfC{$\Gtopt{\Gamma};\Gtopt{\Delta_1 \cap \wtd x'} \proves 
\wtd y' \hastype \wtd M'$} 
\LeftLabel{\scriptsize \textsc{(PolySend)}}
\BinaryInfC{$\Gtopt{\Gamma};\Gtopt{\Delta \cap \wtd x}
\cat 
\wtd s : \Rtopt{S} \cat \wtd {\dual s}: \Rtopt{\dual S}
\proves \propout{k+1}{\wtd y'}\inact
\hastype \Proc$}
\DisplayProof
\end{align}
\noindent where we may notice that 
$$\Gtopt{\Delta \cap \wtd x}
\cat 
\wtd s : \Rtopt{S} \cat \wtd {\dual s}: \Rtopt{\dual S} = 
\Gtopt{\Delta_1 \cap \wtd x'}$$

The following tree proves this case: 
\begin{align}
\AxiomC{\eqref{pi:pt:resrec1}} 
\AxiomC{} 
\LeftLabel{\scriptsize \textsc{(PolyVar)}}
\UnaryInfC{$\Gtopt{\Gamma};\Gtopt{\Delta \cap \wtd x} \proves \wtd y \hastype \wtd M$} 
\LeftLabel{\scriptsize \textsc{(PolyRcv)}}
\BinaryInfC{$\Gtopt{\Gamma \setminus \wtd x};
\Theta' \cat 
\wtd s : \Rtopt{S} \cat \wtd {\dual s}: \Rtopt{\dual S}
\proves 
\propinp{k}{\wtd y}\propout{k+1}{\wtd y'}\inact
\hastype \Proc$} 
\AxiomC{\eqref{pi:eq:recresIh}} 
\LeftLabel{\scriptsize \textsc{(Par)}}
\BinaryInfC{$\Gtopt{\Gamma \setminus \wtd x};\Gtopt{\Delta \setminus \wtd x} \cat \Theta \cat 
\wtd s : \Rtopt{S} \cat \wtd {\dual s}: \Rtopt{\dual S}
\proves 
 \propinp{k}{\wtd y}\propout{k+1}{\wtd y'}\inact \Par \Bopt{k}{\tilde y'}{P'} \hastype \Proc$} 
\LeftLabel{\scriptsize \textsc{(PolyResS)}}
\UnaryInfC{$	\Gtopt{\Gamma \setminus \wtd x};\Gtopt{\Delta \setminus \wtd x} \cat \Theta \proves 
\news{\widetilde{s}:\Rtopt{S}}
 (\propinp{k}{\wtd y}\propout{k+1}{\wtd y'}\inact \Par \Bopt{k}{\tilde y'}{P'})
\hastype \Proc$}
\DisplayProof
\end{align}
\end{enumerate}
This concludes the proof of \Cref{pi:t:typerecur}.
\end{proof}

\thesisalt{
\subsection{Proof of \Cref{pi:t:decompcore} (Minimality Result, Optimized)}
}
{ 
  \subsection{Proof of \Cref{pi:t:thmdecompcore}}
}
\thmdecompcore*

\begin{proof} 
  \label{pi:app:decompcore}
  \label{pi:app:thmdecompcore}

By assumption, $\Gamma;\Delta \proves P \hastype \Proc$. 
Then by applying \Cref{top:lem:subst} we have: 
\begin{align}
\label{pi:eq:rec-typedec-subst}
  \Gamma\sigma;\Delta\sigma \proves P\sigma \hastype \Proc	
\end{align}

By \Cref{pi:def:decomp}, we
  shall prove the following judgment:
  \begin{align} 
  \Gtopt{\Gamma\sigma};\Gtopt{\Delta\sigma} \proves 	
  \news{\wtd \prop}(
  \propout{k}{\wtd r} \inact \Par \Bopt{k}{\tilde r}{P\sigma})
  \hastype \Proc 
  \end{align}
\noindent where $\widetilde \prop = (\prop_k,\ldots,\prop_{k+\lenHOopt{P}-1})$;
   $k > 0$; and 
  $\wtd r = \bigcup_{v \in \tilde v}\{v_1,\ldots, v_{\len{\Gtopt{S}}} \}$ with $v:S$.
  Since $\wtd v \subseteq \fn{P}$ 
  we know $\indexed{\wtd r}{\wtd v}{\Gamma\sigma, \Delta\sigma}$.
  Since $P\sigma$ is an initialized process, 
we apply \Cref{pi:t:typerecur} to \eqref{pi:eq:rec-typedec-subst} to get:
  \begin{align}
    \label{pi:eq:rec-typedec-bdown}
    \Gtopt{\Gamma\sigma};\Gtopt{\Delta \sigma \setminus \wtd v}\cat\Theta
    \proves \Bopt{k}{\tilde r}{P\sigma} \hastype \Proc
  \end{align}
\noindent where	$\Theta$ is balanced with
  $$\dom{\Theta} = \{\prop_k,\prop_{k+1},\ldots,\prop_{k+\lenHOopt{P}-1} \}
  \cup \{\dual{\prop_{k+1}},\ldots,\dual{\prop_{k+\lenHOopt{P}-1}} \}$$ and
  $\Theta(\prop_k)=\btinp{\wtd M}\tinact$ 
  with $\wtd M =\Gtopt{\Delta}(\wtd r)$.
  
  The following tree proves this case: 
  \begin{align*}
  \AxiomC{} 
  \LeftLabel{\scriptsize \textsc{(Nil)}}
  \UnaryInfC{$\Gtopt{\Gamma\sigma};\emptyset \proves \inact
  \hastype \Proc$} 
  \LeftLabel{\scriptsize \textsc{(End)}}
  \UnaryInfC{$\Gtopt{\Gamma\sigma};\prop_k:\tinact \proves \inact
  \hastype \Proc$} 
  \AxiomC{}
  \LeftLabel{\scriptsize \textsc{(PolyVar)}} 
  \UnaryInfC{$\Gtopt{\Gamma\sigma}; 
  \wtd r:\wtd M \proves \wtd r \hastype \wtd M$} 
  \LeftLabel{\scriptsize \textsc{(Send)}}
  \BinaryInfC{$\Gtopt{\Gamma\sigma};
    \dual{\prop_k}:\btout{\wtd M}\tinact 
    \cat \wtd r : \wtd M
    \proves 	
  \propout{k}{\wtd r} \inact   \hastype \Proc$} 
  \AxiomC{\eqref{pi:eq:rec-typedec-bdown}} 
  \LeftLabel{\scriptsize Par}
    \BinaryInfC{$\Gtopt{\Gamma\sigma};\Gtopt{\Delta\sigma} 
    \cat \dual{\prop_k}:\btout{\wtd M}\tinact \cat \Theta
    \proves 	
  \propout{k}{\wtd r} \inact \Par \Bopt{k}{\tilde r}{P\sigma}
  \hastype \Proc$}
    \LeftLabel{\scriptsize \textsc{(ResS)}}
    \UnaryInfC{$\Gtopt{\Gamma\sigma};
    \Gtopt{\Delta\sigma} \proves 	
  \news{\wtd \prop}(
  \propout{k}{\wtd r} \inact \Par \Bopt{k}{\tilde r}{P\sigma})
  \hastype \Proc$}
  \DisplayProof
  \end{align*}
\end{proof}

\thesisalt{
  \subsection{Proof of \Cref{pi:l:lemms}}
}{
  \subsection{Proof of \Cref{pi:l:lemms}}
}

% \begin{lemma}
% \label{pi:l:processrelate-mst}
% Relation $\processrelate$ is a MST bisimulation. 
% \end{lemma}
% \begin{proof}[Sketch]
% Straightforward by the transition induction and \Cref{pi:def:valuesprelation}. 
% \end{proof} 

For convenience, we define the function $\Cbs{-}{-}{\cdot}$,  
relying on $\Cbpi{-}{-}{\cdot}$, as follows: 
%\Cbs
\begin{definition}[Function  $\Cbs{-}{-}{\cdot}$ ]
\label{pi:d:cbs}
Let $P$ be a process, $\rho$ be a name substitution, and $\sigma$ be an indexed 
name substitution. We define $\Cbs{\rho}{\sigma}{P}$ as follows: 
\begin{align*}
  \Cbs{\rho}{\sigma}{P_1}  &= \Cbpi{\bns{u}}{\bns{x}}{P\sigma} \\
      & \quad \text{ with } 
      \rho = \subst{\tilde u}{\tilde x}, 
  % \wtd n = \fn{P} \cup \wtd u \cup \wtd x, \sigma \in \indices{\wtd n}, \\
    \wbns{u} = \bn{\wtd u \sigma : \wtd C}, 
  \wbns x = \bn{\wtd x \sigma : \wtd C} 
\end{align*}
\noindent where $\wtd u : \wtd C$. 
\end{definition}

Recall that $\relS$ has been defined in \Cref{pi:d:relation-s}.

\lemms* 

\begin{proof} 
\label{pi:app:lemms}
By transition induction. 
%  Then, we analyze inductive cases. 
Let $\rho_1 = \subst{\tilde u}{\tilde x}$. 
By inversion of $P_1 \,\relS\, Q_1$ 
we know there is $\sigma_1 \in \indices{\fn{P_1} \cup \wtd u \cup \wtd x}$, 
% $\wbns{u}=\bn{\wtd u \sigma_1}$, $\wbns{x}=\bn{\wtd x \sigma_1}$, 
% such that $Q_1 \in \Cbpi{\bns u}{\bns x}{P_1\sigma_1}$.
such that $Q_1 \in \Cbs{\rho_1}{\sigma_1}{P_1}$. 
We need the following assertion on index substitutions. 
\begin{newaddenv}
If $P_1\rho_1 \by{\ell} P_2\rho_2$ and $\subj{\ell}=n$ then there exists $Q_2$
such that $Q_1 \by{\iname \ell} Q_2$ 
with $\subj{\iname \ell} = n_i$ and $Q_2 \in \Cbs{\rho_2}{\sigma_2}{P_2}$
such that $\sigma_2 \in \indices{P_2\rho_2}$, 
$\nextn{n_i} \in \sigma_2$, 
and 
$\sigma_1 \cdot (\sigma_2 \setminus \nextn{n_i}) = 
(\sigma_2 \setminus \nextn{n_i}) \cdot \sigma_1$. 
\end{newaddenv}
% such that $\sigma_2 \in \indices{P_2\rho_2}$
% with $\nextn{n_i} \in \sigma_2$. 
\begin{comment} 
If $P_1\rho_1 \by{\ell} P_2\rho_2$ and $\subj{\ell}=n$ then there exists $Q_2$
such that $Q_1 \by{\iname \ell} Q_2$ 
with $\subj{\iname \ell} = n_i$ and $Q_2 \in \Cbs{\rho_2}{\sigma_2}{P_2}$
such that $\sigma_2 \in \indices{P_2\rho_2}$
with $\nextn{n_i} \in \sigma_2$.
\end{comment} 
%  and 
%  $Q_2 \in \Cbs{\rho_2}{\sigma_2}{P_2}$
%  % $Q_2 \in \Cbpi{\bns u'}{\bns x'}{P_2\sigma_2}$ 
%  for some $\rho_2$, and for some $\sigma_2'$ we have 
%  % $\sigma_2 = \sigma_1 \cdot \incrname{n}{i}$
%  $\sigma_2 = \sigma_2' \cdot \sigma_1 \cdot \incrname{n}{i}$
%  with $n_i = \subj{\iname \ell}$.  

We proceed as follows.
\begin{itemize}
	\item 
First, we consider two base cases: Rules~\ltsrule{Snd} and 
\ltsrule{Rv}. 
\item Then, we consider two inductive cases: 
Rules~\ltsrule{Par${}_L$} and \ltsrule{Tau}. 
We omit the inductive cases $\ltsrule{New}$ and $\ltsrule{Res}$
as they follow directly by the inductive hypothesis and the 
definition of the restriction case in $\Cbpi{-}{-}{\cdot}$
(\Cref{pi:t:tablecd}). 
\item Finally, we separately treat cases when a process is recursive.
We show two cases (\ltsrule{Rv} and \ltsrule{Par${}_L$}) 
emphasizing the specifics of the recursion breakdown. 
\end{itemize}
%  Let $\rho = \subst{\tilde u}{\tilde x}$ 
%  in every case. 

\paragraph{Non-recursive cases} We detail the four cases mentioned above.
\begin{enumerate}
 \item Case \ltsrule{Snd}. 
 \begin{newaddenv}
  % We first comment on the case when $P_1$ is a trigger collection. 
  We note that we only consider the case when $P_1$  is a pure process, 
  as the case when $P_1$ is a trigger collection follows directly by \Cref{pi:l:processrelate-mst}. 
  % We first note that in the case when $P_1$ is a trigger collection 
  % follows directly by ... . 
  % Next, we discuss the case when $P_1$  is a pure process. 
 \end{newaddenv}

 We distinguish two sub-cases: \rom{1} 
 $P_1 = \bout{n}{v}P_2$
 and \rom{2} $P_1 = \bout{n}{y}P_2$ with 
 $\subst{v}{y} \in \rho_1$. 
 In both sub-cases we know there is 
  $\rho_2 = \subst{\tilde u_2}{\tilde x_2}$ such that
 $$P_1\rho_1 = \bout{n\rho_1}{v}P_2\rho_2$$ 
 We have the following transition: 
 \begin{align*} 
  \AxiomC{} 
  \LeftLabel{\scriptsize \textsc{Snd}}
  \UnaryInfC{$P_1\rho_1
  \by{\bactout{n\rho_1}{v}} 
  P_2 \rho_2
  $}
  \DisplayProof
\end{align*}

Let $\sigma_1 \in \indices{\wtd p}$ where 
$\wtd p = \indices{\fn{P_1} \cup \wtd u \cup \wtd x}$ such that 
$\subst{n_i}{n} \in \sigma_1$ and $\sigma_2' = \sigma_2 \cdot \subst{t_1}{t} \cdot \subst{t_1}{t}$. 
%  Also, let $\sigma_2 = \sigma_1 \cdot \nextn{n_i}$. 
%  
%  Let $\sigma_1 = \sigma' \cdot \initname{n}{i}$  where 
%  $\sigma' \in \indices{\fn{P_1} \cup \wtd u \cup \wtd x}$ 
%  and $\sigma_2 = \sigma_1 \cdot \sigma$ where $\sigma = \nextn{n_i}$. 
\begin{newaddenv}
  Here we take $\sigmav = \sigma_1$, as we know $v \in \fn{P_1}$.
\end{newaddenv}
% Here we take $\sigmav = \sigma_2$. 
Further, let $\wbns{u} =\bname{\wtd u \sigma_1 : \wtd C}$
and $\wbns{x} =\bname{\wtd x \sigma_1 : \wtd C}$ with $\wtd u : \wtd C$. 
Also, let $\wtd z = \fnb{P_2}{\wbns{x} \setminus \wtd w}$ 
where $\wtd w = \{n_i\}$ if $\lin{n_i}$ otherwise $\wtd w = \epsilon$. 
Then, in both sub-cases, there are two possibilities for the shape of $Q_1$, namely: 
\begin{align*}
  Q^1_1 &= \news{\wtd \prop_k}(\apropout{k}{\wbns u} \Par
  \Bopt{k}{\bns x}{P_1  \sigma_1}) \\
  Q^2_1 &= \news{\wtd \prop_{k+1}} \bout{n_i \rho_*}{\wtd v} 
  \apropout{k+1}{\wtd z \rho_*}
 \Par \Bopt{k+1}{\tilde z}{P_2\sigma_2}
\end{align*}

\noindent 
% where $v_j \namesrelate \wtd v$.   
where $v\sigmav \namesrelate \wtd v$ 
and $\rho_* = \subst{\wtd u_*}{\wtd x_*}$
By \Cref{pi:l:c-prop-closed} we know that $Q^1_1 \by{\tau} Q^2_1$. 
Thus, we only consider how $Q^2_1$ evolves. 
We can infer the following: 
$$Q^2_1 \by{\bactout{n_i\rho_*}{\tilde v}} Q_2$$
\noindent where $Q_2 =\news{\wtd \prop_{k+1}}\apropout{k+1}{\wtd z \rho_*}
\Par \Bopt{k+1}{\tilde z}{P_2\sigma_2}$. Then, we should show the following: 
\begin{align}
  \label{pi:eq:snd-goal1}
{(P_2 \parallel \ftrigger{t}{v}{C_1})\rho_2 }
  \ \relS \ 
{(Q_2 \parallel \ftriggerm{t_1}{v\sigmav}{C_1})}
\end{align}

% First, we can see that if $

By \Cref{pi:t:tablecd} we can see that 
\begin{align} 
  \label{pi:eq:snd-c-r1}
  Q_2 \in \Cbpi{\tilde z \rho_*}{\tilde z}{P_2\sigma_2}
\end{align} 

Here we remark that our assertion holds by the definition as we have
$\sigma_2=\sigma_1\cdot \nextn{n_i}$.

\begin{newaddenv}
  Next, by \Cref{pi:lemm:processrelate-trigger} we have
  $$(\ftrigger{t}{v}{C_1})\sigma_2 \processrelate (\ftriggerm{t_1}{v\sigma_v}{C_1})$$ 

  That is, we have 
  \begin{align} 
    \label{pi:eq:snd-c-r1-2}
    (\ftriggerm{t_1}{v\sigma_v}{C_1}) \in 
    \Cbpi{\tilde z \rho_*}{\tilde z}{(\ftrigger{t}{v}{C_1})\sigma_2}
  \end{align} 
  % $$(\ftriggerm{t}{v\sigma_v}{C_1}) \in \Cbpi{\tilde z \rho_*}{\tilde z}{(\ftrigger{t}{v}{C_1})\sigma_2}$$ 

\end{newaddenv}

Thus, by \eqref{pi:eq:snd-c-r1} and \eqref{pi:eq:snd-c-r1-2} we have 
\begin{align}
  \label{pi:eq:snd-c-r1-5}
{(Q_2 \parallel \ftriggerm{t_1}{v\sigmav}{C_1})}
\in \Cbpi{\tilde z \rho_*}{\tilde z}{(P_2 \parallel \ftrigger{t}{v}{C_1})\sigma_2}
\end{align}

Now, by \Cref{pi:d:fnb} and \Cref{pi:d:name-breakdown} we know that 
% there are $\wtd u_2$ and $\wtd x_2$ such that 
% \begin{align*}
%   P_2\subst{\tilde u}{\tilde x} = P_2\subst{\tilde u_2}{\tilde x_2}
% \end{align*}
% and 
% \begin{align*}
%   \wtd z = \fnb{P_2}{\bname{\wtd x : \wtd C} \setminus \wtd w} = 
%   \bname{\wtd x_2 \sigma_1 \cdot \sigma : \wtd C_2}
% \end{align*}
\begin{align*}
  \wtd z = \fnb{P_2}{\bname{\wtd x : \wtd C} \setminus \wtd w} = 
  \bname{\wtd x_2 \sigma_2 : \wtd C_2}
\end{align*}
\noindent with $\wtd x_2 : \wtd C_2$. 
 Similarly, we have 
$$\wtd z\rho_* = \bname{\wtd x_2 \rho_2 \sigma_2: \wtd C_2}$$ 
% $$\wtd z\rho_* = \bname{\wtd x_2 \rho_2 \sigma_1 \cdot \sigma : \wtd C_2}$$ 

Now, by the assumption $\dual n \not\in \wtd v$ 
we know $\subst{\dual n_i}{\dual n} \not\in \sigma_2$.  
Thus, by $\sigma_2 = \sigma_1 \cdot \nextn{n_i} \cdot \subst{t_1}{t}$
and \Cref{pi:d:indexedsubstitution}, we have 
% $$\sigma_2 \in \indices{\wtd p}$$ 

$$\sigma_2 \in \indices{\fn{P_2 \parallel \ftrigger{t}{v}{C_1}} \cup \wtd u_2 \cup \wtd x_2}$$
\begin{comment}
Further, we know there is 
$$\sigma_2' \in \indices{\fn{P_2} \cup \wtd u_2 \cup \wtd x_2}$$ 
%  $$\sigma_2' \in \indices{\fn{P_2} \cup \wtd u_2 \cup \wtd x_2}$$ 
%      $$\sigma_2' \in \indices{\fn{P_2} \cup \wtd u_2 \cup \wtd x_2}$$ 

such that 
$P_2\rho_2 \sigma_2 = P_2\rho_2\sigma_2'$. 
% Thus, by this and \eqref{pi:eq:snd-c-r1} we have that $P_2\rho_2 \relS Q_2$. 
% \begin{newaddenv}
%   Next, by LEMMA [??] we have
%   $$(\ftrigger{t}{v}{C_1}) \processrelate (\ftriggerm{t}{v\rho\sigma}{C_1})$$ 
% \end{newaddenv}
\end{comment} 
\begin{comment} 
Next, by the definition we have 
$\ftrigger{t}{v}{C_1} \relS \ftriggerm{t}{v\rho\sigma}{C_1}$. 
\end{comment} 
By this and \eqref{pi:eq:snd-c-r1-5} the goal \eqref{pi:eq:snd-goal1} follows. 
This concludes \ltsrule{Snd} case. 

\item Case \ltsrule{Rv}. 
\begin{newaddenv}
  % We first comment on the case when $P_1$ is a trigger collection. 
  % As in the above case \ltsrule{Snd}, we only consider the case when $P_1$ is a pure process. 
  As in above case (\ltsrule{Snd}), we only consider the case when $P_1$  is a pure process, 
  as the case when $P_1$ is a trigger collection follows directly by \Cref{pi:l:processrelate-mst}. 
  % We first note that in the case when $P_1$ is a trigger collection 
  % follows directly by ... . 
  % Next, we discuss the case when $P_1$  is a pure process. 
 \end{newaddenv}

Here we know $P_1 = \binp{n}{y}P_2$. 
We know there is $\rho_2 = \subst{\tilde u_2}{\tilde x_2}$
such that 
$$P_1\rho_1 = \binp{n\rho_1}{y}P_2\rho_2$$

The transition is as follows: 
\begin{align*}
  \AxiomC{} 
  \LeftLabel{\scriptsize \textsc{Rv}}
  \UnaryInfC{$\binp{n\rho_1}{y}P_2\rho_2 \by{\abinp{n}{v}} 
  P_2 \rho_2 \cdot \subst{v}{y}$}
  \DisplayProof  	
\end{align*}

Let 
$\sigma_1 \in \indices{\fn{P_1} \cup \wtd u \cup \wtd x}$ 
and $\sigma_2 = \sigma_1 \cdot \nextn{n_i} \cdot \initname{y}{1}$.
% Let $\sigma_1 = \sigma' \cdot \initname{n}{i}$  where 
% $\sigma' \in \indices{\fn{P_1} \cup \wtd u \cup \wtd x}$ 
% and $\sigma_2 = \sigma_1 \cdot \nextn{n_i} \cdot \initname{y}{1}$. 
% $\sigma = \nextn{n_i} \cdot \initname{y}{1}$.
Further, let $\wbns u =\bname{\wtd u \sigma_1 : \wtd C}$
and $\wbns x =\bname{\wtd x \sigma_1 : \wtd C}$ with $\wtd u : \wtd C$. 
Also, let 
$\wtd y = (y_1,\ldots,y_{\len{\Gtopt{S}}})$ and 
$\wtd z = \fnb{P_2}{\wbns x \wtd y \setminus \wtd w}$ 
where $\wtd w = \{n_i\}$ if $\lin{n_i}$ otherwise $\wtd w = \epsilon$. 
Then, there are two possibilities for the shape of $Q_1$, namely: 
\begin{align*}
  Q^1_1 &= \news{\wtd \prop_k}(\apropout{k}{\wbns u} \Par
  \Bopt{k}{\bns x}{P_1  \sigma_1}) \\
  Q^2_1 &= \news{\wtd \prop_{k+1}} \binp{n_i \rho}{\wtd y} 
  \apropout{k+1}{\wtd z \rho}
 \Par \Bopt{k+1}{\tilde z}{P_2\sigma_2}
\end{align*}

By \Cref{pi:l:c-prop-closed} we know that $Q^1_1 \by{\tau} Q^2_1$. 
Thus, we only consider how $Q^2_1$ evolves. 
We can infer the following: 
\begin{align*}
  Q^2_1 \by{\bactinp{n_i\rho}{\wtd v}} Q_2 
\end{align*}
\noindent where 
$Q_2 = \news{\wtd \prop_{k+1}}\apropout{k+1}{\wtd z \rho \cdot \subst{\tilde v}{\tilde y}}
\Par \Bopt{k+1}{\tilde z}{P_2\sigma_2}$ 
and $v\sigmav \namesrelate \wtd v$ for some $\sigmav \in \indices{v}$.
Now, we should show the following: 
\begin{align}
  \label{pi:eq:fo-rv-goal1}
  P_2\rho_2 \cdot \subst{v}{y} \,\relS\, Q_2 
\end{align}

By \Cref{pi:t:tablecd} we can see that 
\begin{align} 
  \label{pi:eq:rv-c-r1}
  Q_2 \in 
  \Cbpi{\tilde z \rho_2 \cdot \subst{\tilde v}{\tilde y}}{\tilde z}{P_2\sigma_2 \cdot \sigmav}
\end{align} 

% We remark that 
% $\sigma_2 \cdot \sigmav= \sigmav \cdot \sigma_1 \cdot \nextn{n_i}$. 
We may notice that 
$$\sigma_2 \cdot \sigmav \in \indices{\fn{P_2} \cup \wtd u_2 \cup \wtd x_2}$$

Futher, by the definition we have 
$(\sigma_2 \cdot \sigmav) \setminus \nextn{n_i} = \sigma_1 \cdot \sigmav$. 
As $v \not\in (\fn{P_1} \cup \wtd u \cup \wtd x) = \codom{\sigma_1}$, we have 
$\sigma_1 \cdot \sigmav = \sigmav \cdot \sigma_1$. That is, our assertion 
on index substitutions holds.

By \Cref{pi:d:fnb} and \Cref{pi:d:name-breakdown} we have that 
\begin{align*}
  \wtd z &= \fnb{P_2}{\bname{\wtd x : \wtd C}\cdot \wtd y \setminus \wtd w} \\
   &= 
  \bname{\wtd x y (\sigma_1 \cdot \nextn{n_i} \cdot \initname{y}{1} ): \wtd C S} 
\end{align*}
\noindent with $y : S$. Similarly, we have 
\begin{align*}
  \wtd z \rho_2 \cdot \subst{\tilde v}{\tilde y} =  
   \bname{\wtd x y \rho_2 \cdot \subst{v}{y}(\sigma_2 \cdot \sigmav ): \wtd C S} 
\end{align*}

% Further, we may notice that 
% $$\sigma_2 \cdot \sigmav \in \indices $$

Thus, by this and \eqref{pi:eq:rv-c-r1} the goal 
\eqref{pi:eq:fo-rv-goal1} follows. 
This concludes \ltsrule{Rv} case (and the base cases).
We remark that base cases concerning triggers collection processes 
follow by  \Cref{pi:l:processrelate-mst}. 
Next, we consider inductive cases. 

\item Case \ltsrule{Par${}_L$}. 
Here we know  $P_1 = P'_1 \Par P''_1$.
Let $\rho_1'$ and $\rho_1''$ be such that 
$$P_1 \rho_1 = 
   P_1' \rho_1' \Par P''_1 \rho''_1$$
% We distinguish two sub-cases: 
% \rom{1} $P_1 = P'_1 \Par P''_1$ and \rom{2} $P_1 = H_1 \parallel P''_1$ 
% where $H_1$ is a triggers collection. 
% We only consider sub-case \rom{1} as sub-case \rom{2} is similar using 
% \Cref{pi:l:processrelate-mst}. 
The final rule in the inference tree is 
as follows: 
\begin{align*} 
  \AxiomC{$P_1' \rho_1' \by{\ell} P_2' \rho_2' $} 
  \AxiomC{$\bn{\ell} \cap \fn{P_1''} = \emptyset$} 
  \LeftLabel{\scriptsize \ltsrule{Par${}_L$}}
  \BinaryInfC{$P_1' \rho_1' \Par P_1'' \rho_1''
  \by{\ell} 
  P_2' \rho_2' \Par P_1'' \rho_1''$} 	
\DisplayProof 
\end{align*}

Let $\sigma_1 \in \indices{\fn{P_1}\cup \wtd u \cup \wtd x}$. 
Further, let $\sigma'_1$ and $\sigma_1''$
   such that 
   $$P_1 \rho_1 \sigma_1 = 
   P_1' \rho_1' \sigma'_1 \Par P''_1 \rho''_1 \sigma''_1$$

% Let $\sigma_1 \in \indices{\fn{P_1}\cup \wtd u \cup \wtd x}$. 
Further, let $\wtd y = \fnb{P_1'}{\wbns x}$, 
$\wtd z = \fnb{P_1''}{\wbns x}$, $\wbns{u} =\bname{\wtd u \sigma_1 : \wtd C}$,
 $\wbns{x} =\bname{\wtd x \sigma_1 : \wtd C}$ with $\wtd u : \wtd C$, and 
 $\rhom=\subst{\bns u}{\bns x}$. 
In sub-case \rom{1}, by the definition of $\relS$
(\Cref{pi:t:tablecd}), there are following possibilities for $Q_1$: 
  \begin{align*}
    Q^1_1 &= \news{\wtd \prop_k}(\apropout{k}{\wbns u} \Par
    \Bopt{k}{\bns x}{P_1  \sigma_1'}) \\
    Q^2_1 &= \propout{k}{\wtd y \rhom}
    \apropout{k+\degree}{\wtd z \rhom}   \Par
    \Bopt{k}{\tilde y}{P'_1\sigma_1'} \Par \Bopt{k+\degree}{\tilde z}{P''_1\sigma_1''} \\
    N^3_1 &= \left\{(R'_1 \Par R''_1) :
    \Cbpi{\tilde y\rhom}{\tilde y}{P'_1\sigma_1'}, R''_1 \in 
    \Cbpi{\tilde z\rhom}{\tilde z}{P''_1 \sigma_1''}\right\}
  \end{align*}
  \begin{comment} 
  By \Cref{pi:l:c-prop-closed} there exist 
  \begin{align}
    Q'_1 &\in \Cbpi{\tilde y\rhom}{\tilde y}{P_1'\sigma_1'}
  \\
    Q''_1 &\in \Cbpi{\tilde z\rhom}{\tilde z}{P_1''\sigma_1''}
    \label{pi:eq:fo-parl-q''_1}
  \end{align}
  such that 
\begin{align*}
  Q^1_1 \By{\tau} Q^2_1 \By{\tau} Q'_1 \Par Q''_1
\end{align*}
\end{comment} 
% \todo[inline]{Check this, $\rho$ substitution here}
By \Cref{pi:l:c-prop-closed} there exist 
\begin{align}
  Q'_1 &\in \Cbs{\rho_1'}{\sigma_1'}{P_1'}
\\
  Q''_1 &\in \Cbpi{\rho_1''}{\sigma_1''}{P_1''}
  \label{pi:eq:fo-parl-q''_1}
\end{align}
such that 
\begin{align*}
Q^1_1 \By{\tau} Q^2_1 \By{\tau} Q'_1 \Par Q''_1
\end{align*}

Thus, in both cases we consider how $Q'_1 \Par Q''_1$
evolves. 
By the definition of $\relS$ we have 
\begin{align} 
&P_1' \rho_1' \,\relS\, Q'_1 
\label{pi:fo-parl-q'1}
\\
&P_1'' \rho_1'' \,\relS\, Q''_1
\label{pi:fo-parl-q''1}
\end{align}

% By unfolding \Cref{pi:d:relation-s} we know 
% % there are $\sigma_1' \in \indices{\fn{P_1'} \cup \rho_1'}$ and 
% % $\sigma_1'' \in \indices{\fn{P_1''} \cup \rho_1''}$
% % such that  
% \begin{align}
%   Q_1' &\in \Cbs{\rho_1'}{\sigma_1}{P_1'} 
%   \label{pi:fo-parl-q'1-2}
%   \\
%   Q_1'' &\in \Cbs{\rho_1''}{\sigma_1}{P_1''}
%   \label{pi:fo-parl-q''1-2}
% \end{align}

To apply IH we do a case analysis on the action $\ell$. There are three sub-cases:
\begin{itemize}
\item Sub-case $\ell \equiv \bactinp{n}{v}$. 
If $v \in \wtd u$ then we take 
$\sigmav = \sigma_1$, otherwise 
$\sigmav = \initname{v}{j}$ for  $j > 0$. 
Then,  by applying IH to \eqref{pi:fo-parl-q'1} we know there is 
$Q'_2$ such that $Q'_1 \By{\bactinp{n_i}{\tilde v}} Q'_2$ and 
\begin{align}
  \label{pi:eq:fo-parl-ih2-inp}
  P_2' \subst{\tilde u'_2}{\tilde y_2} \ \relS \ Q_2' 	
  \end{align}
  \noindent where $v \sigmav \namesrelate \wtd v$ 
  and $\wtd u'_2 =\wtd u'_1 \cdot \wtd v$. 
  We should show that 
  \begin{equation}
    \label{pi:eq:fo-parl-inp-goal}
    P_2' \rho_2' \Par 
    P_1'' \rho_1'' \,\relS\, Q_2' \Par Q_1''   	
  \end{equation}
  % \noindent where $\rho_2' = \subst{\tilde u'_2}{\tilde y_1}$ 
  % and $\rho_1'' = \subst{\tilde u''_1}{\tilde z_1}$. 
  \begin{newaddenv}
Now, by the assertion on index substitutions, 
we have $\sigma_2' \in \indices{\fn{P'_2} \cup \rho'_2}$
such that $\nextn{n_i} \in \sigma_2'$, 
$\sigma_1' \cdot (\sigma_2' \setminus \nextn{n_i}) 
=  (\sigma_2' \setminus \nextn{n_i}) \cdot \sigma_1' $ , and 
\begin{align}
  \label{pi:eq:fo-parl-ih2-inp-2}
  Q_2' \in \Cbs{\rho'_2}{\sigma_2'}{P_2'}
\end{align}

Now, as by the definition we have 
$\sigma_1' \cdot \sigma_1'' = \sigma_1'' \cdot \sigma_1'$, it follows 
$\sigma_1'' \cdot (\sigma_2' \setminus \nextn{n_i}) 
=  (\sigma_2' \setminus \nextn{n_i}) \cdot \sigma_1'' $. 

  % Now, by the assertion we have $\sigma_2' = \sigma_1' \cdot \sigma_v \cdot \nextn{n_i}$ 
  % such that 
\end{newaddenv}
  % ...
%   By  \eqref{pi:eq:fo-parl-ih2-inp} and the 
% assertion on the index substitution we know  
%   there is  
%   % $\sigma_2 = \sigma_1 \cdot \nextn{n_i}$ 
%   $\sigma_2' \in \indices{P_2'\rho_2'}$
%   such that $\nextn{n_i} \in \sigma_2'$ such that 
  % Then, by
  %  \eqref{pi:eq:fo-parl-ih2-inp} we know 
  % \begin{align}
  %   \label{pi:eq:fo-parl-ih2-inp-2}
  %   Q_2' \in \Cbs{\rho'_2}{\sigma_2'}{P_2'}
  % \end{align}

%  \begin{newaddenv}
    Thus, the following holds 
    % $$\Cbs{\rho_1''}{\sigma_1'\cdot \sigma_2'}{P_1''} = 
    % \Cbs{\rho_1''}{\sigma_1'}{P_1''}$$
    \begin{align*}
      &\Cbs{\rho_1'' \cdot \rho_2'}
      {\sigma_2' \cdot \sigma_1''}{P_1''} = 
    \Cbs{\rho_1''}{\sigma_1''}{P_1''} 
    \\
    & 
    \Cbs{\rho_1'' \cdot \rho_2'}
    {\sigma_2' \cdot \sigma_1''}{P_2'} = 
    \Cbs{\rho'_2}{\sigma_2'}{P_2'}
    \end{align*}
 % \end{newaddenv}
By this, \eqref{pi:eq:fo-parl-ih2-inp-2}, and \eqref{pi:eq:fo-parl-q''_1} we have 
  \begin{align} 
    \label{pi:eq:fo-parl-inp-3}
   Q_2' \Par Q_1'' \in 
   \Cbs{\rho_1'' \cdot \rho_2'}
   {\sigma_1''\cdot \sigma_2'}{P_1'' \Par P_2'}
  \end{align} 

  Further, we may notice that  
  $\sigma_1'' \cdot \sigma_2' \in \indices{\fn{P_2' \Par P''_1} \cup \rho'_2 \cup \rho''_1}$ and $\nextn{n_i} \in \sigma_1'' \cdot \sigma_2'$ as by the assumption  we have 
  $\bar n \not\in \fn{P_2' \Par P''_1} \cup \rho'_2 \cup \rho''_1$. Thus, our assertion holds. 
  Hence, by this and \eqref{pi:eq:fo-parl-inp-3} the goal 
    \eqref{pi:eq:fo-parl-inp-goal} follows. 

  % Now, we 
  % ...

  % We may notice that $\sigma_1 = \sigma_1' \cdot \sigma_2'$ 
  % where $\sigma_1'$ is such that 
  % $$\Cbs{\rho_1''}{\sigma_1}{P_1''} = \Cbs{\rho_1''}{\sigma_1'}{P_1''}$$

  % By the definition of $\sigmav$ and $\sigma_2$ we may see that 
  % $$\Cbs{\rho_1''}{\sigma_2 \cdot \sigmav}{P_1''} = \Cbs{\rho_1''}{\sigma_1}{P_1''}$$
  % So, by \eqref{pi:eq:fo-parl-ih2-inp-2} and \eqref{pi:eq:fo-parl-q''_1} we have 
  % \begin{align} 
  %   \label{pi:eq:fo-parl-inp-3}
  %  Q_2' \Par Q_1'' \in \Cbs{\rho'_2 \cdot \rho'_1}{\sigma_2 \cdot \sigmav}{P_2' \Par P_1''}
  % \end{align} 
  % Further, we may notice that  
  % $\sigma_2 \cdot \sigma_v \in \indices{\fn{P_2' \Par P''_1} \cup \rho'_2 \cup \rho''_1}$ and $\nextn{n_i} \in \sigma_2 \cdot \sigma_v$. Thus, our assertion holds. 
  % So, by this and \eqref{pi:eq:fo-parl-inp-3} the goal 
  %   \eqref{pi:eq:fo-parl-inp-goal} follows. 

\item Sub-case $\ell = \tau$. 
By applying IH to \eqref{pi:fo-parl-q'1} we know there is 
$Q'_2$ such that $Q'_1 \By{\ell} Q'_2$ and 
\begin{align}
  \label{pi:eq:fo-parl-tau-ih2}
  P_2' \rho'_2 \ \relS \ Q_2' 	
  \end{align} 
  \noindent where $\rho'_2 = \subst{\tilde u'_2}{\tilde y_2}$. 
  We should show that 
  \begin{align}
    \label{pi:eq:fo-parl-tau-goal}
    P_2' \rho_2' \Par 
    P_1'' \rho_1'' \,\relS\, Q_2' \Par Q_1'' 
  \end{align}
  By \eqref{pi:eq:fo-parl-tau-ih2}, we know there is 
  $\sigma_2 \in \indices{\fn{P_2} \cup \rho'_2}$ such that  
  \begin{align*}
    Q_2' \in \Cbs{\rho'_2}{\sigma_2}{P_2'}
  \end{align*}
  Further, by remark we know that either 
$\sigma_2' = \sigma_1 \cdot \incrname{n}{i} \cdot \incrname{\dual n}{i}$
for some $n_i, \dual n_i \in \fn{P'_2 \rho'_2}$ and 
$n_i, \dual n_i \not\in \fn{P''_1 \rho''_1}$ or 
$\sigma_2' = \sigma_1$
such that 
$$\Cbs{\rho'_2}{\sigma_2}{P_2'} = \Cbs{\rho'_2}{\sigma_2'}{P_2'}$$ 

By this  we have that 
$$\Cbs{\rho'_1}{\sigma_2'}{P_1'} = \Cbs{\rho'_1}{\sigma_1}{P_1'}$$
So, we know that 
$$Q_2' \Par Q''_1 \in \Cbs{\rho'_2 \cdot \rho''_1}{\sigma_2'}{P'_2 \Par P''_1}$$
By \eqref{pi:eq:fo-parl-tau-ih2} and  \eqref{pi:eq:fo-parl-q''_1} and the definition of $\sigma_2'$
the goal \eqref{pi:eq:fo-parl-tau-goal} follows. 

\item Sub-case  $\ell \equiv \news{\wtd m_1}\about{n}{v:C_1}$. 
Here we omit details on substitutions as they are similar 
to the first sub-case. 
By applying IH to 
\eqref{pi:fo-parl-q'1} we know there is 
$Q'_2$ such that $Q'_1 \By{\ell} Q'_2$ and 
\begin{align}
  \label{pi:eq:fo-parl-out-ih1}
  &{\newsp{\wtd{m_1}}{P'_2 \parallel \ftrigger{t}{v}{C_1}}}
  \ \relS \\
  &{\newsp{\widetilde{m_2}}{Q'_2 \parallel \ftriggerm{t_1}{v\sigma_1}{C_1}}}
\end{align}

We should show that 
\begin{align}
  \label{pi:eq:fo-parl-out-goal}
  &{\newsp{\wtd{m_1}}{P'_2 \Par P''_1 \parallel \ftrigger{t}{v}{C_1}}}
  \ \relS \\
  &{\newsp{\widetilde{m_2}}{Q'_2 \Par P''_1 \parallel \ftriggerm{t_1}{v\sigma_1}{C_1}}}
\end{align}

By \Cref{pi:d:relation-s} and \Cref{pi:t:tablecd} from 
\eqref{pi:eq:fo-parl-out-ih1} we can infer the following: 
\begin{align}
  \label{pi:eq:fo-parl-out-ih2}
  P_2' \subst{\tilde u'_1}{\tilde y'} \ \relS \ Q_2' 	
  \end{align}
  So, by \eqref{pi:fo-parl-q''1} and \eqref{pi:eq:fo-parl-out-ih2} 
the goal   \eqref{pi:eq:fo-parl-out-goal} follows. 
\end{itemize}
This concludes the case \ltsrule{Par${}_L$} case. 

\item Case \ltsrule{Tau}. 
Here we know $P_1 = P'_1 \Par P''_1$. 
Without the loss of generality, we assume
$\ell_1 = \news{\wtd m_1}\about{\dual n}{v_1}$
and $\ell_2 = \abinp{n}{v_1}$. 
Let  $\rho_1' = \subst{\tilde u'_1}{\tilde y}$ and 
$\rho_1'' = \subst{\tilde u''_1}{\tilde z}$ 
such that 
$$P_1\rho_1 = P_1'\rho_1' \Par P''_1 \rho''_1$$ 

Then, the final rule in the inference tree is as follows: 
\begin{align*}
\AxiomC{$P_1'\rho_1' \by{\ell_1} P_2' \rho_2'$} 
\AxiomC{$P_1'' \rho_1'' 
\by{\ell_2} P_2'' \rho_2'$} 
\AxiomC{$\ell_1 \asymp \ell_2$} 
\LeftLabel{\scriptsize \ltsrule{Tau}}
\TrinaryInfC{$P_1' \rho_1' \Par P_1'' \rho_1''
\by{\tau} 
\news{\wtd m_1}(P_2'  \rho_2' 
\Par P_2'' \rho''_2)$}	
\DisplayProof
\end{align*}

Let $\sigma_1$, $\wtd y$, $\wtd z$, $\wbns u$, 
$\wbns x$, and $\rhom$ be defined as in the previous case. 
Further, $Q_1$ can have shapes: 
$Q^1_1$, $Q^2_1$, and $N^3_1$ as in the previous case. 
As in the previous case, we know that there are
  \begin{align}
    Q'_1 &\in \Cbpi{\tilde y\rhom}{\tilde y}{P_1'\sigma_1}
    \label{pi:eq:fo-tau-q'_1} \\
    Q''_1 &\in \Cbpi{\tilde z\rhom}{\tilde z}{P_1''\sigma_1}
    \label{pi:eq:fo-tau-q''_1}
  \end{align}
  such that 
\begin{align*}
  Q^1_1 \By{\tau} Q^2_1 \By{\tau} Q'_1 \Par Q''_1
\end{align*}

Thus, in both cases we consider how $Q'_1 \Par Q''_1$
evolves. By the definition of $\relS$ we have 
\begin{align} 
&P_1' \rho_1' \,\relS\, Q'_1 
\label{pi:fo-tau-q'1}
\\
&P_1'' \rho_1'' \,\relS\, Q''_1
\label{pi:fo-tau-q''1}
\end{align}
We apply the IH component-wise: 
\begin{itemize}
\item By applying IH to \eqref{pi:fo-tau-q'1} we know there is 
$Q'_2$ such that 
\begin{align} 
  \label{pi:eq:fo-tau-out-ih1-1}
  Q'_1 \By{\news{\wtd m_2}\about{\dual n_i}{\tilde v}} Q'_2
\end{align} 
\noindent and 
\begin{align}
  \label{pi:eq:fo-tau-out-ih1}
  &{\newsp{\wtd{m_1}}{P'_2 \parallel \ftrigger{t}{v}{C_1}}\rho'_2}
  \ \relS 
  % \\&
  {\newsp{\widetilde{m_2}}{Q'_2 \parallel \ftriggerm{t}{v\sigma_1}{C_1}}}
\end{align}
\noindent where $v\sigma_v \namesrelate \wtd v$ 
\newadd{such that $\sigma_v \subseteq \sigma_1$.}
% with $\sigma_v $.

% \begin{newaddenv}
%   Now, by~\eqref{pi:eq:fo-tau-out-ih1-1} we know there is $R'$ such that 
%   \begin{align*}
%     Q'_1 \By{\tau} R' \by{\news{\wtd m_2}\about{\dual n_i}{V_2}} Q'_2
%   \end{align*}
% \end{newaddenv}

Now, by \eqref{pi:eq:fo-tau-out-ih1} we can infer: 
\begin{align*}
P_2'\rho'_2 \,\relS\, Q_2'
\end{align*}

%\begin{newaddenv}
Now, by the assertion on index substitutions, 
we have $\sigma_2' \in \indices{\fn{P_2'} \cup \rho_2'}$
such that $\nextn{\dual n_i} \in \sigma_2'$, 
$\sigma_1' \cdot (\sigma_2' \setminus \nextn{\dual n_i}) 
=  (\sigma_2' \setminus \nextn{\dual n_i}) \cdot \sigma_1' $, and 
%\end{newaddenv}

% By this we know there is $\sigma_2' \in \indices{\fn{P_2'} \cup \rho'_2}$
% such that 
\begin{align*}
Q_2' \in \Cbs{\rho'_2}{\sigma_2'}{P_2'}
\end{align*}

More precisely, here we know that 
% $\sigma_2'' = \sigma_1'$
$\sigma_2' \subseteq \sigma_1' \cdot \nextn{\dual n_i}
\subseteq \sigma_1 \nextn{\dual n_i}$. 
So, we have 
\begin{align}
  \label{pi:eq:fo-tau-ih2-3}
Q_2' \in \Cbs{\rho'_2}{\sigma_1 \cdot \nextn{\dual n_i}}{P_2'}
\end{align}

\item By applying IH to \eqref{pi:fo-tau-q''1}  we know there is
$Q_2''$ such that 
\begin{align} 
  \label{pi:eq:fo-tau-ih2-1}
 Q''_1 \By{\bactinp{n_j}{\tilde v}} Q''_2
\end{align}
\noindent and 
\begin{align}
  \label{pi:eq:fo-tau-ih2}
  P_2'' \rho''_2 \ \relS \ Q_2'' 	
  \end{align}

  \begin{newaddenv}
    Now, by the assertion on index substitutions, 
    we have $\sigma_2'' \in \indices{\fn{P_2''} \cup \rho_2''}$
    such that $\nextn{n_j} \in \sigma_2''$, 
    $\sigma_1' \cdot (\sigma_2'' \setminus \nextn{n_j}) 
    =  (\sigma_2'' \setminus \nextn{n_j}) \cdot \sigma_1' $, and 
    \end{newaddenv}
% By~\eqref{pi:eq:fo-tau-ih2}, we know there is 
%  $\sigma_2'' \in \indices{\fn{P_2''} \cup \rho''_2}$
% and $\sigma_2'' \subseteq \sigma_1 \cdot \nextn{\dual n_j}$, so that 
% we have the following
\begin{align}
  \label{pi:eq:fo-tau-ih2-2-1}
Q_2'' \in \Cbs{\rho''_2}{\sigma_2''}{P_2''}
\end{align}

More precisely, we know that 
$\sigma_2'' =\sigma_1'' \cdot \sigma_v \cdot \nextn{n_j} 
\subseteq \sigma_1 \cdot \nextn{n_j}$. So, we have 
\begin{align}
  \label{pi:eq:fo-tau-ih2-2}
Q_2'' \in \Cbs{\rho''_2}{\sigma_1 \cdot \nextn{n_j}}{P_2''}
\end{align}

% \begin{align}
%   \label{pi:eq:fo-tau-ih2-2}
% Q_2'' \in \Cbs{\rho''_2}{\sigma_1 \cdot \nextn{n_j}}{P_2''}
% \end{align}

\end{itemize}

\begin{comment} 
Now, by \eqref{pi:eq:fo-tau-out-ih1} we can infer: 
\begin{align*}
P_2'\rho'_2 \relS Q_2'
\end{align*}

By this we know there is $\sigma_2' \in \indices{\fn{P_2'} \cup \rho'_2}$
such that 
\begin{align*}
Q_2' \in \Cbs{\rho'_2}{\sigma_2'}{P_2'}
\end{align*}

Moreover, we know that it must be 
$\sigma_2' \subseteq \sigma_1 \cdot \nextn{\dual n_i}$. 
So, we have 
\begin{align*}
Q_2' \in \Cbs{\rho'_2}{\sigma_1 \cdot \nextn{\dual n_i}}{P_2'}
\end{align*}
\end{comment} 
\begin{comment} 
Similarly, there is $\sigma_2'' \in \indices{\fn{P_2''} \cup \rho''_2}$
and $\sigma_2'' \subseteq \sigma_1 \cdot \nextn{\dual n_i}$, so 
we have the following  
\begin{align*}
Q_2'' \in \Cbs{\rho''_2}{\sigma_1 \cdot \nextn{n_i}}{P_2''}
\end{align*}
\end{comment} 

% \begin{newaddenv}
%   Now, we know there is $R'$ such that 
%   \begin{align*}
%     Q'_1 \By{\tau} R' \by{\news{\wtd m_2}\about{\dual n_i}{V_2}} Q'_2
%   \end{align*}
% \end{newaddenv}

% \begin{newaddenv}
  Now, by~\eqref{pi:eq:fo-tau-out-ih1-1} we know there is $R'$ such that 
  \begin{align*}
    Q'_1 \By{\tau} R' \by{\iname \ell_1} Q'_2
    % Q'_1 \By{\tau} R' \by{\news{\wtd m_2}\about{\dual n_i}{\tilde v}} Q'_2
  \end{align*}
  \noindent where $\iname \ell_1 = \news{\wtd m_2}\about{\dual n_i}{\tilde v}$. 
  Similarly, by~\eqref{pi:eq:fo-tau-ih2} there is $R''$ such that 
  \begin{align*}
    Q''_1 \By{\tau} R'' \by{\iname \ell_2} Q''_2
  \end{align*}
  \noindent where $\iname \ell_2 =\abinp{n_j}{\tilde w}$. 
% \end{newaddenv}

% \begin{newaddenv}
To proceed we must show $\iname \ell_1 \asymp \iname \ell_2$, 
  which boils down to showing that indices of $\dual n_i$ and $n_j$ match.   
  For this, we distinguish two sub-cases: \rom{1} $\neg \traux{\dual n_i}$ and 
  $\neg \traux{n_j}$ 
  and \rom{2} $\traux{\dual{n_i}}$ and $\traux{n_j}$. In the former sub-case,  
    we have 
  $\subst{\dual n_i}{n} \in \sigma_1$ and $\subst{n_j}{n} \in \sigma_1$, 
  where $\sigma_1 = \indices{\wtd u}$. 
  Further, by this and 
   \Cref{pi:d:indexedsubstitution} 
  we know that $i =j$. Now, we consider the latter case. 
  By assumption that $P_1\subst{\tilde u}{\tilde x}$ is well-typed, 
  we know there $\Gamma_1, \Lambda_1$, and $\Delta_1$ such that 
   $\Gamma_1;\Lambda_1;\Delta_1 \proves P_1\subst{\tilde u}{\tilde x} 
   \hastype \Proc$ with $\balan{\Delta_1}$, Thus,  we have 
   $n:S \in \Delta_1$ and $\dual n:T \in \Delta_1$ such that    $S \dualof T$. 
% \end{newaddenv}

% \begin{newaddenv} 
  Hence, we can infer the following transition: 
  \begin{align*}
  \AxiomC{$R' \by{\iname \ell_1} Q_2'$} 
  \AxiomC{$R'' \by{\iname \ell_2} Q_2''$} 
  \AxiomC{$\iname \ell_1 \asymp \iname \ell_2$} 
  \LeftLabel{\scriptsize  \ltsrule{Tau}}
  \TrinaryInfC{$(R' \Par R'') \by{\tau} \news{\wtd m_2}(Q_2' \Par Q_2'')$}
  \DisplayProof	
  \end{align*}  
% \end{newaddenv}

% \begin{newaddenv}
  Now, we should show that 
  \begin{align}
    \news{\wtd m_1}(P_2' \rho_2' \Par P_2'' \rho_2'')  
    \ \relS \ 
    \news{\wtd m_2}(Q_2' \Par Q_2'')
    \label{pi:eq:tau-goal2}
   \end{align} 
% \end{newaddenv}

% \todo[inline]{TODO}
% \begin{comment} 
Now, by~\eqref{pi:eq:fo-tau-ih2-3}, \eqref{pi:eq:fo-tau-ih2-2}, 
and  $(P_2' \Par P_2'') \rho'_2 \cdot \rho''_2 = P_2' \rho'_2 \Par P_2'\rho''_2$ we have 
\begin{align*}
Q_2' \Par Q_2'' \in \Cbs{\rho'_2 \cdot \rho_2''}{\sigma_1 \cdot \nextn{n_i} 
\cdot \nextn{\dual n_i}}
{P_2' \Par P_2''}
\end{align*}
% \end{comment} 

% \begin{align*}
%   Q_2' \Par Q_2'' \in \Cbpi{\bns u'_2 \cdot \bns u''_2}{\bns y' \cdot \bns z'}
%   {(P_2' \Par P_2'')\sigma_1 \cdot \nextn{n_i}}
% \end{align*}

\begin{comment} 
Futher, we need to show that there is $\sigma_2$ such that 
$\sigma_2 
\in  \indices{\fn{P_2' \Par P_2''} \cup  \rho'_2 \cup \rho''_2}$ 
and $\sigma_2 \subseteq \sigma_1 \cdot \nextn{n_i} \cdot \nextn{\dual n_i}$. 
This follow directly by the definition of $\sigma_1$, 
that is $\sigma_1 \in \indices{\fn{P_1}\cup \wtd u \cup \wtd x}$,
and \Cref{pi:d:indexedsubstitution}. 
\end{comment} 

Further, we have 
$\sigma_1 \cdot \nextn{n_i} 
\cdot \nextn{\dual n_i} \in 
\indices{\fn{P_2' \Par P_2''} \cup  \rho'_2 \cup \rho''_2}$ 
This follow directly by the definition of $\sigma_1$, 
that is $\sigma_1 \in \indices{\fn{P_1}\cup \wtd u \cup \wtd x}$,
and \Cref{pi:d:indexedsubstitution}. 

% Now, we may notice that there is $\sigma_2 \in \indices{\fn{P_2' \Par P_2''} \cup  \rho'_2 \cup \rho''_2}$ such that 
% $\sigma_2 \subseteq \sigma_1 \cdot \nextn{n_i} \cdot \nextn{\dual n_i}$. 

% By \Cref{pi:d:indexedsubstitution}, there is 
% $\sigma_2 \in \indices{\fn{P_2' \Par P_2''} \cup  \rho'_2 \cup \rho''_2}$ such that 
% $\sigma_2 \subseteq \sigma_1 \cdot \nextn{n_i} \cdot \nextn{\dual n_i}$. 
\begin{comment} 
Thus, we know 
\begin{align*}
\Cbs{\rho'_2 \cdot \rho_2''}{\sigma_2}
{P_2' \Par P_2''} = \Cbs{\rho'_2 \cdot \rho_2''}{\sigma_1 \cdot \nextn{n_i} 
\cdot \nextn{\dual n_i}}
{P_2' \Par P_2''}
\end{align*}
\end{comment} 
% we have that $\sigma_1 \cdot \nextn{n_i} 
% \cdot \nextn{\dual n_i} \in \indices{\fn{P_2' \Par P_2''} \cup \cup \wtd u \cup \wtd x}$. 
Finally, by this and the definition of $\relS$ (\Cref{pi:d:relation-s}) the goal~\eqref{pi:eq:tau-goal2} follows. 
% \begin{align*}
% &{\newsp{\wtd{m_1}}{P'_2 \Par P''_2}\rho_2' \cdot \rho''_2}
% \ \relS 
% {\newsp{\widetilde{m_2}}{Q'_2 \Par Q''_2}}
% \end{align*}
This concludes case \ltsrule{Tau}. 

% \item Case \texttt{News}. 
 
\end{enumerate}

\paragraph{Recursive cases}
Now, we consider cases where $P'_1 \equiv \recp{X}P^*_1$ 
is a parallel component of $P_1$. 
We focus on two cases, which highlight the specifics of 
the breakdown of recursive processes, 
and omit details that are similar to the corresponding 
non-recursive cases. 
\begin{enumerate}
 \item Case \ltsrule{Rv}. 
Here we know $P_1 = \binp{n}{y}P_2$. Further, we know there exist $P^*$ 
such that 
 $\jdepth{P}{P^*}{d}{X}$ (\Cref{pi:d:jdepth}). 
 Now, we unfold this definition. 
% such that $P_1\equiv\recp{X}P^*_1$. 
Let 
$$
    P^1_1 = \alpha_d.\alpha_{d-1}. \ldots .\alpha_{1}.(X \Par R)
$$
\noindent where $R$ is some processes, 
and $\alpha_d = \abinp{n}{y}$.
%  As $P_1$ is a recursive process, 
  We know that there is $\recp{X}P^*$ such that $P^1_1$ is its sub-processes and
$$
      P_1 \equiv P^1_1\subst{\recp{X}P^*}{X}
$$
%
%    where $\recp{X}P^*$ is subprocess of $P_X$, $\alpha_d = \abinp{n}{y}$, and 
%    $\alpha_d.\alpha_{d-1}. \ldots .\alpha_{d-p}$  and  . 
  Here we can distinguish two sub-cases: \rom{1} $p > 0$ and \rom{2} $p = 0$ and 
  $R \equiv \inact$. 
 We know there is $\rho_2 = \subst{\tilde u_2}{\tilde x_2}$
 such that 
 $$P_1\rho_1 = \binp{n\rho_1}{y}P_2\rho_2$$

  The transition is as follows: 
  \begin{align*}
    \AxiomC{} 
    \LeftLabel{\scriptsize \ltsrule{Rv}}
    \UnaryInfC{$
         P^1_1\subst{\recp{X}P^*}{X}
        \by{\abinp{n}{v}} 
        P_2 \rho_2 \cdot \subst{v}{y}
    $}
    \DisplayProof  	
  \end{align*}

  Let $\sigma_1 = \sigma' \cdot \initname{n}{i}$  where 
$\sigma' \in \indices{\fn{P_1} \cup \wtd u \cup \wtd x}$ 
and $\sigma_2 = \sigma_1 \cdot \sigma$ where 
$\sigma = \nextn{n_i} \cdot \initname{y}{1}$.
Further, let $\wbns u =\bname{\wtd u \sigma_1 : \wtd C}$
and $\wbns x =\bname{\wtd x \sigma_1 : \wtd C}$ with $\wtd u : \wtd C$. 
Also, let 
$\wtd y = (y_1,\ldots,y_{\len{\Gtopt{S}}})$ and 
$\wtd z = \fnb{P_2}{\wbns x \wtd y \setminus \wtd w}$ 
where $\wtd w = \{n_i\}$ if $\lin{n_i}$ otherwise $\wtd w = \epsilon$. 
%    As  $P_1\equiv\recp{X}P^*$ 
%    we know that $P_1 \equiv \alpha_d,\alpha_{d-1},\ldots,\alpha_1.\recp{X}P^*_1 \Par R$ 
%    where $R$ is some process and $\alpha_d,\alpha_{d-1},\ldots,\alpha_1$ is a subprocess 
%    of $P^*_1$ with $\alpha_d = \abinp{n}{y}$.
%    Here we can distinguish two sub-cases: \rom{1} $d > 1$ and \rom{2} $d = 1$. 
%    We consider sub-case \rom{1} first. 
  
   We could see that 
  $d = \rvardepth{\alpha_d,\alpha_{d-1},\ldots,\alpha_1.\recp{X}P^*_1}$. 
  So, by  \Cref{pi:t:tablecd-rec}
  there are following possibilities for the shape of $Q_1$, namely elements in $N$
  defined as: 
  \begin{align*}
      N = \{\news{\wtd \prop}R : R \in \Cbrecpi{\bns u}{\bns x, \rho}{\recp{X}P^*}^d \}
  \end{align*}
  
  Let 
  \begin{align*}
      Q^1_1 = \news{\wtd \prop}
          B_1
          \Par 
           \binp{n_l}{\wtd y} 
          \propoutrec{k+1}{\wtd z\rho}
          \recp{X}\binp{\prop^r_k}{\wtd x}
            \binp{n_l}{\wtd y} 
            \propoutrec{k+1}{\wtd z}
            \rvar{X}
           \Par
%      		\\
      \Brecpi{k+1}{\tilde z}{\alpha_d,\alpha_{d-1}.P^2_1\sigma}_g
  \end{align*}
  \noindent where $B_1$ is, intuitively, trios mimicking  
  prefixes before $\alpha_d$ and $P^2_1$ is such that 
  $P^1_1 = \alpha_d.\alpha_{d-1}.P^2_1$. 
  By unfolding the definition of $\Cbrecpi{\bns u}{\bns x, \rho}{\recp{X}P^*}^d$ 
     we have that $Q^1_1 \in N$. Further, 
     if $R \in \Cbrecpi{\bns u}{\bns x, \rho}{\recp{X}P^*}^d$
     then $R \By{\tau} Q^1_1$. So, we only analyze 
     how $Q^1_1$ evolves. 
  We can infer the following: 
  \begin{align*}
    Q^1_1 \by{\bactinp{n_i\rho}{\wtd v}} Q_2 
  \end{align*}
  where 
   \begin{align*}
       Q_2=
          B_1 
          \Par
          \propoutrec{k+1}{\wtd z \rho \cdot \subst{\tilde v}{\tilde y}}
          \recp{X}\binp{\prop^r_k}{\wtd x}
           \binp{n_l}{\wtd y} 
           \propoutrec{k+1}{\wtd z}
           \rvar{X} \Par \Brecpi{k+1}{\tilde z}{\alpha_{d-1}.P^2_1\sigma}_g
   \end{align*}
 \noindent  with $v\sigmav \namesrelate \wtd v$ for some $\sigmav \in \indices{v}$.
  Now, we should show that 
  \begin{align}
    \label{pi:eq:fo-rv-rec-goal1}
    P_2\rho_2 \cdot \subst{v}{y} \relS Q_2 
  \end{align}

  Let $\wtd u_z =\tilde z \rho_2 \cdot \subst{\tilde v}{\tilde y}$. 
  In sub-case \rom{1} we should show that 
  $Q_2 \in \Cbpi{\tilde u_z}{\tilde z}{\alpha_{d-1}.P^2_1}$.
  So, we should show that 
  $$
    Q_2 \in  \Cbrecpi{\tilde u_z}
    {\tilde z, \rho_2\cdot \subst{\tilde v}{\tilde y}}{P^*}^{d-1}_g
$$
  This follows by inspecting the definition 
  $\Dbrecpi{k}{\tilde z, \rho_2 \cdot \subst{\tilde v}{\tilde y}}{P^*}^{d-1}_g$. 
  That is, we can notice that 
  $$Q^1_2 = B_1 \Par B_2$$
  where 
  $B_2 \in \Dbrecpi{k}{\tilde z, \rho_2 \cdot \subst{\tilde v}{\tilde y}}
  {\alpha_{d-1}.P^2_1\sigma}^{d-1}_g$ 
  as $(d-1) + 1= \rvardepth{\alpha_d.\alpha_{d-1}.P^2_1\sigma}$. 
  So, \eqref{pi:eq:fo-rv-rec-goal1} follows. 
  This concludes sub-case \rom{1}. 
  
  In sub-case \rom{2} we know $P_2 \equiv \alpha_{d-1}.P^2_1 \Par R$.  
  Now, by \Cref{pi:t:tablecd-rec} we have $Q_2 \by{\tau} Q^2_2$ where 
  \begin{align*} 
  Q^2_2=& \news{\wtd \prop} 
     V_1 \Par 
              \apropoutrecsh{k+\degree+1}{\wtd u_{z_2}}
           \Par 
       V_2
  \end{align*}
  \noindent where 
  \begin{align*}
    V_1 = & B_1 
    \Par
    \recp{X}\binp{\prop^r_k}{\wtd x}
     \binp{n_l}{\wtd y} 
     \propoutrec{k+1}{\wtd z}
     \rvar{X} \Par 
     \proproutk{r}{k+1}{\wtd u_{z_1}}
    \recp{X}\proprinpk{r}{k}{\wtd x} (\proproutk{r}{k+1}{\wtd z_1}\rvar{X} \Par 
    \apropoutrecsh{k+\degree+1}{\wtd z_2}) \\
    V_2 = & \Brecpi{k+1}{\tilde z_1}{P_X}_{g_1} \Par 
    \Brecpi{k+\degree+1}{\tilde z_2}{R}_{\es}
  \end{align*}

  Now, we should show that 
 $Q^2_2 \in \Cbpi{\tilde u_z}{\tilde z}
 {\alpha_{d-1}.P^2_1 \Par R}$.
   By \Cref{pi:t:tablecd-rec} we have that 
  \begin{align}
   \{R_1 \Par R_2 : R_1 \in \Cbrecpi{\tilde u_{z_1}}{\tilde z_1}
   {\alpha_{d-1}.P^2_1},~ R_2 \in \Cbpi{\tilde u_{z_2}}{\tilde z_2}{R}\}
   \subseteq	\Cbpi{\tilde u_{z}}{\tilde z}
  {\alpha_{d-1}.P^2_1 \Par R} 
  \label{pi:eq:fo-rv-rec-r1r2}
  \end{align}
  
  Further, we know that 
   \begin{align}
       & \news{\wtd \prop_{k+l+1}} \apropoutrecsh{k+l+1}{\wtd u_z} 
    \Par \Brecpi{k+l+1}{\tilde z_2}{R}_\es
        \in \Cbpi{\tilde u_{z_2}}{\tilde z_2}{R}
    \label{pi:eq:fo-rv-rec-v1}
\\
    & V_1 \Par V_2 \in 
      \Cbrecpi{\tilde u_{z_1}}{\tilde z_1}{\alpha_{d-1}.P^2_1}
      \label{pi:eq:fo-rv-rec-v2}
  \end{align}

  Thus, by \eqref{pi:eq:fo-rv-rec-r1r2}, \eqref{pi:eq:fo-rv-rec-v1}, 
  and \eqref{pi:eq:fo-rv-rec-v2} 
  we have the following: 
  \begin{align*}
    V_1 \Par V_2 \Par 
    \news{\wtd \prop_{k+l+1}} \apropoutrecsh{k+l+1}{\wtd u_z} 
    \Par \Brecpi{k+l+1}{\tilde z_2}{R}_\es
    \in \Cbpi{\tilde u_{z}}{\tilde z}{\alpha_{d-1}.P^2_1 \Par R}
  \end{align*}

  Now, we can notice that 
  $\wtd \prop_{k+l+1} \cap \fpn{V_1} = \es$ and  
  $\wtd \prop_{k+l+1} \cap \fpn{\Brecpi{k+1}{\tilde z_1}{P_X}_{g_1}} = \es$. 
  Further, we have 
  $$\news{\wtd \prop_{k+l+1}} \Brecpi{k+\degree+1}{\tilde z_2}{R}_{\es} 
  \fwb \inact$$
  \noindent (where $\fwb$ is as in \Cref{top:d:fwb}) 
  as first shared trigger $\prop_{k+l+1}$ in the 
  breakdown of $R$ is restricted so it could not get activated. 

  % $$\news{\wtd \prop} 
  % V_1 \Par 
  %    \apropoutrecsh{k+\degree+1}{\wtd u_{z_2}}
  %   \Par 
  %   V_2$$

  Thus, we have  
  $$Q^2_2 \equiv V_1 \Par V_2 \Par 
  \news{\wtd \prop_{k+l+1}} \apropoutrecsh{k+l+1}{\wtd u_z} 
  \Par \Brecpi{k+l+1}{\tilde z_2}{R}_\es$$
  
  This concludes the  case \ltsrule{Rv}. Now, we consider the inductive case. 
\item Case \ltsrule{Par${}_L$}. 
    Here we know $P_1 = P'_1 \Par P''_1$. 
    Further, we know there exist $P^*$ 
such that 
 $\jdepth{P}{P^*}{d}{X}$ (\Cref{pi:d:jdepth}). 
%  Now, we unfold this definition. 
% % such that $P_1\equiv\recp{X}P^*_1$. 
% Let 
% \begin{align*}
%     P^1_1 = \alpha_d.\alpha_{d-1}. \ldots .\alpha_{1}.(X \Par R)
% \end{align*}
% \noindent where $R$ is some processes, 
% and $\alpha_d = \abinp{n}{y}$.
% %  As $P_1$ is a recursive process, 
%   We know that there is $\recp{X}P^*$ such that $P^1_1$ is its sub-processes and
%   \begin{align*}
%       P_1 \equiv P^1_1\subst{\recp{X}P^*}{X}
%   \end{align*} 
    Similarly to the previous case, let 
    \begin{align*}
      P^1_1 = \alpha_d.\alpha_{d-1}. \ldots .\alpha_{1}.(X \Par R)
    \end{align*}
    We know 
    there is $\recp{X}P^*$ such that $P^1_1$ is its subprocess 
    and 
    $$P'_1 \equiv P^1_1\subst{\recp{X}P^*}{X}$$

    % Here, we can distinguish sub-cases depending 
    % on if $P''_1 \equiv R$ and $p > 0$ or $p = 0$. 

    Here, we can distinguish two sub-cases: \rom{1} 
    $P''_1 \equiv R$ and \rom{2} $P''_1 \not\equiv R$. 
    Here, we consider sub-case \rom{1} as it is an interesting case. 
    The sub-case \rom{2} is similar to the corresponding 
    case of non-recursive process. 
    As in the previous case, we can further distinguish 
    cases in which $p = 0$ and $p > 0$. 
    We consider $p = 0$ and $\ell = \bactinp{n}{v}$.

    The final rule in the inference tree is as follows: 
    \begin{align*} 
      \AxiomC{$P_1' \subst{\tilde u'_1}{\tilde y_1} \by{\ell} 
      P_2' \subst{\tilde u'_2}{\tilde y_2} 
      \Par 
      R\subst{\tilde w_1}{\tilde z_1}\cdot 
       \subst{\tilde u'_R}{\tilde w_1}$} 
      \AxiomC{$\bn{\ell} \cap \fn{P_1''} = \emptyset$} 
      \LeftLabel{\scriptsize \ltsrule{Par${}_L$}}
      \BinaryInfC{$P_1' \subst{\tilde u'_1}{\tilde y_1} \Par 
      R \subst{\tilde u''_1}{\tilde z_1}
      \by{\ell} P_2' \subst{\tilde u'_2}{\tilde y_2} \Par 
      R \subst{\tilde u''_1}{\tilde z_1} \Par
       R\subst{\tilde w_1}{\tilde z_1}\cdot 
       \subst{\tilde u'_R}{\tilde w_1}$} 	
      \DisplayProof 
      \end{align*}

    Let $\sigma_1 \in \indices{\fn{P_1}\cup \wtd u \cup \wtd x}$. 
    Further, let $\wbns y_1 = \fnb{P_1'}{\wbns x}$, 
    $\wbns z_1 = \fnb{R}{\wbns x}$, $\wbns{u} =\bname{\wtd u \sigma_1 : \wtd C}$ 
    where 
      $\wbns{x} =\bname{\wtd x \sigma_1 : \wtd C}$ with $\wtd u : \wtd C$, and 
      $\rhom=\subst{\bns u}{\bns x}$. 
      By the definition of $\relS$
      ( \Cref{pi:t:tablecd-rec}), there are following possibilities for $Q_1$:
      \begin{align*}
        Q^1_1 =& \news{\wtd \prop_k}(\apropout{k}{\wbns u} \Par
        \Bopt{k}{\bns x}{P_1  \sigma_1}) \\
        Q^2_1 =& \news{\wtd \prop_k}\propout{k}{\wbns u'_2}
        \apropout{k+\degree}{\wbns u''_1}   \Par
        \Bopt{k}{\bns y_1}{P'_1\sigma_1} \Par 
        \Bopt{k+\degree}{\bns z_1}{R\sigma_1}  \\
        N^3_1 =& \{(R'_1 \Par R''_1) :
        \Cbrecpi{\bns u'_2}{\bns y_1}{\recp{X}P^*\sigma_1}^d_g,~ R''_1 \in 
        \Cbpi{\bns u''_1}{\bns z_1}{R \sigma_1}\}
      \end{align*}

      By \Cref{pi:l:c-prop-closed} there exist
      \begin{align}
        Q'_1 &\in \Cbrecpi{\bns u'_2}{\bns y_1}{\recp{X}P^*\sigma_1}^d_g
      \\
        Q''_1 &\in \Cbpi{\bns u''_1}{\bns z_1}{R\sigma_1}
        \label{pi:eq:fo-parl-rec-q''_1}
      \end{align}
      such that 
    $$
      Q^1_1 \By{\tau} Q^2_1 \By{\tau} Q'_1 \Par Q''_1
   $$

  The interesting case is to consider  a process $Q^3_1$ defined as: 
  \begin{align*} 
    Q^3_1=& \news{\wtd \prop} 
       V_1 \Par 
          \apropoutrecsh{k+\degree+1}{\bns u''_{1}}
         \Par 
         V_2
    \end{align*}
    \noindent where 
    \begin{align*}
      V_1 & =  B_1 
      \Par
       \proproutk{r}{k+1}{\wbns u'_{2}}
      \recp{X}\proprinpk{r}{k}{\wbns y_1} (\proproutk{r}{k+1}{\wbns y_2}\rvar{X} \Par 
      \apropoutrecsh{k+\degree+1}{\wbns z_2}) \\
      V_2 & =  \Brecpi{k+1}{\bns y_2}{P_X}_{g_1} \Par 
      \Brecpi{k+\degree+1}{\bns w_2}{R\subst{\tilde w_2}{\tilde z_2}}_{\es}
    \end{align*}

    We have $Q^3_1 \fwb Q'_1 \Par Q''_1$ (where $\fwb$ is as in \Cref{top:d:fwb}). We may notice that  
  % equivalent to $Q'_1 \Par Q''_1$ which 
  $Q^3_1$ can be a descendent of the recursive process 
  (following a similar  
  reasoning as in the previous case). 
  So, we consider how $Q^3_1$ evolves. As in the corresponding
  case of non-recursive processes, we do the case analysis on $\ell$. 
  % We consider case $\ell = \bactinp{n}{v}$.  
  If $v \in \wtd u$ then we take $\sigmav = \sigma_1$, otherwise $\sigmav =
  \initname{v}{j}$ for  $j > 0$.  
  Now, we could see that 
  $$Q^3_1 \By{\bactinp{n}{\tilde v}} Q_2$$ 

  \noindent where 
  \begin{align*}
    Q_2 =  \news{\wtd \prop} 
    V_1 \Par 
    \proproutk{r}{k+1}{\wbns u'_{2}}
    \recp{X}\proprinpk{r}{k}{\wbns y_1} (\proproutk{r}{k+1}{\wbns y_2}\rvar{X} \Par 
    \apropoutrecsh{k+\degree+1}{\wbns z_1})
      \Par 
      \apropoutrecsh{k+\degree+1}{\wbns u'_{R}}
      \Par 
      V_2
  \end{align*}

  We define $Q^1_2$ as follows
  \begin{align*}
    Q^1_2 &= \news{\wtd \prop} V_1 \Par    
    \proproutk{r}{k+1}{\wbns u'_{2}}
    \recp{X}\proprinpk{r}{k}{\wbns y_1} (\proproutk{r}{k+1}{\wbns y_2}\rvar{X} \Par 
    \apropoutrecsh{k+\degree+1}{\wbns z_1})
    \Par V_2 \Par \\
    & \quad \news{\wtd \prop_{k}} (\apropoutrecsh{k}{\wbns u''_1} 
    \Par \Brecpi{k}{\bns z_1}{R}_\es) 
   \Par \news{\wtd \prop_{k}} ( \apropoutrecsh{k}{\wbns u'_R} 
    \Par \Brecpi{k}{\bns w_1}{R\subst{\tilde w_1}{\tilde z_1}}_\es ) 
  \end{align*}

  By definition, we could see that 
  $Q^1_2 \in \Cbpi{\bns u_2}{\bns z_1}{P_2' \Par R \Par R\subst{\tilde w_1}{\tilde z_1}}$.
  Now, 
  by the definition of $\Brecpi{-}{-}{\cdot}_\es$ we have
  \begin{align*}
    &\apropoutrecsh{k+\degree+1}{\wbns u''_{1}}
    \Par 
    \apropoutrecsh{k+\degree+1}{\wbns u'_{R}}
    \Par 
    \Brecpi{k+\degree+1}{\bns z_1}{R}_{\es} 
    \fwb \\
    &\quad \news{\wtd \prop_{k}} (\apropoutrecsh{k}{\wbns u''_1} 
    \Par \Brecpi{k}{\bns z_1}{R}_\es) 
   \Par \news{\wtd \prop_{k}} ( \apropoutrecsh{k}{\wbns u'_R} 
    \Par \Brecpi{k}{\bns w_1}{R\subst{\tilde w_1}{\tilde z_1}}_\es
  \end{align*}
  % \noindent (where $\fwb$ is defined in Definition 18~\cite{KouzapasPY17}). 
  As each trio in  
  $\Brecpi{k+\degree+1}{\bns z_1}{R}_{\es}$ makes a replica of itself 
  when triggered along a propagator.  
  % and by assumption we have $\wtd z_1 = \wtd z_2$, 
  % that is $\bns z_1 = \bns z_2$. 
  So, finally we have 
  $$Q_2 \fwb Q^1_2$$
\end{enumerate}
  This concludes the proof of \Cref{pi:l:lemms}.

\end{proof}

\end{document}